\documentclass[11pt,a4paper]{article}
\usepackage{amsmath, amsfonts,amssymb,mathtools,nicefrac,nccmath,cases}
\usepackage{microtype,booktabs,multirow}
\usepackage[usenames,dvipsnames,table]{xcolor}
\usepackage{dsfont}
\usepackage{colortbl}
\usepackage{tikz}
\usepackage{verbatim}
\usetikzlibrary{shapes,arrows}
\usetikzlibrary{calc,shapes.callouts,shapes.arrows,bending,intersections}
\usetikzlibrary{shapes.geometric}
\usetikzlibrary{arrows.meta,arrows}
\usetikzlibrary{decorations.pathmorphing}
\usetikzlibrary{decorations.markings}
\usetikzlibrary{shapes.multipart}
\usetikzlibrary{tikzmark}
\usepackage{arydshln}
\usepackage{caption}
\usepackage{subcaption}
\usepackage{bm}
\usepackage[nosort]{cite}
\usepackage{colortbl}
\usepackage{amsthm}
\usepackage{soul}

\usepackage{xcolor}
\usepackage{lipsum}

\usepackage[percent]{overpic}

\usepackage[linktocpage=true]{hyperref}
\hypersetup{colorlinks=true,linkcolor=nicecolor,citecolor=nicecolor,urlcolor=nicecolor}
\definecolor{nicecolor}{rgb}{0.1, 0.3, 0.4}

\definecolor{blue}{rgb}{0.06, 0.3, 0.57}
\definecolor{Gray}{gray}{0.4}

\definecolor{nicecolor}{rgb}{0.1, 0.3, 0.4}
\definecolor{blue}{rgb}{0.06, 0.3, 0.57}
\definecolor{Gray}{gray}{0.4}
\colorlet{tableheadcolor}{gray!15} 
\colorlet{tablerowcolor}{gray!7} 

\def\hybrid{\topmargin -20pt    \oddsidemargin 0pt
	\headheight 0pt \headsep 0pt
	\textwidth 6.5in        
	\textheight 9in         
	\textwidth 6.25in       
	\textheight 9 in       
	\marginparwidth .875in
	\parskip 5pt plus 1pt 
	\jot = 1.5ex
}
\hybrid
\numberwithin{equation}{section}
\numberwithin{table}{section}
\usepackage{float}
\usepackage{tabularx}
\usepackage{multirow}
\usepackage{hhline}
\usepackage{stackengine}

\newcolumntype{D}{>{\centering\arraybackslash}X}
\newcolumntype{L}{>{$}l<{$}}
\newcolumntype{R}{>{$}r<{$}}
\newcolumntype{C}{>{$}c<{$}}
\usepackage{IEEEtrantools}

\newcommand{\beq}{\begin{equation}}  \newcommand{\eeq}{\end{equation}}
\newcommand{\bal}{\begin{aligned}}   \newcommand{\eal}{\end{aligned}}
\newcommand{\bea}{\begin{eqnarray}}  \newcommand{\eea}{\end{eqnarray}}
\def\beqa{\begin{eqnarray}}
\def\eeqa{\end{eqnarray}}

\newcommand{\bmat}{\left(\begin{array}}
\newcommand{\emat}{\end{array}\right)}

\usepackage{cancel}

\usepackage[section]{placeins} 

\newcommand{\be}{\begin{equation}}
\newcommand{\ee}{\end{equation}}

\definecolor{Gray}{gray}{0.95}
\definecolor{darkspringgreen}{rgb}{0.09, 0.45, 0.27}
\definecolor{darkseagreen}{rgb}{0.56, 0.74, 0.56}
\definecolor{darkmouthgreen}{rgb}{0.05, 0.5, 0.06}
\definecolor{darkcyan}{rgb}{0.0, 0.55, 0.55}

\def\d {{\rm d}}
\def\del          {\partial}

\def\ii           {{\rm i}}

\def\tr           {\mathop{\rm tr}}

\def\Im           {{\rm Im\hskip0.1em}}

\def\cala         {{\cal A}}

\def\calb         {{\cal B}}
\def\calc         {{\cal C}}
\def\cald         {{\cal D}}
\def\cale         {{\cal E}}
\def\calf         {{\cal F}}
\def\calg         {{\cal G}}

\def\calh         {{\cal H}}
\def\cali         {{\cal I}}
\def\calj         {{\cal J}}
\def\calk         {{\cal K}}
\def\call         {{\cal L}}
\def\calm         {{\cal M}}
\def\caln         {{\cal N}}
\def\calo         {{\cal O}}
\def\calp         {{\cal P}}

\def\calr         {{\cal R}}
\def\cals         {{\cal S}}
\def\calt         {{\cal T}}

\def\calw         {{\cal W}}

\usepackage{xcolor}
\definecolor{colorloc1}{RGB}{0,0,102}  
\definecolor{colorloc2}{RGB}{0,125,253} 
\usepackage[framemethod=default]{mdframed}
\newmdenv[skipabove=10pt,
skipbelow=7pt,
rightline=false,
leftline=true,
topline=false,
bottomline=false,
linecolor=colorloc1,
backgroundcolor=colorloc2!5,
innerleftmargin=4pt,
innerrightmargin=0pt,
innertopmargin=0pt,
leftmargin=2pt,
rightmargin=0pt,
linewidth=2pt,
innerbottommargin=0pt]{lbBox}

\begin{document}

\baselineskip=14pt
\parskip 5pt plus 1pt

\vspace*{-1.5cm}
\begin{flushright}    
  {\small 
  }
\end{flushright}

\vspace{2cm}
\begin{center}        

\textbf{\textsc{\huge 
Wormholes in the axiverse,\\\smallskip and the species scale
}}\\[.3cm]

\end{center}

\vspace{0.5cm}
\begin{center}        
{\large   Luca Martucci$^{1,2}$, Nicol\`o Risso$^{1,2}$, Alessandro Valenti$^{1,2,3}$ and Luca Vecchi$^2$}
\end{center}

\vspace{0.15cm}
\begin{center}  
${}^1$ \emph{Dipartimento di Fisica e Astronomia ``Galileo Galilei",  Universit\`a degli Studi di Padova} \\ 
${}^2$\emph{INFN Sezione di Padova, Via F. Marzolo 8, 35131 Padova, Italy}\\
${}^3$\emph{Department of Physics, University of Basel, Klingelbergstrasse 82, CH-4056 Basel, Switzerland}
\\
\end{center}

\vspace{2cm}


\begin{abstract}

\noindent We analyze a large class of four-dimensional $\caln=1$ low-energy realizations of the axiverse satisfying various quantum gravity constraints. We propose a novel upper bound on the ultimate UV cutoff of the effective theory, namely the species scale, which only depends on data available at the two-derivative level. Its dependence on the moduli fields and the number $N$ of axions matches expectations from other independent considerations. After an assessment of the regime of validity of the effective field theory, we investigate the non-perturbative gravitational effects therein. We identify a set of axionic charges supported by extremal and non-extremal wormhole configurations. We present a universal class of analytic wormhole solutions, explore their deformations, and analyze the relation between wormhole energy scales and the species scale. The connection between these wormholes and a special subclass of BPS fundamental instantons is discussed, and an argument in favor of the genericity of certain axion-dependent effective superpotentials is provided.
We find a lower bound increasing with $N\gg 1$ on the Gauss-Bonnet coefficient, resulting in an exponential suppression of non-extremal wormhole effects. Our claims are illustrated and tested in concrete string theory models.

\end{abstract}

\thispagestyle{empty}
\clearpage

\setcounter{page}{1}


\newpage

  \tableofcontents


\section{Introduction}

From a purely bottom-up perspective, light axions can provide a solution to several open problems in particle phenomenology: they are well known to offer an elegant solution to the strong CP problem \cite{Peccei:1977hh,Weinberg:1977ma,Wilczek:1977pj}, to be interesting dark matter candidates \cite{Preskill:1982cy, Abbott:1982af,Dine:1982ah,Hu:2000ke} and possibly drive inflation \cite{Frieman:1995pm}.  
Remarkably, since the early eighties it was recognized that string theory offers a natural framework for such particles, see for instance \cite{Svrcek:2006yi} for an overview and references to the older literature. In fact, the low-energy description of string theory generically features a number of moduli fields; some of these are axions, spinless particles carrying shift symmetries that are broken solely by non-perturbative effects. In generic string theory models the number $N$ of low-energy axions can easily be in the hundreds, if not thousands. A setup with $N\gg 1$ axions can lead to an array of interesting phenomenological signatures   and is often referred to as axiverse \cite{Arvanitaki:2009fg}. A recent review and an updated list of references on the string theory axiverse can be found in \cite{McAllister:2023vgy}.

Despite their genericity, stringy axions can be phenomenologically relevant only if their potential interactions are very small. The non-perturbative effects, expected to break the axionic shift symmetries in generic string constructions, must somehow be more suppressed than naively expected. Non-perturbative effects can belong to two qualitatively different classes. The first class has a genuinely UV nature and, in string theory models, typically includes worldsheet and brane instantons. We will refer to them as {\em fundamental} instantons. These would appear to any  effective field theory (EFT) observer simply as bare symmetry-breaking local operators. Yet, non-perturbative effects may also emerge within the EFT itself. Besides the familiar gauge theory instantons, in the presence of axions the second class includes axionic wormholes  \cite{Giddings:1987cg} (for a review and more references on  aspects relevant for the present paper, see
\cite{Hebecker:2018ofv}). Given the variety of non-perturbative contributions available in these theories, and their genericity, we should ask: How likely is it to find light axions in axiverse models compatible with quantum gravity? More generally, how and under which conditions are non-perturbative corrections to their potentials suppressed?

The answer to the first question might simply turn out to be a mere probabilistic argument: the number of axions in the axiverse is so large that it becomes statistically likely that at least a few of them remain light. A more interesting possibility is perhaps that some structural property of quantum gravity lies behind it, though. Yet, to assess the viability of this option one must at least partially address the second question. In particular, we must improve our understanding of the string axiverse and all non-perturbative effects therein.  Encouraging results have been obtained in particular in \cite{Demirtas:2018akl,Demirtas:2021gsq}, where it was shown that in a large class of IIB models with $N\gg 1$ axions  non-perturbative stringy corrections  are  suppressed in an unexpectedly strong way.

The present paper makes a further contribution to this subject. We adopt the formalism of \cite{Lanza:2020qmt,Lanza:2021udy,Lanza:2022zyg} and consider four-dimensional effective field theories with an arbitrary, possibly large, number $N$ of  axions and minimal $\caln=1$ supersymmetry. For simplicity we take the minimal field content that is sufficient to describe the low-energy limit of an axiverse, but our framework can be easily generalized by including additional degrees of freedom, if needed. We focus on  fundamental axions, which originate directly from the UV completion of the EFT, rather than arising as (approximate) Nambu-Goldstone bosons of some linearly realized accidental compact symmetry within the EFT. The structure of our theories is constrained by quantum gravity and string theory in a number of non-trivial ways; we will make heavy use of those constrains. Two are the main objectives of our work. The first is a clear identification of the regime of validity and of the relevant quantum gravity scales of such effective theories.  The second is a characterization of the non-perturbative gravitational effects within that setup.

After an introduction of our EFTs (Section \ref{sec:N=1models}), in Section \ref{sec:regimesEFT} we present a careful analysis of their regime of validity both in energy and field space. An interesting result  is a lower bound of order $N$ on the (field dependent) coefficient of the Gauss-Bonnet operator. We also present a detailed discussion of the highest possible UV cutoff of our theory, namely the {\em species scale} $M_{\rm sp}$ \cite{Dvali:2007hz,Dvali:2007wp,Dvali:2009ks,Dvali:2010vm}. The determination of the species scale is by itself a very active area of research especially in the context of the Swampland program (see \cite{Palti:2019pca,vanBeest:2021lhn,Grana:2021zvf,Agmon:2022thq} for reviews). Our contribution is the proposal of a new upper bound on that quantity:
\be\label{boundINTRO}
M_{\rm sp}\leq \sqrt{2\pi\calt}\,,
\ee
where $\calt$ is the field-dependent tension of the lightest EFT string \cite{Lanza:2020qmt,Lanza:2021udy,Lanza:2022zyg}, see Section \ref{sec:UVcutoff} for more details. 
Our upper bound is fully determined by EFT data already available at the two-derivative level and has the advantage of having a clear physical interpretation and being radiatively stable. We subsequently test our lower bound on the coefficient of the Gauss-Bonnet operator and our upper bound on the species scale in a number of explicit string theory models in Section \ref{sec:stringtheorymodels}.

Having firmly established our framework and its perturbative regime, we can next move to  wormholes.  We begin in Section \ref{sec:WHs} recalling the derivation of $O(4)$-invariant wormhole configurations in Euclidean space, and presenting an analysis of their regime of validity. In particular, we discuss the relation between extremal wormholes and fundamental BPS instantons, and subsequently introduce the notion of EFT instanton \cite{Lanza:2021udy}. 
Section \ref{sec:WHAxiverse} is dedicated to non-extremal wormholes. We identify a universal class of {\emph{homogeneous}} wormhole  solutions, which involve all $N$ axions and their supersymmetric saxionic partners, taking advantage of crucial inputs from string theory. Their perturbative domain is analyzed, and  a universal constraint relating the minimal radius of the wormhole throat to the species scale is pointed out. We later show how these homogeneous solutions are instrumental in understanding some properties of more general solutions.  We argue that a peculiar role is played by  {\emph{marginally degenerate}}  non-extremal wormholes, unveiling a suggestive analogy between this type of wormholes and the fundamental EFT instantons carrying the same charges, which will be explored in more detail subsequently. The general discussion is illustrated and tested, also through numerical simulations,  in concrete  string theory models.

The physical implications of the non-perturbative effects associated to the configurations discussed in Section \ref{sec:WHAxiverse} are analyzed in Section \ref{sec:phys}. We argue that extremal wormholes can induce potentially relevant symmetry-breaking effective superpotentials, as well as higher derivative F-terms \cite{Beasley:2004ys,Beasley:2005iu}, at low energies. On the other hand, non-extremal non-degenerate wormholes can at most induce corrections to the K\"ahler potential. However, such D-terms come with associated Coleman's $\alpha$-parameters \cite{Coleman:1988cy}, which are undetermined in the EFT, and so their physical relevance cannot be firmly established within our framework. String theory experience and quantum gravity arguments strongly suggest \cite{Marolf:2020xie,McNamara:2020uza} that the $\alpha$-parameters should be determined by the UV completion, possibly in terms of other dynamical fields, but the actual mechanism by which such determination might occur remains a mystery. Interestingly, the physics of a specific subclass of non-extremal wormholes, which we call {\emph{marginally degenerate}}, might shed light on this puzzling open problem. Indeed, various considerations indicate that marginally degenerate wormholes represent the low-energy manifestation of fundamental EFT instantons and, as such, should be capable of inducing effective superpotentials at low energies, generalizing a mechanism first pointed out in \cite{Giddings:1989bq,Park:1990ep}. If correct, this conclusion in turn implies that the $\alpha$-parameters of marginally-degenerate wormholes must necessarily be fixed by the UV-complete description of the fundamental EFT instantons. Perhaps something similar might happen to the $\alpha$-parameters of non-degenerate wormholes as well. In Section \ref{sec:phys} we also discuss how our results are inherited by non-supersymmetric scenarios UV-completed by an $\caln=1$ axiverse. Our conclusions are presented in Section \ref{sec:conclusions}.

Our work is complemented by a few appendices. Naive Dimensional Analysis is proposed in Appendix \ref{app:2pis} as a guide to estimate the factors of $2\pi$ appearing in some of the formulas of the paper. Some details on dual heterotic/F-theory models, which we use in Section \ref{sec:stringtheorymodels}, are presented in Appendix \ref{app:hetFtheory}. Additional evidence of the validity of our new bound on the species scale is given in Appendix \ref{app:SCtests}. Appendix \ref{app:WHsusy} discusses in some detail the wormhole fermionic zero-modes and their implications on the structure of the wormhole-induced low-energy  effective operators.

\section{${\caln}=1$ axiverse models}
\label{sec:N=1models}

The basic assumption underlying our work is the existence of an effective four-dimensional field theory (EFT) with a possibly large number $N$ of light axions. We will focus on {\emph{fundamental}} axions, namely periodic axions like those that typically  originate from string theory, and that cannot be regarded as angular components of some elementary field in four dimensions.

Because in quantum gravity (QG) global symmetries are expected to be at most approximate, the global shift symmetries that prevent our axions to acquire large masses must ultimately be broken. It is then crucial to identify a concrete and realistic general  framework in which the axionic shift symmetries can be considered exact up to small corrections dictated by QG. Such a framework is provided by the ${\cal N}=1$ setup outlined in \cite{Lanza:2020qmt,Lanza:2021udy,Lanza:2022zyg}. The associated EFTs emerge from large classes of string theory models and naturally take into account the relevant QG constraints.

In Section \ref{sec:N=1EFT} we review the basic setup of \cite{Lanza:2020qmt,Lanza:2021udy,Lanza:2022zyg}. We focus on the leading two-derivative approximation but, for reasons that will become clearer later, also keep an eye on the (semi-)topological higher-derivative couplings to gravity, and in particular on the Gauss-Bonnet interaction (see Section \ref{sec:semitop}). 
Section \ref{sec:GBcontribution} reformulates the EFT in a dual 2-form language in preparation of the subsequent sections.

\subsection{Axions in $\caln=1$ EFTs}
\label{sec:N=1EFT}

Let us start recalling the minimal structure of an $\caln=1$ effective theory involving $N$ {\emph{periodic}} axions $a^i$, $i=1,\ldots,N$ and their bosonic partners. Without loss of generality we normalize the $a^i$'s so as to have unit periodicity: 
\be\label{unitperiodicity}
a^i\simeq a^i+1\,.
\ee
The  ``angular" variables $\theta^i$ often used to denote axions are related to our fields via $\theta^i=2\pi a^i$. By supersymmetry the axions combine with {\em saxions} $s^i$ into complex fields
\be
t^i=a^i+\ii s^i\simeq t^i+1\,,
\ee
which represent the bottom components of corresponding $\caln=1$ chiral multiplets. The EFT in general contains other fields. For simplicity we will ignore them and just consider $t^i$ plus gravity. Our main conclusions do not depend on this assumption.

The exact gauge symmetry \eqref{unitperiodicity} combined with supersymmetry constrains significantly the EFT. The only manifestly supersymmetric non-derivative couplings of the axions can be either semi-topological or functions of $e^{2\pi iq_it^i}$ with $q_i\in {\mathbb Z}$, and of their complex conjugate, exponentially suppressed by the saxions. We will discuss the semi-topological couplings shortly and postpone an analysis of the exponentially suppressed instanton-like corrections to the following subsection, where we also provide a quantitative definition of the perturbative regime in which such corrections can be considered small.

Up to semi-topological couplings and instanton-like effects, the EFT is  invariant under arbitrary constant shifts of the axions. At the two-derivative level and in Lorentzian signature, the contribution of the terms involving only gravity and $t^i$ to the most general $\caln=1$ shift-symmetric  action  is
\be\label{kinetic} 
\frac{1}{2}M_{\text{\tiny P}}^2\int R*1-\frac12M_{\text{\tiny P}}^2\int \calg_{ij}(s)\left(\d s^i\wedge *\d s^j+\d a^i\wedge *\d a^j\right)
\ee
where $\calg_{ij}(s)$ is a symmetric positive matrix function of the saxion fields and we have omitted  appropriate Gibbons-Hawking boundary term. By supersymmetry, the kinetic terms of the scalars are specified by a K\"ahler potential $K$. Within our perturbative regime, $K$ depends on the complex fields $t^i$ only through their saxionic component $s^i=-\frac\ii2(t^i-\bar t^i)$, i.e. $K=K({s})$ (a possible dependence on additional spectator multiplets is ignored), via the relation
\be\label{calg1}
\calg_{ij}\equiv\frac12\frac{\del^2 K}{\del 
 s^i\del s^j}.
\ee
We can actually say more about our EFT if we take into account additional non-trivial inputs from the UV completion. Indeed, within the perturbative regime we are considering (to be more precisely defined later), for a large class of string theory models the K\"ahler potential reads
\be\label{KP(s)} 
K({s})=-\log P({s})\,,
\ee
where $P({s})$ is a positive homogeneous function, $P(\lambda {s})=\lambda^nP({s})$ and $n$ is an integer ranging from $1$ to $7$ --  see Sections \ref{sec:stringtheorymodels} and \ref{sec:WHAxiverse} for explicit examples. As discussed in \cite{Lanza:2020qmt,Lanza:2021udy} the perturbative structure \eqref{KP(s)} conforms with various formulations of the weak gravity conjecture \cite{ArkaniHamed:2006dz} and the distance conjecture \cite{Ooguri:2006in} in the present setting. The homogeneity of $P$ and the relation \eqref{KP(s)} will play a crucial role in some of the subsequent sections.

\subsection{Semi-topological couplings to gravity}
\label{sec:semitop}

Eq.~\eqref{kinetic} just represents the leading two-derivative term in our EFT. In general one should also allow the presence of higher dimensional interactions suppressed by some mass scale  $M_{\text{\tiny UV}}$ that depends on the UV completion of the EFT and in general on the EFT scalar fields. On the other hand, from the Wilsonian viewpoint  any EFT is associated with a  (field independent)  cutoff scale $\Lambda$, which defines the upper bound on the allowed momentum scales ($p\leq \Lambda$) and must obey  $\Lambda\ll M_{\text{\tiny UV}}$. Note that  $M_{\text{\tiny UV}}$ in general depends on saxions $s^i$, $M_{\text{\tiny UV}}=M_{\text{\tiny UV}}(s)$, and then the condition $\Lambda\ll M_{\text{\tiny UV}}(s)$ will in general restrict the field space region in which the EFT is valid.  But how can we know what $M_{\text{\tiny UV}}(s)$ is without knowing the UV completion of the EFT? We will come back to this important question in Section \ref{sec:UVcutoff}. For the moment we observe that 
  imposing $\Lambda\ll M_{\text{\tiny UV}}$ one may naively presume that the effect of all higher dimensional operators can be safely neglected.
  However, a very special class of higher-dimensional operators may be unsuppressed at low scales. These are the {\emph{topological terms}} which can be obtained as integrals of total derivative operators. Because of their nature, they do not affect the equations of motion nor induce particle vertices. Nevertheless, they can contribute to the on-shell action and therefore impact semiclassical calculations.  

  In a four-dimensional gravitational context, a well-known example of such a topological term is provided by the integral of the Gauss-Bonnet (GB) operator
\be\label{GBdensity}
E_{\text{\tiny GB}}\equiv \frac1{32 \pi^2}\left(R_{abcd}R^{abcd}-4 R_{ab}R^{ab}+R^2\right)\,,
\ee
which can indeed be locally written as a total covariant derivative. 
In an ${\caln}=1$ SUSY framework, this operator originates from a superspace combination of the form \cite{Townsend:1979js,Cecotti:1985mf,Cecotti:1987mr,Bonora:2013rta}
\be\label{superGB} 
\left(\int\d^4x\,\d^2\theta\,2\cale\,f\, \calw^{\alpha\beta\gamma}\calw_{\alpha\beta\gamma}+\text{c.c.}\right)+\text{D-terms}\,,
\ee
where $\calw^{\alpha\beta\gamma}$ is the Weyl chiral superfield,  $f$ is a holomorphic function of chiral superfields and the D-terms take a specific form which will not be relevant in the following -- see \cite{Martucci:2022krl} for more details in the present context. If $f$ is constant then \eqref{superGB} is topological, but in our context can in general depend on the chiral fields $t^i$.
More precisely, the F-term appearing in \eqref{superGB} includes both a coupling of
${\rm Im}f$ to the bosonic Weyl density and of ${\rm Re}f$ to the Pontryagin form $\tr(\calr\wedge\calr)$, while the D-terms provide the $R^{mn}R_{mn}$ and $R^2$ counterterms which combine with the Weyl density to give the GB operator \eqref{GBdensity}.

Consistency with \eqref{unitperiodicity} implies that, in addition to a constant,   $f(t)$ can contain a linear combination $\tilde C_it^i$ for some real constants $\tilde C_i$, plus possible  exponentially suppressed instanton-like corrections which will be ignored. Adopting the same normalization conventions of  \cite{Martucci:2022krl}, the $\tilde C_it^i$ contribution in \eqref{superGB} gives a GB term 
\be\label{GRtheta}
\int\d^4x\,\sqrt{-g}\,\gamma(s)\,E_{\text{\tiny GB}}\,,
\ee
with 
\be\label{gammas}
\gamma(s)\equiv \frac{\pi}{6}\tilde C_is^i\,,
\ee
and a Pontryagin term proportional to $\tilde C_i\int a^i\tr(\calr\wedge\calr)$. Taking into account the precise numerical factors and imposing various consistency conditions on the Pontryagin operator, one finds that the constants $\tilde C_i$ must be integrally quantized \cite{Martucci:2022krl}: $\tilde C_i\in\mathbb{Z}$.\footnote{In complete analogy, supersymmetry fixes linear couplings of (s)axions to vector fields to take the form
\be\label{QCDtheta} 
-\frac{1}{8\pi}\int C_is^i \tr(F\wedge *F)- \frac{1}{8\pi}\int C_ia^i \tr(F\wedge F)\,.
\ee
We will study only solutions with trivial gauge configurations, and so the above couplings are not of primary interest here. See section 2 of \cite{Martucci:2022krl} for a more detailed discussion of these quasi-topological terms in the present setting.} Note that in presence of boundaries \eqref{GRtheta} must be supplemented  by Gibbons-Hawking-like boundary terms, which will be explicitly discussed in Section \ref{sec:WHs}.

The coefficients of the D-terms appearing in \eqref{superGB} are instead not protected by holomorphy and hence supersymmetry is not enough to provide robust information about them. In particular, they can in principle have a more  complicated dependence on $s^i$, and moreover be affected by radiative corrections.  Fortunately, Ricci squared terms are also basis-dependent, in the sense that re-defining the metric one can always trade them for operators involving derivatives of  $s^i$. Hence, without loss of generality, we can choose a field basis in which the non-derivative saxions couplings to curvature squared operators reduce to the GB term \eqref{GRtheta}. 

The coefficient of the GB operator receives non-perturbative as well as perturbative corrections. The former are negligible in our setup (see Section \ref{sec:perturbative}). The latter are of two types. By supersymmetry, radiative contributions to $\gamma$ are exhausted by a constant 1-loop correction $\propto N\log\Lambda$. Yet, a more subtle correction to the GB appears in our scenario. This is because, strictly speaking, the standard manifestly supersymmetric formulation  \cite{Wess:1992cp} is not automatically in the Einstein frame, which instead we used in \eqref{kinetic}. In order to pass to the Einstein frame a Weyl rescaling $\Phi\to e^{\beta K}\Phi$ of all fields, with $\beta$ some number, is necessary. Such transformation is anomalous and brings a non-manifestly supersymmetric correction to the coefficient of the GB term of the parametric form $\sim N\ln e^{\beta K}=\beta NK$.~\footnote{Similarly, the re-definition of the metric necessary to remove the saxion couplings to $R_{ab}R^{ab}$ and $R^2$ may induce a Weyl anomaly, but that does not carry an $\sim N$ enhancement and is hence parametrically smaller.} Despite the $\propto N$ nature of these two perturbative corrections, however, we will see in Section \ref{sec:QGbounds} that they are both subleading compared to \eqref{gammas} in any tractable framework. Therefore \eqref{GRtheta} and \eqref{gammas} provide an accurate approximation of the GB term.

The above GB term is singled out from the infinite set of higher-derivative interactions by its quasi-topological nature.  For example, after stabilization of the saxions \eqref{GRtheta} (and supplemented by the boundary terms discussed in Section \ref{sec:WHs}), it becomes a purely topological term that does not alter the axions' equations of motion but nevertheless contributes to the on-shell action of topologically non-trivial space-times. In particular it plays an important role in wormhole physics, as we will see. At the perturbative level, the topological nature of GB is connected to the absence of ghosts, which is why string theory effective actions of any dimension  seem to favor it, so to speak, over other higher curvature terms \cite{Zwiebach:1985uq}.

\subsection{Dual formulation and EFT strings}
\label{sec:GBcontribution}

In order to make contact with the QG structures highlighted in \cite{Lanza:2020qmt,Lanza:2021udy,Martucci:2022krl}, it is convenient to recall the basic features of the dual formulation, in which the axions are traded for two-form potentials $\calb_{2,i}$ with corresponding field-strengths $\calh_{3,i}=\d\calb_{2,i}=-M_{\text{\tiny P}}^2\calg_{ij}*\d a^i$. This duality transformation can be completed into a full supersymmetric duality which trades the $t^i$ chiral multiplets for corresponding linear multiplets \cite{Lindstrom:1983rt}. Following \cite{Lanza:2019xxg}, with the notation of \cite{Lanza:2020qmt,Lanza:2021udy}, the linear multiplets have as bottom components the {\em dual saxions} $\ell_i$, which are related to the saxions $s^i$ by  
\be\label{elldef} 
\ell_i=-\frac12\frac{\del K}{\del s^i}\,.
\ee
The kinetic terms are specified by the kinetic potential
\be\label{kinF} 
\calf=K+2\ell_i s^i\,,
\ee
which must be considered as a function of the dual saxions $\ell_i$ (and of the spectator fields). Note that $K$ and $\calf$ are defined up to an arbitrary constant. The leading order action \eqref{kinetic} is equivalently re-written as
\be\label{kinetic2} 
\frac{1}{2}M_{\text{\tiny P}}^2\int R*1-\frac12M_{\text{\tiny P}}^2\int \calg^{ij}\d \ell_i\wedge *\d \ell_j-\frac1{2M_{\text{\tiny P}}^2}\int\calg^{ij}\calh_{3,i}\wedge *\calh_{3,j}\,,
\ee
where 
\be\label{calg2}
\calg^{ij}\equiv-\frac12\frac{\del^2\calf}{\del\ell_i\del\ell_j}
\ee
is the inverse matrix of \eqref{calg1}. Furthermore the inverse of the relation \eqref{elldef} is given by
\be\label{elldef2} 
s^i=\frac12\frac{\del\calf}{\del\ell_i}\,.
\ee
The dualization from the axions $a^i$ to the two-forms $\calb_{2,i}$ produces also a boundary term, to be added to \eqref{kinetic2},
\be\label{aHboundary} 
-\ii\int_{\del\calm}a^i\calh_{3,i}\,,
\ee
which may be relevant in evaluating the on-shell actions.

The field-strengths $\calh_{3,i}$ satisfy the Bianchi identity 
\be\label{H3BI} 
\d\calh_{3,i}=-\frac1{96\pi}\tilde C_i\tr(\calr\wedge \calr)\,.
\ee 
In fact, for our purposes one can consistently neglect the Pontryagin four-form appearing in \eqref{H3BI}, since it will be identically vanishing in all the configurations that we will explore. Because in the following discussions  gauge fields will play no role,  in \eqref{H3BI} we have not included their  contributions, which is dual to the axionic terms appearing in \eqref{QCDtheta} (see for instance Section 3.2 of \cite{Martucci:2022krl}).  Eq.\ \eqref{H3BI} can also be corrected by the localized  contribution of fundamental instantons, of the type  considered in Section \ref{sec:extremalBPS}.

Note that by the homogeneity of $P({s})$ (see Eq.\ \eqref{KP(s)}) we have $\ell_is^i=\frac{n}2$ and then, omitting an  irrelevant additional constant, the dual saxion  kinetic potential \eqref{kinF} takes the form 
\be\label{calfP}
\calf({\ell})=\log\tilde P({\ell})\,,
\ee
  where
\be 
\tilde P({\ell})\equiv \frac{1}{P({s}({\ell}))}\,
\ee 
is a homogeneous function of degree $n$: 
\be\label{tildePhom} 
\tilde P(\lambda{\ell})=\lambda^n\tilde P({\ell})\,.
\ee
Similarly, the (semi-)topological couplings to gravity can be written as in \eqref{GRtheta}-\eqref{gammas} provided we interpret $s^i$ as a function of the dual saxions, as dictated by \eqref{elldef2}.

In the present setting it is natural to consider strings  carrying magnetic axionic charges $e^i\in\mathbb{Z}$, around which $a^i\rightarrow a^i+e^i$. If such strings are BPS, their tension is completely fixed by supersymmetry \cite{Lanza:2019xxg}:
\be\label{strten} 
\calt_{\bf e}=M^2_{\text{\tiny P}}e^i\ell_i\equiv M^2_{\text{\tiny P}}\langle {\bm \ell},{\bf e}\rangle\,,
\ee
where we have introduced the index-free notation  $\langle {\bm \ell},{\bf e}\rangle\equiv \ell_ie^i$, which will be largely used in the following.\footnote{See \cite{Lanza:2020qmt,Lanza:2021udy} for a thorough discussion on the precise interpretation of the field-dependence of \eqref{strten}.} 
The formula  \eqref{strten} imposes a non-trivial constraint $\langle {\bm \ell},{\bf e}\rangle\geq 0$ on the charges as well as the dual saxions. We will elaborate on this constraint in the next section, when a domain for the saxions and dual saxions is identified.

\section{EFT regime of validity}
\label{sec:regimesEFT}

The EFT described in the previous section has a limited domain of validity. First, as any low-energy description, it has an associated derivative expansion and thus a maximal UV cutoff. Second, any such EFT can reliably  describe the low-energy limit of string theory models  only in a limited domain in field space. Furthermore, in the spirit of the Swampland Program, compatibility with QG/string theory imposes some non-trivial constraints on the EFT structure, in addition to the standard QFT ones. In this section we will discuss  these aspects in some detail. Specifically, in Section \ref{sec:perturbative} we quantify the size of the QG effects that violate the axionic shift-symmetry and introduce the concept of saxionic cone. In Section \ref{sec:saxconvhull} we provide an unambiguous definition of the domain of validity of our EFT {\emph{in field space}}. In  Section \ref{sec:QGbounds} it is then shown that, within this perturbative domain and in the large $N$ limit, the coefficient of the Gauss-Bonnet interaction is subject to an interesting lower bound. The domain of validity of our EFT {\emph{in momentum space}} is finally analyzed in Section \ref{sec:UVcutoff}.

Before embarking in our detailed analysis it is useful to make some simple consideration based on dimensional analysis, which is sufficient to qualitatively understand how the EFT domain of validity may be  controlled by the saxions $s^i$. If we momentarily restore the powers of $\hbar$ and insist that $\tilde C_i$ in \eqref{GRtheta} -- see also $C_i$ in \eqref{QCDtheta} -- be truly dimensionless integers, one infers that $s^i$ and $a^i$ have dimension  $[s^i]=[a^i]=[\hbar]$. Because quantum corrections come with powers of $\hbar$, one realizes that 
\be\label{EFTcoupling}
\alpha_*\equiv \frac{1}{s_*},
\ee
with $s_*$ denoting some appropriate positive linear combination of the saxions $s^i$,
represents a sort of ``fine structure constant" of our theory.
 Accordingly, in order for our EFT to make sense one should require the dimensionless ``loop counting parameter" in four-dimensional theories to be small, namely 
\be\label{saxions=couplings} 
\frac{\alpha_*\hbar}{2\pi}\equiv\frac\hbar{2\pi s_*}\ll1~~~~~~~~~~~~~~~~\left(\frac{\hbar\ell_*}{2\pi}\ll1\right)\,,
\ee 
where $(2\pi)^{-1}$ arises from the usual four-dimensional loop factor.\footnote{Recall that the structure constant is a coupling squared divided by $2\pi$. Throughout the paper we will keep track of the ``geometric" factors of $2\pi$ but ignore factors of order unity (see Appendix \ref{app:2pis}), and $2\pi\alpha_*$ will represent our ``coupling squared". The analogous expansion in $\hbar/(2\pi a_*)$ simply cannot appear because of the approximate shift symmetry.}  As a simple example, which is obvious from the effective field theorist's viewpoint,   the gauge coupling squared appearing in \eqref{QCDtheta} takes the form $2\pi\alpha_*=2\pi/s_*$ with $s_*= C_is^i$, and so \eqref{saxions=couplings} represents the standard perturbative regime for the gauge theory. What the effective field theorist cannot know, however, is that conditions of the form \eqref{saxions=couplings} are in fact instrumental in computing EFTs from string theory models and, for instance,   disguise  large volume or weak string coupling expansions. As an example, of crucial importance for the present paper is the form \eqref{KP(s)} of the K\"ahler potential: such form holds only to first approximation in an appropriate large-saxion expansion and is expected to receive both perturbative and non-perturbative corrections. In view of our dimensional analysis argument, it should not come as a surprise that perturbative corrections are controlled by $\alpha_*/(2\pi)$ whereas the non-perturbative effects that break the axionic shift symmetries are of order $~e^{-\frac{2\pi}{\alpha_*\hbar}}=e^{-\frac{2\pi s_*}{\hbar}}$, and are hence exponentially suppressed by a requirement of the form \eqref{saxions=couplings}. 

Non-perturbative effects may be due to fundamental instantons beyond the EFT or by physics within the EFT, e.g.\ gauge instantons or wormholes.  Wormholes  will be discussed at length in the following sections while gauge instantons, not being directly related to QG aspects, will not be considered in the present paper. In the following  subsection we will instead  focus on fundamental instantons, since they encode non-trivial information on the UV completion of the theory and turn out to strongly characterize the EFT structure. From now on we will go back to the more conventional $\hbar=1$ units.

\subsection{Saxionic cones, fundamental instantons and   strings}
\label{sec:perturbative}

The contribution of point-like fundamental instantons are ubiquitous in string theory compactifications, in which they are typically associated to Euclidean branes wrapping internal cycles -- see for instance \cite{Blumenhagen:2009qh} for a review. Their effects show up in the four-dimensional EFT defined at its highest possible UV cutoff as shift-symmetry breaking local operators. Imposing that these are sufficiently small is what defines our perturbative regime. Among all fundamental instantons, the BPS ones --- preserving $\frac12$ of the bulk supersymmetry --- are expected to be the most relevant.\footnote{Experience with supersymmetric instantons suggests that the action $S_{\rm inst}$ of a possible non-BPS fundamental instanton carrying charges $q_i$ obeys a BPS bound $S_{\text{non-BPS}}>S_{\text{BPS}}$. On the other hand, the axion form of the weak gravity conjecture \cite{ArkaniHamed:2006dz} suggests the possible existence of non-BPS instantons violating such bound -- see for instance  \cite{Demirtas:2019lfi,Long:2021lon} and appendix B of \cite{Demirtas:2021gsq}  for related discussions  in string theory contexts. We nevertheless expect such possible violations not to affect the following considerations.} These carry a set of quantized axionic charges $q_i\in\mathbb{Z}$ and contribute to the effective action by terms proportional to $e^{2\pi\ii q_i a^i}$. By holomorphy, BPS instantons generate terms proportional to $e^{2\pi \ii q_i t^i}$. For each BPS instanton of charges $q_i$ there exists an {\em anti}-instanton of charges $-q_i$ preserving the opposite $\frac12$ supersymmetry and contributing by terms proportional to $e^{-2\pi \ii q_i \bar t^i} $.  Hence, BPS and anti-BPS instantons combine and contribute to the effective action via operators proportional to 
\be\label{expms} 
e^{-2\pi \langle {\bf q}, {\bm s}\rangle}\,,
\ee
where we are again using the index-free pairing introduced in \eqref{strten}:
\be 
\langle {\bf q}, {\bm s}\rangle\equiv q_i s^i.
\ee
In our notation $\langle\,.\,,\,.\,\rangle$ is the canonical pairing between the elements of dual vector spaces $V_{\mathbb{R}}$ and $V^*_{\mathbb{R}}$, and corresponding dual lattices $V_{\mathbb{Z}}\subset V_{\mathbb{R}}$ and $V^*_{\mathbb{Z}}\subset V^*_{\mathbb{R}}$. One can  introduce an integral basis $\{{\bf v}_i\}^N_{i=1}$  of generators of $V_{\mathbb{Z}}$ and the dual basis $\{{\bf w}^i\}^N_{i=1}$ of generators of $V^*_{\mathbb{Z}}$,  such that $\langle {\bf w}^i,{\bf v}_j\rangle=\delta^i_j$. The saxions $s^i$ and the charges $q_i$ are the components of the vectors ${\bm s}=s^i{\bf v}_i\in V_{\mathbb{R}}$ and ${\bf q}=q_i{\bf w}^i\in V^*_{\mathbb{R}}$ respectively. The set of all BPS instanton charges is denoted by
 \be\label{BPSinstcharges} 
 \calc_{\rm I}=\{\text{set of BPS instanton charges ${\bf q}$}\}\subset V^*_{\mathbb{Z}}\,.
 \ee 
 Given two BPS instantons of charge vectors ${\bf q}_1$ and ${\bf q}_2$, being mutually BPS, they can be superimposed to form a BPS instanton of charge vector ${\bf q}_1+{\bf q}_2$. Hence  $\calc_{\rm I}$ can be regarded as discrete convex cone, generated by a set of ``elementary'' BPS instanton charges.

The combination $2\pi \langle {\bf q}, {\bm s}\rangle$ appearing in \eqref{expms} represents the real part of the BPS instanton Euclidean action, and must be positive. Hence the saxions necessarily take values in the {\em saxionic cone}:\footnote{The definition of saxionic cone of \cite{Lanza:2020qmt,Lanza:2021udy} slightly differs from \eqref{sCone}, in that it does not include the boundary faces at which  $\langle {\bf q}, {\bm s}\rangle= 0$ for some ${\bf q}$, and then some instanton action degenerates. We include these faces since  we will anyway  restrict ourselves to more interior regions -- see Section \ref{sec:saxconvhull}.}
\be\label{sCone} 
\Delta\equiv \{{\bm s}\in V_{\mathbb{R}}|\langle {\bf q}, {\bm s}\rangle\geq 0,\ \forall {\bf q}\in \calc_{\rm I}\}\,.
\ee
This is a convex cone, whose prototypical example is provided by the K\"ahler cone in heterotic or type IIA string compactifications on Calabi-Yau spaces, where $\calc_{\rm I}$ represents the cone of effective curves which can be wrapped by world-sheet instantons. These and other  string theory realizations will be more explicitly discussed  in  the following. 

The magnetic axionic string charges $e^i\in\mathbb{Z}$  introduced around \eqref{strten} specify an element ${\bf e}=e^i{\bf v}_i$ of $V_\mathbb{Z}$. We call {\em EFT strings} those strings associated to charge vectors ${\bf e}$ belonging to the set
\cite{Lanza:2020qmt,Lanza:2021udy,Lanza:2022zyg}
\be\label{EFTstrcharges}
\calc_{\rm S}^{\text{\tiny EFT}}=V_\mathbb{Z}\cap\Delta-\{{\bf 0}\}\,.
\ee 
According to this definition, one may regard the saxionic cone $\Delta$ as being generated by the EFT string charges. We will assume $\Delta$ to be polyhedral -- see \cite{Lanza:2021udy} for a more detailed discussion on this assumption -- and hence to be generated by a finite number of {\em elementary} EFT string charges, i.e.\ charges ${\bf e}\in\calc_{\rm S}^{\text{\tiny EFT}}$ that cannot be written as a sum with positive integral coefficients of other elements of $\calc_{\rm S}^{\text{\tiny EFT}}$. (We will extend this terminology to BPS strings and BPS instantons in an obvious way.) As we will see,  EFT strings will play a key role in  the following discussions.

We can now define the  dual saxionic domain $\calp$ as  the closure of the image of $\Delta$ under the transform \eqref{elldef}:
\be\label{Pdomain}
\calp=\overline{\Big\{{\bm \ell}=\ell_i{\bf w}^i\in V^*_{\mathbb{R}}\Big|\ell_i=-\frac12\frac{\del K}{\del s^i}\Big|_{{\bm s}\in\Delta}\Big\}}.
\ee
While the general structure of $\calp$ can be a priori complicated, if $K$ takes the form \eqref{KP(s)} then $\calp$ becomes conical. Indeed, if ${\bm \ell}\in\calp$ is the image of ${\bm s}\in\Delta$ then $\lambda{\bm \ell}$ is the image of $\lambda^{-1}{\bm s}$,  for any $\lambda>0$. Since also $\lambda^{-1}{\bm s}$ belongs to $\Delta$, then  $\lambda{\bm \ell}$ belongs to $\calp$. (On the other hand, $\calp$ is not necessarily convex.) Note that by consistency any BPS (non-necessarily EFT)  string tension \eqref{strten} must be positive in the interior of $\calp$, and the condition $\langle {\bm \ell},{\bf e}\rangle\geq 0$ for any $ {\bm \ell}\in\calp$ can be taken as defining condition of the BPS string charges  ${\bf e}\in\calc_{\rm S}$, that is:\footnote{We recall that, given two dual vector spaces $V_{\mathbb{R}}$ and $V^*_{\mathbb{R}}$ and any subset $I\subset V_{\mathbb{R}}$, its dual cone $I^\vee$ is by definition the set of elements ${\bm a}\in V^*_{\mathbb{R}}$ such that $\langle {\bm a},{\bm b}\rangle \geq 0$ for any ${\bm b}\in I$.} 
\be\label{def:Cs} 
\calc_{\rm S}\in  V_{\mathbb{Z}}\cap\calp^\vee-\{{\bf 0}\}\,.
\ee  
This implies that different boundary components of  $\del\calp$ can be associated with the possible vanishing of different BPS string tensions \eqref{strten}.  In particular, components of $\del\calp$ which are at infinite field distance are detected by the vanishing of some EFT string tensions, i.e.\ corresponding to some  ${\bf e}\in\calc^{\text{\tiny EFT}}_{\rm S}$  \cite{Lanza:2021udy}. On the other hand, finite distance components of  $\del\calp$ could be associated with (classically) tensionless non-EFT BPS strings, i.e.\ with ${\bf e}\in\calc_{\rm S}-\calc^{\text{\tiny EFT}}_{\rm S}$. These tensionless strings naturally identify a rational polyhedral part of the boundary of $\calp$. There may also be more general finite distance boundaries of $\calp$, not directly associated with tensionless strings. In any case, one should keep in mind that finite distance boundaries are not so sharply defined, since around them non-perturbative corrections can a priori become relevant and may for instance generate strong corrections  to the formula \eqref{strten}.  Other strongly coupled regions are reached by radially moving away from the tip of $\calp$ along different directions. If one  insists in using the ``bare'' kinetic potential \eqref{calfP}, these appear as infinite distance limits in which all the BPS tensions \eqref{strten} diverge. However in these limits the above description breaks down since it assumes that 
\be\label{TMP}
\calt_{\bf e}< 2\pi M^2_{\text{\tiny P}}\,,
\ee 
in order for the string to have a weak gravitational backreaction \cite{Lanza:2020qmt}.\footnote{The $2\pi$ factor is introduced to match the NDA arguments of Appendix \ref{app:2pis}, but can also be understood   recalling that a string of constant tension $\calt$ generates a deficit angle $\Delta\theta=\calt/M^2_{\text{\tiny P}}$ (see e.g.\ \cite{Vilenkin:2000jqa}) and imposing $\Delta\theta<2\pi$.}  Therefore, non-perturbative physics   may again completely change the nature of these limits. In any case, we see how the behavior of the  BPS tension \eqref{strten} can be a useful proxy to qualitative characterize the different boundary components of $\calp$.

\subsection{Perturbative domain and  saxionic convex hull}
\label{sec:saxconvhull}

The realization that saxions must necessarily belong to $\Delta$ is bringing us closer to a precise definition of the domain of validity of our EFT. Unfortunately, ${\bm s}\in\Delta$ is not sufficient to suppress instantons nor to ensure the perturbativity requirement suggested in  \eqref{saxions=couplings} holds.  In this subsection we will more precisely identify a perturbative domain with a subset of $\Delta$, controlled by a single perturbative coupling $\alpha< 2\pi$, analogous to \eqref{EFTcoupling}. We will consider two possible  perturbative domains: the $\alpha$-{\em saxionic convex hull} $\hat\Delta_\alpha$ and the   $\alpha$-{\em stretched} saxionic cone  $\tilde\Delta_\alpha$, the latter being defined in analogy to  the stretched K\"ahler cones introduced in \cite{Demirtas:2018akl}. Since $\hat\Delta_\alpha\subseteq \tilde\Delta_\alpha$, for simplicity in the rest of the paper we will adopt the more conservative $\hat\Delta_\alpha$ as our main definition perturbative domain, though most of our  conclusions would clearly hold for $\tilde\Delta_\alpha$ as well. 

Consider the set of all the  elementary EFT string charges  $\{{\bf e}_A\}_{A\in\calj}\subset \calc^{\text{\tiny EFT}}_{\rm S}$, which generate the entire $\calc^{\text{\tiny EFT}}_{\rm S}$, where $\calj$ denotes the corresponding set of indices. 
Take any subset $\calj_\sigma\subset \calj$ of $N$ elements, such that the corresponding elementary charges  $\{{\bf e}_A\}_{A\in\calj_\sigma}$ are linearly independent. Each of these subsets is associated to a regular simplicial cone  
\be\label{simpsigma}
\sigma=\{\sum_{A\in\calj_\sigma}\lambda^A{\bf e}_A|\lambda^A\geq 0\}.
\ee 
 For each of these cones we construct a corresponding ``$\alpha$-stretched" cone 
 \be
 \tilde\sigma_\alpha=\{\sum_{A\in\calj_\sigma}\lambda^A{\bf e}_A|\lambda^A\geq \frac1\alpha\}
 \ee
 with $\alpha>0$ some small number that represents the largest possible value of the couplings of the form  \eqref{EFTcoupling}. The
$\alpha$-{\em saxionic convex hull} $\hat\Delta_\alpha$, anticipated at the beginning of this subsection, is defined as the convex hull of all the stretched sub-cones $\tilde\sigma_\alpha$, that is
\be\label{sconvhull} 
\hat\Delta_\alpha=\Big\{ {\bm s}=\sum_\sigma \lambda^\sigma{\bm s}_\sigma\in \Delta\Big|\  \lambda^\sigma\geq 0,\ \sum_\sigma \lambda^\sigma=1,\ {\bm s}_\sigma\in \tilde\sigma_\alpha\Big\}.
\ee
Intuitively, any element of $\hat\Delta_\alpha$  can be considered as a linear average  of saxions  whose components satisfy $s^i\geq 1/\alpha$ in some basis of elementary EFT strings charges.
The perturbative regime  can now be precisely identified by the requirement ${\bm s}\in\hat\Delta_\alpha$. With a sufficiently small $\alpha$, this definition simultaneously formalizes the perturbativity requirement \eqref{saxions=couplings} as well as guarantees the suppression of non-perturbative corrections.  

In order to better understand this latter point, let us relate $\hat\Delta_\alpha$ to the $\alpha$-{\em stretched saxionic cone}\,\footnote{We are adapting the terminology of \cite{Demirtas:2018akl} which may be slightly misleading, since $\tilde\Delta_{\alpha}$ is generically not a cone, but rather a convex polyhedron.} 
\be
\tilde\Delta_\alpha\equiv\{{\bm s}\in \Delta|\langle {\bf q},{\bm s}\rangle\geq \frac{1}{\alpha},\ \forall {\bf q}\in{\cal C}_{\rm I} \}\,.
\ee
This definition is more directly motivated  by the non-perturbative corrections \eqref{expms} (rather than the perturbative ones), since in $\tilde\Delta_\alpha$ all such  corrections are bounded from above by $e^{-2\pi/\alpha}$. Now, 
because our $\tilde\sigma_\alpha$ appearing in \eqref{sconvhull} are clearly subsets of $\tilde\Delta_\alpha$, by convexity the same inclusion extends to the entire saxionic convex hull, so that: 
\be 
\hat\Delta_\alpha\subset \tilde\Delta_\alpha\subset\Delta\,.
\ee
Hence the condition ${\bm s}\in \hat\Delta_\alpha$ is stronger, though qualitatively similar, to the condition ${\bm s}\in \tilde\Delta_\alpha$. (Clearly $\hat\Delta_\alpha=\tilde\Delta_\alpha$ 
 if $\Delta$ is a simplicial cone.) As a result, the non-perturbative corrections to our EFT are at least suppressed by $e^{-2\pi/\alpha}$ when ${\bm s}\in \hat\Delta_\alpha$.\footnote{Note  that, given a saxionic cone $\Delta$, its boundary can be regarded as the union of conical faces $\Delta'\subset\del\Delta$ of various codimensions. These may be associated with corresponding perturbative domains  $\tilde\Delta'_\alpha$ or $\hat\Delta'_\alpha$ (which are {\em not} subsets  of $\tilde\Delta_\alpha$ or $\hat\Delta_\alpha$).}

The perturbative saxionic regions $\tilde\Delta_\alpha,\hat\Delta_\alpha\subset \Delta$ are associated to corresponding dual saxionic regions $\tilde\calp_\alpha,\hat\calp_\alpha\subset\calp$ through the map \eqref{elldef}.  In the case of K\"ahler potentials of the form \eqref{KP(s)} the regions dual to $\tilde\Delta_\alpha,\hat\Delta_\alpha$ are concentrated around the dual saxion origin ${\bm \ell}=0$ -- see Fig.~\ref{fig:model1domainDeltaP} below for a simple but non-trivial concrete example.

To summarize, by assuming ${\bm s}\in \hat\Delta_\alpha$ (or ${\bm \ell}\in \hat\calp_\alpha$) we are certain that our EFT \eqref{kinetic} with \eqref{KP(s)} provides a reliable low-energy description of a large class of string theory models up to controllable powers of $\alpha/2\pi\ll1$ and $e^{-2\pi/\alpha}$. In particular, the latter exponential factor suppresses any explicit breaking of the axion shift symmetries. Our expansion parameter $\alpha$ is nothing but an upper bound on the quantity $\alpha_*$ alluded to in Eq.\ \eqref{EFTcoupling}. In the following we will therefore assume that ${\bm s}\in\hat\Delta_\alpha$ and for concreteness have in mind $\alpha\simeq 0.1$ as benchmark value.  In fact, $\alpha\simeq 1$ might already be enough to sufficiently suppress both perturbative and non-perturbative effects. However, such values of $\alpha$ do not necessarily ensure the reliability of the leading order expression \eqref{KP(s)}. Concretely, in the case of the K\"ahler cone of heterotic compactifications, setting $\alpha\simeq 1$ allows for string-size internal cycles, which for instance cast doubts on the geometric formula corresponding to \eqref{KP(s)} -- see Section \ref{sec:stringtheorymodels} for more details. From these considerations $\alpha\simeq 0.1$ seems a more reassuring choice. Much smaller values face another problem, though, which will be analyzed in Section \ref{sec:UVcutoff}: in the  limit $\alpha\rightarrow 0$ the maximal possible UV cutoff of the EFT tends to zero!

\subsection{Quantum gravity bounds}
\label{sec:QGbounds}

In \cite{Martucci:2022krl} it was shown how quantum consistency in the presence of EFT strings imposes strong constraints on the structure of the bulk theory. These constraints crucially involve the constants $\tilde C_i$ appearing in \eqref{gammas}. In particular $\tilde C_i$ must satisfy the quantization condition $\langle\tilde {\bf C}, {\bf e}\rangle\equiv \tilde C_ie^i\in\mathbb{Z}$, for any string charge vector ${\bf e}\in V_{\mathbb{Z}}$. More importantly for us, in \cite{Martucci:2022krl} it was argued that $\langle\tilde {\bf C}, {\bm s}\rangle$ enter some positivity bounds which, in their weakest form, imply that 
\be\label{Ctildebound} 
\langle \tilde{\bf C},{\bf e}\rangle\geq 0\quad~~~~~~~~~~~~\forall {\bf e}\in\calc^{\text{\tiny EFT}}_{\rm S}\,.
\ee
Recalling \eqref{gammas} and the fact that the EFT string charges generate the saxionic cone, \eqref{Ctildebound} has as a consequence that $\gamma({s})\geq  0$ for any ${\bm s}\in\Delta$. 

In order to get a stronger lower bound for $\gamma({s})$ we observe that $\langle \tilde{\bf C},{\bf e}\rangle$ is proportional to the two-dimensional gravitational anomaly of the EFT string of charge vector ${\bf e}$. Since such strings break half of the bulk supersymmetry and support a chiral  $(0,2)$ world-sheet, they generically have a chiral spectrum with non-vanishing gravitational anomaly. This means that \eqref{Ctildebound} generically translates into the stricter bound 
\be\label{gammabound1} 
\langle \tilde{\bf C},{\bf e}\rangle\in \mathbb{Z}_{\geq 1}\quad~~~~\forall\text{ generic }{\bf e}\in\calc^{\text{\tiny EFT}}_{\rm S}\,.
\ee
Note that in many models \eqref{gammabound1} can be strengthened to $\langle \tilde{\bf C},{\bf e}\rangle\in 3\mathbb{Z}_{\geq 1}$. This stronger bound holds if the world-sheet  normal bundle $U(1)_{\rm N}$ symmetry is classically preserved (though generically anomalous at the quantum world-sheet level). This is a conceivable expectation, which is  indeed realized in large classes of string theory models, such as the F-theory/type IIB ones of Section \ref{sec:FIIB}.  On the other hand, in \cite{Martucci:2022krl} it was pointed out that in addition to the standard quantum $U(1)_{\rm N}$ anomaly there could be  classical Green-Schwarz-like terms on the world-sheet, which signal the existence of an intermediate microscopic  description in terms of a five dimensional $\caln=1$ supergravity (in presence of possible supersymmetry breaking defects). The five-dimensional arguments of \cite{Katz:2020ewz} hence lead to \eqref{Ctildebound}, and then also \eqref{gammabound1}. For instance, this weaker bound  can hold in the $E_8\times E_8$ heterotic models of Section \ref{sec:het}. 

Let us now assume that in our models there are $N\gg 1$ (s)axions and an even larger number of elementary EFT string charge vectors $\{{\bf e}_A\}_{A\in\calj}$ (since these generate the generically non-simplicial $\Delta$). We can then assume \eqref{gammabound1} to be satisfied for {\em all} ${\bf e}_A$, since for $N\gg 1$  any non-generic violation of this assumption would affect our conclusions by negligible corrections. Hence we will assume that
\be\label{tildeCvA} 
\langle\tilde{\bf C},{\bf e}_A\rangle\in\mathbb{Z}_{\geq 1}\quad\,,
\ee 
for any elementary  ${\bf e}_A\in\calc_{\rm S}^{\text{\tiny EFT}}$. Take now any regular simplicial sub-cone \eqref{simpsigma} and an element ${\bm s}_\sigma$ of the corresponding $\alpha$-stretched cone: ${\bm s}_\sigma\in\tilde\sigma_\alpha$.  We can write ${\bm s}_\sigma=\frac1\alpha\sum_{A\in\calj_\sigma}{\bf e}_A+{\bm v}_\sigma$, where the first contribution represents the tip of ${\bm s}_\sigma$ and   ${\bm v}_\sigma$ is an element of $\sigma$. As a consequence, the lower bounds \eqref{gammabound1} and \eqref{tildeCvA} imply that 
\be\label{Ctildebsigma}
\langle\tilde{\bf C},{\bm s}_\sigma\rangle=\frac1\alpha\sum_{A\in\calj_\sigma}\langle\tilde{\bf C},{\bf e}_A\rangle+\langle\tilde{\bf C},{\bm v}_\sigma\rangle\geq \frac1\alpha\sum_{A\in\calj_\sigma}\langle\tilde{\bf C},{\bf e}_A\rangle\geq \frac{N}\alpha\,.
\ee
Consider next a more general point ${\bm s}$ of the saxionic convex hull \eqref{sconvhull}. By definition, we can write it as ${\bm s}=\sum\lambda^\sigma{\bm s}_\sigma$, with $\sum_\sigma\lambda^\sigma=1$ and $\lambda^\sigma\geq 0$. According to \eqref{Ctildebsigma} we thus have 
\be\label{tildeCs} 
\langle\tilde{\bf C},{\bm s}\rangle=\sum_\sigma\lambda^\sigma\langle\tilde{\bf C},{\bm s}_\sigma\rangle\geq \frac{N}{\alpha}\sum_\sigma\lambda^\sigma=\frac{N}{\alpha}\,.
\ee 
Hence we conclude that for ${\bm s}\in\hat\Delta_\alpha$ the coefficient \eqref{gammas} of the GB term \eqref{GRtheta} satisfies the lower bound
\be\label{gammabound}
\gamma({s})|_{\hat\Delta_\alpha}\geq\frac{N\pi}{6\alpha}\,.
\ee
This bound may receive $1/N$ corrections due to non-generic violations of \eqref{tildeCvA}, but in the large $N$ regime these can be safely neglected. We also observe that, as stressed above, in many models we could alternatively adopt $\langle \tilde{\bf C},{\bf e}_A\rangle\in 3\mathbb{Z}_{\geq 1}$ instead of \eqref{tildeCvA}, obtaining a slightly stronger lower bound $\gamma({s})|_{\hat\Delta_\alpha}\geq\frac{N\pi}{2\alpha}$.

At the end of Section \ref{sec:semitop} we discussed the possible corrections to the GB coefficient \eqref{gammas}. Within the perturbative regime $\alpha\ll 2\pi$ the radiative effects not included in \eqref{gammas} are parametrically smaller than the one in \eqref{gammabound}, as anticipated there. We can thus conclude with confidence that in this regime the bound \eqref{gammabound} is not significantly spoiled by such corrections. Note that while this result was derived by restricting ${\bm s}$ to the $\alpha$-stretched saxionic convex hull, we expect it to qualitatively hold (up to a possible overall constant) also if we consider the stretched saxionic cone. We will provide some evidence of this claim in Section \ref{sec:stringtheorymodels}, where we will show that \eqref{gammabound} is actually very conservative in a set of concrete string theory models.

\subsection{UV mass scales}
\label{sec:UVcutoff}

Any EFT is associated with a cutoff energy $\Lambda$ above which it is no longer valid. As in the Wilson's view of the renormalization group, $\Lambda$ is not in general a physical scale, but rather a conventional definition of the regime of validity of the description. As anticipated in Section \ref{sec:semitop}, the EFT cutoff must satisfy $\Lambda< M_{\text{\tiny UV}}$, where  $M_{\text{\tiny UV}}$ is some UV  mass scale suppressing the irrelevant operators appearing in the effective action. In the present context, the UV  physical mass scales beyond the EFT do depend on the moduli fields of \eqref{kinetic}, and specifically on $s^i$. As a result, $M_{\text{\tiny UV}}$ not only represents the highest possible energy above which the momentum expansion ceases to be effective, but also an implicit constraint on the domain of the saxions through the condition $M_{\text{\tiny UV}}(s)>\Lambda$.

In this section we would like to identify a proxy for $M_{\text{\tiny UV}}$ in the specific scenarios introduced in Section \ref{sec:N=1models}. We will see that these models are characterized by two relevant scales: the {\emph{tower scale}} \cite{Ooguri:2006in} and the {\emph{species scale}} \cite{Dvali:2007hz,Dvali:2007wp,Dvali:2009ks,Dvali:2010vm}. The former identifies the energy threshold beyond which a four-dimensional EFT should be replaced by a more fundamental description whereas the latter scale sets the absolute maximum UV cutoff at which {\emph{any}} EFT inevitably breaks down. In the process of discussing these two relevant scales we will also propose a novel upper bound on the species scale defined solely in terms of EFT data. In the reminder of the paper we will then conservatively take the species scale as our proxy for $M_{\text{\tiny UV}}$.

In the perturbative framework outlined in the  Subsection \ref{sec:saxconvhull}, the weak coupling limit $\alpha\rightarrow 0$ pushes the entire domain $\hat\Delta_\alpha$ to infinite field space  distance. According to the Distance Conjecture \cite{Ooguri:2006in} (see also \cite{Agmon:2022thq} for a recent review) this signals the appearance of towers of massive single-particle states. More precisely, denoting by $M_{\rm t}$ the {\em tower scale}, i.e.\ the mass of the lightest particle of such towers, the Distance Conjecture implies that $M_{\rm t}/M_{\text{\tiny P}}$ decreases exponentially with the field space distance -- see also \cite{Etheredge:2022opl}. Furthermore, the possible nature of such towers, and with it of the UV completion of our EFT, is significantly restricted by the Emergent String Conjecture (ESC) \cite{Lee:2019wij}, which states that (in appropriate duality frames) those UV towers can be formed by either  Kaluza-Klein modes or by excitation modes of a weakly coupled critical superstring. The former modes become massless in  limits involving decompactifications to higher-dimensional EFTs, whereas the latter in limits in which the critical string coupling goes to zero. In the following we will assume the validity of the ESC conjecture and take
\be
M_{\rm t}\equiv{\rm min}\left\{M_{\text{\tiny KK}},M_{\rm s}\right\}, 
\ee
where $M_{\text{\tiny KK}}$ and $M_{\rm s}$ are the  KK and superstring scales (as measured by the four-dimensional observer). 

In general it is not known how to precisely determine $M_{\rm t}$ without knowing the details of the UV completion, that is, of the higher dimensional EFT and/or string theory. Yet, in the present context, useful information is encoded in the tension \eqref{strten} of EFT strings, as indicated by the Integral Weight Conjecture (IWC) \cite{Lanza:2021udy,Lanza:2022zyg}. Namely,
each EFT string charge ${\bf e}\in\calc_{\rm S}^{\text{\tiny EFT}}$ identifies an infinite distance saxionic flow ${\bm s}={\bm s}_0+{\bf e}\sigma$, with $\sigma\rightarrow\infty$, and is associated with an integral {\em scaling weight} $w_{\bf e}\in\{1,2,3\}$. Along this EFT string flow  $\calt_{\bf e}\sim M^2_{\text{\tiny P}}/\sigma$ and the scaling weight relates this behavior with the asymptotic scaling of  $M_{\rm t}$ as follows:
\be\label{ISW} 
M^2_{\rm t}\sim \left(\frac{\calt_{\bf e}}{M^2_{\text{\tiny P}}}\right)^{w_{\bf e}}M^2_{\text{\tiny P}}\,. 
\ee
Note that this relation reveals the scaling behavior in $\sigma$, but does not allow one to precisely determine $M^2_{\rm t}$. Nevertheless, combined  with the ESC, \eqref{ISW} contains important pieces of information on the UV nature of the EFT strings and of the corresponding infinite distance limits. First of all, any $w_{\bf e}=1$ EFT string in the corresponding EFT string limit must uplift to a weakly coupled critical superstring. This in turn implies that any $w_{\bf e}=1$ EFT string flow\footnote{For instance, assuming a dual $E_8\times E_8$ string model of the type discussed below in Section \ref{sec:het},  this limit corresponds to $s^0\sim \sigma\rightarrow\infty$, with $s^0$ as in \eqref{hets0} and fixed $s^a$.} is dual to a ten-dimensional weak string coupling limit $g_{\rm s}\rightarrow 0$, along which the tower scale $M_{\rm t}$ can be identified with the mass of the first excited string mode. Thus, for any elementary EFT string charge ${\bf e}$ of scaling weight $w_{\bf e}=1$, \eqref{ISW} can  actually be promoted to the identity $M_{\rm t}^2|_{w_{\bf e}=1}=M_{\rm s}^2=2\pi \calt_{\bf e}$. On the other hand, EFT strings with $w_{\bf e}\geq 2$ {\em cannot} be identified with critical strings, and the corresponding infinite distance limits must correspond to decompactification limits along which $M_{\rm t}$ corresponds to a Kaluza-Klein (KK) mass scale.

There is another important mass scale that controls the transition away from the EFT: the {\em species scale} $M_{\text{sp}}$ \cite{Dvali:2007hz,Dvali:2007wp,Dvali:2009ks,Dvali:2010vm} -- see also the recent review \cite{Agmon:2022thq}. Various definitions of species scale have been given. In this paper, by $M_{\text{sp}}$ we mean the highest possible scale at which our gravitational setup admits a reliable (possibly higher dimensional) EFT description. 
The ESC allows us to make this concept more concrete: $M_{\text{sp}}$ is either the lightest string excitation  mass or the scale at which the KK excitations become strongly-coupled, depending on which of the two is lower. If a weakly coupled string description exists, then the KK excitations are never strongly coupled and the scale $M_{\text{sp}}$ can be identified with the critical superstring mass, properly converted to the four-dimensional frame: $M_{\text{s}}=\sqrt{2\pi\calt_{\rm F1}}$.  On the other hand, if no perturbative stringy description exists then $M_{\text{sp}}$ corresponds to a quantum gravity scale $M_{\text{\tiny QG}}$, at which the gravitational interactions become strong,  which roughly coincides with the higher-dimensional Planck mass (properly converted to the four-dimensional frame). In the latter case, a quantitative criterion for determining the 
quantum gravity scale, based on ``Naive Dimensional Analysis'' (NDA), is discussed in Appendix \ref{app:2pis} and tested on some concrete string theory models in Section \ref{sec:stringtheorymodels}.

We will refer to $M_{\text{sp}}={\rm min}\left\{M_{\text{s}},M_{\text{\tiny QG}}\right\}$ as ``species scale" to conform to the terminology adopted in most of the literature. Such terminology originates from more general models with a large number $N_{\rm sp}\gg 1$ of species \cite{Dvali:2007hz,Dvali:2007wp}, in which perturbative as well as non-perturbative arguments indicate that the gravitational interactions become strong at a scale of order
\be\label{axiSC} 
\frac{2\pi M_{\text{\tiny P}}}{\sqrt{N_{\rm sp}}}\,.
 \ee 
In our setup, identifying  $N_{\rm sp}$  with the number of (massless and massive) KK  modes   of mass smaller than $M_{\text{\tiny QG}}$, one can show that $M_{\text{\tiny QG}}$ is indeed consistent with \eqref{axiSC},
see for instance \cite{Castellano:2022bvr} and Appendix \ref{app:2pis} for a systematic discussion.  Yet, $M_{\text{\tiny QG}}$ ceases to be a reliable measure of the species scale when $M_{\rm s}<M_{\text{\tiny QG}}$. In such circumstances black-hole arguments \cite{Dvali:2009ks,Dvali:2010vm} (see also \cite{Agmon:2022thq}) lead to the identification $M_{\text{sp}}=M_{\rm s}$ adopted above.\footnote{Applying \eqref{axiSC} to string excitation modes, one gets $M_{\rm s}$ up to logarithmic corrections \cite{Marchesano:2022axe,Castellano:2022bvr} -- see also related discussions in \cite{Blumenhagen:2023yws,Basile:2023blg}. One may adopt that ``stringy species scale" as a definition of species scale, but that would not significantly affect our conclusions.}

As the tower scale, also the species scale generically depends on the details of the EFT UV completion. However, it was recently proposed \cite{vandeHeisteeg:2022btw} that information on the species scale is captured by the coefficient of certain higher-derivative gravitational interactions -- see also \cite{vandeHeisteeg:2023ubh,Cribiori:2023ffn,Calderon-Infante:2023uhz,vandeHeisteeg:2023dlw,Castellano:2023aum,Castellano:2023stg,Castellano:2023jjt} for related discussions. By applying the proposal of  \cite{vandeHeisteeg:2022btw} to our context one gets a relation of the form  $
\gamma\sim {\cal O}(1){M_{\text{\tiny P}}^2}/{M_{\text{sp}}^2}$ between the species scale and GB coefficient appearing in \eqref{GRtheta}.
This relation can be understood as follows. 

The four-dimensional gravitational theories we consider in this paper represent the low-energy description of some UV complete theory. The lowest threshold that characterizes the latter theory has been denoted by $M_{\text{sp}}$. It is therefore natural to imagine deriving our EFT by matching it to its UV completion precisely at that scale. On general grounds, the resulting QFT will contain operators of arbitrary dimensions with coefficients set by powers of $M_{\text{sp}}$ times dimensionless coefficients $c$. Perturbativity at the matching scale demands that $|c|\lesssim1$. One would thus be naively tempted to claim that $M_{\text{\tiny UV}}$ should be identified with $M_{\text{sp}}$. However, a further step is usually needed when $M_{\text{\tiny KK}}<M_{\text{sp}}$. In order to arrive at our four-dimensional EFT one should first integrate out the KK excitations within the intermediate higher-dimensional description. In carrying out this last step some of the higher-dimensional operators of our four-dimensional EFT will inevitably receive corrections proportional to powers of $M_{\text{sp}}/M_{\text{\tiny KK}}$. In particular, any operator of the form ``current squared" can in principle receive tree-level corrections from the integration of KK resonances of the appropriate spin. This is for instance the case for $R^2$ and $R_{ab}R^{ab}$, unless protected by extended supersymmetries, which may a priori be mediated by scalar and spin-2 KK modes. On the other hand, there is no KK excitation with the quantum numbers appropriate to couple linearly to $R_{abcd}$ in any known low-energy description of string theory, and so the coefficient of  $R_{abcd}R^{abcd}$ is not expected to be renormalized at tree-level. The same tree-level non-renormalization property thus extends to the GB operator \eqref{GBdensity}. Moreover, by power-counting the coefficient of the latter operator can only receive logarithmic corrections and cannot depend on inverse powers of $M_{\text{\tiny KK}}$. This is particularly clear in the $\caln=1$ context we are considering, see Section \ref{sec:semitop}.  Taking into account our normalization conventions, which are discussed in more detail in Appendix \ref{app:2pis}, it follows that 
\be\label{GBNDA}
{\gamma}= {4\pi^2}\, c_{\text{GB}}\frac{M_{\text{\tiny P}}^2}{M_{\text{sp}}^2},
\ee
where $c_{\text{GB}}$ is some dimensionless coefficient that, up to logarithmic radiative corrections, directly arises from some higher-dimensional description. The consistency condition $|c_{\text{GB}}|\lesssim1$ implies an upper bound on the GB coefficient, which combined with \eqref{gammabound} gives  
\be\label{axiSCus}
\frac{\pi N}{6\alpha}\leq{\gamma}\lesssim {4\pi^2}\,\frac{M_{\text{\tiny P}}^2}{M_{\text{sp}}^2}.
\ee
A particular implication of this relation is an upper bound on the species scale, as pointed out in \cite{vandeHeisteeg:2023dlw}: 
\be\label{vafaSC} 
M_{\text{sp}}\lesssim M_{\gamma}\equiv \frac{2\pi M_{\text{\tiny P}}}{\sqrt{\gamma(s)}} \,.
\ee 
We emphasize the moduli-dependence of $M_{\gamma}$, which expectedly vanishes as $\alpha\to0$.  Note that \eqref{vafaSC} is defined in terms of a Wilson coefficient, and therefore in general the mass scale $M_{\gamma}$  has no direct  physical interpretation, which may for instance provide some sharper criterion to fix its normalization. Furthermore, in non-supersymmetric as well as our $\caln=1$ context,  the EFT coefficient $\gamma(s)$  receives scheme-dependent renormalization corrections (see Section \ref{sec:semitop}). It could therefore be useful to identify an alternative and more  physical proxy for the species scale. This is what we will do next.

We propose that an alternative upper bound on the species scale  can be derived from the physics of EFT strings.
Take the set of EFT string charges \eqref{EFTstrcharges}. As emphasized above,  if an elementary EFT charge ${\bf e}$ has scaling weight   $w_{{\bf e}}= 1$, then there exists an asymptotic regime, defined by the associated EFT string limit, in which this EFT string uplifts to a critical superstring at weak string coupling. Hence, according to the above definitions,  in this regime we can make the identifications $M^2_{\text{sp}}=M_{\rm t}^2=M^2_{\rm s}=2\pi\calt_{\bf e}$. In all the other cases, in which either $w_{{\bf e}}= 1$ but the theory is not in the corresponding asymptotic regime, or $w_{{\bf e}}\geq 2$, the EFT string does {\em not} uplift to a weakly coupled critical  superstring. Hence it should not be quantizable, and so its would-be excitation masses should be above the species scale, that is  $2\pi\calt_{\bf e}\geq M^2_{\text{sp}}$. This is what was emphasized in  \cite{Cota:2022yjw}, which compared  the species scale with the  EFT string tensions along the corresponding EFT string limits in F-theory models.

The above considerations motivate us to propose the following general upper bound 
\be\label{SPbound} 
M^2_{\text{sp}}\leq  M^2_{\calt}\,,
\ee
where we have introduced the {\em dominant EFT string scale}
\be\label{SPbounddef} 
M^2_{\calt}(\ell)\equiv\min\big\{2\pi\calt_{\bf e}(\ell) \,|\,{\bf e}\in \calc^{\text{\tiny EFT}}_{\rm S}\big\}\,,
\ee
with $\calt_{\bf e}(\ell)=\langle {\bm\ell},{\bf e}\rangle M^2_{\text{\tiny P}}$, as in \eqref{strten}. This is the bound anticipated in \eqref{boundINTRO}. The strict equality holds only when $w_{\bf e}=1$ and the saxions are in the asymptotic regime identified by the corresponding EFT string flow. We stress that the simple formula \eqref{strten} is fixed by $\caln=1$ supersymmetry and is thus {\em protected} against perturbative corrections. Hence, 
as a function of the dual saxions $\ell_i$, the dominant EFT string scale \eqref{SPbounddef} gives an explicit and robust upper bound on the species scale, valid for instance also if classical or quantum perturbative corrections  to  \eqref{KP(s)} and \eqref{calfP} cannot be neglected anymore (while non-perturbative corrections continue to be negligible). In other words, it enjoys a sort of non-renormalization theorem. By expressing \eqref{SPbounddef} in terms of the saxions $s^i$ by means of \eqref{elldef}, we would get a formula that formally depends on the K\"ahler potential $K$, which itself is sensitive to perturbative corrections. Despite that, such corrections get all ``resummed" in the $\ell_i$, as manifest in the dual saxionic formulation.

Note that the dominant EFT string scale \eqref{SPbounddef} depends on the dual saxions $\ell_i$ in such a way that, under an overall constant rescaling of the saxions, it behaves precisely as $M^2_{\gamma}$  in \eqref{vafaSC}. However, \eqref{SPbounddef} is fully determined by data available within the two-derivative EFT, namely the EFT string tensions $\calt_{\bf e}$. Once the K\"ahler potential and the saxionic cone are given, or analogously the dual saxions $\ell_i$ and the set of EFT string charges \eqref{EFTstrcharges}, \eqref{SPbounddef} is determined at each point of the perturbative region. It is  sufficient to restrict to the elementary generators of $\calc^{\text{\tiny EFT}}_{\rm S}$, compute the corresponding tensions and identify the lowest one. Of course, the charge corresponding to the lowest tension generically changes as we move in the saxionic domain. Hence $M^2_{\calt}$ is a continuous but possibly non-smooth function of the dual saxions. 

In Section \ref{sec:stringtheorymodels} we will verify the bound  \eqref{SPbound} in explicit string theory models and compare it to \eqref{vafaSC}. Other checks are provided in Appendix \ref{app:SCtests}. In fact, $M_{\calt}$ turns out to provide a good estimate of the species scale, not only when  $M_{\rm sp}=M_{\rm s}$ (in which case by construction $M_{\rm sp}=M_{\calt}$), but also more generically, at least for ``not-too-large" saxions $s^i$. Instead, a large hierarchy $M_{\text{sp}}\ll M_{\calt}$ can  occur in extreme limits in asymptotic field  space regions where the species scale is set by $M_{\text{\tiny QG}}$, a typical example being realized in the strong string coupling limit of M-theory.

We conclude this section by stressing that one of the basic assumptions that underlie our analysis is that supersymmetry is exact and in particular that no perturbative stabilizing mechanism for axions and saxions is present. Clearly, if supersymmetry gets broken at low energies a potential for the saxions is generically induced. In that case some of the results obtained using our formulation would be qualitatively wrong. For example one could not reliably identify the vacuum configuration of a realistic string theory model using Eq.\ \eqref{kinetic}.
Nevertheless, the considerations presented in our paper are short-distance in nature and, therefore, largely insensitive to IR deformations like supersymmetry breaking. To guarantee this we will restrict our attention to $\Lambda$'s satisfying
\be\label{IRcutoff}
\Lambda> M_{{\text{\tiny IR}}}
\ee
where  $M_{{\text{\tiny IR}}}$ is some physical IR mass scale below which the long-distance modifications of  \eqref{kinetic} can no longer be ignored.

\section{String theory models}
\label{sec:stringtheorymodels}

In order to make the general discussion of Sections \ref{sec:N=1models} and \ref{sec:regimesEFT} more concrete, we now describe two broad classes of string theory models, namely the F-theory and heterotic models in the large volume regime. These have the advantage that can be described quite easily in our general framework, and will allow us to provide a few explicit examples thereof to better illustrate and check our main points. 
As in \cite{Lanza:2021udy,Martucci:2022krl}, our general claims also apply to other string theory models or perturbative regimes -- e.g.\ type I, type IIA and M-theory -- which are however either very similar/dual to the heterotic and F-theory cases, or admit a less explicit EFT description. Hence, for concreteness and clarity, in Section \ref{sec:FIIB} we focus on the F-theory and in Section \ref{sec:het} on heterotic models. We will encounter M-theory models on $G_2$ manifolds in Section \ref{sec:WHst3}.

For clarity, we collect here the conventions we adopt on the relevant scales in string/M-theory. The ten-dimensional Ricci scalar for type I, IIA, IIB appears in the Einstein and string frame actions  as
\begin{align}\label{S-10-theory}
   \frac{2\pi}{l_{(10)}^8}\int R_{(10)}= \frac{2\pi}{l_{(10)}^8}\int e^{-2\phi}R^{(\rm s)}_{(10)}\,,
\end{align}
where $l_{(10)}=2\pi\sqrt{\alpha'}$ represents the ten-dimensional Planck length in the Einstein frame, and  the string length in the string frame.   The string and Einstein frame metrics are related by $\d s^2_{\rm (s)}=e^\frac{\phi}{2}\d s^2_{\text{\tiny(E)}}$. Concerning M-theory, the eleven-dimensional Einstein-Hilbert term is 
\begin{align}\label{S-M-theory}
    \frac{2\pi}{l_{(11)}^9}\int R_{(11)},
\end{align}
where $l_{(11)}$ is the Planck length in M-theory. 
Choosing 
\be\label{11to10} 
\d s_{11}^2=e^{-\frac16\phi}\d s_{10}^2+l_{(11)}^2e^{\frac43\phi}\d y^2\,,
\ee 
and compactifying on the interval  $y\in[0,1]$, \eqref{S-M-theory}  reduces to the Einstein frame action in \eqref{S-10-theory} with $l_{(10)}=l_{(11)}$.

In both cases,  the four-dimensional metric is embedded in the higher-dimensional one according to the ansatz
  \be\label{lineelement}
  \d s^2_{d}=e^{2A}\d s^2_4+\d s^2_{X}. 
  \ee
The Weyl rescaling factor $e^{2A}=M^2_{\text{\tiny P}}l_{(d)}^2/(4\pi V_X)$, where $V_X$ denotes the volume of the $(d-4)$-dimensional compactified space in $l_{(d)}$ units, is necessary to identify $\d s^2_4$ with the four-dimensional Einstein frame metric. Explicit expressions of this quantity for our models will be provided below. Note that the appropriate dimension $d$ generically depends on the saxions.

According to Appendix \ref{app:2pis}, the strong coupling scales for ten-dimensional string theory and M-theory are, respectively, 
\be\label{strPS1011} 
\widehat M_{(10)}=\frac{(2\pi)^\frac34}{l_{(10)}},\quad\quad \widehat 
 M_{(11)}=\frac{(2\pi)^{\frac23}}{l_{(11)}}\,.
\ee
The quantum gravity scale introduced in Section \ref{sec:UVcutoff} is then given by 
 $M_{\text{\tiny{QG}}}=e^A\widehat M_{(d)}$.

\subsection{F-theory/type IIB orientifold models}
\label{sec:FIIB}

An important large class of examples is provided by the F-theory compactifications  -- see e.g.\ \cite{Denef:2008wq,Weigand:2018rez} for reviews. An F-theory  model corresponds to a type IIB compactification on a K\"ahler space $X$ in presence of 7-branes. The space $X$ can be regarded as the base of an elliptically  fibered Calabi-Yau four-fold, whose fiber's complex structure can be identified with the type IIB axio-dilaton. In particular, this requires the base $X$ to have an effective anti-canonical divisor $\overline K_X$.\footnote{We adopt the quite common usage of denoting holomorphic line bundles and corresponding  divisors by the same symbol.} In the following we will for simplicity assume that the elliptically  fibered Calabi-Yau four-fold  has vanishing third Betti number,  so to avoid technical complications associated with  moduli 
 of the M-theory gauge three-form.

These models admit a natural perturbative regime corresponding to the large volume limit. Let us pick a basis of divisors $D^a\in H_4(X,\mathbb{Z})$ of $X$ and a dual  basis of two-cycles $\Sigma_a\in H_2(X,\mathbb{Z})$, such that $D^a\cdot \Sigma_b=\delta^a_b$, $a=1,\ldots,b_2(X)$. The K\"ahler moduli $v_a$ are obtained by expanding the (Einstein frame) K\"ahler form $J$ of $X$ in the Poincar\'e dual basis $[D^a]\in H^2(X,\mathbb{Z})$. Keeping Poincar\'e duality implicit, we can write 
\be\label{JexpF} 
J=v_a D^a\quad~~~~\Leftrightarrow\quad~~~~ v_a=\int_{\Sigma_a}J=\Sigma_a\cdot J\,. 
\ee 
The  corresponding saxions $s^a$ are then defined as follows: 
\be 
s^a=\frac12\kappa^{abc}v_bv_c\,,
\ee
where we have introduced the triple intersection numbers $\kappa^{abc}=D^a\cdot D^b\cdot D^c$. Hence in this case $N=b_2(X)$.

As discussed in \cite{Lanza:2021udy}, one can identify the saxionic cone with the cone ${\rm Mov}_1(X)$ generated by {\em movable} curves (see e.g.\ \cite{fulger2016zariski}). We can then write
\be\label{sF}
{\bm s}=s^a\Sigma_a\in  \Delta_{\text{\tiny K}}\simeq {\rm Mov}_1(X)\,.
\ee
Note that the string charge vectors ${\bf e}$ can be identified with effective curves $\Sigma_{\bf e}=e^a\Sigma_a$ and, in particular, the EFT string charges correspond to movable curves
\be 
\calc^{\text{\tiny  EFT}}_{\rm S}\simeq   {\rm Mov}_1(X)_{\mathbb{Z}}\,.
\ee
Physically, EFT strings are realized by  D3-branes wrapping  movable curves. The corresponding BPS instantons are instead realized by Euclidean branes wrapping effective divisors $D\in {\rm Eff}^1(X)_{\mathbb{Z}}$, so that we can make the identification $\calc_{\rm I}\simeq {\rm Eff}^1(X)_{\mathbb{Z}}$.

The constants $\tilde C_a$ defining $\gamma(s)$ as in \eqref{gammas} admit a nice geometrical interpretation \cite{Martucci:2022krl}:
\be\label{FtildeC} 
\tilde C_a=6 \overline K_X\cdot \Sigma_a\,.
\ee
In particular, the pairing appearing in \eqref{Ctildebound} corresponds to the intersection number 
\be\label{tildeCeF} 
\langle\tilde{\bf C},{\bf e}\rangle=6 \overline K_X\cdot \Sigma_{\bf e}\,.
\ee
Recalling that $\overline K_X$ is an effective divisor, the bound \eqref{Ctildebound} is always satisfied, since movable curves can be precisely characterized as those curves that have non-negative intersection with all effective divisors \cite{boucksom2013pseudo}. In order to test the  bound \eqref{gammabound}, which is expected to hold up to subleading corrections in $1/N\ll 1$, let us focus on the large class of models with  toric $X$ --  see for instance  \cite{cox2011toric}. Then the anti-canonical divisor is given by
\be\label{toricantiK} 
\overline K_X=\sum_{I\in \text{toric div.}}\cald_I\,,
\ee
where the sum is over the set of prime toric divisors $\cald_I$, $I=1,\ldots, N+3$. All these divisors are effective, and in fact generate the whole cone of effective divisors. So any movable curve $\Sigma_{\bf e}$ has strictly positive intersection number with at least one toric divisor $\cald_I$, and then
\be
\overline K_X\cdot\Sigma_{\bf e}\geq 1\,.
\ee
Combined with \eqref{toricantiK}, this implies that
\be 
\langle\tilde{\bf C},{\bf e}\rangle\geq 6\,,\quad~~~~~~~~~~~~\forall {\bf e}\in\calc^{\text{\tiny EFT}}_{\rm S}\,.
\ee
We then see that, in this large class of models, the condition \eqref{tildeCvA} is indeed satisfied, strengthened by a factor of 6, and then  the bound \eqref{gammabound} is realized  in the  stronger form:
\be\label{gammaboundF} 
\gamma(s)\geq \frac{N\pi}{\alpha}\,.
\ee

The effective theory is more easily described in the dual saxionic formulation 
\be\label{ellJrel}
\ell_a=\frac{3v_a}{\kappa(J,J,J)}\,,
\ee
where $\kappa(J,J,J)\equiv J\cdot J\cdot J=\kappa^{abc}v_av_bv_c$. The dual saxionic cone $\calp_{\text{\tiny K}}$ can  be identified with an ``extended'' K\"ahler cone $\calk(X)_{\rm ext}$ obtained by gluing different spaces connected by  flop transitions, in which curves collapse or blow-up:
\be\label{calpdec}
\calp_{\text{\tiny K}}=\calk_{\rm ext}(X)=\bigcup_{X'\sim X} \overline{\calk(X')}\,. 
\ee
Here $X'\sim X$ means that $X'$ can be obtained from $X$ by a chain of flops (which may also be  trivial, corresponding to $X'=X$).  Hence ${\bm \ell}\in \calp_{\text{\tiny K}}$ if there exists one chamber of $\calk_{\rm ext}(X)$, associated with a  compactification space $X'\sim X$, in which ${\bm \ell}=\ell_a\,D^a$ is a {\em nef} $\mathbb{R}$-divisor, that is ${\bm \ell}\in \overline{\calk(X')}$.\footnote{
\label{foot:movable} In the following we will often focus on spaces $X$ which are toric or orientifold quotients of Calabi-Yau three-folds. In these cases  $\calk(X)_{\rm ext}$ can be identified with the space of the so-called movable divisors. Hence we can write ${\bm \ell}=\ell_a\,D^a\in \calp_{\text{\tiny K}}\simeq {\rm Mov}^1(X)$.
This identification can actually hold more generically  -- see \cite{Lanza:2021udy} for more details.} 

At large volume, the kinetic potential $\calf(\ell)$ takes the form \eqref{calfP}:
\be\label{FFK} 
\calf_{\text{\tiny K}}(\ell)=\log\kappa({\bm\ell},{\bm\ell},{\bm\ell})\,.
\ee
Hence $\tilde P(\ell)=\kappa({\bm\ell},{\bm\ell},{\bm\ell})$, which is clearly homogeneous  as in \eqref{tildePhom}, with $n=3$.

If one can take Sen's orientifold limit, the space $X$ can be regarded as the $\mathbb{Z}_2$-orientifold quotient of a Calabi-Yau three-fold $\hat X$. A new saxion  $ \hat s\equiv e^{-\phi}$ appears, detected by $D(-1)$-instantons,  where $\phi$ is the standard type IIB dilaton, so that we now have $N=b_2(X)+1$. The corresponding dual saxion is
\be\label{dualdilaton} 
\hat \ell=\frac12 e^\phi\,.
\ee
 In the perturbative regime described by the dual saxions $\ell_i=(\hat\ell,\ell_a)$, the leading contribution to the K\"ahler potential is  given by
\be\label{Fcalf} 
\calf=\log \hat\ell+\calf_{\text{\tiny K}}(\ell)=\log \hat \ell+\log \kappa({\bm\ell},{\bm\ell},{\bm\ell})
\ee
and can then be written as in \eqref{calfP} with $\tilde P=\hat\ell\, \kappa({\bm\ell},{\bm\ell},{\bm\ell})$, which has homogeneity $n=4$.\footnote{In fact,  the saxionic and  dual saxionic cones are expected to receive corrections coming from higher derivative terms.  This type of effect has been discussed in some detail for heterotic models in \cite{Martucci:2022krl} and we will  encounter it in subsection \ref{sec:het} -- see e.g.\ \eqref{hets0}. In particular, if we choose $\ell_0$ so that $\ell_0M^2_{\rm P}$  gives the tension of the lightest D7-string, we generically have $\ell_0=\hat\ell+c^a\ell_a$, where 
$c^a\in\mathbb{Q}$ accounts for possible world-volume curvature/bundle corrections. This means that  in \eqref{Fcalf} we should set $\hat\ell=\ell_0-c^a\ell_a$, which induces also a shift $s^a\rightarrow \hat s^a=s^a+c^a s^0$ in the K\"ahler potential. These subtleties will be  studied in more detail elsewhere, but since they are not crucial to our purposes for simplicity we will just ignore them, tacitly keeping  them in mind.}

The conversion factor  from ten- to four-dimensional scales, appearing in \eqref{lineelement}, is given by
\be\label{Fconv} 
e^{2A}=\frac{M^2_{\text{\tiny P}}l^2_{(10)}}{4\pi V(X)}=\frac{M^2_{\text{\tiny P}}l^2_{(10)}}{2\pi}\sqrt\frac{\kappa({\bm\ell},{\bm\ell},{\bm\ell})}3\,,
\ee
where $V(X)=\kappa(J,J,J)/6$ is the compactification volume in $l_{(10)}$-units. 
By applying this  conversion factor to $\widehat M_{(10)}$ in \eqref{strPS1011} we get 
\be\label{FtheorySC} 
M^2_{\text{\tiny QG}}=e^{2A}\widehat M_{(10)}^2=\sqrt{\frac{2\pi\kappa({\bm\ell},{\bm\ell},{\bm\ell})}{3}}\, M^2_{\text{\tiny P}}\,.
\ee
It is interesting to explicitly check how the powers of $2\pi$'s precisely combine so that $M^2_{\text{\tiny QG}}$ is  the square of $2\pi M_{\text{\tiny P}}$ --- the scale a naive low-energy observer would identify as the quantum gravity scale --- times a non-trivial suppression $\sim (\ell/2\pi)^{3/2}$ controlled by the ``loop parameter" \eqref{saxions=couplings}.

\subsubsection{Model 1: ${\mathbb{P}}^3$}
\label{sec:FtheoryP3}
For illustrative purposes, it is useful to describe a couple of simple explicit models (though with a small number of (s)axions) and their relevant energy scales. The first and easiest example is  obtained by choosing $X=\mathbb{P}^3$.

In this case, the set of effective divisors $\operatorname{Eff}^1(X)_{\mathbb{Z}}$ is spanned by a single element, the hyperplane divisor $D\equiv H$, which has triple self-intersection $\kappa(D,D,D)=1$. Hence, in the large volume perturbative regime, ${\bm\ell}=\ell D$, the dual saxionic cone is just given by $\ell\geq 0$ (including the degenerate boundaries) and \eqref{FFK} reduces to
\be 
\calf_{\text{\tiny K}}=3\log\ell \,.
\ee
The corresponding saxionic cone is spanned by the curve $\Sigma\equiv D\cdot D$. The only saxion $s$ of this model encodes the volume of the hyperplane divisor, has  K\"ahler potential 
$K = -3 \operatorname{log} s$
and is related to the  dual saxion by $\ell = \frac{3}{2 s}$, see \eqref{elldef}. The saxionic cone is simply given by $s \geq 0$. Moreover the $\alpha$-saxionic convex hull is just $\hat\Delta_\alpha=\{s\geq 1/\alpha\}$, and then $\hat\calp_\alpha=\{\ell\leq \frac32\alpha\}$. 

The hyperplane divisor $D$ generates the set of BPS instanton charges $\calc_{\rm I}=\{{\bf q}=q D|q\in\mathbb{Z}_{\geq 0}\}$, while 
the  curve $\Sigma$ defines the elementary EFT string charge, which  generates the set of EFT string charges $\calc^{\text{\tiny EFT}}_{\rm S}=\{{\bf e}=e\Sigma|e\in\mathbb{Z}_{\geq 0}\}$. All these EFT string charges have scaling weight $w=2$ \cite{Lanza:2021udy}, and  the tension of the elementary string is given by $\calt=M^2_{\text{\tiny P}}\,\ell $. The anti-canonical divisor in this setting is just $\overline K_X=4 H$ and, by using \eqref{gammas} and \eqref{FtildeC}, it yields
\be 
\gamma(s) = 4\pi s \, .
\ee 
This is manifestly positive in the saxionic cone and we have also $\gamma(s)|_{\hat\Delta_\alpha}\geq \frac{4\pi}{\alpha}$, stricter than \eqref{gammabound} with $N=1$ by a factor of 24.

Let us now discuss the species scale. The upper bounds given by \eqref{vafaSC} and \eqref{SPbounddef} become 
\be 
M^2_{\gamma}=\frac{\pi M^2_{\text{\tiny P}}}{s}\,,\quad\quad M^2_{\calt}=2\pi M^2_{\text{\tiny P}}\ell=\frac{3\pi M^2_{\text{\tiny P}}}{s}\,.
\ee
We see that $M_{\gamma}\sim M_{\calt}$ in the entire perturbative domain. Because the dominant EFT string scale $M_{\calt}$ is associated to a $w=2$ string and there are no $w=1$ EFT strings (and, correspondingly, no weak string coupling), we expect $M_{\rm sp}=M_{\text{\tiny QG}}< M_{\calt}$, realizing the bound \eqref{SPbound} in its  strict form. We can verify this by using the  quantum gravity scale \eqref{FtheorySC}, which reads
\be 
M^2_{\text{\tiny QG}}=\sqrt{\frac{2\pi\ell^3}{3}}\,M^2_{\text{\tiny P}}
=\frac32\sqrt\frac{\pi}{s^3}\,M^2_{\text{\tiny P}}\,. 
\ee
It follows that
$M_{\calt}/M_{\rm sp}=(4\pi s)^{\frac14}\geq (4\pi/ \alpha)^{\frac14}$, and the strict form of the bound \eqref{SPbound} is always satisfied in the perturbative regime $\alpha/2\pi\ll1$. Nevertheless, $M_{\calt}$ provides a good proxy for $M_{\rm sp}$, say up to an $\calo(10)$ factor, for $s\lesssim 10^3$.

\subsubsection{Model 2: $\mathbb{P}^1$ fibration over $\mathbb{P}^2$}
\label{sec:Ftheory1}

In our context this model has already been discussed  
 in \cite{Lanza:2021udy}, which we can then follow. The internal space $X$ is a $\mathbb{P}^1$ fibration over  $\mathbb{P}^2$, and the fibration is specified by the integer $p\geq 0$. 

The cone $\operatorname{Eff}^1(X)_{\mathbb{Z}}$ of effective divisors, which can be identified with the cone of BPS instanton charges $\calc_{\rm I}$,  is simplicial and is generated by two effective divisors $E^1,E^2$: $E^1$ is the divisor obtained by restricting the  $\mathbb{P}^1$ fibration over  $\mathbb{P}^1\subset\mathbb{P}^2$, while $E^2$ corresponds to a global section of the $\mathbb{P}^1$ fibration. One can then identify a basis of nef divisors $D^1=E^1$ and $D^2=E^2+pE^1$, which generate the K\"ahler cone: $J=v_1 D^1+v_2 D^2$, with $v_{1,2}>0$. The triple intersection numbers are given by the coefficients of the formal object $\cali(X)=(D^1)^2D^2+pD^1(D^2)^2+p^2(D^2)^3$. 
Hence, by using the expansion  ${\bm \ell}=\ell_aD^a$ the kinetic potential \eqref{FFK} becomes
\be\label{m1FK}
\calf_{\text{\tiny K}}=\log\kappa\left({\bm\ell},{\bm\ell},{\bm\ell}\right)=\log\left(3\ell_1^2\ell_2+3p\ell_1\ell_2^2+p^2\ell_2^3\right)
\ee
In this model the dual saxionic cone  coincides with the closure of  the K\"ahler cone: $\calp_{\text{\tiny K}}=\{{\bm\ell}=\ell_1D^1+\ell_2D^2|\ell_1\geq  0,\ell_2\geq 0 \}$. From \eqref{elldef2} one can obtain the corresponding saxions:
\be\label{mod2sl}
s^1=\frac{6 \ell_1 + 3 p \ell_2 }{6 \ell_1^2 + 6p \ell_1 \ell_2  + 2p^2\ell_2^2 }\quad,\quad s^2=\frac{3 (\ell_1 + p\ell_2 )^2 }{6\ell_1^2\ell_2+6p\ell_1\ell_2^2+2p^2\ell_2^3 }\,.
\ee

The (Mori) cone of effective curves is generated by $\Sigma_1=E^1\cdot E^2$ and $\Sigma_2=(E^1)^2$, which are dual to the nef divisors $D^{a}$: $D^a\cdot\Sigma_b=\delta^a_b$. The cone of movable curves is instead generated by $\hat\Sigma_1=D^1\cdot D^2=\Sigma_1+p\Sigma_2$ and $\hat\Sigma_2=(D^1)^2=\Sigma_2$, which are dual to the effective divisors $E^a$: $E^a\cdot\hat\Sigma_b=\delta^a_b$. Hence ${\bm s}=s^1\Sigma_1+s^2\Sigma_2=s^1\hat\Sigma_1+ (s^2-ps^1)\hat\Sigma_2$, and  the saxionic cone is $\Delta=\{{\bm s}=s^a\Sigma_a|s^1\geq 0\,,\ s^2\geq ps^1\}$. One can also invert the relation between saxions and dual saxions:
\be\label{sellinv1} 
\ell_1=\frac{3p\sqrt{s^2-ps^1}}{2\left[(s^2)^{\frac32}-(s^2-ps^1)^{\frac32}\right]}\quad,\quad \ell_2=\frac{3\left(\sqrt{s^2}-\sqrt{s^2-ps^1}\right)}{2\left[(s^2)^{\frac32}-(s^2-ps^1)^{\frac32}\right]}
\ee

\begin{figure}[!htb]
\centering
\hfill
\begin{subfigure}[t]{.49\textwidth}
  \centering
\includegraphics[width=\textwidth]{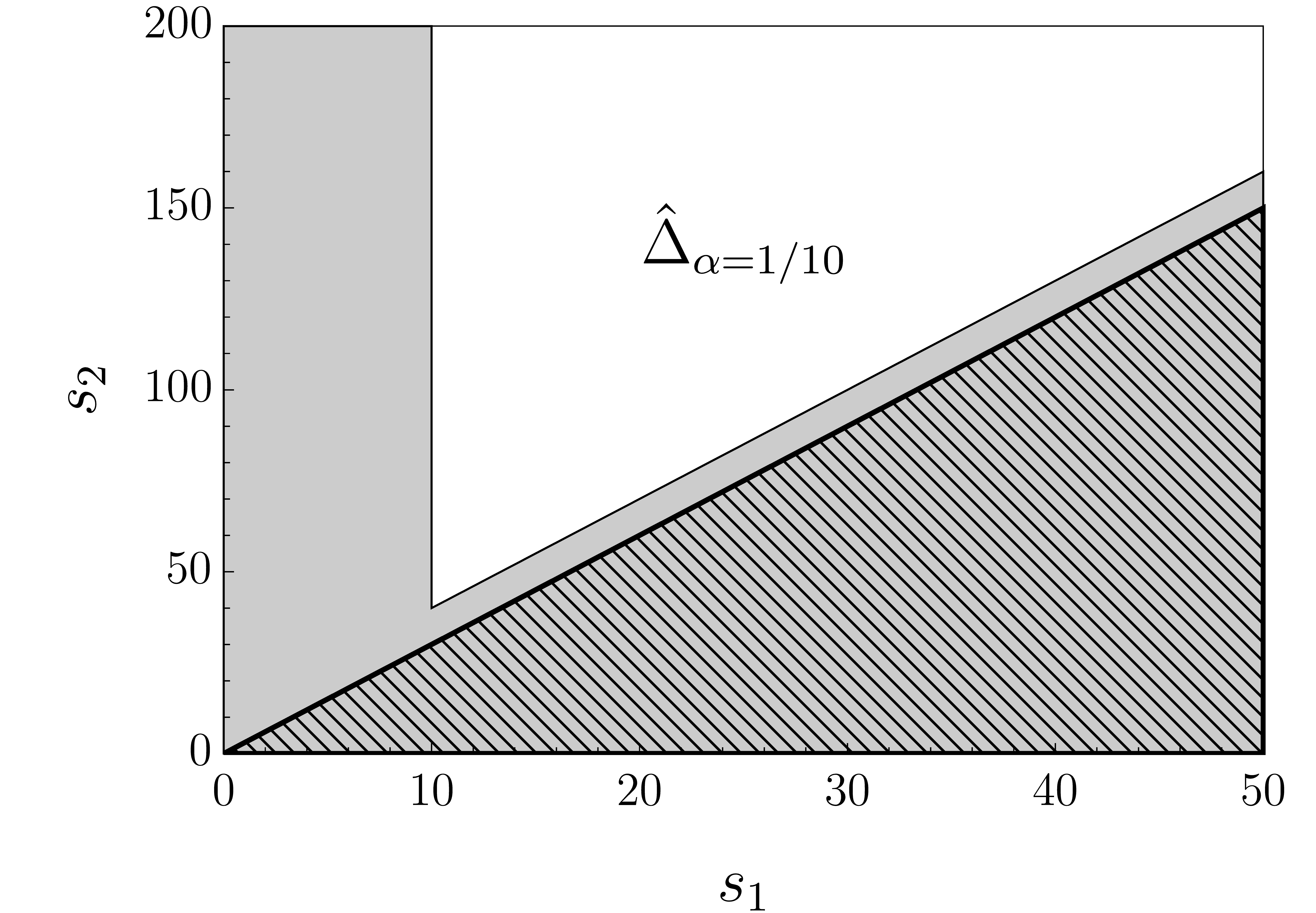}
  \caption{\small Saxionic convex hull $\hat{\Delta}_\alpha$.}
  \label{fig:model1DomainDelta}
\end{subfigure}%
\hfill
\begin{subfigure}[t]{.49\textwidth}
  \centering
\includegraphics[width=\textwidth]{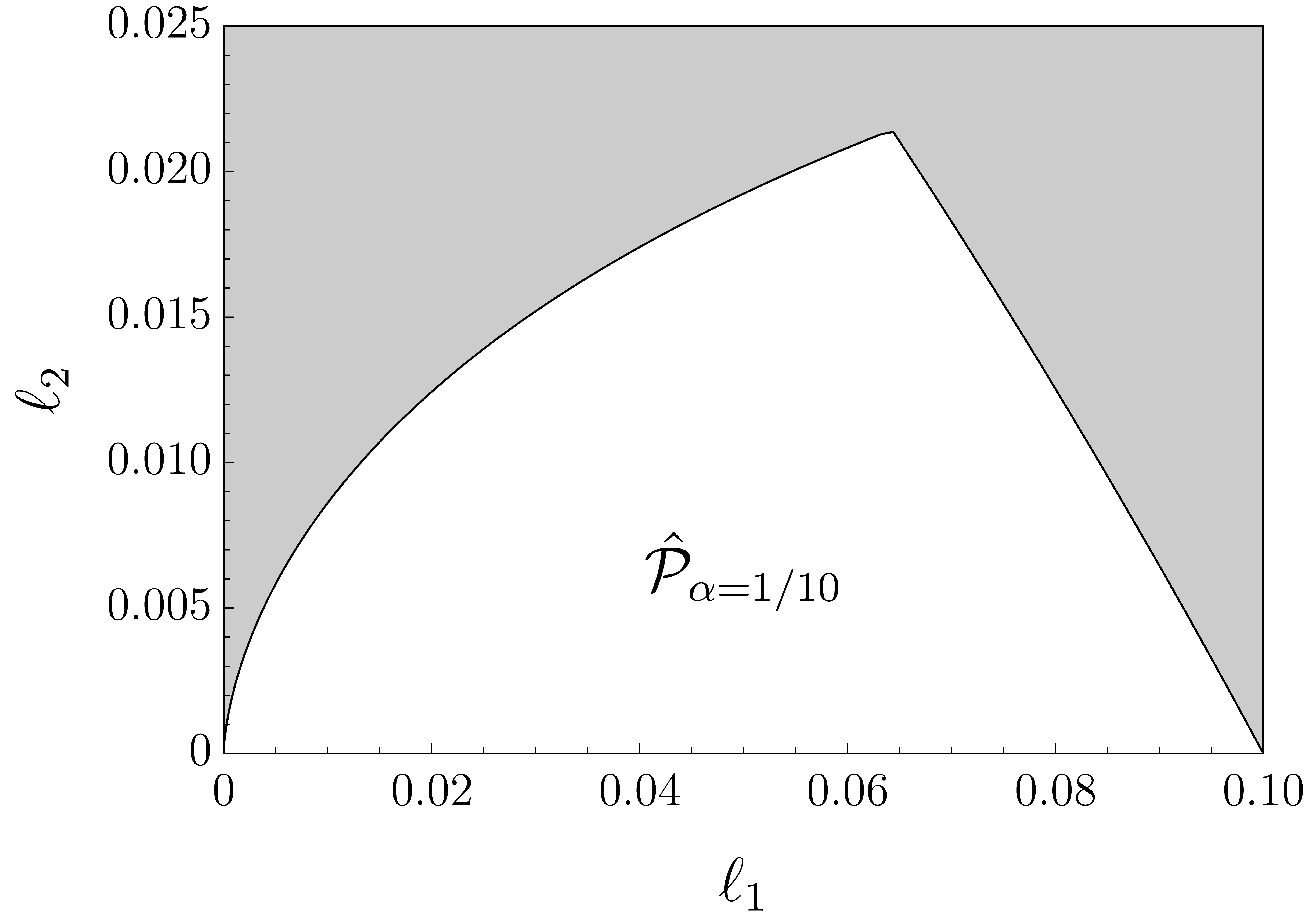}
  \caption{\small Dual saxionic convex hull $\hat{\mathcal{P}}_\alpha$.}
  \label{fig:model1domainP}
\end{subfigure}
\hfill
\caption{\small Saxionic convex hull $\hat{\Delta}_\alpha$ and dual saxionic convex hull $\hat{\mathcal{P}}_\alpha$ for the F-theory model 2. The plot has been drawn with the reference values $\alpha=1/10$ and $p=3$. The hatched area in figure \ref{fig:model1DomainDelta} is outside the saxionic cone $\Delta$, which comprise the gray and white regions.}
\label{fig:model1domainDeltaP}
\end{figure}

As in our general discussion, we can characterize the boundaries of $\calp_{\text{\tiny K}}$ in terms of tensionless strings. 
The set of EFT string charges 
\be\label{model2EFTs} 
\calc^{\text{\tiny EFT}}_{\rm S}=\{{\bf e}=e^1\Sigma_1+e^2\Sigma_2|(e^1,e^2)\in\mathbb{Z}^2\,,\ e^1\geq 0\,,\ e^2\geq pe^1\}\,
\ee 
is generated by ${\bf e}_{(1)}=\hat\Sigma_1=\Sigma_1+p\Sigma_2$ and ${\bf e}_{(2)}=\hat\Sigma_2=\Sigma_2$, which have tensions $\calt_{(1)}=M^2_{\text{\tiny P}}(\ell_1+p\ell_2)$ and $\calt_{(2)}=M^2_{\text{\tiny P}}\ell_2$. We notice that $\calt_{(2)}$ vanishes at $\ell_2=0$, while  $\calt_{(1)}$ vanishes at the tip $\ell_1=\ell_2=0$. These are infinite distance boundary components of $\calp_{\text{\tiny K}}$. On the other hand, on the boundary component $\ell_1=0$ no EFT string tension vanishes. This is instead characterized by the vanishing of the tension  $M^2_{\text{\tiny P}}\ell_1$ associated with the non-EFT string charge $\Sigma_1$, which together with $\Sigma_2$ generates the set of BPS charges  $\calc_{\rm S}$. This implies that, even if the saxionic convex hull is simply given by $\hat\Delta_\alpha=\{s^1\geq \frac1{\alpha},s^2-ps^1\geq \frac1\alpha\}$, the corresponding dual saxionic image  $\hat\calp_\alpha$ is more complicated -- see figure \ref{fig:model1domainDeltaP}.

Remember that the GB coefficient is determined by the anti-canonical divisor.
In the present examples, the latter is given by 
\be
\begin{aligned}
\overline K_X&=(3-p)D^1+2D^2=(3+p)E^1+2E^2\,.
\end{aligned}
\ee
From \eqref{gammas} and \eqref{FtildeC} we then get
\be\label{gammasF1} 
\gamma(s)=\pi\left[(3-p)s^1+2s^2\right]\,, 
\ee
which is positive since $s^2\geq ps^1$.
Furthermore, we see that 
$\gamma(s)|_{\hat\Delta_\alpha}\geq \frac{(5+p)\pi}{\alpha}$, which  is stronger than \eqref{gammabound} with $N=2$.
It is for instance sufficient to take $\alpha\leq \frac1{10}$ and $p\geq 1$ to get $\gamma(s)|_{\hat\Delta_\alpha}>188$. 

Now let us turn our attention to the relevant energy scales at play. It is easy to check that $K\simeq -\log\sigma$ asymptotically along the EFT string flow associated with ${\bf e}_{(2)}$, while $K\simeq -3\log\sigma$ along the EFT string flow associated with ${\bf e}_{(1)}$. This is consistent with the fact that only ${\bf e}_{(2)}$ has scaling weight $w=1$, while ${\bf e}_{(1)}$ has $w=2$ \cite{Lanza:2021udy}.  Indeed, the string obtained by wrapping a D3 on $\hat\Sigma_2$ is dual to a fundamental heterotic superstring via F-theory/heterotic duality -- see Appendix \ref{app:hetFtheory} for a more general discussion.
We can then distinguish different  regimes set by the two elementary EFT string tensions $\calt_{{(1)}}=M^2_{\text{\tiny P}}(\ell_1+p\ell_2)$ and $\calt_{(2)}=M^2_{\text{\tiny P}}\ell_2$. Let us start assuming that $p>0$, the particular case $p=0$ will be discussed at the end. 

For $p>0$ we have $\calt_{(2)}\leq \calt_{{(1)}}$ and for any value of the saxions the dominant EFT string scale \eqref{SPbounddef} is given by
\be\label{Fboundmax} 
M^2_{\calt}
=2\pi
M^2_{\text{\tiny P}}\ell_2\,.
\ee
Furthermore, the condition \eqref{TMP} requires that  $\ell_2<2\pi/p$, and so $M_{\calt}<2\pi M_{\text{\tiny P}}/\sqrt{p}$. We can distinguish two regimes, namely $\calt_{(2)}\ll \calt_{{(1)}}$ or $\calt_{(2)}\simeq \calt_{{(1)}}$. 
If $\ell_2\ll\ell_1$ we are in the first regime, where $\calt_{(2)}$ corresponds to the tension of a dual  weakly coupled critical string. Here \eqref{SPbound} is actually saturated.  The second regime is defined by $\calt_{(2)}\simeq\calt_{(1)}$. Since $\calt_{(1)}/\calt_{(2)}=p+\ell_1/\ell_2> p$, this can be reached only if $p\sim \calo(1)$ and $\ell_1/\ell_2\lesssim 1$.  In this second regime there should not exist a controlled dual weakly coupled string theory description. The species scale should then be identified with the quantum gravity scale \eqref{FtheorySC}.  By combining \eqref{FtheorySC} and \eqref{m1FK} we  get  
\be 
M^2_{\text{\tiny QG}}= \sqrt{2\pi\left(\ell_1^2\ell_2+p\ell_1\ell_2^2+\frac{1}{3}p^2\ell_2^3\right)}\, M^2_{\text{\tiny P}}\lesssim\sqrt{2\pi\ell_2^3}\, M^2_{\text{\tiny P}}=\sqrt{\frac{\ell_2}{2\pi}}\,M^2_{\calt}\,,
\ee 
where in the second step we have used $\ell_1\lesssim\ell_2<\frac{2\pi}{p}$ and $p\sim \calo(1)$, and we have neglected an $\calo(1)$ overall constant. Consistently with the bound \eqref{SPbound} we find that $M_{\text{\tiny QG}}\lesssim M_{\calt}$ whereas $M_{\text{\tiny QG}}\ll M_{\calt}$ for $\ell_2\ll 2\pi$, that is, far away from the tip of the saxionic domain. Other regimes can  be better studied  through the dual  heterotic M-theory description, which will be discussed in subsection \ref{sec:hetscales} and will confirm that $M^2_{\text{sp}}$ is still bounded by \eqref{Fboundmax}.

It is instructive to also discuss the bound  \eqref{vafaSC} for this model. Recalling \eqref{gammasF1} and \eqref{mod2sl} we get 
\be\label{SCboundex1} 
M^2_{\gamma}
=\frac{4\pi M^2_{\text{\tiny P}}}{(3-p)s^1+2s^2}=\frac{8\pi\ell_2(3\ell_1^2+3p\ell_1\ell_2+p^2\ell_2^2)}{6\ell_1^2+(3+p)(6\ell_1\ell_2+3p\ell_2^2)}M^2_{\text{\tiny P}}.
\ee
A comparison between the two mass scales \eqref{Fboundmax} and \eqref{SCboundex1} gives:
 \be
\frac{M^2_{\calt}}{M^2_{\gamma}}=\frac12\left[1+\frac{18\ell_1\ell_2+(9p+p^2)\ell_2^2}{4(3\ell_1^2+3p\ell_1\ell_2+p^2\ell_2^2)}\right]\,.
 \ee
Since we are assuming $p>0$, we have 
\be 
\frac12 \leq\ \frac{M^2_{\calt}}{M^2_{\gamma}} \leq   \frac58 + \frac{9}{8p} \leq \frac74\,,
\ee 
where the two extrema correspond to $\ell_2=0$ and $\ell_1=0$, respectively. This shows that $M_{\calt}$ and $M_{\gamma}$ are always of the same order, and then the upper bound \eqref{vafaSC} is satisfied too.

The case $p=0$, which we ignored so far, is characterized by $M^2_{\calt}=2\pi M^2_{\text{\tiny P}}\min\{\ell_1,\ell_2\}$ and $M^2_{\gamma}=4\pi\frac{\ell_1\ell_2}{\ell_1+3\ell_2}M^2_{\text{\tiny P}}$, so that $\frac12\leq M^2_{\calt}/M^2_{\gamma}\leq \frac32$. Again, the two upper bounds on the species scale parametrically agree. Invoking \eqref{TMP} we require $\ell_1,\ell_2\leq 2\pi$ and obtain the inequality $M^2_{\calt}\leq M^2_{\text{\tiny QG}}$ for $\ell_2\leq\ell^2_1/2\pi$, consistently with the identification $M^2_{\text{sp}}=M^2_{\calt}=M^2_{\text{\tiny P}}\ell_2$, while $M^2_{\text{sp}}=M^2_{\text{\tiny QG}}\leq M^2_{\calt}$ for $\ell_2\geq\ell^2_1/2\pi$.

A similar discussion can be carried out for the models where $X$ is $\mathbb{P}^1$ fibration over an Hirzebruch surface $\mathbb{F}_p$, and is presented in Appendix \ref{app:Ftheory2}.

\subsection{Heterotic models}
\label{sec:het}

Our second class of models is given by $E_8\times E_8$ heterotic compactifications on Calabi-Yau spaces, and their M-theory counterpart, at large volume. (The $SO(32)$ case is completely analogous.) As discussed in \cite{Martucci:2022krl}, the relevant saxionic  cone is affected by ten- and eleven-dimensional higher derivative terms. Here we summarize only the necessary information.

We consider a perturbative regime associated to $N=b_2(X)+1$  saxions  $s^i=(s^0,s^a)$, which include the K\"ahler moduli $s^a$ of the Calabi-Yau compactification space $X$. These are obtained  by expanding the string frame K\"ahler form $J=s^a D_a$ in (a basis Poincar\'e dual to) a basis of divisors $D_a$, $a=1,\ldots,b_2(X)$. The remaining saxion $s^0$ combines the dilaton and the K\"ahler moduli:
\be\label{hets0} 
s^0=\frac16e^{-2\phi}\kappa_{abc}s^as^bs^c+\frac12p_as^a\,.
\ee
where $\kappa_{abc}\equiv D_a\cdot D_b\cdot D_c$ and 
\be\label{padef} 
p_a\equiv -\int_{D_a}\left[\lambda(E_2)-\frac12 c_2(X)\right]\in\mathbb{Z}\,,
\ee
where  $E_1$ and $E_2$ denote the two $E_8$ internal bundles, and  
\be 
\lambda(E)\equiv-\frac1{16\pi^2}\tr(F\wedge F)\,.
\ee
The tadpole cancellation condition imposes the topological constraint $\lambda(E_1)+\lambda(E_2)=c_2(X)$. Let us also introduce the integer  
\be 
n_a\equiv \frac12\int_{D_a}c_2(X)\,.
\ee
Notice that $n_as^a=\frac12\int_X c_2(X)\wedge J\geq  0$  \cite{miyaoka1985chern} and that, by supersymmetry, the internal $E_8$ bundles must satisfy  $\int_X \lambda(E_{1,2})\wedge J\geq  0$. Combining these positivity conditions with \eqref{padef} and the tadpole condition, one gets 
\be\label{upperpasa} 
\left|p_as^a\right|\leq n_as^a\,.
\ee
One  could also include NS5/M5-branes wrapping internal curves (see \cite{Martucci:2022krl}), but  for simplicity here we will not do that. 

The saxionic cone is given by
\be\label{hetDelta} 
\Delta=\left\{(s^0,s^a)\ \big|\  {\bm s}=s^aD_a\in\overline{\calk(X)}\,,\ s^0\geq 0\,,\ s^0\geq p_as^a\right\}\,,
\ee
and the GB coupling \eqref{gammas} takes the form
\be\label{GBhet} 
\gamma(s)=\pi\left(2s^0-p_as^a+\frac16n_as^a\right)\,.
\ee

Let us also recall that in the M-theory realization \cite{Horava:1995qa,Horava:1996ma}, the Calabi-Yau $X$ three-fold is fibered over an interval, representing the 11-th M-theory direction. Then  $s^0$ and $s^0-p_as^a$ can be interpreted as the volume of $X$ at the two endpoints of this interval \cite{Martucci:2022krl}. For simplicity we will henceforth  assume that $p_as^a\geq 0$. (By \eqref{upperpasa}, we can actually have $0\leq p_as^a\leq n_as^a$.) In this case  the saxionic cone \eqref{hetDelta} reduces to 
\be\label{npDelta} 
\Delta=\left\{(s^0,{\bm s})\ \big|\  {\bm s}=s^aD_a\in\overline{\calk(X)}\,,\  s^0\geq p_as^a\right\}\,.
\ee

The K\"ahler potential can be in principle obtained by dimensionally reducing the ten- and eleven-dimensional heterotic (M-)theory. One must  take into account the corrections discussed in \cite{Witten:1996mz}  (see also \cite{Blumenhagen:2006ux}). These affect the choice of saxionic variables and induce the tilting of the saxionic cone \eqref{npDelta} due to the constants $p_a$, which encode the effect of ten- and eleven-dimensional higher derivative terms. Moreover, these terms may induce additional corrections to the K\"ahler potential. Fully determining these corrections is beyond the scope of the present paper, and so we will content ourselves with considering contributions coming from the leading heterotic M-theory terms \cite{Horava:1996ma},  while taking into account the tilting of the saxionic cone \eqref{npDelta} and possible additional information coming from heterotic/F-theory duality. 

Under these working assumptions and using the above saxionic parametrization, the K\"ahler potential takes the form  
\be\label{hetK} 
K=-\log(s^0-\frac12p_a s^a)-\log\kappa({\bm s},{\bm s},{\bm s})\,,
\ee
where $\kappa({\bm s},{\bm s},{\bm s})\equiv \kappa_{abc}s^as^bs^c$, and we ignore  irrelevant additional constants. 
By \eqref{elldef}, the corresponding dual saxions are then given by 
 \be\label{ellshet} 
\ell_0=\frac{1}{2(s^0-\frac12p_b s^b)}\quad,\quad \ell_a=\frac{3\kappa_{abc}s^bs^c}{2\kappa({\bm s},{\bm s},{\bm s})}-\frac{p_a}{4(s^0- \frac12p_b s^b)}\equiv \hat\ell_a-\frac12p_a\ell_0\,,
 \ee
and their kinetic potential takes the form
\be\label{hetKinpot} 
\calf=\log\ell_0+\log\hat P({\hat\ell})\,. 
\ee
Here $\hat\ell_a\equiv \ell_a+\frac12p_a\ell_0$ and $\log\hat {P}({\hat\ell})$  are the dual saxions and the kinetic potential that one would obtain by ignoring the $s^0$ saxion and starting from a K\"ahler potential $\hat K=-\log\kappa({\bm s},{\bm s},{\bm s})$. 

Note that the Calabi-Yau volume changes along the M-theory interval \cite{Witten:1996mz} and that $s^0-p_as^a$ represents its smallest value in $l_{(11)}$ units.  Hence the assumed  validity of the geometric heterotic M-theory regime of \cite{Horava:1996ma} requires that $s^0-p_as^a\gg 1$, which in turn implies that the $p_as^a$ contribution  in \eqref{hetK} may be considered as a subleading contribution, potentially of the same order of other neglected corrections coming from higher-derivative M-theory terms. More information can be obtained by looking at the models that admit a dual F-theory description, whose perturbative regime described in Section \ref{sec:FIIB} should correspond to freezing one of the heterotic saxions $s^a$.  In Appendix \ref{app:hetFtheory} we show how, in the regime $\ell_0\ll|\ell_a|$, the corresponding  restriction of \eqref{ellshet} matches the F-theory K\"ahler potential  \eqref{FFK}, up to $\ell_0^2$ corrections. We then expect the F-theory K\"ahler potential \eqref{FFK} to capture possible corrections to the corresponding restricted version of  \eqref{ellshet}.  

These uncertainties clearly affect the identification of the dual saxionic cone.  First focus on the modified dual saxionic vector  $\hat{\bm\ell}={\bm\ell}+\frac12\ell_0{\bf p}=\hat\ell_a\Sigma^a$, where  $\Sigma^a$ is the basis of curves dual to $D_a$ ($D_a\cdot\Sigma^b=\delta_a^b$). The second relation in \eqref{ellshet} implies that $\hat{\bm\ell}\in\calp^{\text{\tiny het}}_{\text{\tiny K}}$, where $\calp^{\text{\tiny het}}_{\text{\tiny K}}\subset H_2(X,\mathbb{R})$ is the Poincar\'e dual of the closure of the image of $\calk(X)$ under the map $J\rightarrow J\wedge J$. Note that $\calp^{\text{\tiny het}}_{\text{\tiny K}}$  is a subcone of the cone Mov$_1(X)$ introduced in \eqref{sF}. However, this condition may not precisely represent the dual saxionic domain, as we may be missing modifications of the dual saxionic cone which are negligible only if $\ell_0\ll |\ell_a|$. Indeed, as discussed in more detail in Appendix \ref{app:hetFtheory}, in models admitting a dual F-theory description  ${\bm\ell}$,  rather than $\hat{\bm\ell}$, should belong to $\calp^{\text{\tiny het}}_{\text{\tiny K}}$. This suggests the following possible refinement of the dual saxionic cone $\calp$: 
\be\label{hetellcond} 
 \calp \simeq\{(\ell_0,{\bm\ell})\ |\  \ell_0\geq 0\,,\ {\bm\ell}\in\calp^{\text{\tiny het}}_{\text{\tiny K}}\}\,.
\ee 
While $\calp$ may receive further corrections and it would certainly be more satisfying to have a more precise derivation thereof, for concreteness we will henceforth assume  \eqref{hetellcond}, keeping in mind that its  reliability is more robust for models admitting an F-theory dual in the regime  $\ell_0\ll |\ell_a|$.

Going back to the saxionic coordinates, the saxionic convex hull is now given by
\be\label{hethatdelta} 
\hat\Delta_\alpha=\left\{(s^0,{\bm s})\ \big|\ {\bm s}=s^aD_a\in \hat\calk_\alpha(X)\,,\ s^0\geq \frac1\alpha+ p_as^a\right\}\,.
\ee
where $\hat\calk_\alpha(X)$ is the K\"ahler convex hull defined as in the generic saxionic case, which is  contained in the stretched  K\"ahler cone introduced in \cite{Demirtas:2018akl}. If one restricts to $\hat\Delta_\alpha$,  \eqref{GBhet} satisfies 
\be 
\gamma(s)|_{\hat\Delta_\alpha}\geq \pi \left[ \frac{2}{\alpha}+\big(p_a+\frac16 n_a\big)s^a|_{\hat\calk_\alpha(X)}  \right]\,.
\ee

\begin{figure}[!htb]
    \centering
    \includegraphics[width=0.8\textwidth]{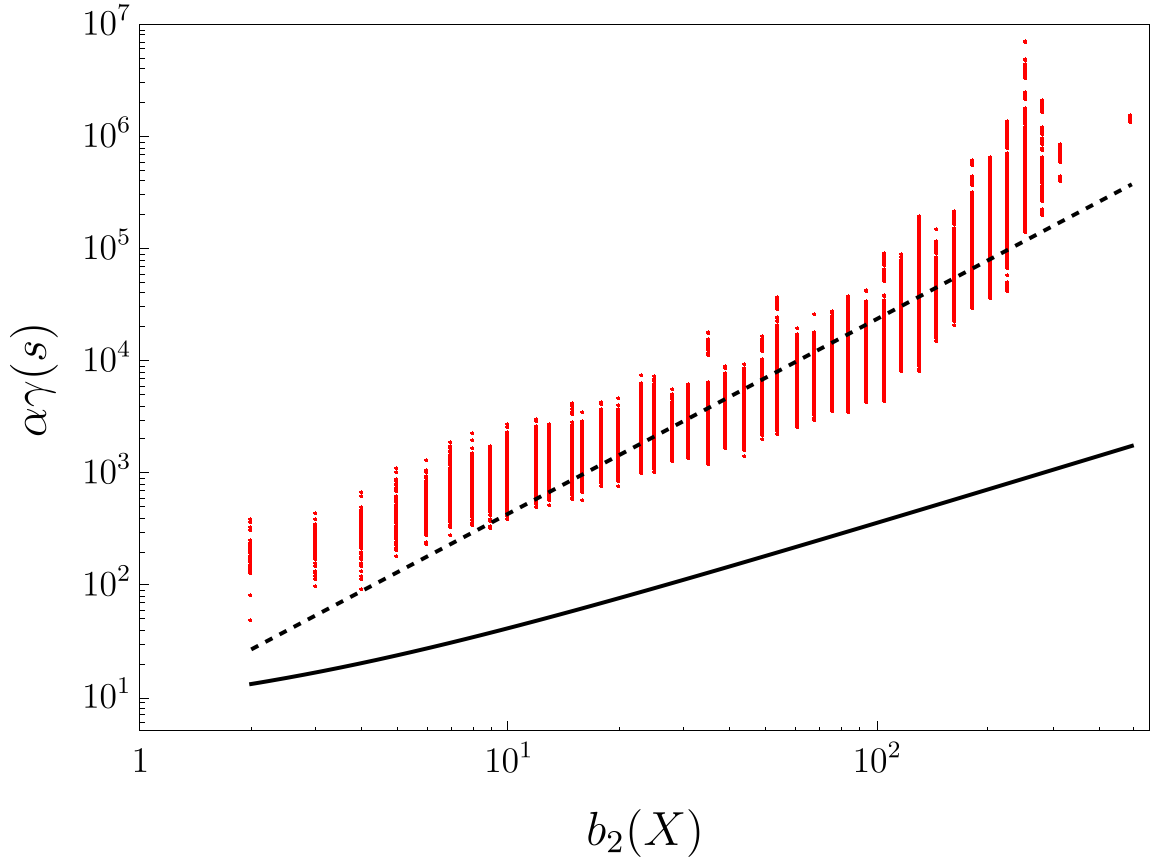}
    \caption{\small Value of $\alpha \gamma(s)$ at the tip of the $\alpha$-stretched saxionic cone as a function of $b_2(X)=N-1$. This quantity, which is $\alpha$-invariant, has been computed in an explicit set of Calabi-Yau compactifications analyzed with \texttt{CYtools} \cite{Demirtas:2022hqf}. The set consists of 71908 Calabi-Yau manifolds with $b_2(X)$ ranging from 2 to 491, with up to 100 polytopes per fixed $b_2(X)$ and with 25 triangulations per polytope obtained with the \texttt{random\_triangulation\_fast} method. The quantity $n_a s^a$ of equation \eqref{GBhet} has been obtained with the \texttt{second\_chern\_class} method and evaluated at the tip of the $\alpha$-stretched Kähler cone. The black solid line refers to the lower bound of equation \eqref{nsbound}. The best fit of our numerical result, obtained assuming a power-law scaling, is $\alpha\gamma\simeq10^{0.9}b_2^{1.7}$, and is shown by the black dashed line.}
    \label{fig:GBvsN}
\end{figure}

For concreteness, consider for instance the models with $\lambda(E_2)=0$. In this case $p_a=n_a$ and by \eqref{upperpasa} $p_as^a$ takes its highest possible value, i.e. $n_as^a\geq 0$. (Smaller values of $p_as^a\geq 0$ lead to similar conclusions.)
 By applying the same arguments that led us to \eqref{gammabound}, we expect $\big(n_as^a\big)|_{\hat\calk_\alpha(X)}\geq  b_2(X)/\alpha$,  which implies that 
\be\label{nsbound} 
\gamma(s)|_{\hat\Delta_\alpha}\geq\pi \left( \frac{2}{\alpha}+\frac76 n_as^a|_{\hat\calk_\alpha(X)}  \right)\geq  \frac{(5+7N)\pi}{6\alpha} \,.
\ee
Note that this lower bound is stronger than \eqref{gammabound}. We numerically tested this bound in a set of explicit Calabi-Yau compactifications with \texttt{CYtools} \cite{Demirtas:2022hqf}. The result is reported in figure \ref{fig:GBvsN}, in which we plot the value of $\alpha\gamma(s)$ evaluated at the tip of stretched Kähler cone against $b_2(X)=N-1$.  Because the saxionic convex hull $\hat\Delta_\alpha$ is contained in the stretched K\"ahler cone, the numerical analysis provides an important non-trivial check of our general bound \eqref{nsbound}, as well as of \eqref{gammabound},  which relied on certain non-trivial quantum gravity constraints.

\subsubsection{Energy scales in heterotic models}
\label{sec:hetscales}

Finally, let us consider the scales characterizing the heterotic models  discussed in this section. By discretizing \eqref{npDelta} one gets  $\calc_{\rm S}^{\text{\tiny EFT}}$, which is generated by  $(1,{\bf 0})$ and vectors of the form $(e^ap_a,{\bf e})$, where ${\bf e}\equiv e^aD_a$ are generators of the cone of nef divisors.  These EFT strings have tensions
\begin{subequations}
\label{eq:HetTensions}
\begin{align}
\calt_*&=
  M^2_{\text{\tiny P}}\ell_0=\frac{M^2_{\text{\tiny P}}}{2(s^0-\frac12 p_a s^a)}=\frac{3M^2_{\text{\tiny P}}e^{2\phi}}{\kappa({\bm s},{\bm s},{\bm s})}
  \,,\label{eq:HetTensionsa}\\
  \calt_{\bf e} 
 &=M^2_{\text{\tiny P}}(e^a\ell_a+e^ap_a\ell_0)=\left(\frac{3\kappa({\bf e},{\bm s},{\bm s})}{2\kappa({\bm s},{\bm s},{\bm s})}+\frac{e^ap_a}{2(s^0-\frac12 p_b s^b)}\right)M^2_{\text{\tiny P}}\,,\label{eq:HetTensionsb}
\end{align}
\end{subequations}
where we have used \eqref{ellshet}. By looking at the behavior of \eqref{hetK} under the corresponding EFT string flows we can check that $K\sim -\log\sigma$ under the $(1,{\bf 0})$ flow, and $w=1$, consistently with the fact that $(1,{\bf 0})$ indeed represents a critical heterotic string. On the other hand, under the flow of the EFT string charges $(e^ap_a,{\bf e})$ we have  $K\sim -n\log\sigma$, with integer $n$ determined by the self-intersections of ${\bf e}$. If $e^ap_a\geq 1$, one has $n=2,3,4$, and consistently  these strings have scaling weights $w=2$ or $w=3$ \cite{Martucci:2022krl}. If instead $e^ap_a=0$, one can also have $n=1$ and $w=1$ \cite{Lanza:2021udy}.

It is now interesting to compare the dominant EFT string scale $M_{\calt}$ introduced in \eqref{SPbounddef}  with $M_{\gamma}$, see \eqref{vafaSC}, and  the species scale  $M_{\text{sp}}$. We will approach this task first analytically, in certain controllable limits, and then numerically in more general setups.

Consider first  the asymptotic regime identified by the EFT string flow generated by $(1,{\bf 0})$, that is,  $s^0\gg 1$ with $s^a$ fixed within \eqref{hethatdelta}.  Inspecting \eqref{hets0} and \eqref{eq:HetTensions}, one finds that this limit  corresponds to  the weak string coupling limit $e^\phi\ll 1$, and that in this limit $\calt_*\ll  \calt_{\bf e}$, for any nef divisor ${\bf e
 }$.  Hence in this regime $M^2_{\calt}=2\pi\calt_*$ and the bound \eqref{SPbound} is saturated, since $\calt_*$ corresponds to the critical string tension: $M_{\rm sp}=M_{\rm s}=M_{\calt}$.  In order to move away from this specific regime,  we have to distinguish whether there exists or there does not exist a nef divisor ${\bf e}=e^aD_a$ such that $e^ap_a=0$.

Assume first that $e^ap_a\geq 1$ for any nef divisor ${\bf e}$, which also implies the strict positivity $p_as^a>0$. In this case, the equations \eqref{eq:HetTensions} clearly show that $\calt_*\leq \calt_{\bf e}$, and therefore that $M^2_{\calt}=2\pi\calt_*$,  at any point of \eqref{hethatdelta}. 
We can now compare compare $M^2_{\calt}$ to $M^2_{\gamma}$:
\be\label{rspport}
\frac{M^2_{\calt}}{M^2_{\gamma}}=\frac12\left[1+\frac{n_as^a}{12(s^0-\frac12p_as^a)}\right] \,,
\ee
where $M^2_{\gamma}$ can be obtained by using  \eqref{GBhet} in the definition \eqref{vafaSC}.
Since $s^0\geq p_as^a$ and $n_as^a\geq 0$, we can then conclude that
 \be\label{rspportop>0}  \frac12\leq\frac{M^2_{\calt}}{M^2_{\gamma}} \leq \frac12\left(1+\frac{n_as^a}{6p_as^a}\right)\,. 
\ee

\begin{figure}[!htb]
    \centering
    \includegraphics[width=0.8\textwidth]{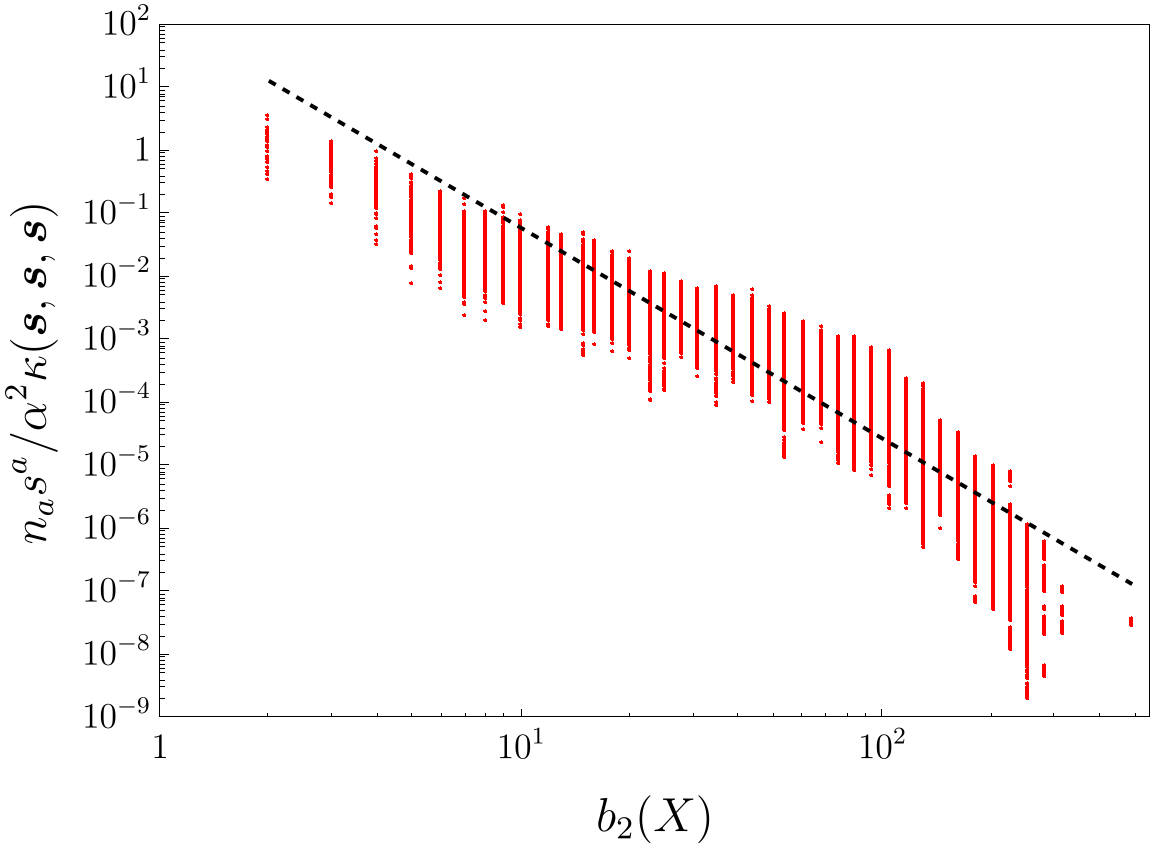}
    \caption{\small Scaling of the ratio $n_a s^a/\alpha^2 \kappa ({\bm s},{\bm s},{\bm s})$, appearing in \eqref{rspport2}, as a function of $b_2(X) = N-1$. The ratio has been evaluated at the tip of the $\alpha$-stretched Kähler cone in the same set of Calabi-Yau compactifications analyzed in figure \ref{fig:GBvsN}. The best fit for the approximate scaling with $N$, again assuming a power-law behavior, is given by $n_a s^a/\alpha^2 \kappa ({\bm s},{\bm s},{\bm s})\simeq 10^{2.1}\,b_2^{-3.3}$ and is qualitatively compatible with the scaling of $n_a s^a$ from Fig. \ref{fig:GBvsN} and with that of $\kappa({\bm s},{\bm s},{\bm s})$ from \cite{Demirtas:2018akl} (modulo a small difference due to the algorithms used to generate Calabi-Yau manifold, which are different in the latter reference and lead to manifolds with qualitatively different properties).
    }
    \label{fig:ratiovsN}
\end{figure}

By \eqref{upperpasa} the smallest possible value of the upper bound in \eqref{rspportop>0} is $\frac{7}{12}$ and is obtained by picking a trivial $E_2$ bundle, that is $p_a=n_a$. Therefore, in this case
the two scales clearly agree within a factor of order one. Even though for more general $p_a$  the upper bound appearing in \eqref{rspportop>0} is a priori larger than $\frac{7}{12}$, by our assumption that $p_ae^a\geq 1$ for any nef divisor ${\bf e}$ we expect $p_as^a$ not to be   much smaller than $n_as^a$ in our perturbative domain. Verifying this expectation would require a thorough investigation  of the internal bundle structure, which enters the definition \eqref{padef}, but this  is beyond the scope of the present paper. We can however get some qualitative information by rewriting \eqref{rspport} in the form
 \be\label{rspport2}
\frac{M^2_{\calt}}{M^2_{\gamma}}=
 \frac{1}{2}\left[1+\frac12 e^{2\phi}\frac{n_as^a }{\kappa({\bm s},{\bm s},{\bm s})}\right]\,.
\ee
By using  \texttt{CYtools} \cite{Demirtas:2022hqf} we have evaluated numerically the ratio $n_as^a/\kappa({\bm s},{\bm s},{\bm s})$  at the tip of the $\alpha$-stretched K\"ahler cone of a large number of models. The result is shown in Fig.~\ref{fig:ratiovsN}, which presents the value of the $\alpha$-invariant combination $n_as^a/[\alpha^2\kappa({\bm s},{\bm s},{\bm s})]$. If for instance $\alpha= 0.1$,  $n_as^a/\kappa({\bm s},{\bm s},{\bm s})$ is roughly given by $N^{-3}$ for large $N$. 
Moving away from the tip of the  stretched K\"ahler cone we expect an even larger suppression. This suggests  that $M_{\calt}$ and $M_{\gamma}$ basically agree also in more general $N\gg 1$ models, at least if $e^\phi$ is not unnaturally large.

\begin{figure}[!htb]
    \centering
    \includegraphics[width=0.8\textwidth]{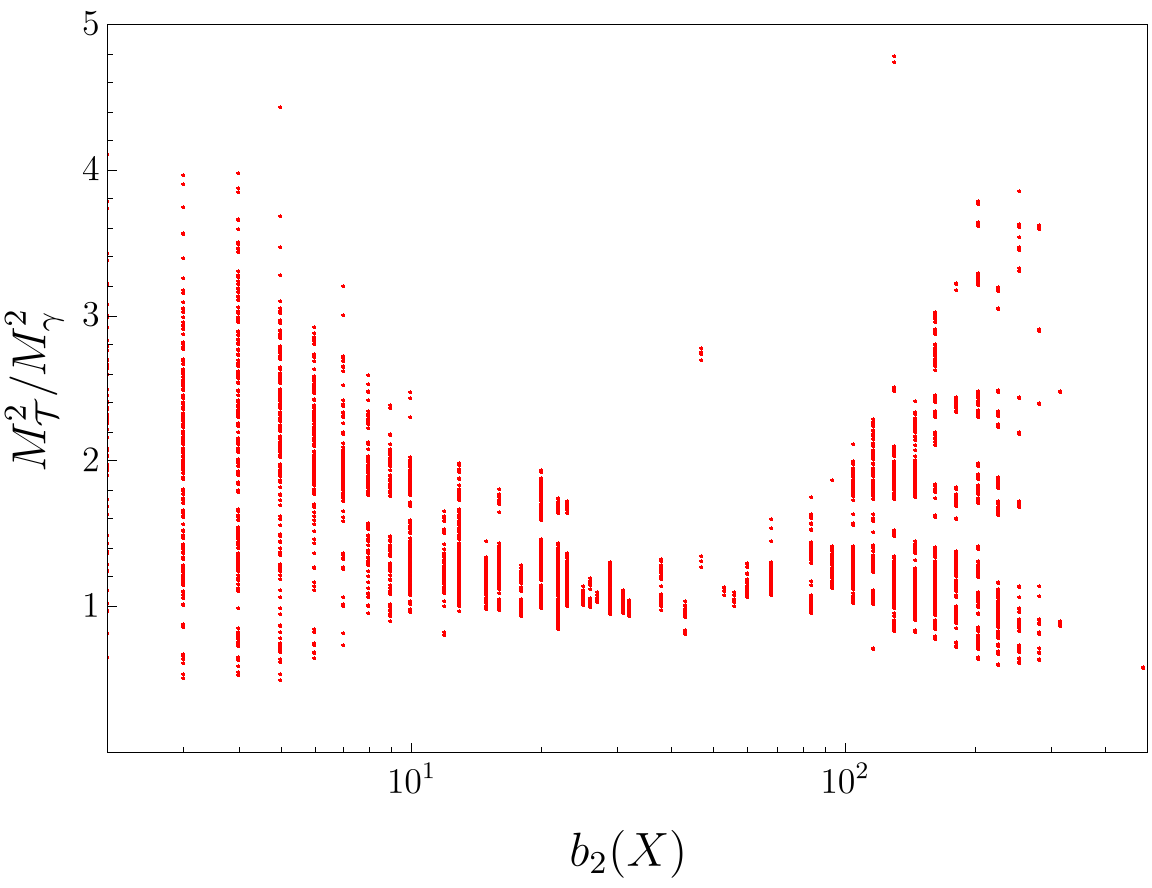}
    \caption{\small Sample of values of the ratio between  $M^2_{\calt}$ and $M^2_{\gamma}$ in a large class of heterotic models, assuming $p_a \equiv 0$.  This ratio, which is $\alpha$-invariant, is evaluated at the tip of the $\alpha$-stretched saxionic cone, that is, at the tip of of the 
    $\alpha$-stretched Kähler cone and with $s^0=1/\alpha$, in an explicit set of 11962 distinct Calabi-Yau compactifications analyzed with \texttt{CYtools} \cite{Demirtas:2022hqf}. The manifolds have been obtained with the \texttt{random\_triangulation\_fast} method with up to 100 polytopes per fixed $b_2(X)$ and 25 triangulations per polytope.  $M^2_{\gamma}$ can be obtained by using  \eqref{GBhet} in the definition \eqref{vafaSC},  with the numerical values of $n_a$ obtained with the \texttt{second\_chern\_class} function. $M^2_{\calt}$ has been estimated using \eqref{eq:HetTensionsb},  in which we employed the minimum volume among the nef divisors that we found in each compactification. These have been individuated requiring $e^a M_{\alpha a} \geq 0 \, \forall \, \alpha$, where $M_{\alpha a}$ is the matrix of Mori cone generators associated to the basis of the inherited $b_2 (X)+4$ prime toric divisors obtained using the \texttt{toric\_mori\_cone} function. Their volumes have been then computed at the tip of the Kähler cone with \texttt{compute\_divisor\_volumes}. As typically only a small fraction of the basis divisors are also nef divisors, the available statistic is reduced compared to the analysis of figure \ref{fig:GBvsN}.
    }
    \label{fig:speciesScalesHet}
\end{figure}

Let us now allow for the existence of one or more  nef divisors $\hat{\bf e}=\hat e^aD_a$ such that $p_a\hat e^a=0$. We will restrict to the elementary charges $\hat{\bf e}$ of this type. The corresponding tensions  \eqref{eq:HetTensionsb} reduce to
\be\label{hatetension} 
\calt_{\hat{\bf e}}=M^2_{\text{\tiny P}}\hat e^a\ell_a=M^2_{\text{\tiny P}}\frac{3\kappa(\hat{\bf e},{\bm s},{\bm s})}{2\kappa({\bm s},{\bm s},{\bm s})}\,,
\ee
and control $M^2_{\calt}$ in the region of the saxionic domain such that $e^{2\phi}\geq \frac12\kappa(\hat{\bf e},{\bm s},{\bm s})$ for some of these $\hat{\bf e}$. By recalling \eqref{hets0}, this condition corresponds to $s^0\leq \frac{M^2_{\text{\tiny P}}}{2\calt_{\hat{\bf e}}}+\frac12p_as^a$. Unfortunately, in this case it is not easy to draw general   conclusions about the ratio $M^2_{\calt}/M^2_{\gamma}$. In order to get some quantitative understanding, we pragmatically assume that $p_a\equiv 0$, and we again use   \texttt{CYtools} \cite{Demirtas:2022hqf} to  numerically investigate $M^2_{\calt}/M^2_{\gamma}$.  
The result is shown in Fig.\ \ref{fig:speciesScalesHet}. Again we see that, up to irrelevant numerical factors of order one, the two scales agree at the tips of the stretched K\"ahler cones of the entire set of geometries that we explored. Note that these results hold also  for the more general case with non-vanishing $p_a$,  as long as we go  far enough along the EFT string flows ${\bm s}={\bm s}_0+\sigma \hat{\bf e}$, $\sigma\gg 1$. Indeed, precisely in this limit we can neglect the contribution $p_as^a\simeq p_as^a_0$, which does not scale with $\sigma$, and $\calt_{\hat{\bf e}}$ is expected to identify the lightest EFT string tension, and then to determine $M_{\calt}$. Furthermore,  the same conclusions immediately apply to ${\cal N}=2$ type IIA models. This clearly indicates  that $M^2_{\calt}$ agrees (in the geometric regime) with the estimate of the species scale proposed  for these models in \cite{vandeHeisteeg:2022btw}. 

Let us now turn to the verification of the bound \eqref{SPbound} beyond the weak string coupling limit, that is, in the regime in which $M_{\rm sp}$ is given by the quantum gravity scale $M_{\text{\tiny QG}}$, rather than the string scale $M_{\rm s}$. Let us again first assume that $e^ap_a\geq1$ for any nef divisor ${\bf e}=e^aD_a$, so that $M^2_{\calt}=2\pi\calt_*$ 
 as already pointed out.
 We can check \eqref{SPbound} analytically in the M-theory supergravity regime. This requires $e^\phi$ and  $ e^{-\frac{2\phi}{3}}{\bm s}$ to be large enough, since the internal six-dimensional M-theory and heterotic string frame  metrics are related by $\d  s^2_{\text{\tiny M}}(X)=e^{-\frac{2\phi}{3}}\d  s^2_{\text{st}}(X)$. For instance we may require that $e^{-\frac{2\phi}{3}}{\bm s}\in \hat\calk_{\bf \alpha}$. Recalling the ansatz \eqref{11to10} (which implies $l_{(11)}=l_{(10)}$), the scaling factor appearing in \eqref{lineelement} is given by
\be\label{scalingMth}
e^{2A}=\frac{M^2_{\text{\tiny P}}l^2_{(11)}}{4\pi V(X)}e^{-\frac23\phi}\,,
\ee
where $V(X)=\frac16e^{-2\phi}\kappa({\bm s},{\bm s},{\bm s})$ is the average value of the Calabi-Yau volume, as measured in M-theory and in $l_{(11)}$ units, along the M-theory interval. The quantum gravity scale is then given by 
 \be\label{hetMQG}
 M^2_{\text{\tiny QG}}=e^{2A}\widehat M^2_{(11)}= \frac12\left[\frac{12\pi}{(s^0-\frac12p_as^a)^2\kappa({\bm s},{\bm s},{\bm s})}\right]^\frac13\,M^2_{\text{\tiny P}}\,,
 \ee
where $\widehat M^2_{(11)}$ is as in  \eqref{strPS1011}, and we have used \eqref{hets0}. Since $M^2_{\calt}=2\pi\calt_*$, with $\calt_*$ as in \eqref{eq:HetTensionsa}, we conclude that  
\be\label{MhetTQG} 
\frac{M^2_{\calt}}{M^2_{\text{\tiny QG}}}=  \left[\frac{(2\pi)^2\kappa({\bm s},{\bm s},{\bm s})}{6(s^0-\frac12p_as^a)}\right]^{\frac13}= \left(2\pi e^{\phi}\right)^{\frac23}\quad~~~~~\text{(for $2\pi e^\phi>1$)}\,.
\ee
(This formula could have been derived more directly  by identifying $M^2_{\calt}/2\pi=\calt_*$ with the tension of a string  corresponding to an open M2-brane stretching along the M-theory interval, which has length $l_{(11)}e^{2\phi/3}$.)  
 
 Note that  the ratio \eqref{MhetTQG} is controlled by the expansion parameter $2\pi e^\phi$, which more precisely distinguishes the ten- and eleven-dimensional regimes, as also discussed in Appendix \ref{app:2pis}.  The bound \eqref{SPbound} is satisfied almost tautologically, since the M-theory regime we are working in requires $2\pi e^\phi>1$. If instead $2\pi e^\phi<1$, the M-theory description is not trustable anymore, and one should rather compute $M^2_{\text{\tiny QG}}$ starting from $\widehat M_{(10)}$ in \eqref{strPS1011}. Since $\calt_*$ corresponds to an F1-string tension, the ratio \eqref{MhetTQG} is more easily obtained by comparing  $2\pi \calt_{\rm F1}=(2\pi)^2l^{-2}_{(10)}e^{\frac12\phi}$ and $\widehat M^2_{(10)}$ directly in ten dimensions. This gives (see also Appendix \ref{app:2pis})
\be\label{MhetTQG2} 
\frac{M^2_{\calt}}{M^2_{\text{\tiny QG}}}= \frac{2\pi \calt_{\rm F1}}{\widehat M^2_{(10)}}= \left(2\pi e^{\phi}\right)^{\frac12}\quad~~~~~\text{(for $2\pi e^\phi<1$)}\,.
\ee
Clearly  $M_{\calt}< M_{\text{\tiny QG}}$ if $2\pi e^\phi<1$, and hence the bound \eqref{SPbound} is saturated: $M_{\rm sp}=M_{\rm s}=M_{\calt}$.

\begin{figure}[!htb]
    \centering
    \includegraphics[width=0.8\textwidth]{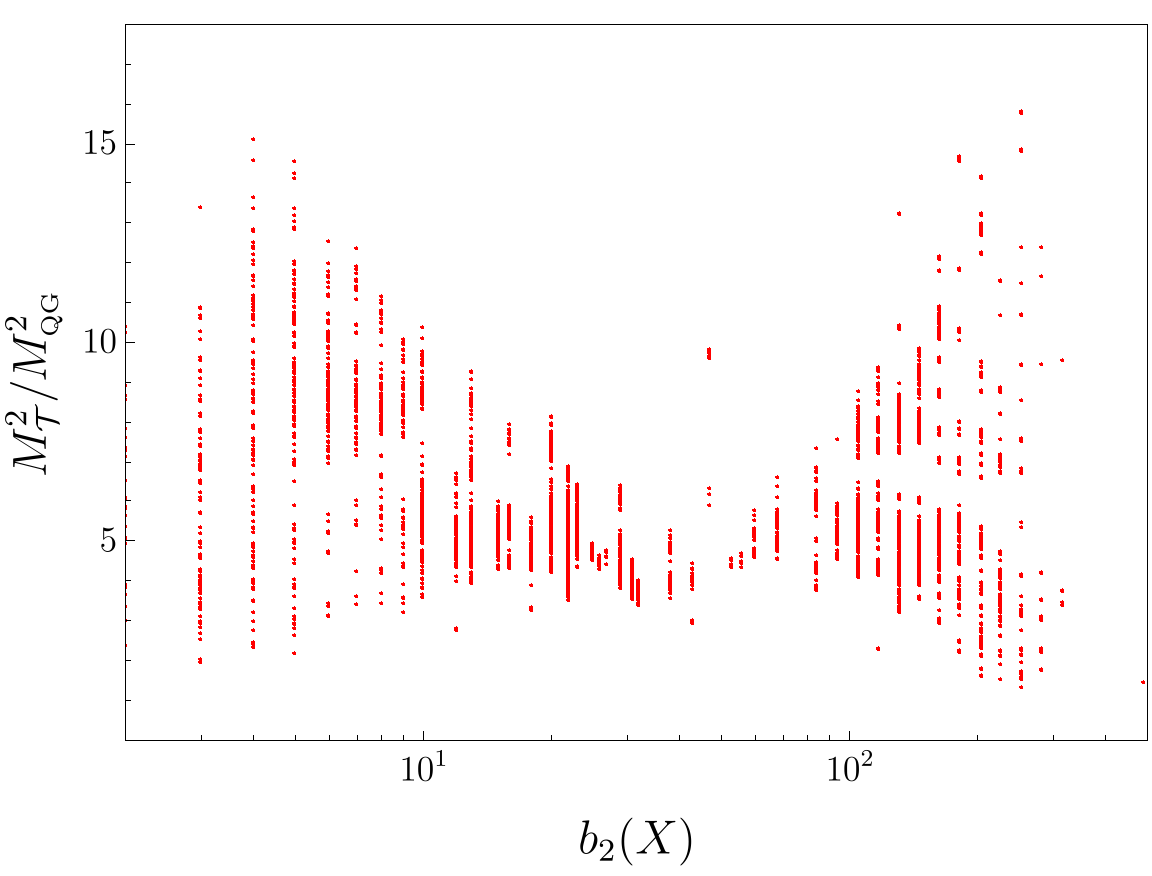}
    \caption{\small Plot of the lower bound \eqref{MTMQG} evaluated  evaluated at the tip of the $\alpha$-stretched saxionic cone with $\alpha=1$ in the same set of Calabi-Yau manifolds of figure \ref{fig:speciesScalesHet}. Since the r.h.s.\ of \eqref{MTMQG} is invariant under an overall rescaling of the saxions $s^i$, the result for more general $\alpha$ can be obtained through a  rescaling  by $\alpha^{-\frac23}$.
    }
    \label{fig:QGscaleHet}
\end{figure}

The case in which there is some nef divisor $\hat{\bf e}=\hat e^aD_a$  such that $p_a\hat e^a=0$ again requires a separate discussion. In this case, the corresponding lightest possible tension \eqref{hatetension} determines $M^2_{\calt}$ if $(2\pi e^{\phi})^2\geq \frac12\kappa(\hat{\bf e},2\pi{\bm s},2\pi{\bm s})$. Clearly, in our perturbative saxionic regime this can happen only if $2\pi e^{\phi}>1$, namely in the M-theory regime, since ${\bm s}\in\hat\calk_\alpha$ and then we certainly have $\frac12\kappa(\hat{\bf e},2\pi{\bm s},2\pi{\bm s})>1$. As discussed above, while we will not attempt to make exact statements, we can nevertheless get non-trivial information assuming that we can set $p_a\equiv 0$, either exactly or approximately. Hence, picking the charge $\hat{\bf e}$ with lightest tension, which thus determines $M^2_{\calt}$,  combining \eqref{hatetension} and \eqref{hetMQG} we get
\be 
\frac{M^2_{\calt}}{M^2_{\text{\tiny QG}}}=\frac{\kappa(\hat{\bf e},2\pi{\bm s},2\pi{\bm s})}{2(2\pi e^\phi)^{\frac43}}\,.
\ee
Applying the perturbative condition \eqref{hethatdelta} to \eqref{hets0} with $p_a=0$ we get
\be 
e^{2\phi}\leq \frac{\alpha}{6}\kappa({\bm s},{\bm s},{\bm s})
\ee
and hence
\be\label{MTMQG} 
\frac{M^2_{\calt}}{M^2_{\text{\tiny QG}}}\geq \left(\frac{2\pi}{\alpha}\right)^{\frac23}\frac{V(\hat{\bf e})}{V(X)^\frac23}=\frac12\left(\frac{12\pi}{\alpha}\right)^{\frac23}\frac{\kappa(\hat{\bf e},{\bm s},{\bm s})}{\kappa({\bm s},{\bm s},{\bm s})^\frac23}\quad~~~~~~~~~\text{(for $p_a=0$)}\,.
\ee
We have numerically evaluated $\kappa(\hat{\bf e},{\bm s},{\bm s})/\kappa({\bm s},{\bm s},{\bm s})^\frac23$ at the tip of the stretched K\"ahler cone of a large sample of models using again  \texttt{CYtools} \cite{Demirtas:2022hqf}. The result is shown in Fig.~\ref{fig:QGscaleHet} and, since $2\pi/\alpha\gg1$ and the combination $\kappa(\hat{\bf e},{\bm s},{\bm s})/\kappa({\bm s},{\bm s},{\bm s})^\frac23$ is invariant under an overall saxionic rescaling, it clearly indicates that  \eqref{MTMQG} should  always be  greater than one, compatibly with the bound \eqref{SPbound} with $M_{\rm sp}=M_{\text{\tiny QG}}$. 

One can further check the bound \eqref{SPbound} in other limits. These are discussed in Appendix \ref{app:N=2tensions}, which partly apply also to $\caln=2$ models.

\section{$SO(4)$-symmetric wormhole configurations}
\label{sec:WHs}

In this section we show that the broad class of models described by the two-derivative action \eqref{kinetic}, or equivalently \eqref{kinetic2}, admits non-extremal and extremal wormhole configurations with $SO(4)$ symmetry, after continuation to Euclidean space. These solutions can be considered as generalizations of the ones provided in the seminal paper \cite{Giddings:1987cg} and encompass several other generalizations already appeared in the literature -- see for instance \cite{Hebecker:2018ofv} for a review and \cite{Loges:2022nuw, Jonas:2023qle,Hertog:2024nys} for recent discussion on the stability of these types of wormholes. In Sections \ref{sec:eucliact} and \ref{sec:effd=1} we present the Euclidean formulation of our models, derive the equations of motion associated to the $SO(4)$-symmetric ansatz, and finally calculate the on-shell action for non-extremal wormholes. This part of our work is not new but serves to set the stage for the original results presented in the subsequent sections. Extremal wormholes are discussed in Section \ref{sec:extremalBPS} and their relation to fundamental instantons is clarified. 
The concept of EFT instantons is introduced in Section \ref{sec:EFTinst}. A puzzling BPS bound on the on-shell action of non-extremal wormholes is briefly commented upon in Section \ref{sec:BPSbound}, a possible interpretation of which is suggested in later sections.

\subsection{Euclidean action}
\label{sec:eucliact}

When discussing non-perturbative effects such as wormholes we have to pass to the Euclidean formulation. The way in which one identifies charged saddles in the axionic Euclidean action  has been clarified in \cite{Coleman:1989zu} (see also \cite{Arkani-Hamed:2007cpn}) and involves a continuation to imaginary axion fields as well as the introduction of boundary terms. No such subtleties appear if one works with the dual formulation, which is what we do. One just has to Wick rotate \eqref{kinetic2}, giving 
\be
\begin{aligned}
\label{kinetic3} 
S=&\, -\frac{1}{2}M_{\text{\tiny P}}^2\int_\calm\sqrt{g}\, R-\frac{1}{2}M_{\text{\tiny P}}^2\int_{\partial\cal M}\sqrt{h}(K-K_0)\\
&\,+\frac12M_{\text{\tiny P}}^2\int_\calm \calg^{ij}\d \ell_i\wedge *\d \ell_j+\frac1{2M_{\text{\tiny P}}^2}\int_\calm\calg^{ij}\calh_{3,i}\wedge *\calh_{3,j}\,,
\end{aligned}
\ee
where the Gibbons-Hawking term \cite{Gibbons:1976ue} for asymptotically flat vacua has been included. In this dual formulation  $\calh_{3,i}$ must satisfy \eqref{H3BI},  up to possible localized terms corresponding to fundamental instantons.\footnote{\label{foot:fluxsup}Apart from the additional contribution to $\d\calh_3$ of standard gauge fields (see for instance \cite{Martucci:2022krl}), which we set to zero,  $\d\calh_3$ can also get a contribution from the field-strengths of three-forms potentials. As discussed in \cite{Farakos:2017jme,Bandos:2018gjp,Lanza:2019xxg}, in $\caln=1$ supersymmetry this effect is dual to the presence of special multi-branched superpotentials which are often realized in flux compactifications \cite{Bielleman:2015ina}. In this paper we assume that such superpotential terms are not present or do not affect the (s)axionic sector under discussion.} 
The GB term \eqref{GRtheta} will play an important role. Rotated to Euclidean signature, this reads 
\be\label{GBterm2}
S_{\text{\tiny GB}}\equiv-\int_\calm\sqrt{g}\,\gamma(\ell) \,E_{\text{\tiny GB}}-\int_{\del\calm}\sqrt{h}\,\gamma(\ell)\, (Q-Q_0)\,,
\ee
where in $\gamma(\ell)=\frac{\pi}{6}\tilde C_is^i\equiv\frac{\pi}{6}\langle \tilde{\bf C}, {\bm s}\rangle$ (see \eqref{gammas})  we regard $s^i$ as functions of the dual saxions $\ell_i$ as defined by \eqref{elldef2}, and we included the appropriate boundary term \cite{Myers:1987yn} necessary to make the variational principle well defined. This  term is analogous to the Gibbons-Hawking term in \eqref{kinetic3}, and  we will not need its precise form. Suffice it to say that it allows one to make the identification \cite{Chern45} (see also \cite{Eguchi:1980jx,Gibbons:1979xm})
\be\label{Chern} 
\int_\calm\sqrt{g}\, \,E_{\text{\tiny GB}}+\int_{\del\calm}\sqrt{h}\,Q=\chi(\calm)\,,
\ee 
where $\chi(\calm)=\sum_{k}(-)^kb_k(\calm)$ is the Euler characteristic of $\calm$. On the other hand, the counterterm $\int_{\del\calm}\sqrt{h}\,\gamma(\ell)Q_0$ in \eqref{GBterm2} subtracts the contribution of a flat vacuum configuration so as to have a vanishing action in flat space, analogously to what done with the Gibbons-Hawking term.

\subsection{Effective one-dimensional action and equations of motion}
\label{sec:effd=1}

In deriving the relevant equations of motion we will follow the approach of \cite{Arkani-Hamed:2007cpn}, but will work with the formulation in terms of gauge two-forms $\calb_{2\,i}$ and dual linear multiplets.

We are looking for Euclidean wormholes preserving $SO(4)$ rotational symmetry. This means that we can restrict the metric to take the form
\be\label{genmetric}
\d s^2=\frac{1}{M^2_{\text{\tiny P}}}\left[e^{2A(\rho)}\d\rho^2+e^{2B(\rho)}\d\Omega^2\right]\,.
\ee
Here $\rho\in I$ is an arbitrary dimensionless radial  coordinate taking values in some interval $I\subset \mathbb{R}$, to be defined below, and $\d\Omega^2$ is the line element of a three-sphere of unit radius, which has volume $2\pi^2$. One can of course remove the arbitrariness of  $\rho$ by gauge-fixing $e^{A(\rho)}$ or $e^{B(\rho)}$.  
A particularly convenient choice is given by
\be\label{rcoordin} 
e^B=\rho= M_{\text{\tiny P}}r
\ee
where $r$ represents the radius of the three-sphere. This gauge fixing corresponds to the following line element
\be\label{gauge1} 
\d s^2=e^{2A(r)}\d r^2+r^2\d\Omega^2.
\ee 
Note that, however, in this case $r$ can smoothly parametrize only half wormhole.\footnote{This issue is avoided if instead we use the `geodesic'  radial coordinate $\chi$ defined by $M_{\text{\tiny P}}\d\chi= e^A\d\rho$. 
With this coordinate the metric takes the form $\d s^2=\d \chi^2+M_{\text{\tiny P}}^{-2}e^{2B(\chi)}\d\Omega^2$, 
corresponding to the gauge-fixing $e^{A}=1$ and the identification $\rho=\chi M_{\text{\tiny P}}$.} Another useful parametrization is discussed below, but for the moment we keep $\rho$ arbitrary.  

In addition to the metric, we will allow the fields $\ell_i,\calh_{3,i}$  to have non-trivial profiles. By $SO(4)$ symmetry, they  can only depend on the radial coordinate: $\ell_i=\ell_i(\rho)$, or equivalently $s^i=s^i(\rho)$. Instead the field-strengths $\calh_{3\,i}$ must necessarily take the form 
\be\label{H3ansatz}
\calh_{3\,i}=\frac1{\pi}q_i\,{\rm vol}_{S^3}\,,
\ee
with $q_i$ constants. Notice that such a choice satisfies $\d\calh_{3,i}\equiv 0$, and this is consistent with \eqref{H3BI} since the $SO(4)$ symmetry implies that $\tr(\calr\wedge\calr)\equiv0$. 
Observing that $*\calh_{3\,i}=M_{\text{\tiny P}}^2\calg_{ij}\d a^i$ corresponds to the 1-form current associated to the axion shift symmetry, the quantities $q_i$ are to be interpreted as the wormhole charges 
\be\label{H3quant}
\frac1{2\pi}\int_{S^3}\calh_{3\,i}=q_i\in\mathbb{Z}\,.
\ee
Charge quantization can be either seen as a consequence of the axion periodicity \eqref{unitperiodicity} (i.e. the momentum conjugate to a periodic variable is quantized) or, in the dual language, to the quantization of the  $\calh_{3\,i}$ field-strengths.

The $SO(4)$ symmetry ensures that our ansatz can be regarded as a consistent truncation of the full theory. Inserting it in Eq.\ \eqref{kinetic3}, the action reduces to 
\be\label{1dS}
\begin{aligned}
\left.S\right|_{\rm ansatz}=&\, -6\pi^2\int_I\d \rho\left[e^{A+B}+\left(\frac{\d B}{\d \rho}\right)^2e^{3B-A}\right]+6\pi^2\left[e^{2B}\right]_{\del I}\\
&+\int_I\d\rho\,\left[e^{3B-A}\pi^2\calg^{ij}\,\frac{\d \ell_i}{\d \rho}\frac{\d \ell_j}{\d \rho}+e^{A-3B}\calg^{ij}\,q_iq_j\right]\,.
\end{aligned}
\ee
Extremizing \eqref{1dS} with respect to $A$ we get the constraint
\be\label{Aconst}
e^{6B-2A}\pi^2\calg^{ij}\,\frac{\d \ell_i}{\d \rho}\frac{\d \ell_j}{\d \rho}-\calg^{ij}\,q_iq_j=6\pi^2\left(\frac{\d B}{\d \rho}\right)^2 e^{6B-2A}-6\pi^2\,e^{4B}\,.
\ee
The equations of motion for $\ell_i$ are derived from the second line of \eqref{1dS} and will be shown shortly. The variation of \eqref{1dS} with respect to $B$ is on the other hand redundant once the former two conditions are satisfied, and hence will not be discussed.

To write the field equations for $\ell_i$ it is convenient to introduce \cite{Arkani-Hamed:2007cpn} a `proper' radial coordinate $\tau$ such that 
\be\label{taurho} 
\d\tau=\pm\frac1\pi e^{A-3B}\d\rho\,,
\ee
where the two signs correspond to the two possible relative orientations between $\tau$ and $\rho$. With such a choice, the second line of \eqref{1dS} resembles the action for a particle with non-canonical kinetic term moving in a non-trivial potential: 
\be\label{tau1dS}
2\pi\int \d\tau \left[\frac12 \calg^{ij}(\ell)\,\dot\ell_i\dot\ell_j-V_{\bf q}({\ell})\right]
\ee
with $\dot\ell_i\equiv \frac{\d\ell_i}{\d \tau}$. The potential is 
\be\label{potVq}
V_{\bf q}({\ell})\equiv -\frac12\calg^{ij}({\ell})\,q_iq_j\equiv -\frac12\|{\bf q}\|^2\,
\ee
where we have introduced the norm defined by the metric \eqref{calg2}:
\be
\|{\bf q}\|^2\equiv \calg^{ij}(\ell)\,q_iq_j\,.
\ee
The (dimensionless) ``particle'' energy
\be\label{Econs}
E \equiv \frac12 \calg^{ij}({\ell})\,\dot\ell_i\dot\ell_j+V_{\bf q}({\ell})
\ee
is thus manifestly conserved. From \eqref{tau1dS} -- or equivalently \eqref{1dS} -- we may now obtain the equations of motion for $\ell_i$ 
\be\label{elleom}
\calg^{ij}\frac{D\dot\ell_j}{\d\tau}=-\frac{\del V_{\bf q}}{\del\ell_i}\quad\Leftrightarrow\quad \calf^{ij}\ddot\ell_j=\frac12\calf^{ijk}\left(q_jq_k-\dot\ell_j\dot\ell_k\right)\,,
\ee
where ${D\dot\ell_j}/{\d\tau}$ is the Levi-Civita covariant derivative associated with the metric $\calg^{ij}$; we have used  \eqref{calg2} and we have introduced the shorthand notation $\calf^{i}\equiv{\del\calf}/{\del\ell_i}$, $\calf^{ij}\equiv{\del^2\calf}/{\del\ell_i\del\ell_j}$, etc.  In addition the dual saxions $\ell_i$ satisfy appropriate boundary conditions, which we assume to be Dirichlet.\footnote{In presence of low-energy supersymmetry breaking, the asymptotic values ${\bm \ell}_\infty$ may ultimately be determined by an hypothetical stabilizing potential for the saxions. We will come back to this in Section \ref{sec:QFTpheno}.}

We are interested in asymptotically flat wormhole configurations manifesting a Euclidean ``time-reversal" symmetry.\footnote{The Euclidean time-reversal symmetry  ensures that by cutting these solutions at the minimal radius one gets half-wormholes which, once analytically continued back to a Lorentzian spacetime, describe the nucleation or absorption of a baby universe,  with real boundary values of the fields and of their time derivatives \cite{Giddings:1987cg}.} We can take a symmetric $\tau$ interval, $\tau\in [-\tau_\infty,\tau_\infty]$, with both endpoints $\tau=\pm\tau_\infty$  corresponding to $r=\infty$. The dual saxions can then flow from some asymptotic value ${\bm \ell}_\infty$ at $\tau=-\tau_\infty$  to some value ${\bm \ell}_*$ at the neck of the wormhole $\tau=0$, and eventually back to the asymptotic value ${\bm \ell}_\infty$ at $\tau=\tau_\infty$. The analogy with the point particle makes it clear that the saxionic flow can be interpreted as a scattering process, in which the particle climbs the potential \eqref{potVq} until it reaches the turning point ${\bm\ell}_*$, at which it stops and then rolls down back to its original position. Since  $V_{\bf q}< 0$, the saxions can bounce  only if $E=-|E|<0$. We will however be a bit more general and consider
\be\label{extrlowerb}
E=-|E|\leq 0\,.
\ee
We will refer to the configurations with $E<0$ and $E=0$  as non-extremal  and extremal wormholes, respectively, even though  strictly speaking the extremal configurations do not describe proper wormholes connecting two asymptotically flat regions, see Section \ref{sec:extremalBPS} below for more details. (Rather, they can be formally considered as half-wormholes in the extreme  $E\rightarrow 0$ limit.) In the following we will concentrate on these cases, without considering the configurations with $E>0$, which have a singular geometry. See for instance the review papers \cite{Hebecker:2018ofv,VanRiet:2020pcn,Harlow:2022ich} for additional discussions on these three cases.

Inserting \eqref{Econs} in the constraint \eqref{Aconst}, and writing the result in the gauge \eqref{rcoordin} one gets 
\be\label{wormwarp}
e^{2A}=\frac{1}{1-\frac{L^4}{r^4}}
\ee
where
\be\label{WHL} 
L^4\equiv \frac{|E|}{3\pi^2M^4_{\text{\tiny P}}}
=\frac{\|{\bf q}\|^2_*}{6\pi^2M^4_{\text{\tiny P}}}\,,
\ee
with $\|{\bf q}\|^2_*\equiv \|{\bf q}\|^2({\ell}_*)$.
Hence the metric takes the form 
\be\label{WHmetric}
\d s^2= \frac{1}{1-\frac{L^4}{r^4}}  \d r^2+r^2\d\Omega_3^2,
\ee
with $r\in [L,\infty)$ and $L$ can be identified with the minimal $S^3$ radius $r_*=L$. In the second equality in \eqref{WHL} we used the fact that, if $E<0$, at the turning point of the wormhole throat the particle stops ($\dot\ell_i|_{\tau=0}=0$) and $E$ reduces to $V_{\bf q}(\ell_*)$. We see that the minimal $S^3$ radius of non-extremal wormholes is controlled by the charge norm $\|{\bf q}\|$ at such radius. If instead $E=0$, then  $L=0$ and the metric becomes flat.

By using \eqref{wormwarp} in \eqref{taurho} (in the gauge \eqref{rcoordin}) one gets a differential relation between $r$ and $\tau$, which can be easily integrated. For non-extremal ($E<0$) wormholes, the integration constant can be fixed by imposing that
\be\label{tauzero}
\tau=0\quad~~~~\Leftrightarrow\quad~~~~ r=L
\,.
\ee
Then \eqref{taurho}, with $+/-$ sign for positive/negative $\tau$, integrates to 
\be\label{rtaurel}
\tau=\pm\frac{1}{2\pi M^2_{\text{\tiny P}}L^2}\left[\frac{\pi}2-\arcsin\left(\frac{L^2}{r^2}\right)\right]
\ee
or similarly ${L^2}/{r^2}=\cos\left(2\pi M^2_{\text{\tiny P}}L^2\tau\right)$. This in particular implies that $e^{-2A}$ as a function of $\tau$ is given by $e^{-2A}={\sin^2\left(2\pi M^2_{\text{\tiny P}}L^2\tau\right)}$. The maximal extension of the $\tau$ interval is $|\tau|\leq\tau_\infty$, with
\be\label{taurange}
\tau_\infty\equiv\frac1{4M^2_{\text{\tiny P}}L^2}
\quad~~~~\Leftrightarrow\quad~~~~ r_\infty=\infty
\,,
\ee
where $\cos\left(2\pi M^2_{\text{\tiny P}}L^2\tau_\infty\right)=0$. In particular, the region 
$0\leq \tau\leq \tau_\infty$ parametrizes the first half-wormhole, while the interval $ -\tau_\infty\leq\tau\leq 0$ 
parametrizes the second half-wormhole. In other words, the particle takes a proper time $\tau_\infty$ to make half of its journey.

The particle interpretation also suggests a simple characterization of regular wormholes. The point particle is at rest at $\tau=0$ and starts rolling down the potential $V_{\bf q}$. If it reaches $\tau_\infty$ remaining inside the perturbative dual saxionic domain, then the corresponding trajectory describes an everywhere regular on-shell wormhole. Whenever along its path the particle exits the allowed domain, the solution develops a singularity and should either be somehow regularized  or discarded.

Finally, the on-shell wormhole action can be obtained by plugging \eqref{Aconst} into \eqref{1dS}, using the gauge \eqref{rcoordin} and the explicit solution \eqref{wormwarp}. Alternatively, we can use the traced Einstein equation
\be 
R\,*1=\calg^{ij}\left(\d\ell_i\wedge * \d\ell_j-\frac{1}{M_{\text{\tiny P}}^4}\calh_{3\,i}\wedge *\calh_{3\,j}\right)\,
\ee
in  \eqref{kinetic3},
observing that the asymptotic Gibbons-Hawking boundary term vanishes for the metric \eqref{WHmetric}. 
In either case, one obtains the wormhole action
\be
\left.S\right|_{\rm w}=2\pi\int\d\tau\, \|{\bf q}\|^2=\frac1{M_{\text{\tiny P}}^2}\int\calg^{ij}\calh_{3,i}\wedge *\calh_{3,j}\,.
\ee
As argued in \cite{Giddings:1987cg} (see also Section \ref{sec:phys}), the physically relevant quantity is $1/2$ of $\left.S\right|_{\rm w}$, namely the on-shell action of half-wormhole:
\be\label{Sw0}
S|_{\text{hw}}=2\pi\int_{0}^{\tau_\infty}\d\tau\, \|{\bf q}\|^2\,.
\ee
The latter can also be written using \eqref{potVq}, \eqref{WHL}, and \eqref{taurange} in an alternative form:
\be\label{Swhsplit} 
\begin{aligned}
S|_{\text{hw}}
&=-4\pi\int_{0}^{\tau_\infty}\d\tau\, V_{\bf q}\\
&=3\pi^3M^2_{\text{\tiny P}}L^2+4\pi\int_{0}^{\tau_\infty}\d\tau\, (E-V_{\bf q})\\
&\geq3\pi^3M^2_{\text{\tiny P}}L^2\,.
\end{aligned}
\ee
 The first term in the second line coincides with what one would find for a purely axionic theory, with dual saxions  ``frozen'' at their throat values $\ell_{i,*}$.\footnote{Up to a different convention for the Planck scale, the semi-wormhole action \eqref{Swhsplit} agrees with the one first derived in \cite{Giddings:1987cg}. It also agrees with \cite{Kallosh:1995hi} provided no Gibbons-Hawking term is added at the inner boundary of the half-wormhole.} The second  contribution can be identified with the integrated ``kinetic'' energy of the dual saxions, see \eqref{Econs}, and is hence always positive. 

The on-shell action \eqref{Swhsplit} does not include the imaginary contribution of the boundary term \eqref{aHboundary}. These are generically non-vanishing and will be taken into account in Section \ref{sec:phys}.

We remind the reader that the results obtained in this section rely on the assumption that a truncation of the action at the 2-derivative level is justified. As emphasized in Section \ref{sec:UVcutoff} this requires $L>1/\Lambda$, where $\Lambda$ is an appropriate UV cutoff. A reasonable bound would be $\Lambda<M_{\rm t}$, since the tower scale $M_{\rm t}$ represents the highest possible energy at which our four-dimensional EFT makes sense. However, in principle a wormhole with $L\lesssim M^{-1}_{\rm t}$ could still make sense in a weakly coupled higher-dimensional EFT. This is certainly not possible if $L^{-1}$ exceeds the species scale $M_{\text{sp}}$. So, we will  impose the less conservative bound 
\be\label{boundonL}
L^2>M^{-2}_{\text{sp}}
\ee
as the most fundamental consistency condition. Note that this requirement can always be satisfied by rescaling ${\bf q}$, so to increase $\|{\bf q}\|$.

The analysis of this section may also require the introduction of an IR cutoff, as discussed in general at the end of Section \ref{sec:UVcutoff}. Suppose for example that a more complete description includes a small saxion potential  $V_{\rm s}(s)$, see for instance  \cite{Andriolo:2022rxc,Jonas:2023ipa} for recent discussions and more references, as well as Section \ref{sec:QFTpheno}. This would contribute to the equations of motion and to the on-shell action. A sufficient condition for these corrections  to be negligible and not to significantly affect our results  is that $r^6V_{\rm s}\ll M_{\text{\tiny P}}^2L^4$. If the overall size of the potential is $V_{\rm s}\ll M_{\text{\tiny P}}^2/L^2$, such condition is certainly satisfied sufficiently close to the wormhole throat. In the same regime our equations of motion are reliable as well. Yet, for $r> M^{-1}_{\text{\tiny IR}}\sim (M_{\text{\tiny P}}^2L^4/V_{\rm s})^{1/6}$ both the classical configuration and its action can significantly depart from what we have found. 
We should thus either stay away from the asymptotic region, or properly take into account the non-trivial potential.

\subsection{Extremal wormholes and fundamental BPS instantons}
\label{sec:extremalBPS}

So far in this section we have not really exploited supersymmetry, and what we have found would hold even if the kinetic matrices of $\ell_i$ and $\calh_{3,i}$ were different from each other (see for instance the recent work  \cite{Cheong:2023hrj}). On the other hand supersymmetry forces such matrices to be identical, see \eqref{kinetic3}, and this allows for a particularly simple class of BPS extremal $E=0$ configurations. These solutions can be considered a generalization of the extremal wormholes found in \cite{Rey:1989xj} -- see also \cite{Gibbons:1995vg} for their ten-dimensional counterpart.

Given the form of the potential \eqref{potVq}, we see that taking $\dot \ell_i=-q_i$ obviously satisfies \eqref{Econs} with $E=0$ (another possible choice is $\dot \ell_i=q_i$ and can be viewed as being associated to conjugate charges). The equations of motion \eqref{elleom}
are manifestly solved as well. The condition can be easily integrated to ${\bm\ell}(\tau)={\bm\ell}(\tau_\infty)+{\bf q}(\tau_\infty-\tau)$. The wormhole metric \eqref{gauge1} is flat in the extremal case $E=0$, and in the gauge \eqref{rcoordin} our solution has $e^{2A}= 1$ with the coordinate $r\geq0$ spanning the entire space. In flat space \eqref{H3ansatz} implies that $\calh_{3,i}=-M^2_{\text{\tiny P}}*\d\ell_i$ and $\d\calh_{3,i}=2\pi\delta_4(0)$. As discussed in Appendix \ref{app:WHsusy1}, these relations can be generalized to  BPS multi-centered extremal wormhole solutions.

When considering extremal wormholes it is more convenient introduce a new proper radial coordinate 
\be\label{tauhat} 
\hat\tau\equiv \tau_\infty -\tau\,,
\ee
which has opposite orientation with respect to $\tau$ and is such that $r=\infty$ corresponds to $\hat\tau_\infty=0$, while $r=0$ corresponds to $\hat\tau_*=\tau_\infty$. This is motivated by the fact that in the extremal solution $r=0$ will soon be shown to correspond to a singular point, which is then moved to infinite $\hat\tau$-distance, and this simplifies our presentation. By solving \eqref{taurho} one obtains 
 \be 
 \hat\tau=\frac{1}{2\pi M^2_{\text{\tiny P}} r^2}\,
 \ee 
 and
\be\label{extrwh} 
{\bm \ell}(r)={\bm\ell}_\infty +{\bf q}\hat\tau={\bm\ell}_\infty +\frac{{\bf q}}{2\pi  M^2_{\text{\tiny P}}r^2}\,.
 \ee 
As promised, the solution is singular at $r=0$. The nature of the singularity will be explored shortly. Similarly, the conditions on ${\bm \ell}_\infty$ and ${\bf q}$ necessary for \eqref{extrwh} to lie within the allowed saxion domain identified in Section \ref{sec:perturbative} will be analyzed in detail below.

For the moment let us point out a suggestive fact. By recalling \eqref{elldef}, one can easily derive the identity $\calg^{ij}q_iq_j=\calg^{ij}q_i\frac{\d\ell_j}{\d\hat\tau}=-q_i\frac{\d s^i}{\d\hat\tau}$. Thus, paying attention to the relation \eqref{tauhat}, the on-shell action \eqref{Sw0} of our extremal solution reads\footnote{The coordinate $r$ here describes the whole space. Hence the notation ``half extremal wormhole" may be a bit misleading.}
\be\label{extrwhS}
S|_{\text{hw-extr.}}=2\pi\int_{\hat\tau_\infty}^{\hat\tau_*}\d\hat\tau\, \calg^{ij}q_iq_j=-2\pi\int_{\hat\tau_\infty}^{\hat\tau_*}\d\hat\tau\, q_i\frac{\d s^i}{\d\hat\tau}=2\pi q_is^i_\infty-2\pi q_is^i_*\,.
\ee
This action reminds us of that of a BPS instanton, which appears in the exponential of \eqref{expms} (with $s^i$ replacing $s^i_\infty$), but differs from it because of the term $-2\pi q_is^i_*$. Yet, inspecting \eqref{extrwh} we see that the dual saxions diverge as $r\rightarrow 0$. If the K\"ahler potential is as in \eqref{KP(s)}, this implies that ${\bm s}_*\to0$ in this limit. So, apparently, the two actions do in fact coincide. Note that this in particular implies that the charge vector ${\bf q}=\{q_i\}$ should belong to the set $\calc_{\rm I}$ of BPS instanton charges introduced in \eqref{BPSinstcharges}. These observations suggest that extremal wormholes be somehow related to fundamental BPS instantons. In order to clarify this relation we need to understand how to interpret the singularity.  

The singularity at $r=0$ indicates that the expression \eqref{extrwh} is not fully reliable. From a genuinely EFT perspective, that solution should be interpreted as viable only in a region $r\geq 1/\Lambda$ away from the origin, where $\Lambda$ lies in the range \eqref{IRcutoff} and a derivative expansion is meaningful. Below the UV cutoff the singularity should be regularized by some local counterterm placed around $r=0$. Restricting the integral in \eqref{extrwhS} to the reliable region $r\geq 1/\Lambda$ we obtain the on-shell contribution 
\be
S^{\Lambda}_{\rm bulk}=2\pi q_i(s^i_\infty-s^i_{\Lambda})\,,
\ee
where we defined
\be
s^i_{\Lambda}\equiv s^i(r=\Lambda^{-1})\,.
\ee
That action should then be supplemented by a cutoff-dependent localized term $S_{\rm loc}^\Lambda$ accounting for the effect of the unknown physics at scales above $\Lambda$. Such a localized term cannot be determined by sole considerations of the low-energy observer. Fortunately we have crucial information about the UV. The analysis presented in Appendix E of \cite{Lanza:2021udy} shows that fundamental BPS instantons localized at $r\sim1/\Lambda$ are captured within the EFT precisely by the introduction of a localized term of the form 
\be\label{SlocBPS} 
S^{\Lambda}_{\rm loc}=2\pi q_is^i_{\Lambda}=2\pi\langle {\bf q},{\bm s}_{\Lambda}\rangle \,,
\ee 
along with its supersymmetric counterpart $- 2\pi\ii\langle {\bf q},{\bm a}_{\Lambda}\rangle$. Hence the real part of the complete on-shell action for an extremal wormhole, including the localized term, should be given by 
\be\label{BPSinstaction}
S_{\text{\tiny BPS}}=S^{\Lambda}_{\rm bulk}+S^{\Lambda}_{\rm loc}=2\pi q_is^i_\infty\equiv2\pi\langle {\bf q}, {\bm s}_\infty\rangle
\ee
and exactly reproduces the expression appearing in \eqref{expms}. 

The BPS fundamental instanton acts as a magnetic source for the potentials $\calb_{2,i}$, as codified by the modified Bianchi identity $\d\calh_{3,i}=2\pi\delta_4(0)$. In the dual axionic formulation, this arises from the localized imaginary term written below \eqref{SlocBPS}, see Appendix E of \cite{Lanza:2021udy} for more details. Keeping track of the boundary term \eqref{aHboundary} arising in the duality transformation, one find that the BPS action, including its imaginary part, reads
\be\label{complBPSlocaction}
S_{\text{\tiny BPS}}-2\pi\ii\langle {\bf q},{\bm a}_{\infty}\rangle=-2\pi\ii \langle {\bf q},{\bm t}_{\infty}\rangle\,,
\ee
which is holomorphic, as expected by supersymmetry. Extremal wormholes are therefore a low-energy description of BPS fundamental instantons. 

Not all BPS fundamental instantons can be reliably described by extremal wormholes within the EFT, though. Those that admit such a description will be called {\em EFT instantons}, as in \cite{Lanza:2021udy}. This class is the subject of the next subsection.

\subsection{EFT instantons}
\label{sec:EFTinst}

EFT instantons are BPS instantons whose charges belong to the set 
\be\label{EFTinstc1} 
\calc_{\rm I}^{\text{\tiny EFT}}\equiv \calp\cap \calc_{\rm I}\,,
\ee
where  $\calp$ is the dual saxionic domain  \eqref{Pdomain}. A trivial example is provided by the model of Section \ref{sec:FtheoryP3}, in which $\calc_{\rm I}^{\text{\tiny EFT}}=\calc_{\rm I}$ is generated by the hyperplane divisor $D$. As a less trivial example,  in the model of Section \ref{sec:Ftheory1} $\calc_{\rm I}$ is generated by the effective divisors $E^1,E^2$, while $\calc_{\rm I}^{\text{\tiny EFT}}\subset \calc_{\rm I}$ is generated by the nef divisors $D^1,D^2$. In more general F-theory models satisfying  the conditions of footnote \ref{foot:movable}, the set of EFT instantons corresponds microscopically to D3-branes wrapping  movable divisors. If one instead considers the heterotic models of Section \ref{sec:het}, according to  \eqref{hetellcond} EFT instantons are represented by F1-strings wrapping movable curves in $\calp^{\rm het}_{\text{\tiny K}}$, and to NS5-branes wrapping the entire internal Calabi-Yau and possibly supporting additional F1-charge.

One reason for the requirement \eqref{EFTinstc1} is the following.   In the perturbative models in which the K\"ahler potential takes the form \eqref{KP(s)}, $\calp$ has conical shape. Since \eqref{extrwh} describes a straight dual saxionic line generated by ${\bf q}$, any ${\bf q}\in\calc_{\rm I}^{\text{\tiny EFT}}$ identifies  profiles completely contained in $\calp$. In general $\calp$ may be non-convex, and in that case ${\bf q}\in\calc_{\rm I}^{\text{\tiny EFT}}$ is not enough to ensure that the trajectory \eqref{extrwh} is completely contained in $\calp$ for any ${\bm\ell}_\infty\in\calp$. Yet, if ${\bf q}$ is in the interior of $\calp$ and ${\bm\ell}_\infty\propto{\bf q}$ then the solution certainly exists. By continuity we thus expect the same to be true for nearby choices of ${\bm\ell}_\infty$. We conclude that for any ${\bf q}\in \calc_{\rm I}^{\text{\tiny EFT}}$ one can judiciously adjust ${\bm\ell}_\infty$ in order to get extremal solutions belonging to $\calp$. 

Yet, the discussion of Section \ref{sec:extremalBPS} seems to suggest that any BPS instanton charge ${\bf q}\in \calc_{\rm I}$, either EFT or non-EFT, can be associated with a corresponding extremal wormhole provided $r>1/\Lambda$. As we will now show, for charges in the domain \eqref{EFTinstc1} there is no obvious obstruction to extending our solutions all the way to $r\sim1/M_{\text{sp}}$. Extremal wormholes carrying non-EFT charge can instead reach a finite distance boundary of $\calp$ at some finite radial distance, uncorrelated to the species scale. Hence, in a sense, non-EFT instantons are associated to strong coupling effects of non-gravitational nature. This does not happen if ${\bf q}$ belongs to \eqref{EFTinstc1}. In the following we will offer a few arguments motivating this claim. 

Let us first make some preliminary remark on the tensions of BPS strings. Recall that tensionless strings signal field space boundaries of $\calp$: tensionless EFT strings are associated with boundaries at infinite field space distance, while finite distance boundaries may be detected by tensionless BPS but non-EFT strings. As one approaches either of these two types of boundaries, the EFT becomes strongly coupled, but in a different way. To see this difference, recall that the semiclassical description of BPS strings roughly requires their tension to be larger than the EFT cutoff $\Lambda$: 
\be\label{Tpertcond} 
\Lambda^2<2\pi\calt_{\bf e}\,.
\ee
The violation of \eqref{Tpertcond}
by EFT and non-EFT strings has different implications, though. Indeed, in the former case,  the bound \eqref{SPbound}  would imply that 
\be 
\Lambda^2\geq M^2_{\calt}\geq M^2_{\rm sp}\,.
\ee
That is, the violation of \eqref{Tpertcond}
by an EFT string implies that the theory has entered a phase in which gravity does not admit any semiclassical gravitational description.   On the other hand, the tension of an elementary BPS but non-EFT string is not directly correlated with the species scale, and can be much smaller. So, the violation of \eqref{Tpertcond} by some elementary BPS but non-EFT string signals an EFT break-down of non-gravitational nature. Phases of this type  have been for instance studied in  \cite{Mayr:1996sh}.

Let us now use this consideration to better appreciate the distinction between EFT and non-EFT string instanton charges. Take first a BPS instanton charge ${\bf q}$ that is {\em not} EFT, i.e.\ that does not belong to \eqref{EFTinstc1}. Recalling \eqref{def:Cs}, this means that there may exist a BPS but non-EFT string charge ${\bf e}$ such that $\langle{\bf q},{\bf e}\rangle <0$.\footnote{This certainly happens only if $\calp=\calc_{\rm S}^\vee$. More generically $\calp\subset \calc_{\rm S}^\vee$, and there may be BPS instanton charges ${\bf q}$ in  $\calc_{\rm S}^\vee-\calp$, which would then have $\langle{\bf q},{\bf e}\rangle \geq 0$ for any ${\bf e}\in \calc_{\rm S}$.} In such a case, the corresponding tension {\em decreases} along the flow  \eqref{extrwh}, as one approaches the instanton:
\be 
\calt_{\bf e}(r)=\calt^\infty_{\bf e}-\frac{|\langle{\bf q},{\bf e}\rangle| }{2\pi r^2}\,.
\ee 
Hence $\calt_{\bf e}(r)$ violates \eqref{Tpertcond} at a critical radius  
\be 
r^2_{\rm cr}=\frac{|\langle{\bf q},{\bf e}\rangle|}{2\pi\calt^\infty_{\bf e}-\Lambda^2}\,.
\ee
Since  $2\pi\calt^\infty_{\bf e}-\Lambda^2$ can be much smaller than $M^2_{\rm sp}$, the theory can enter a non-gravitationally strongly coupled phase at distances $r_{\rm cr}\gg M^{-1}_{\rm sp}$. If ${\bf q}$ is not EFT and nevertheless $\langle{\bf q},{\bf e}\rangle \geq 0$ for any ${\bf e}\in\calc_{\rm S}$, we cannot run this argument, but we still expect the radial flow to reach some finite distance boundary of $\calp$ at some finite $r_{\rm cr}$ uncorrelated to $M_{\rm sp}$. It would be very interesting to more thoroughly investigate the possible non-perturbative effects associated with the various non-EFT instantons.

Suppose instead that ${\bf q}$ belongs to \eqref{EFTinstc1}. In this case the condition $\langle{\bf q},{\bf e}\rangle\geq 0$ is guaranteed. Hence, not only can one always find extremal wormholes that never exit $\calp$, but it is also guaranteed that any BPS (either EFT or non-EFT) string tension always increases as one approaches $r=0$,
\be\label{caltevol} 
\calt_{\bf e}(r)=\calt^\infty_{\bf e}+\frac{\langle{\bf q},{\bf e}\rangle }{2\pi r^2}\geq \calt^\infty_{\bf e}\,.
\ee  
Therefore, if \eqref{Tpertcond} is asymptotically satisfied by $\calt^\infty_{\bf e}$, then it will hold true for any $r$ and the theory will never enter a phase 
in which the semiclassical EFT description necessarily breaks down.

While along an EFT instanton flow there do not appear dangerous light string tensions, the theory formally exits the perturbative domain $\hat\calp_\alpha$ as $r\rightarrow 0$, since the dual saxions \eqref{extrwh} diverge in the ${\bf q}$-direction. This is signaled also by the divergence of BPS string tensions, which should instead 
 satisfy the condition \eqref{TMP} in order not to have a too strong backreaction. In particular, \eqref{TMP} should be satisfied by all elementary EFT string charges. (At the same time, \eqref{TMP} determines an upper bound on the string charges which can admit a weakly coupled world-sheet description.) The formal divergence of the string tensions as $r\rightarrow 0$ is not really a problem,  since one should actually restrict the flow to $r\geq   \Lambda^{-1}> M^{-1}_{\rm sp}$.  Combined with \eqref{caltevol}, this restriction implies that 
\be 
\calt_{\bf e}(r)\leq \calt_{\bf e}^\infty+\frac{\langle {\bf q},{\bf e}\rangle }{2\pi}\Lambda^2\,.
\ee
Hence, if \eqref{TMP} is asymptotically satisfied, then it is satisfied for any  $r\geq   \Lambda^{-1}$ provided  
\be\label{qeextcond1}
\langle {\bf q},{\bf e}\rangle
< \frac{(2\pi M_{\text{\tiny P}})^2}{\Lambda^2}\,.
\ee
Since $\Lambda\ll 2\pi M_{\text{\tiny P}}$, this condition can be easily satisfied if the charges are not too large. This shows that the condition \eqref{TMP} is certainly satisfied by the elementary EFT instantons.

So far we have thus shown that, approaching an EFT instanton string, tensions never vanish but rather diverge, and this latter behavior does not signal a loss of calculability. We next proceed our discussion by comparing the dominant EFT string scale \eqref{SPbounddef} to $r^{-1}$, which represents the energy scale probed by the solution \eqref{extrwh}. From \eqref{extrwh} it is easy to see that, as one approaches $r=0$, $M_{\calt}$ is identified by the tension of an EFT string charge ${\bf e}$ with minimal possible paring $\langle {\bf q},{\bf e}\rangle$. We will denote such charge as ${\bf e}_{\bf q}$. If $\langle {\bf q},{\bf e}_{\bf q}\rangle\geq 1$, then $M_{\calt}$ is always larger than $r^{-1}$ along the entire flow:
\be\label{rMsp} 
r^2M^2_{\calt}=2\pi r^2\calt^\infty_{{\bf e}_{\bf q}}+\langle {\bf q},{\bf e}_{\bf q}\rangle> \langle {\bf q},{\bf e}_{\bf q}\rangle\,. 
\ee
(Of course, no weakly coupled description is possible anymore below Planckian radii.) The situation is different if instead $\langle {\bf q},{\bf e}_{\bf q}\rangle=0$. In such a case $M_{\calt}\equiv 2\pi \calt^\infty_{{\bf e}_{\bf q}}$ would remain constant and there would exist a minimal radius at which $r^{-1}$ exceeds $M_{\calt}$. Note that this can happen only if ${\bf q}$ belongs to some infinite distance boundary of $\calp$, and viceversa. Indeed, for any ${\bf e}\in\calc_{\rm S}^{\text{\tiny EFT}}$, $\langle {\bf q},{\bf e}\rangle$ can be identified with an EFT string tension in Planck units evaluated at ${\bm\ell}={\bf q}$. If $\langle {\bf q},{\bf e}\rangle=0$ then this tension vanishes and ${\bm\ell}={\bf q}$ is at infinite field distance. Viceversa, if ${\bm\ell}={\bf q}$ is at infinite field distance, then it should correspond to a tensionless point of some EFT string.  We will see that the same  distinction will be relevant when we discuss regular wormholes in Section \ref{sec:WHAxiverse}. 

Finally, we emphasize  that the existence of fundamental instantons depends on the UV completion of the theory, while our identification of BPS and EFT instanton charges just uses  EFT data. Hence, a priori, one is not guaranteed that for any ${\bf q}\in\calc_{\rm I}$ or ${\bf q}\in\calc^{\text{\tiny EFT}}_{\rm I}$ there actually exists a corresponding fundamental instanton (rather than a multi-instanton configuration carrying total charge ${\bf q}$). Regarding this point,
we notice the analogy between our distinction between EFT and non-EFT BPS  instantons, and the extremal and non-extremal BPS particles, respectively, discussed in \cite{Alim:2021vhs} in the context of five-dimensional $\caln=1$ models. In particular, in our axionic context the counterpart of the BPS tower and sublattice versions of the  weak gravity conjecture (WGC) \cite{Arkani-Hamed:2006emk} proposed in \cite{Alim:2021vhs} may be formulated  as follows:\footnote{The five-dimensional setting  considered in \cite{Alim:2021vhs} is more directly related to an $\caln=2$ version of our four-dimensional framework, in which the BPS fundamental instantons may be regarded as BPS particle world-lines  wrapped along a compactified Euclidean  time in five dimensions. We leave  a more detailed investigation of this relation, and of its connection with wormhole effects in $\caln=2$ models, to future work.}  

\noindent {\bf BPS instanton tower  WGC}.   For any ${\bf q}\in\calc^{\text{\tiny EFT}}_{\rm I}$ there exists  an integer $k\geq 1$ such that there is a fundamental  EFT instanton of charge $k{\bf q}$. 

\noindent {\bf BPS instanton sublattice  WGC}.  There exists  an integer $k\geq 1$ such that for any ${\bf q}\in\calc^{\text{\tiny EFT}}_{\rm I}$ there is a fundamental  EFT instanton of charge $k{\bf q}$.

The latter conjecture is clearly stronger than the former, but both imply that there is an infinite subset of EFT instanton charges  (which coincides with the complete $\calc^{\text{\tiny EFT}}_{\rm I}$ if $k=1$) which is populated by fundamental instantons.    On the other hand, in analogy with the results of \cite{Alim:2021vhs}, we expect no tower or sublattice WGC to hold for BPS but non-EFT instanton charges. We will see how the above conjectures are compatible with some of the physical effects associated with non-extremal wormholes that we will discuss in Section \ref{sec:phys}.

\subsection{A non-standard BPS bound}
\label{sec:BPSbound}

Given the underlying supersymmetric structure of our EFT, one expects some kind of BPS bound relating the action of the (non-BPS) wormhole \eqref{Sw0} and the one of the BPS instanton \eqref{BPSinstaction}. In the following section we will show that regular wormholes solutions are generically expected to exist only for charges  belonging to (a subset of) $\calc^{\text{\tiny EFT}}_{\rm I}$. However, the observations of this subsection do not really rely on that, and we can just assume there exists a regular wormhole carrying a BPS instanton charge ${\bf q}\in\calc_{\rm I}$. Recalling \eqref{Econs}, we observe that the extremality bound \eqref{extrlowerb} is equivalent to $\|{\bf q}\|^2\geq\|\dot{\bm\ell}\|^2$. Employing \eqref{calg2} and \eqref{elldef2} we then have
\be
\|{\bf q}\|^2\geq \|{\bf q}\|\| \dot{\bm\ell}\|\geq |q_i\calg^{ij}\dot\ell_j|=|q_i\dot s^i|\,,
\ee
which is saturated only in the extremal BPS instanton case: $\dot{\bm\ell}=\pm{\bf q}$. The action \eqref{Sw0} is therefore subject to the following bound:
\be\label{BPSbound}
S|_{\text{hw}}\geq 2\pi \int^{\tau_\infty}_0\d\tau\,\left|q_i\dot s^i\right| \geq  2\pi\Big| q_i\int^{\tau_\infty}_0\d\tau\,\dot s^i\Big|=\big|S_{\text{\tiny BPS}}-2\pi \langle {\bf q}, {\bm s}_*\rangle\big|\,,
\ee
where $S_{\text{\tiny BPS}}$ is as in \eqref{BPSinstaction} and  we have again adopted  the  condition  \eqref{tauzero}. The bound involves in a crucial way the neck contribution $\langle {\bf q}, {\bm s}_*\rangle$. In the extremal case, as we have seen in the previous section, that contribution is actually absent once one takes into account the counterterm representing the insertion of the fundamental BPS instanton at the singularity. On the other hand, in the non-extremal wormhole case there is no singularity and no need for a local counterterm. A large enough $\langle {\bf q}, {\bm s}_*\rangle$ in that case may indicate that $S|_{\text{hw}}<S_{\text{\tiny BPS}}$, which would contradict the expectation that supersymmetric solutions saturate inequalities like \eqref{BPSbound}. This puzzling relation has indeed been observed in some explicit wormhole solutions (see e.g. \cite{Arkani-Hamed:2007cpn}) and will be encountered and further discussed in the following section. A possible correction to that puzzling result is suggested at the end of Section \ref{sec:univWH}.

\section{Wormholes in the $\caln=1$ axiverse}
\label{sec:WHAxiverse}

We are now ready to analyze explicit non-extremal (non-supersymmetric) wormhole configurations. We will restrict our focus on the large class of models characterized by  a K\"ahler potential of the form  \eqref{KP(s)}, or equivalently  by a kinetic potential of the form \eqref{calfP}. We would like to show that, under rather general conditions, {\em if} the degree $n$ of homogeneity of $P(s)$ satisfies $n\geq 3$ then the subset of EFT instanton charges
\be\label{calcWH} 
\calc_{\text{\tiny WH}}\equiv \{{\bf q}\in \calc^{\text{\tiny EFT}}_{\rm I}|\ \langle{\bf q},{\bf e}\rangle\geq 1,\ \forall {\bf e}\in \calc^{\text{\tiny EFT}}_{\rm S}\} \ \subset\ \calc^{\text{\tiny EFT}}_{\rm I}\,,
\ee
admits corresponding non-extremal wormhole configurations. Recalling \eqref{EFTinstc1}, the condition $\langle{\bf q},{\bf e}\rangle\geq 1$ for any ${\bf e}\in \calc^{\text{\tiny EFT}}_{\rm S}$ is basically equivalent to the requirement that either ${\bf q}$ belongs to the interior of $\calp$, or to  its finite field distance boundary $\del\calp|_{\text{fin.dist.}}$. Indeed, without imposing this condition   we would have $\langle{\bf q},{\bf e}\rangle\geq 0$ anyway. Hence the condition can be violated only if there exists an EFT string charge ${\bf e}$ such that $\langle{\bf q},{\bf e}\rangle=0$. But, as already discussed below \eqref{rMsp}, this would mean  that ${\bf q}$ should lie on an infinite field distance boundary of $\calp$.\footnote{\label{foot:CWH} 
$\calc_{\text{\tiny WH}}$ depends on the boundary properties  of the (dual) saxionic cone. 
A restriction of the perturbative domain, corresponding to the removal of some generator  ${\bf e}\in\del\Delta$ of $\Delta$, may result in the inclusion in the new $\calc_{\text{\tiny WH}}$ of some ${\bf q}\in \calc^{\text{\tiny EFT}}_{\rm I}$  that were excluded in the initial  
perturbative domain.} Note also that the existence of the wormholes does not depend on the overall sign of ${\bf q}$. Hence, if a wormhole exists for ${\bf q}\in \calc_{\text{\tiny WH}}$, one with $-{\bf q}\in \calc_{\text{\tiny WH}}$ must exist as well. We will refer to these latter wormholes as {\em anti}-wormhole. 
(Of course, the distinction between wormholes and anti-wormholes is only a matter of convention.) 

In the following subsections we will provide evidence that the set \eqref{calcWH} identifies the charges of physically acceptable non-extremal wormholes. The general {\emph{homogeneous solution}} is introduced in Section \ref{sec:univWH} and its reliability is investigated in Section \ref{sec:validityWH}. Considerations applying to more general wormhole solutions are presented in Section \ref{sec:asymptoticBC}. Wormholes arising from kinetic potentials with degree of homogeneity $n=3$ are discussed separately in Section \ref{sec:n=3}. Explicit realizations in string theory models are finally presented in Section \ref{sec:WHst}.

\subsection{The homogeneous solution}
\label{sec:univWH}

The sharpest statements can be made for wormhole charges  ${\bf q}\in\calc_{\text{\tiny WH}}$ belonging  to the {\em interior} of $\calp$.  For each of these charges we will  show that one can construct an explicit wormhole configuration,  no matter how complicated $P( s)$ or $\tilde P( \ell)$  are, as long as they are homogeneous functions of degree $n\geq 3$.  
In a certain sense, our result can be considered as a four-dimensional generalization of the string theory solutions found in \cite{Giddings:1989bq}, which also applies to many more string models. So, we will first focus on charges in the interior of $\calp$, while those belonging to the finite-distance boundary of $\calp$ will be discussed afterwards.

Now, since $\calp$ is conical, if ${\bf q}$ is in its interior the whole ``ray'' generated by ${\bf q}$ is contained in $\calp$. It is then natural to consider {\emph{homogeneous}} wormhole configurations with
\be\label{ellansatz} 
{\bm \ell}(\tau)=\tilde\ell(\tau)\,{\bf q}\,.
\ee
Note that the condition ${\bm \ell}\in\calp$ simply reads $\tilde\ell>0$. Inserting the ansatz \eqref{ellansatz} into \eqref{tau1dS} leads to   
\be\label{1dSsingle}
\begin{aligned}
2\pi \int\d\tau\,\Big[\,\frac{n}{4\tilde\ell^{2}}\Big(\frac{\d \tilde\ell}{\d \tau}\Big)^2+\frac{n}{4{\tilde\ell^{2}}}\Big]\,,
\end{aligned}
\ee
where we used the identity
$\calg^{ij}({\bf q})q_i q_j=\frac{n}2$, 
which follows from degree-$n$ homogeneity of $\tilde P$. Furthermore, one can verify that the homogeneity of $\tilde P$ guarantees that the equations of motion \eqref{elleom} for $\ell_i$ coincide with those derived from the effective action in Eq.\ \eqref{1dSsingle}. This drastically simplifies the study of homogeneous  multi-saxion wormholes.

We have basically reduced the problem to a simple axio-dilaton model with kinetic potential 
\be
\tilde\calf=n\log\tilde\ell\,
\ee
and (effective) unit charge, as the ones studied in \cite{Giddings:1987cg,Giddings:1989bq}. Similar results also follow. In particular, since the reduced effective potential is 
\be\label{potelltilde} 
\tilde V(\tilde\ell)=-\frac{n}{4\tilde\ell^2}\,,
\ee 
the wormhole radius \eqref{WHL} and the maximal proper time \eqref{taurange} are  given respectively by 
\be\label{Ltau} 
L^4=\frac{n}{12\pi^2\tilde\ell_*^2M^4_{\text{\tiny P}}}
\quad~~~\Rightarrow\quad~~~ \tau_\infty=\frac\pi2 \sqrt{\frac{3}{n}}\,\tilde\ell_* \,.
\ee
We note in particular that $L$ explicitly depends on $\tilde\ell_*$, but not on ${\bf q}$.
The explicit profile $\tilde\ell(\tau)$ can be obtained by integrating directly \eqref{Econs} in our reduced one-saxion model. Imposing the boundary conditions $\tilde\ell|_{\tau=0}=\tilde\ell_*$ and $\frac{\d\tilde\ell}{\d\tau}|_{\tau=0}=0$ fixes completely the solution:
\be\label{ellsol}
\tilde\ell(\tau)=\tilde\ell_*\cos\left(\frac{\tau}{\tilde\ell_*}\right)=\tilde\ell_*\cos\left(\frac\pi2 \sqrt{\frac{3}{n}}\,\frac{\tau}{\tau_\infty}\right)\,,
\ee
where the asymptotic value is determined by  
\be\label{linfstar}
\tilde\ell_\infty=\tilde\ell_*\cos\left(\frac\pi2 \sqrt{\frac{3}{n}}\right)\,. 
\ee

Since by definition $\tilde\ell$
must be positive, Eq.~\eqref{ellsol} describes a completely smooth wormhole if and only if
\be 
n\geq 3\,.
\ee
In all string theory models we are aware of, $n$ is an integer taking values from 1 to 7 -- see Sections \ref{sec:stringtheorymodels} and  \ref{sec:WHst} for some examples.  We hence have regular wormholes only in models with $n=3,\ldots,7$.
Note that in the limiting case $n=3$ wormholes are asymptotically degenerate, since the dual saxions reach the infinite field distance point ${\bm\ell}_\infty=0$ at  spatial infinity. They are nevertheless sensible and will be further investigated in Sections \ref{sec:n=3} and \ref{sec:n=3physics}. In Section \ref{sec:n=3}  we will  also more precisely see how wormholes corresponding to $n=1,2$ become singular at a radial distance of order $L$, and should then  be discarded. If instead $n>3$ the homogeneous wormholes are everywhere regular and asymptotically flow to a finite $\tilde\ell_\infty>0$. The condition $n>3$ represents a lower bound on the coefficient of the scalar kinetic term of the type first identified in \cite{Giddings:1987cg,Giddings:1989bq} for the case of simple dilatonic models, and subsequently generalized in \cite{Arkani-Hamed:2007cpn}.

We can next present the on-shell action. Observing that $\|{\bf q}\|^2=\frac{n}{2\tilde\ell^2}$, the integral appearing in \eqref{Sw0} is easily computed:
\be\label{Sw01}
\begin{aligned}
    S|_{\text{hw}}&=\frac{n\pi}{\tilde\ell_*}\tan\left(\frac\pi2 \sqrt{\frac{3}{n}}\right)\\
    &=2\pi\sin\left(\frac\pi2 \sqrt{\frac{3}{n}}\right)\langle{\bf q},{\bm s}_\infty\rangle
\,.
\end{aligned}
\ee
Sticking for now to the everywhere controllable solutions corresponding to $n>3$, we observe that the last line of \eqref{Sw01}, where the relation \eqref{tildelbound} was used, implies $S|_{\text{hw}}<S_{\text{\tiny BPS}}$, explicitly realizing the puzzling possibility mentioned in Subsection \ref{sec:BPSbound}. Perhaps the claim that non-extremal wormholes have actions smaller than the extremal BPS ones is a bit rushed, though. Other contributions, beyond the two-derivative approximation, may not always be negligible, at least for not too large wormhole charges. Consider for example the GB term \eqref{GBterm2}. Treating it as a perturbation and working at the first non-trivial order, it contributes to the on-shell action without modifying the two-derivative solution. This suggests that GB may contribute non-negligibly to the on-shell action of non-extremal wormholes, while of course it does not in the case of (flat) extremal wormholes. Interestingly, for sufficiently small charges we have checked that $S_{\text{\tiny BPS}}<(S+S_{\text{\tiny GB}})|_{\text{hw}}$ in some of the examples discussed in Section \ref{sec:WHst}, despite of $S|_{\text{hw}}<S_{\text{\tiny BPS}}$. What we can infer from this simple exercise is that higher order terms can significantly affect the relation between the on-shell action of non-supersymmetric wormholes and the corresponding BPS actions. Still, we cannot firmly conclude that the puzzle is solved. As soon as the GB starts competing with the leading two-derivative contribution, one should in principle also worry about other higher derivative corrections, which we did not consider because not under theoretical control. Furthermore, for sufficiently large charges we can show that in the case of homogeneous solutions $S_{\rm tot}|_{\text{hw}}\simeq S|_{\text{hw}}<S_{\text{\tiny BPS}}$. In fact, by rescaling ${\bf q}$ (with fixed ${\bm \ell}_*$), we can make $L$ and \eqref{Sw01} arbitrarily large (for fixed ${\bm s}_\infty$) and all higher-derivative corrections, including the GB one, arbitrarily small.\footnote{Note, however, that  large charge non-extremal wormholes may in fact be less relevant, not only because their contribution $e^{-S}$ is strongly suppressed, but also because their very existence is put under question once one has included the corrections due to the smaller-charge ones \cite{Coleman:1989ky,Coleman:1990tz}. }

Let us finally discuss the case in which  ${\bf q}\in\calc_{\text{\tiny WH}}$ belongs to some finite distance boundary  of $\calp$. Our proposal \eqref{calcWH} is based on the expectation  that, being at finite field distance, this boundary region can be identified with a restricted perturbative sector $\calp'\subset \del\calp|_{\text{fin.dist.}}$. Furthermore, it implicitly assumes that this boundary perturbative sector is sufficiently decoupled from any strongly coupled sector that may appear in the limit, so that a description according to our general scheme is reliable. If ${\bf q}$ lies in the interior of $\calp'$, or if $\calp'$ is one-dimensional, one can in such a case run the same arguments followed above for charges in the interior of $\calp$. While we do not have a general proof that these expectations are certainly realized, these are partially supported by the observation that the homogeneous function $\tilde P(\ell)$ appearing in \eqref{calfP} cannot degenerate, either vanishing or diverging, as one approaches  finite distance boundaries.\footnote{Here we are of course using the kinetic potential \eqref{calfP} to compute distances, neglecting corrections that may a priori be present. Again, we expect these corrections not to significantly affect our main points.} This can be understood in the following way. Take a finite distance boundary point  ${\bm \ell}\in\del\calp|_{\text{fin.dist.}}$. By definition, it can be connected  to an interior point ${\bm \ell}_0\in\text{Int}\calp$ by a  path $\gamma$ of finite length ${\rm d}(\gamma)<\infty$. The variation of $\log P$ along this path is always finite, since:
\be
\left|\log \tilde P(\ell)-\log \tilde P(\ell_0)\right|=\Big|\int_\gamma\frac{\d \tilde P}{\tilde P}\Big|=2\Big|\int_\gamma \calg^{ij}\ell_i\d\ell_j\Big|\leq n\int_\gamma\|\d{\bm \ell}\|=n\,\text{d}(\gamma)<\infty\,,
\ee 
where we have again used the degree-$n$ homogeneity of $\tilde P(\ell)$. Since  $\log \tilde P(\ell_0)$ is finite by assumption, $\log \tilde P(\ell)$ is necessarily finite too, and $\tilde P(\ell)$ can neither diverge nor vanish. Note that by continuity the restriction of $\tilde P(\ell)$ to $\del\calp|_{\text{fin.dist.}}$ is still homogeneous of degree $n$.

What said so far holds for charges ${\bf q}$ belonging to \eqref{calcWH}. This does mean that there cannot exist non-extremal wormhole solutions carrying other charges ${\bf q}$. In particular, there may exist wormholes carrying an EFT instanton charge belonging to the infinite distance boundary of $\calp$, for which there exists an EFT string charge ${\bf e}$ such that $\langle{\bf q},{\bf e}\rangle=0$. This for instance happens in presence of decoupled  perturbative sectors, which separately admit regular wormholes. However, as we will discuss in Section \ref{sec:validityWH}, precisely these wormholes appear  not to be consistent with the energetic bounds imposed by the species scale in the original perturbative regime. Nevertheless, wormholes carrying these charges may be compatible with the energetic bounds in appropriate restricted perturbative regimes, which are sufficiently far from the infinite distance  boundary of the original $\calp$.\footnote{\label{foot:modelinfdist} Consider a toy model with $\calp=\{{\bm\ell}=(\ell_1,\ell_2)| \ell_1,\ell_2\in \mathbb{R}_{\geq 0}\}$, so that $\calc_{\rm I}^\text{\tiny EFT}=\{{\bf q}=(q_1,q_2)| q_1,q_2\in \mathbb{Z}_{\geq 0}\}$,  and $\calf=\log \ell_1+4\log\ell_2$. The charges ${\bf q}=(q_1,0)$ and ${\bf q}=(0,q_2)$ are at infinite field distance in $\calp$ and must be excluded from \eqref{calcWH}: $\calc_\text{\tiny WH}=\{{\bf q}=(q_1,q_2)| q_1,q_2\in \mathbb{Z}_{> 0}\}$. On the other hand, the charges ${\bf q}=(0,q_2)$ with $q_2>0$ can still be regarded as elements of the set \eqref{calcWH} associated with the restricted perturbative regime obtained by assuming $\ell_1\simeq 1$, which is  parametrized only by $\ell_2$.}

\subsection{Perturbative conditions and relevant  scales}
\label{sec:validityWH}

In this subsection we verify that the homogeneous wormholes described in Section \ref{sec:univWH} are compatible with the perturbative domains defined in Section \ref{sec:perturbative} and with the energy bounds associated with the species scale (see Section \ref{sec:UVcutoff}). 

Let us first discuss the compatibility with the perturbative regime. Without  loss of generality, we can  assume that ${\bf q}\in\calc_{\text{\tiny WH}}$ belongs to the interior of $\calp$, since as  discussed in Section \ref{sec:univWH} the case of charges ${\bf q}$ belonging to a finite distance boundary of $\calp$ can be treated similarly by restricting the perturbative regime.   We note that, once  expressed in terms of saxions $s^i$,  the profiles \eqref{ellansatz} move along  straight saxionic radial lines as well. Indeed from \eqref{elldef2} and \eqref{tildePhom}, \eqref{ellansatz} is mapped to 
\be\label{siuniv}
s^i(\tau)=\frac{\calf^i({\bf q})}{2\tilde\ell(\tau)}\,,
\ee
where $\calf^i({\bf q})\equiv {\del\calf}/{\del \ell_i}({\bf q})$ is a charge dependent constant. Homogeneity also implies that 
\be\label{tildelbound} 
\langle{\bf q},{\bm s}\rangle=\frac{n}{2\tilde\ell}\ . 
\ee 
Combining \eqref{siuniv} and \eqref{ellsol}, we clearly see that ${\bm s}$ is rescaled by a number $>1$ as we move away from the wormhole throat. Hence, if the ${\bm s}_*$ belongs to the perturbative region $\hat\Delta_\alpha$, ${\bm s}_*(\tau
)$ will remain so for  all other values of $\tau$.  Since ${\bm s}_*$ can be freely chosen to be any point  of the ray in $\calp$ generated by ${\bf q}$, we can always choose  it such that the entire trajectory lies in the perturbative region $\hat\Delta_\alpha$. We are thus confident that our solution is within the ``large saxion'', and ``small dual-saxion'', regions identified in Section \ref{sec:perturbative}. The smallness of the dual saxions can be bounded as follows. If ${\bm s}_*$ belongs to $\hat\Delta_\alpha$, by repeating the same argument leading to \eqref{gammabound} with ${\bf q}$ now playing the role of $\tilde{\bf C}$, we get the lower bound $\langle{\bf q},{\bm s}_*\rangle\geq c_{{\bf q}}N/\alpha$, where $c_{{\bf q}}\geq 1$ is some ${\bf q}$-dependent constant scaling as $c_{k{\bf q}}=k c_{{\bf q}}$. By using \eqref{tildelbound} and the explicit form of the solution \eqref{ellsol}, this can be translated into an upper bound on $\tilde\ell$ along the entire wormhole
$\tilde\ell(\tau)\leq \tilde\ell_*\leq  \frac{n \alpha}{2c_{\bf q}N}$. Note that combined with \eqref{ellansatz}, this sets an upper bound on the value of the dominant EFT string scale  \eqref{SPbounddef} along the wormhole:
\be\label{MThomWH} 
M^2_{\calt}= 2\pi M^2_{\text{\tiny P}}\,\tilde\ell(\tau) \min_{{\bf e}\in\calc_{\rm S}^{\text{\tiny EFT}}}\langle {\bf q},{\bf e}\rangle\,. 
\ee
 For not loo large charges we therefore see that $M^2_{\calt}\lesssim2\pi \alpha M_{\text{\tiny P}}^2/N$. This has  the same scaling of the upper bound on $M_\gamma^2$ obtained by applying \eqref{axiSCus} to \eqref{vafaSC}, indicating that $M_{\calt}$ and $M_\gamma$  again agree parametrically with each other.

We can also get a {\em lower} bound on $M^2_{\calt}$.  Since we are assuming that ${\bf q}\in \calc_{\text{\tiny WH}}$, by our definition \eqref{calcWH} we have 
\be\label{qecond} 
\langle {\bf q},{\bf e}\rangle\geq 1\quad\forall {\bf e}\in\calc^{\text{\tiny EFT}}_{\rm S}\,. 
\ee 
Combined with \eqref{MThomWH}, this shows that  ${\bf q}\in \calc_{\text{\tiny WH}}$ implies
\be\label{Lambdamaxhom} 
M^2_{\calt} \geq 2\pi M^2_{\text{\tiny P}}\,\tilde\ell(\tau) \,,
\ee
which suggests that the species scale increases monotonically as one approaches the wormhole throat. Furthermore, from \eqref{Ltau} and \eqref{Lambdamaxhom} we get the identity
\be\label{lMTLrel}  
L^2 M_{\calt}^2|_{\tau=0}=\sqrt{\frac{n}3} \min_{{\bf e}\in\calc_{\rm S}^{\text{\tiny EFT}}}\langle {\bf q},{\bf e}\rangle\,.
\ee
Therefore, by \eqref{qecond} and the regularity condition $n\geq 3$, we deduce that 
\be\label{LMTcond} 
L M_{\calt}|_{\tau=0}\geq 1\quad~~~~~~~\text{if}\quad {\bf q}\in \calc_{\text{\tiny WH}}\,. 
\ee 
This conclusion   provides an important and non-trivial consistency check on the universal reliability of our homogeneous wormhole solutions, even for small charges ${\bf q}$. It is interesting to compare \eqref{lMTLrel} to the analogous \eqref{rMsp}  for extremal wormholes. In the latter case the charge ${\bf q}$ is bounded from {\emph{above}} by \eqref{qeextcond1}, while in the regular non-extremal case the charge can be arbitrarily large. For them such an upper bound does not exist, and $L M_{\calt}|_{\tau=0}$ can be made arbitrarily large by increasing ${\bf q}$ (and consistently rescaling ${\tilde\ell}$ to keep ${\bm\ell}$ of the same order). Note that these observations hold not only for $n>3$, but also for the marginally degenerate case $n=3$ discussed in Section \ref{sec:n=3}.

Finally, from \eqref{ellsol} and \eqref{Ltau}, we can also easily  compute the field distance  traveled by the saxions, or equivalently the dual saxions, along the half-wormhole. In  Planck units it is given by
\be\label{distell*inf} 
{\rm d}({\bm\ell}_*,{\bm\ell}_\infty)=\int^{\tau_\infty}_{0}\d\tau \sqrt{\calg^{ij}(\ell)\dot\ell_i\dot\ell_j}=\sqrt{\frac{n}2}\,\log\left[\cos\left(\frac\pi2\sqrt{\frac{3}{n}}\right)\right]^{-1}\,.
\ee
This is clearly well defined only for $n>3$.  For instance ${\rm d}({\bm\ell}_*,{\bm\ell}_\infty)|_{n=4}\simeq 2.2$,   while  ${\rm d}({\bm\ell}_*,{\bm\ell}_\infty)|_{n=7}\simeq 1.2$. In those cases the total displacement is moderately super-Planckian, indicating no severe problem associated to the swampland Distance Conjecture \cite{Ooguri:2006in}. This is even more so once we introduce an IR cutoff. Even if, as we will discuss in more detail in Section \ref{sec:n=3}, such an IR regularization is strictly necessary only if $n=3$, it is physically motivated also for $n>3$.

\subsection{Non-homogeneous generalization}
\label{sec:asymptoticBC}

As emphasized in Section \ref{sec:validityWH}, the endpoints  ${\bm s}_*$ and ${\bm s}_\infty$ of the saxionic trajectory  along a homogeneous wormhole   can be freely chosen to be any point  of the ray in $\calp$ generated by ${\bf q}$. In fact, this property comes from a more general scaling symmetry of the wormhole equations. Namely, if ${\bm\ell}(\tau)$ is a solution of the equations of motion then ${\bm\ell}'(\tau')=\lambda{\bm\ell}(\frac{\tau'}{\lambda})$ ($\lambda>0$) is a solution too, which starts from ${\bm\ell}'_{*}=\lambda {\bm\ell}_{*}$ (at $\tau'=0$) and arrives at  ${\bm\ell}'_{\infty}=\lambda {\bm\ell}_{\infty}$ (at $\tau'_\infty=\lambda\tau_\infty$). The corresponding saxionic flow is 
${\bm s}'(\tau')=\frac1\lambda {\bm s}(\frac{\tau'}{\lambda})$, which shows that for any given regular wormhole solution we can always make such a rescaling in order to ensure that the flow is  inside $\hat\Delta_\alpha$.  We would now like to use this property to identify more general wormhole solutions corresponding to an EFT instanton charge ${\bf q}\in\calc_{\text{\tiny WH}}$.  
In order to understand this possibility, it is useful to discuss some properties of the metric $\calg_{ij}$ and the potential $V_{\bf q}$. 

First of all, the saxionic metric admits as particular geodesics the radial directions.  This ``isotropy'' is broken by the presence of the potential $V_{\bf q}(\ell)$ which  in fact identifies ${\bf  q}$ as preferred direction. Indeed, assuming that ${\bf q}$ belongs to the interior of $\calp$, if we restrict along the ${\bf q}$ direction  as in \eqref{ellansatz}, we get $\calg_{ij}\frac{\del V_{\bf q}}{\del\ell_j}=\frac{q_i}{\tilde\ell}$. Hence along the radial direction identified by ${\bf q}$ the ``force'' associated with $V_{\bf q}$ is directed along $-{\bf q}$. The sign can be understood by noticing that
$V_{\bf q}(\lambda\ell)=\frac{1}{\lambda^2}V_{\bf q}(\ell)$ and then, since $V_{\bf q}$ is negative definite, it decreases as one moves radially towards the tip of the dual saxionic cone $\calp$. Correspondingly,  $V_{\bf q}(\lambda s)=\lambda^2 V_{\bf q}(s)$ and then $V_{\bf q}(s)$ decreases as we move radially away from the tip of the saxionic cone $\Delta$.

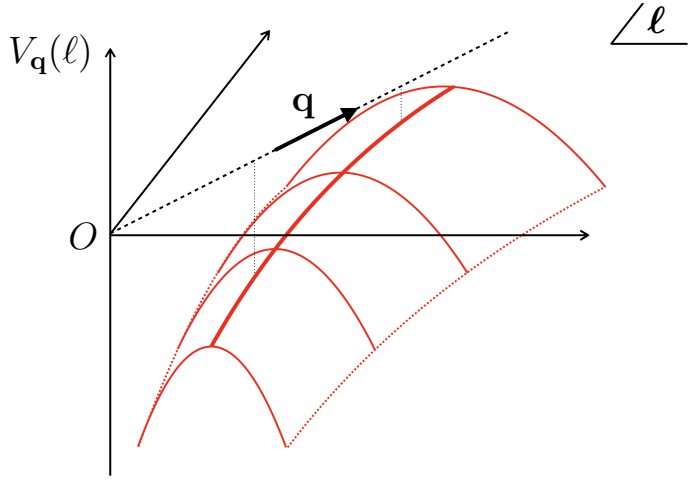
\begin{figure}[!htb]
\centering
\begin{overpic}[width=0.5\textwidth]{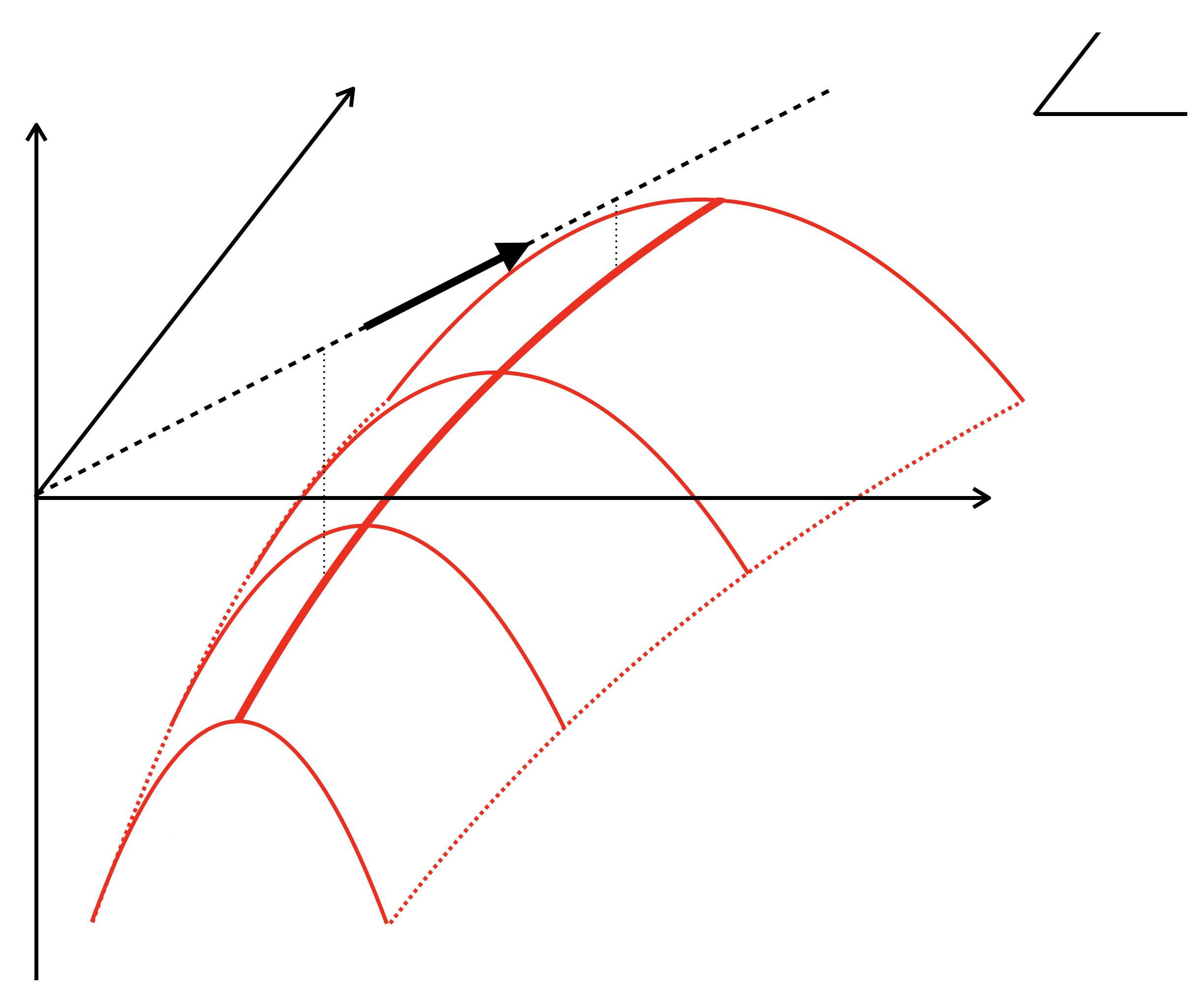}
\put (33,62) {\Large$\displaystyle {\bf q}$}
\put (-14,70) {\Large$\displaystyle V_{\bf q}({ \ell})$}
\put (-4,39) {\Large$O$}
\put (92,75) {\Large$\displaystyle { \bm \ell}$}
\end{overpic}
\caption{\small Schematic representation of the potential $V_{\bf q} (\ell)$. The potential has a hill shape with crest descreasing along the direction $-{\bf q}$. On the homogeneous solution the profile of the dual saxions follow the geodesic identified by the crest.}
\label{fig:potential}
\end{figure}

We can say something more about the shape of the potential $V_{\bf q}$. 
Still assuming that ${\bf q}$ belongs to the interior of $\calp$, take a particular point $\hat{\bm\ell}$ along the ray in $\calp$ generated by ${\bf q}$ and the corresponding point $\hat{\bm s}$ along the ray generated by ${\bm s}|_{{\bm \ell}={\bf q}}=\frac12\frac{\del\calf}{\del{\bm\ell}}({\bf q})$,\footnote{\label{foot:delFq}By $\frac{\del\calf}{\del{\bm\ell}}({\bf q})$ we mean the saxionic vector with components  $\frac{\del\calf}{\del{\ell_i}}({\bf q})$. Since ${\bf q}$ belongs to $\calp$, then by definition $\frac{\del\calf}{\del{\bm\ell}}({\bf q})$ belongs to $\Delta$.}     
and consider the saxionic plane $\cals_{\hat{\bm s}}$ passing through it and
 `orthogonal' to ${\bf q}$:
\be\label{saxhyperplane} 
\cals_{\hat{\bm s}}=\left\{{\bm s}\in\Delta\Big| \langle{\bf q},{\bm s}\rangle= \langle{\bf q},\hat{\bm s}\rangle\right\}\,.
\ee
By using again the homogeneity, one can easily show that $V_{\bf q}(\hat{s})=-\frac1n\left|\langle{\bf q},\hat{\bm s}\rangle\right|^2=-\frac1n\left|\langle{\bf q},{\bm s}\rangle\right|^2|_{\cals_{\hat{\bm s}}}$. On the other hand
$\left|\langle{\bf q},{\bm s}\rangle\right|^2\leq \|{\bf q}\|^2\|{\bm\ell}\|^2=-nV_{\bf q}({s})$. 
We deduce that 
\be\label{Vqbound}
V_{\bf q}(\hat{s})\geq V_{\bf q}({s})|_{\cals_{\hat{\bm s}}}
\ee
and that the potential restricted to $\cals_{\hat{\bm s}}$ has an absolute maximum at $\hat{\bm\ell}$. Hence $V_{\bf q}({s})$ is negative definite and has a hill shape with crest descending along ${\bm s}|_{{\bm \ell}={\bf q}}$ (which belongs to $\Delta$) or, correspondingly, $V_{\bf q}({\ell})$ has a hill shape with crest descending along $-{\bf q}$, as schematically depicted in figure \ref{fig:potential}. An additional confirmation comes by inspecting the Hessian of $V_{\bf q}({\ell})$,
\be 
\frac{\del^2 V_{\bf q} }{\del\ell_i\del\ell_j}|_{\hat\ell}=\frac14\calf^{ijkl}(\hat{\bm\ell})q_k q_l=\frac{1}{4\tilde\ell^4}\calf^{ijkl}({\bf q})q_k q_l=\frac{3}{2\tilde\ell^4}\calf^{ij}({\bf q})=-\frac{3}{\tilde\ell^2}\calg^{ij}(\hat{\bm\ell})\,,
\ee
where we have set $\hat{\bm\ell}=\tilde\ell{\bf q}$. 
As expected, this is negative definite and also reveals that as we move towards the origin of the dual saxionic space the steepness of the hill becomes more and more accentuated. 
 
It is now clear why, if   we start from a point ${\bm s}_*$ initially at rest along the ray identified by ${\bm s}|_{{\bm \ell}={\bf q}}$, the point is radially driven away from the tip of the  cone $\calp$ along the hill's crest down to the point ${\bm s}_\infty$. If $n\geq 3$ one gets precisely the saxionic counterpart of the homogeneous solution  described in Section \ref{sec:univWH} and, as already discussed, we can rescale it in order to make it lie inside $\hat\Delta_\alpha$ (or equivalently $\hat\calp_\alpha$). 
 
Restrict now to $n> 3$, with ${\bf q}\in\calc_{\text{\tiny WH}}$ still in the interior of $\calp$, so that the corresponding homogeneous wormhole is everywhere regular and does not degenerate asymptotically. If we slightly move ${\bm s}_*$ away from the  the  ray generated by ${\bm s}|_{{\bm \ell}={\bf q}}$, the potential will drive it further away the radial direction, but if ${\bm s}_*$ is close enough to the initial position one should still get a sensible wormhole solution, ending at some ${\bm s}_\infty$ inside $\Delta$. Combining this observation with the above scaling symmetry, we then expect to be able to fine-tune the initial value ${\bm s}_*$ to reach any final point  ${\bm s}_\infty$ inside $\hat\Delta_\alpha$. For instance, this is certainly true in the case of a diagonal metric (which corresponds to a completely  factorized  $P(s)=(s^1)^{n_1}(s^2)^{n_2}(s^3)^{n_3}\ldots$, with $n_1+n_2+\ldots>3$), which is a case already discussed in the literature -- see for instance \cite{Arkani-Hamed:2007cpn}.  The above argument suggests that this property holds also for more general models with non-factorizable $P(s)$. While we do not have a general proof, this expectation is confirmed by the following perturbative analysis.  Consider a given  homogeneous solution ${\bm\ell}_0(\tau)$ of the form \eqref{ellansatz}, with $\tilde\ell$ as in \eqref{ellsol}, and take a small deformation ${\bm\ell}(\tau)={\bm\ell}_0(\tau)+\delta{\bm\ell}(\tau)$. By expanding the action \eqref{tau1dS} up to quadratic order, we get the quadratic contribution
\be\label{quadaction}
\begin{aligned}
\pi\int \d\tau~\calg^{ij}({
\bf q
})\left\{\frac{\delta\dot\ell_i\delta\dot\ell_j}{\tilde\ell^2}+\left[1-\sin^2\left(\frac{\tau}{\tilde\ell_*}\right)\right]\frac{\delta\ell_i\delta\ell_j}{\tilde\ell^4}\right\}\,,
\end{aligned}
\ee
where we have ignored all boundary terms since we are only interested in the equations of motion for $\delta\ell_i$. By setting 
$ \delta\ell_i=\tilde\ell f_i\ $, \eqref{quadaction} can be rewritten as 
\be
\pi\int \d\tau~\calg^{ij}({
\bf q
})\left\{\dot f_i\dot f_j+\frac2{\tilde\ell^2}f_if_j\right\}\,.
\ee
from which the equations of motion for $f_i$ follow
\be\label{fisol} 
\ddot f_i=\frac{2}{\tilde\ell^2}f_i=\frac{2}{\tilde\ell^2 _* \cos^2\big(\frac{\tau}{\tilde\ell_*}\big)}f_i\,.
\ee
The general solution of \eqref{fisol} satisfying the initial condition $\dot f_i(0)=0$, corresponding to $\delta\dot\ell_i(0)=0$, is\,\footnote{
Without loss of generality we can impose the transversality condition $ \calg^{ij}({\bf q})f_{*i}q_j=0$, so that the deformation changes the wormhole radius \eqref{WHL} only to second order in $f_{*i}$:
\be\label{L4corr} 
L^4=L^4_{(0)}\left(1+\delta\right)\quad~~~~\text{with}\quad~~ \delta=\frac{6}{n}\calg^{ij}({\bf q})f_{*i}f_{*j}
\ee
Recalling \eqref{taurange}, this implies $\tau_\infty$ is only modified by second order corrections too.}
\be\label{pertsolf} 
f_i(\tau)=f_{*i}\left[1+\frac{\tau}{\tilde\ell_*}\tan\Big(\frac{\tau}{\tilde\ell_*}\Big)\right] = f_{*i}\left[ 1+ \frac{\pi}{2} \sqrt{\frac{3}{n}} \frac{\tau}{\tau_\infty} \tan \left(  \frac{\pi}{2} \sqrt{\frac{3}{n}}\frac{\tau}{\tau_\infty}\right)\right],
\ee
where $f_{*i}=f_i(0)$ is the integration constant representing the value of $f_i(\tau)$ at the throat. For the perturbative expansion to make sense $|f_i(\tau)/q_i|$ must remain small all along the flow, and since the perturbation grows monotonically towards $\tau=\tau_\infty$ this requires $|f_{*i}/q_i|\ll 1$.\footnote{For $n=3$ the perturbation \eqref{pertsolf} diverges at infinite radial distance and thus, for any arbitrarily  small initial values $f_{*i}$, the perturbative expansion  breaks down at some finite radius. Nevertheless, as discussed in Section \ref{sec:n=3}, it can still make sense by introducing an IR cutoff.}
Even though this implies that we cannot fully trust the quadratic approximation, the growing behavior of the deformation suggests that in a complete non-perturbative treatment it may be sufficient to pick a small $|f_i(0)/q_i|$ to allow for a much larger $|f_i(\tau_\infty)/q_i|$. This is confirmed by the numerical study presented in section \ref{sec:WHst1}, in which we inspect non-homogeneous wormhole solutions in some concrete string theory models.

So far we have mostly assumed that ${\bf q}\in\calc_{\text{\tiny WH}}$ belongs to the interior of $\calp$.  However, by reasoning as in Section \ref{sec:univWH}  one can extend the above arguments to the cases with ${\bf q}\in\calc_{\text{\tiny WH}}$ belonging to some finite distance boundary of $\calp$. Hence, at least for $n>3$, all these considerations support the main claim of this section: if the perturbative regime is described by a K\"{a}hler potential \eqref{KP(s)} (or a kinetic potential \eqref{calfP}) with homogeneous $P(s)$ (or $\tilde P(\ell)$) of degree $n\geq 3$, then for each ${\bf q}\in\calc_{\text{\tiny WH}}$  there exists a corresponding smooth wormhole solution, and an anti-wormhole solution carrying charge $-{\bf q}$. The marginally degenerate case $n=3$ will be discussed in Section \ref{sec:n=3}. 

Finally, let us revisit the consistency condition  $LM_{\calt}|_{\tau=0}\geq 1 $, already discussed in Section \ref{sec:validityWH} for homogeneous wormholes. In that case we proved \eqref{lMTLrel}, which shows that the consistency condition is always satisfied. Even though we are not able to derive such a sharp result for the more general wormholes discussed in the present subsection, we still expect no serious issue to show up for the following reasons. From \eqref{WHL} and \eqref{SPbounddef} we get
\be\label{LMT2} 
L^2M_{\calt}^2|_{\tau=0}=\sqrt{\frac{2}{3}}\|{\bf q}\|_*\min_{{\bf e}\in\calc_{\rm S}^{\text{\tiny EFT}}}\langle{\bm\ell}_*,{\bf e}\rangle\,.
\ee
Note that all wormhole solutions related by the scaling symmetry discussed at the beginning of this subsection have the same $LM_{\calt}|_{\tau=0}$, since \eqref{LMT2} is invariant under an overall rescaling of ${\bm\ell}_*$. Therefore, in order to investigate its behavior as we move ${\bm\ell}_*$ away from the radial direction identified by ${\bf q}$, with no loss of generality we can impose   ${\bm s}_*$ to lie in a saxionic hyperplane $\cals_*$ of the form \eqref{saxhyperplane}. Let  $\hat{\bm s}_*$ be the point of $\cals_*$ corresponding to $\hat{\bm\ell}_*\propto {\bf q}$. At this point \eqref{lMTLrel} holds. Recalling \eqref{potVq}, the inequality \eqref{Vqbound} translates into
\be 
\|{\bf q}\|_{*}\geq \|{\bf q}\|_{\hat{\bm\ell}_*}\,. 
\ee
Hence the first factor appearing on the right-hand side of  \eqref{LMT2} increases as ${\bm\ell}_*$ moves away from $\hat{\bm\ell}_*$. Of course, this may be compensated by a faster decrease of $\min_{{\bf e}\in\calc_{\rm S}^{\text{\tiny EFT}}}\langle{\bm\ell}_*,{\bf e}\rangle$. However, since $\langle{\bm\ell}_*,{\bf e}\rangle$ represents an EFT string tension in Planck units, its decrease  should correspond to approaching an infinite distance boundary of $\calp$. Qualitatively, we therefore expect the wormhole to exist only for  ${\bm\ell}_*$ sufficiently close to $\hat{\bm\ell}_*$ and away from the infinite distance boundaries, since otherwise there would be ``no room" for accommodating the corresponding flow. This suggests that an hypothetical decrease of $\langle{\bm\ell}_*,{\bf e}\rangle$ should not overwhelm the increase of $\|{\bf q}\|_{*}$, so that overall we expect no violation of $LM_{\calt}|_{\tau=0}\geq 1 $. In Section \ref{sec:WHst} we will provide numerical support to this expectation.

\subsection{Marginally degenerate case}
\label{sec:n=3}

In this section we will have a closer look at the  wormhole solutions corresponding to K\"ahler and kinetic potentials \eqref{KP(s)} and \eqref{calfP} with $n=3$. But before discussing that specific case, it is instructive to quantify how much  homogeneous wormholes with $n=1,2,3$ fail to be globally non-degenerate. This will reveal that the ``marginally'' degenerate case $n=3$ is in fact special and, as we will see in Section \ref{sec:phys}, also  physically relevant. 

We begin observing that, for general $n$, \eqref{Ltau} implies 
\be\label{Ln=3} 
\tilde\ell_*=\sqrt{\frac{n}{3}}\frac1{2\pi  M^2_{\text{\tiny P}}L^2}
\ee 
whereas \eqref{rtaurel} reads
\be\label{rtaurel1} 
\frac{\tau}{\tilde\ell_*}=\sqrt{\frac{3}{n}}\left[\frac{\pi}{2}-\arcsin\left(\frac{L^2}{r^2}\right)\right].
\ee
The homogeneous solution \eqref{ellsol} may thus be rewritten as 
\be\label{tildellr} 
\tilde\ell(r)=\tilde\ell_*\cos\left[\sqrt{\frac3{n}}\left(\frac{\pi}{2}-\arcsin\left(\frac{L^2}{r^2}\right)\right)\right]\,.
\ee
This equation  shows that if $n\leq 3$ then $\tilde\ell(r)$ vanishes at the radius
\be\label{degradius} 
r_{\rm deg}=\frac{L}{\sqrt{\sin\left[\frac{\pi}{2}\left(1-\sqrt{\frac{n}3}\right)\right]}}\,.
\ee
At the degeneration point the theory reaches the  tip ${\bm\ell}=0$ of the dual saxionic cone, which is at infinite field distance. This means that all BPS string tensions \eqref{strten} and the dominant EFT string scale $M_{\calt}$ defined in \eqref{SPbounddef}  vanish, and with them $M_{\rm t}$ and $M_{\rm sp}$ vanish as well. When this happens the solution can no longer be trusted. From \eqref{degradius} one gets $r_{\rm deg}|_{n=1}\simeq 1.27 L$ and $r_{\rm deg}|_{n=2}\simeq 1.88 L$. Therefore, for $n=1,2$ already at $r_{\rm deg}\sim L$ the solution degenerates and does not appear to make any physical sense.  So, if the $n=1,2$ the charges belonging to the set \eqref{calcWH} can only be regarded as charges of fundamental EFT instantons of the type discussed in Section \ref{sec:extremalBPS}. 

The story is very different for the $n=3$ case. First of all, in that case $r_{\rm deg}|_{n=3}\to \infty$. So the solution is everywhere well defined and degenerates only asymptotically at spatial infinity. The degeneration  ${\bm \ell}(r\to\infty)\rightarrow 0$ is of course associated to an infinite field distance limit, as it was for the $n=1,2$ cases, but here this degeneracy occurs at infinite radius. This is  not a real concern since it is natural to introduce an IR cutoff which sets a finite maximal radial distance below which the solution is required to be non-degenerate. The IR cutoff may correspond to a physical mass scale $M_{{\text{\tiny IR}}}$, as mentioned at the end of Section \ref{sec:UVcutoff}. Alternatively, as we will presently  elaborate upon, it may correspond to a new lower Wilsonian cutoff $\Lambda_{\text{\tiny IR}}\ll L^{-1}$ associated with an infra-red EFT. In any event, the IR cutoff allows us to make physical sense of the marginally degenerate solution. On the contrary, there is no way to interpret the $n<3$ configurations as wormholes connecting asymptotically flat spaces because $r_{\rm deg}|_{n=1,2}\sim L$ would force us to take $M_{{\text{\tiny IR}}},\Lambda_{\text{\tiny IR}}\gtrsim1/L$, leaving essentially no room for such an interpretation. For completeness we emphasize that there is instead no urgent reason to introduce a regulator for $n>3$, because those solutions are regular everywhere. Nevertheless, physically we should expect a non-trivial $M_{{\text{\tiny IR}}}$ in those cases as well, or  we may still be interested in introducing a lower Wilsonian cutoff $\Lambda_{\text{\tiny IR}}$, as we will do below -- see also the comments below \eqref{boundonL} and \eqref{distell*inf}.

There is another peculiarity of the marginally degenerate wormhole. Indeed, for $n=3$ Eq.\ \eqref{tildellr} reduces to
\be\label{homn=3} 
\tilde\ell(r)=\frac{1}{2\pi M^2_{\text{\tiny P}}r^2} \quad\Rightarrow\quad {\bm \ell}(r)=\frac{{\bf q}}{2\pi M^2_{\text{\tiny P}}r^2}\,.
\ee
We then see that, as a function of $r$, the dual saxionic profile is the same as the extremal BPS profile \eqref{extrwh} with vanishing (and hence degenerate) asymptotic saxionic value ${\bm\ell}_\infty=0$. While this observation is strictly valid  only  for homogeneous solutions, it remains approximately true also for more general wormholes corresponding to $n=3$. To see this we look for new $n=3$ solutions via a perturbative expansion around the homogeneous configuration as we did in the previous section. At leading order $\delta\ell_i=\ell_i-\ell_i^0 \equiv \tilde \ell f_i$ with $f_i$ as in  \eqref{pertsolf}. At the same order, specifying $n=3$ and employing \eqref{rtaurel1}, the result is 
\begin{align}
{\bm\ell}\label{moregenn=3}
&\simeq{\bf q}\tilde\ell_*\,\cos\Big(\frac{\tau}{\tilde\ell_*}\Big)+{\bm f}_{*}\,\tilde\ell_*\Big[\cos\Big(\frac{\tau}{\tilde\ell_*}\Big)+\frac{\tau}{\tilde\ell_*}\sin\Big(\frac{\tau}{\tilde\ell_*}\Big)\Big]\\\nonumber
&=\frac{{\bf q}}{2\pi M^2_{\text{\tiny P}}r^2}+\frac{{\bm f_*}}{4 M^2_{\text{\tiny P}}L^2}\left[1+\calo(L^4/r^4)\right],
\end{align}
where $f_{*i}$ are arbitrary constants satisfying $|f_{*i}/q_i|\ll1$ and in the second line we assumed $r/L\gg 1$. Strictly speaking the perturbative expansion breaks down at very large distances $r^2/L^2>|q_i/f_{*i}|$, as shown in section \ref{sec:asymptoticBC}. Nevertheless, we already emphasized that an IR cutoff is necessary. We can thus safely conclude that, for $n=3$ and within the regime of validity of our EFT, there exist more general solutions  that have approximately the same saxionic profile as a BPS extremal wormhole, including a (small but non-vanishing) ${\bm\ell}_\infty={{\bm f_*}}/{(4 M^2_{\text{\tiny P}}L^2)}$. In the following subsection we will provide numerical evidence that the same conclusion generalizes to non-perturbative deformations of the homogeneous solution. The only qualitative difference between the $n=3$ non-extremal half-wormholes and the extremal ones is in the metric (which is flat in the extremal case) and in the  range or the radial coordinate $r$, which is restricted to $r\in[L,\infty)$ in the half-wormhole case. Yet, the two solutions are {\emph{completely indistinguishable}} to an observer at distances $r\gg L$ from the wormhole throat, up to corrections of order $\calo(L^4/r^4)$.

This observation extends to the  on-shell action as well.  Let us start from the homogeneous solution first, whose complete on-shell action \eqref{Sw01} diverges, since  the solution degenerates asymptotically. However, as already emphasized, it is natural to introduce an IR regulator. To compute the action we hence introduce an IR cutoff $\Lambda_{\text{\tiny IR}}$ as above.  We thus assume $\Lambda_{\text{\tiny IR}} \ll L^{-1}$ and remove the region $r>\Lambda_{\text{\tiny IR}}^{-1}$, for some $\Lambda_{\text{\tiny IR}}$ comfortably within the region \eqref{IRcutoff}, from the $n=3$ version of \eqref{Sw0}.  This gives 
\begin{align}\label{apprSn=3}
S|_{\text{hw}}^{\Lambda_{\text{\tiny IR}}}
=&2\pi\int_{\tau_*}^{\tau_{\text{\tiny IR}}}\d\tau\, \|{\bf q}\|^2=\frac{3\pi}{\tilde\ell_*}\tan\frac{\tau_{\text{\tiny IR}}}{\tilde\ell_*}\\\nonumber
=&{2\pi}\langle{\bf q},{\bm s}_{\text{\tiny IR}}\rangle\left[1+{\calo}(\Lambda_{\text{\tiny IR}}^4L^4)\right]
\end{align}
where $\tau_{\text{\tiny IR}}\equiv\tau(r=\Lambda^{-1}_{\text{\tiny IR}})$ and ${\bm s}_{\text{\tiny IR}}\equiv {\bm s}(\tau_{\text{\tiny IR}})$. Moreover, we used \eqref{rtaurel1} and \eqref{tildelbound} with $n=3$ and  $\tilde\ell_{\text{\tiny IR}}=\tilde\ell_*\cos(\tau_{\text{\tiny IR}}/\tilde\ell_*)$ -- see \eqref{ellsol}. Up to $\calo(\Lambda_{\text{\tiny IR}}^4L^4)$ corrections, the cutoff on-shell action \eqref{apprSn=3} is therefore equivalent to the localized operator \eqref{SlocBPS} accounting for the insertion of a BPS fundamental instanton in an EFT with Wilsonian cutoff $\Lambda_{\text{\tiny IR}}$. The cutoff on-shell action for the more general solutions \eqref{moregenn=3} is again given by \eqref{apprSn=3} within the quoted uncertainty. Analogously, one can verify that the Gauss-Bonnet term evaluated on \eqref{homn=3} vanishes with $\Lambda_{\text{\tiny IR}}^2L^2\to0$. 

We also note that for $n=3$ \eqref{Ln=3} reduces to
\be\label{n=3Larb}
\tilde\ell_*=\frac{1}{2\pi M^2_{\text{\tiny P}} L^2}\,,
\ee
and that this formula is not affected  by the above perturbations up to corrections of order $|f_{*i}|^2$ -- see section \ref{sec:asymptoticBC} for more details. We emphasize that one can freely change $\tilde\ell_*$  and $L$ without affecting the profile \eqref{homn=3} and its perturbation \eqref{moregenn=3}, provided $f_*$ is appropriately chosen. This implies that we can consider $L$ and ${\bm\ell}_\text{\tiny{IR}}$
as independent quantities. This is in sharp contrast with what happens in the  $n>3$  wormholes, in which  ${\bm \ell} (r)$ depends explicitly on $L$, which is then fixed by  ${\bm\ell}_\infty$. 

We finally show that  the $n=3$ IR-regularized version of \eqref{distell*inf} reads
\be 
{\rm d}({\bm\ell}_*,{\bm\ell}_{\text{\tiny IR}})=-\sqrt{\frac{3}2}\log( {\Lambda_{\text{\tiny IR}}^2L^2})\,,
\ee
which e.g.\ for  $\Lambda_{\text{\tiny IR}}^2L^2=10^{-1}$ or $\Lambda_{\text{\tiny IR}}^2L^2=10^{-2}$ gives the moderately super-Planckian distances ${\rm d}({\bm\ell}_*,{\bm\ell}_{\text{\tiny IR}})\simeq 2.8$ or ${\rm d}({\bm\ell}_*,{\bm\ell}_{\text{\tiny IR}})\simeq 5.6$, respectively. This suggests that marginally degenerate wormholes (and their deformations) may have potential control issues related to the Distance Conjecture \cite{Ooguri:2006in}.  Nevertheless,  because of the almost-BPS nature noticed above,  
we expect their physical effects to be protected by supersymmetry and hence to be reliable. This expectation will be strengthened by the discussion of  Section \ref{sec:n=3physics}, where we will clarify the physical meaning of the IR regularization leading to \eqref{apprSn=3} and further elaborate on the almost-BPS nature of the marginally degenerate wormholes.

\subsection{Examples in string theory models}
\label{sec:WHst}

Let us now discuss explicit realizations of our wormholes in string theory models.

\subsubsection{Wormholes in F-theory models}
\label{sec:WHst1}

Consider first the F-theory models  discussed in subsection  \ref{sec:FIIB}. 
In this case case, we recall that the set $\calc_{\rm I}$  of BPS instanton charges can be identified with the cone of effective divisors. Microscopically, these instantons correspond to Euclidean D3-branes wrapping  effective divisors. In order to identify the possible wormhole configurations, we must first identify the set \eqref{EFTinstc1} of possible EFT instanton charges. Recalling the discussion  around \eqref{calpdec}, we immediately conclude that ${\bf q}=q_aD^a\in\calc^{\text{\tiny EFT}}_{\rm I}$ is a nef divisor in some space $X'$ which can be obtained from $X$ through small transitions. In the following 
we will for simplicity take ${\bf q}=q_aD^a$ to be  nef already in $X$. Note that the cone generated by nef divisors is contained in the cone generated by movable divisors \cite{kawamata1988crepant}.  Any movable effective divisor $D\subset X$ admits a multiple $mD$,  for some integer $m\geq 1$, that can be freely deformed along the entire $X$. Hence for any EFT instanton of charge ${\bf q}=q_aD^a\in\calc^{\text{\tiny EFT}}_{\rm I}$, there exists an EFT instanton of charge $m{\bf q}$, with $m\geq 1$, corresponding to a non-isolated Euclidean D3-brane that can freely explore the entire internal space.  Even though if $m>1$ a single Euclidean D3-brane of charge ${\bf q}$ does not have such a property,  $m$  such D3-branes can recombine into a single non-isolated one that does. This shows the intrinsically gravitational nature of all these EFT instantons (for both $m=1$ or $m>1$), which would not exist if we decoupled gravity by decompactifying the internal space. We expect these qualitative features to characterize the UV completion of  EFT instantons also in other models. We will provide other examples in the following subsections. Note also that, if correct,  the BPS instanton tower and sublattice WGCs of Section \ref{sec:EFTinst}  imply the existence of an infinite set of Euclidean D3-brane wrapping irreducible movable divisors  populating a (possibly non-strict) infinite subset of $\calc^{\text{\tiny EFT}}_{\rm I}$.     

As already mentioned below \eqref{FFK},  the $\ell_a$-sector alone  corresponds to $n=3$. This means that, in addition to the extremal wormholes corresponding to EFT instantons, it only allows for the marginally degenerate wormholes of Section \ref{sec:n=3}. As anticipated therein and further elaborated in Section \ref{sec:phys}, it makes perfect sense to regularize these wormholes by introducing an IR cutoff. However, for illustrative purposes, in this section we prefer to first discuss everywhere regular wormholes, which can be obtained by extending the $\ell_a$-sector in order to get an overall $n>3$. One such extension is provided by Sen's weak coupling limit, which adds the IIB axio-dilaton, and hence  one  extra dual saxion $\hat\ell$ as defined in \eqref{dualdilaton}, getting an overall $n=4$.\footnote{This is just one simple possibility. For instance, we may  also turn on charges corresponding to the (s)axionic sector appearing in some asymptotic region of the complex structure moduli space.} In this case we know that the family of homogeneous wormholes certainly exists for any charge vector $(\hat q,{\bf q})$ belonging to the set $\calc_{\text{\tiny WH}}$ defined in \eqref{calcWH}. Since the set of EFT string charges take the form $(\hat e,{\bf e})$, where $\hat e\geq 0$ and ${\bf e}=e^a\Sigma_a$  is a (possibly trivial) movable curve,
this means that  $\hat q>0$ and  ${\bf q}=q_aD^a$ represents a  nef divisor in $X$ with non-vanishing (positive) intersection with any non-trivial movable curve. In these wormholes the dual saxions have a profile of the form $\hat\ell(\tau)=\hat q\, \tilde\ell(\tau)$  and ${\bm\ell}(\tau)=\ell_a(\tau)D^a={\bf q}\,\tilde\ell(\tau)$, where  $\tilde\ell(\tau)$ is as in \eqref{ellsol} with $n=4$. Note that, since we will consider the two sectors as decoupled as in \eqref{Fcalf}, by ``forgetting'' the $\hat\ell$ direction, one gets information on the profile of the marginally degenerate wormhole of charge ${\bf q}$ as well. 

\begin{figure}[!htb]
\centering
  \centering
  \includegraphics[width=0.6\textwidth]{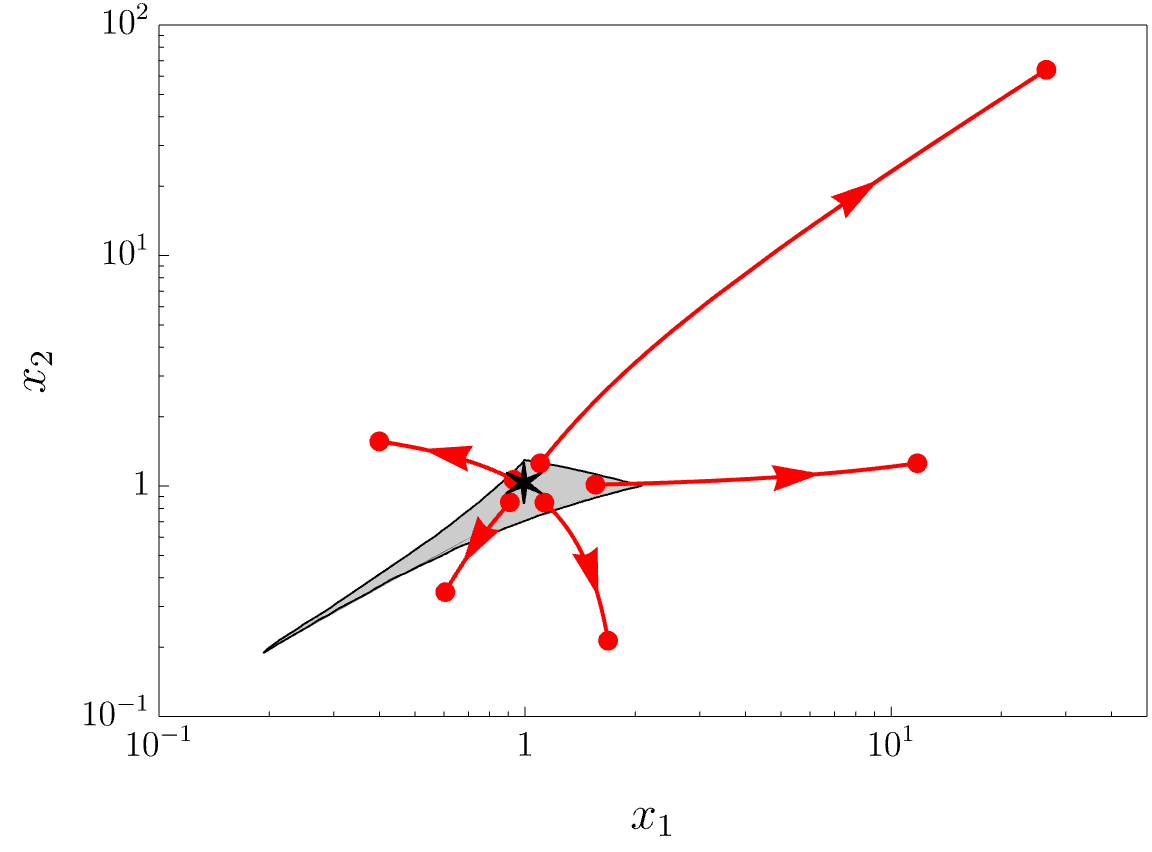}
 \caption{\small Sample of $\bm{x}(\tau)$ trajectories corresponding to the charge vector $\mathbf{q}=(1,1,1)$  in the F-theory model 2 of Section \ref{sec:Ftheory1} (with $p=3$),  with throat position $\bm{x}_*$ displaced from the homogeneous solution, indicated by a black star in the plot. The gray region is an interpolation of the red $\bm{x}_*$ acceptable domain of figure \ref{fig:model1domain}. Small displacements around the star lead to completely different trajectories. }
  \label{fig:model1trajectories}
\end{figure}

According to the general arguments of Section \ref{sec:asymptoticBC}, we also expect that there exist other wormhole solutions whose initial position   $(\hat\ell_*,{\bm \ell}_*)$ is non-aligned but sufficiently closed  to the direction identified by the charge vector $(\hat q,{\bf q})$. 
Note that the scaling symmetry discussed at the beginning of Section \ref{sec:asymptoticBC}  identifies equivalence classes of solutions. It is hence convenient to rephrase our discussion in the projectivized dual saxionic cone, obtained by modding out their overall rescaling. In the present case, it is conveniently parametrized  by the vector 
\be 
{\bm x}(\tau)\equiv x_a(\tau)D^a\equiv\frac{\hat q\,{\bm\ell}(\tau)}{\hat\ell(\tau)}\,,
\ee
which characterizes wormholes that are not equivalent under scaling symmetry. Note that the homogeneous  solution corresponds to the constant profile ${\bm x}(\tau)\equiv {\bf q}$ (hence with ${\bm x}_*={\bm x}_\infty= {\bf q}$), while other equivalence classes of solutions correspond to any other profile ${\bm x}(\tau)$ completely contained in $\calp_{\text{\tiny K}}$ and with  ${\bm x}_*\neq {\bf q}$. Not only do we expect to find admissible solutions for any  ${\bm x}_*$ in a sufficiently small neighborhood of ${\bf q}$, but we also expect that we can tune  ${\bm x}_*$ in this neighborhood to get any ${\bm x}_\infty\in \calp_{\text{\tiny K}}$. Note also that by the scaling symmetry, for  any admissible solution ${\bm x}(\tau)$,  we can always find a non-empty subset of  corresponding  flows $(\hat\ell(\tau),{\bm \ell}(\tau))$ that lie inside the $\alpha$-saxionic convex hull.    

\begin{figure}[!htb]
\centering
\begin{subfigure}[t]{.49\textwidth}
  \centering
  \includegraphics[width=\textwidth]{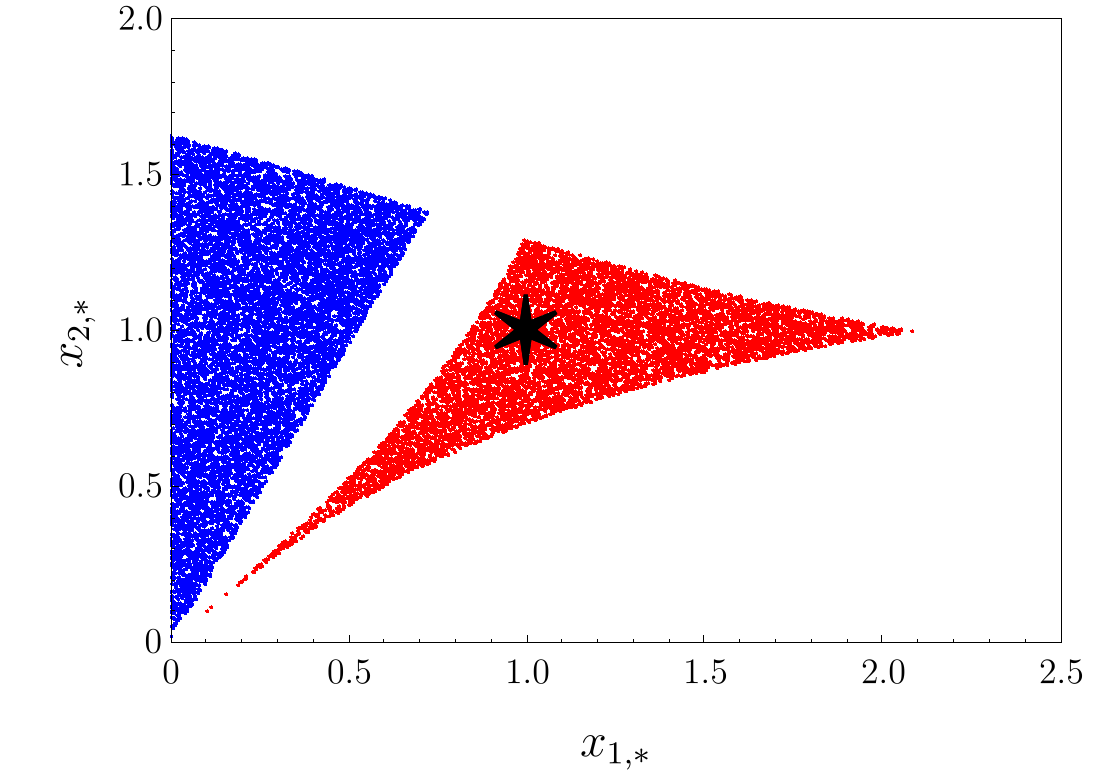}
  \caption{\small Scanned throat values ${\bm x}_*\equiv{\bm x}|_{r=L}$.}
  \label{fig:model1domain}
\end{subfigure}%
\hfill
\begin{subfigure}[t]{.49\textwidth}
  \centering
  \includegraphics[width=\textwidth]{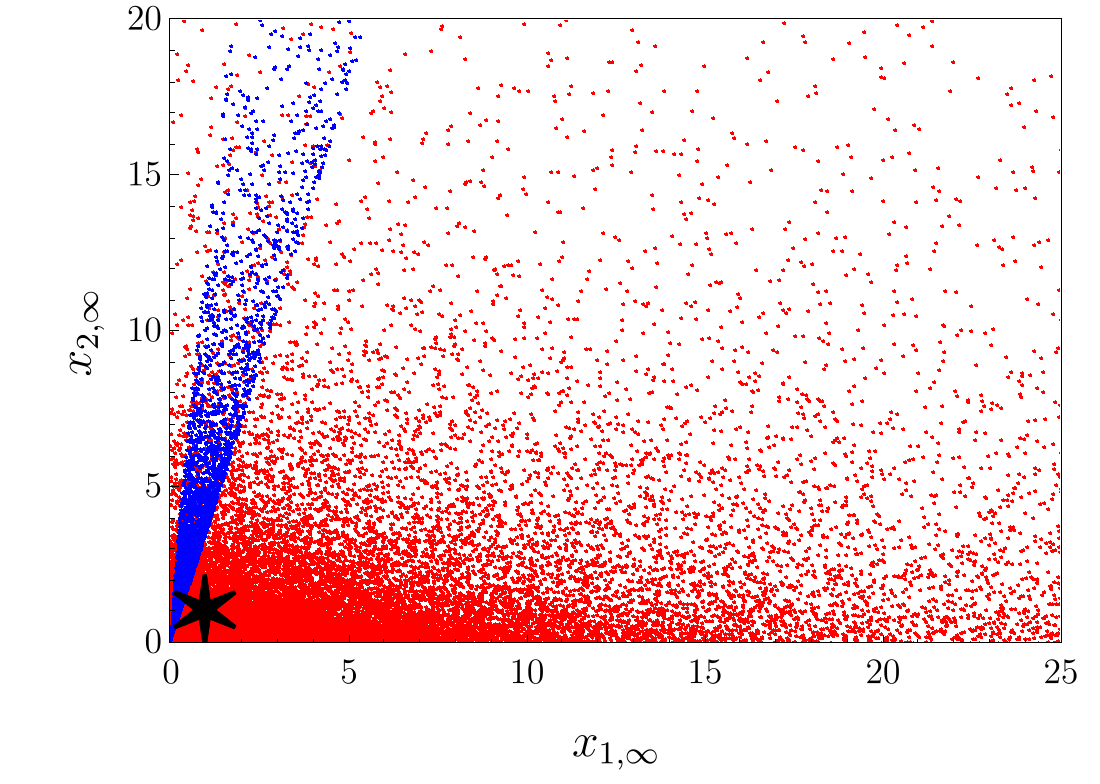}
  \caption{\small Corresponding asymptotic values ${\bm x}_\infty\equiv{\bm x}|_{r=\infty}$.}
      \label{fig:model1inf}
\end{subfigure}
\caption{\small Numerical scan obtained by integrating the equations of motion \eqref{elleom} with charge vector $\mathbf{q}=(1,1,1)$ in the F-theory model of Section \ref{sec:Ftheory1} with $p=3$ and with random initial conditions for the dual saxions. The first plot contains the allowed throat points ${\bm x}_*$ identified by the scan: all the points in the box have been randomly scanned, but we just show the ones that lead to acceptable solutions, i.e.\ with $x_1,x_2>0$ all along their trajectory. The shape of the domains of acceptable solutions are clearly distinguished and visible and have been indicated with different colors. The black star denotes the point corresponding to the homogeneous solution and the red dots in its neighborhood can be identified with the solutions predicted by our general arguments, while the blue set on the left represents ``accidental'' solutions, not predicted by our argument. The second plot shows the asymptotic values ${\bm x}_\infty$ associated to the acceptable solutions. We can again  clearly distinguish two sets of points, corresponding to the two 
disjoint sets  of the first plot: the set corresponding to the red neighborhood of the black star in the first plot, which is spread out all over $\mathbb{R}^2_{>0}$, as predicted by our general argument; the set corresponding to the accidental  solutions,  which identifies the blue thin denser cone which is visible in the second plot.  }
\label{fig:model1}
\end{figure}
\begin{figure}[!h]
\centering
\begin{subfigure}[t]{.49\textwidth}
  \centering
  \includegraphics[width=\textwidth]{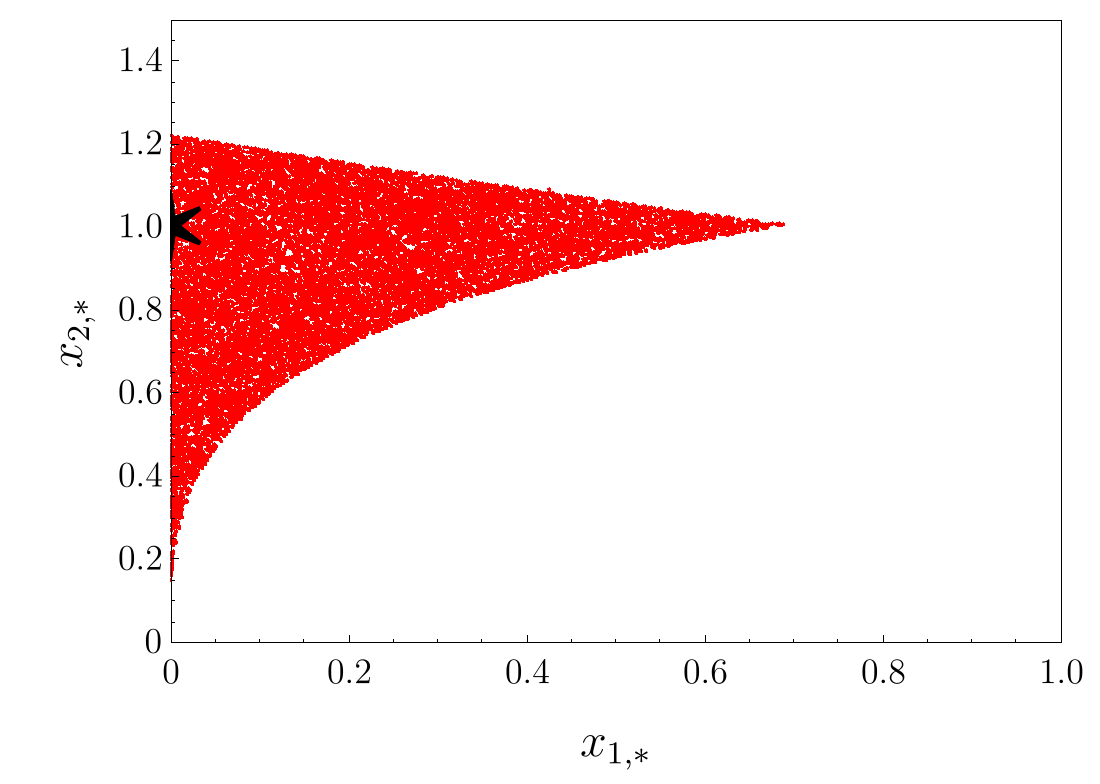}
  \caption{\small Throat values ${\bm x}_*\equiv{\bm x}|_{r=L}$ of solutions with boundary charge vector $\mathbf{q}=(1,0,1)\in\calc_{\text{\tiny WH}}$.}
  \label{fig:model1domain101}
\end{subfigure}%
\hfill
\begin{subfigure}[t]{.49\textwidth}
  \centering
  \includegraphics[width=\textwidth]{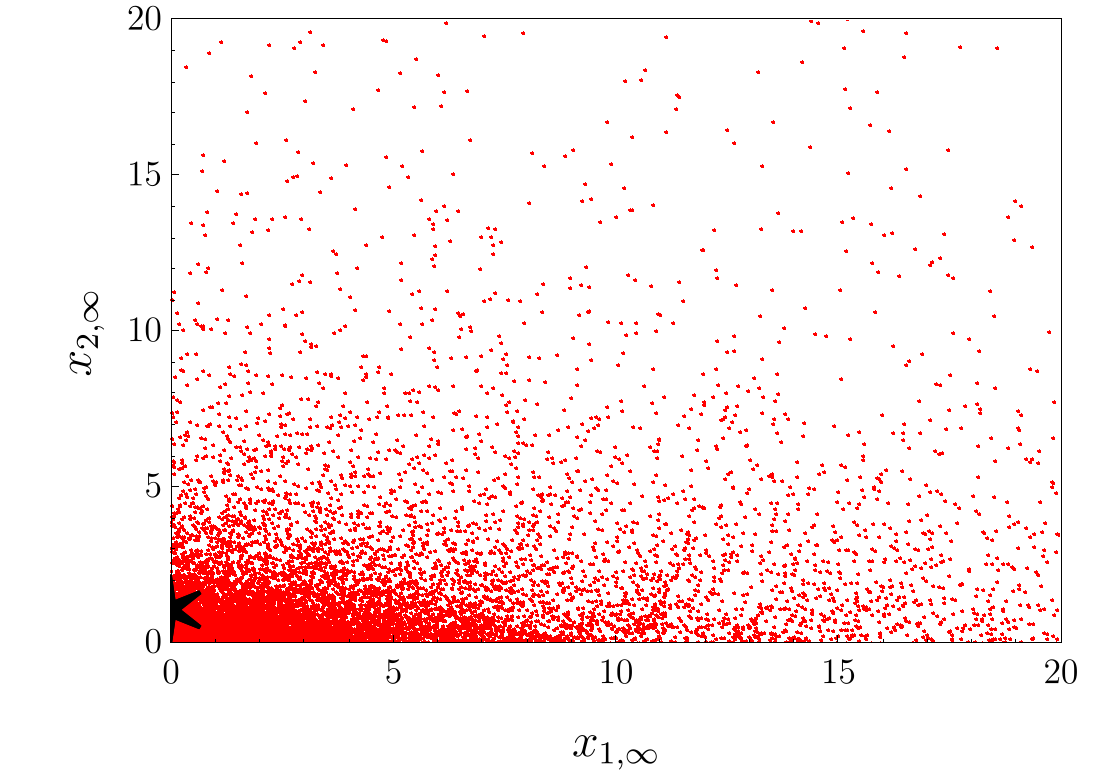}
  \caption{\small Asymptotic values ${\bm x}_\infty\equiv{\bm x}|_{r=\infty}$ of solutions with boundary charge vector $\mathbf{q}=(1,0,1)\in \calc_{\text{\tiny WH}}$.}
  \label{fig:model1inf101}
  \end{subfigure}
  \hfill
  \begin{subfigure}[t]{.49\textwidth}
  \centering
  \includegraphics[width=\textwidth]{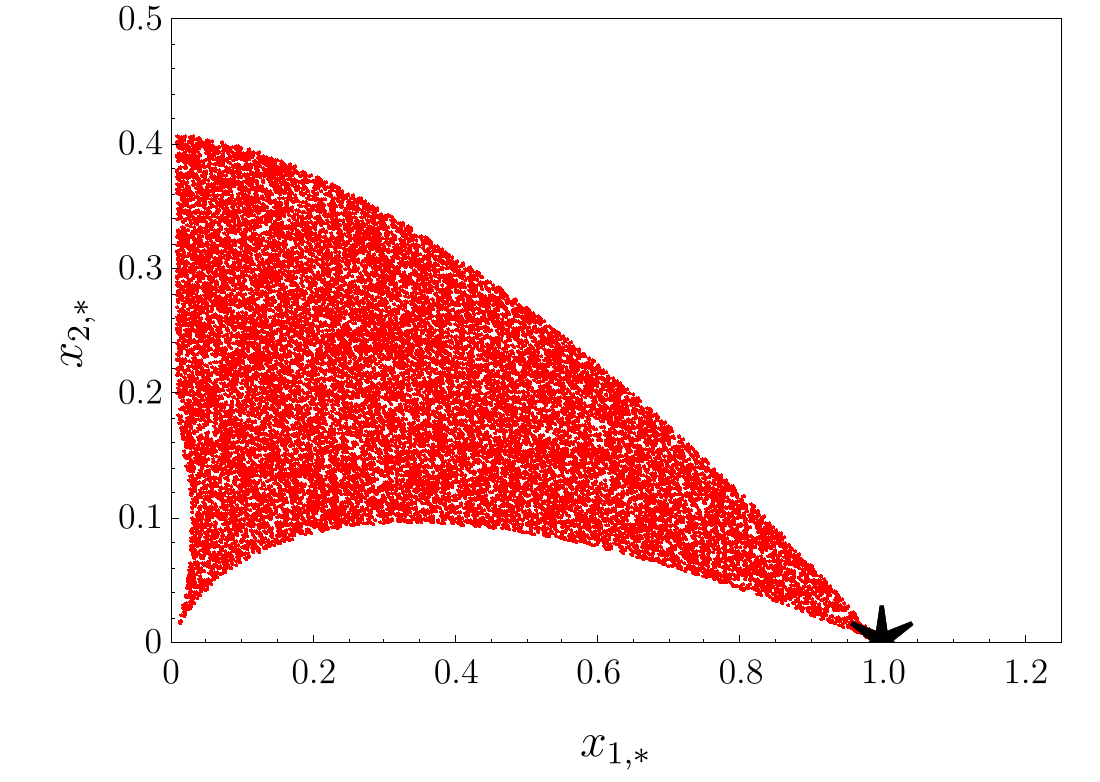}
  \caption{\small Throat values ${\bm x}_*\equiv{\bm x}|_{r=L}$ of solutions with boundary charge vector $\mathbf{q}=(1,1,0)\notin\calc_{\text{\tiny WH}}$.}
  \label{fig:model1domain110}
    \end{subfigure}%
    \hfill
  \begin{subfigure}[t]{.49\textwidth}
  \centering
  \includegraphics[width=\textwidth]{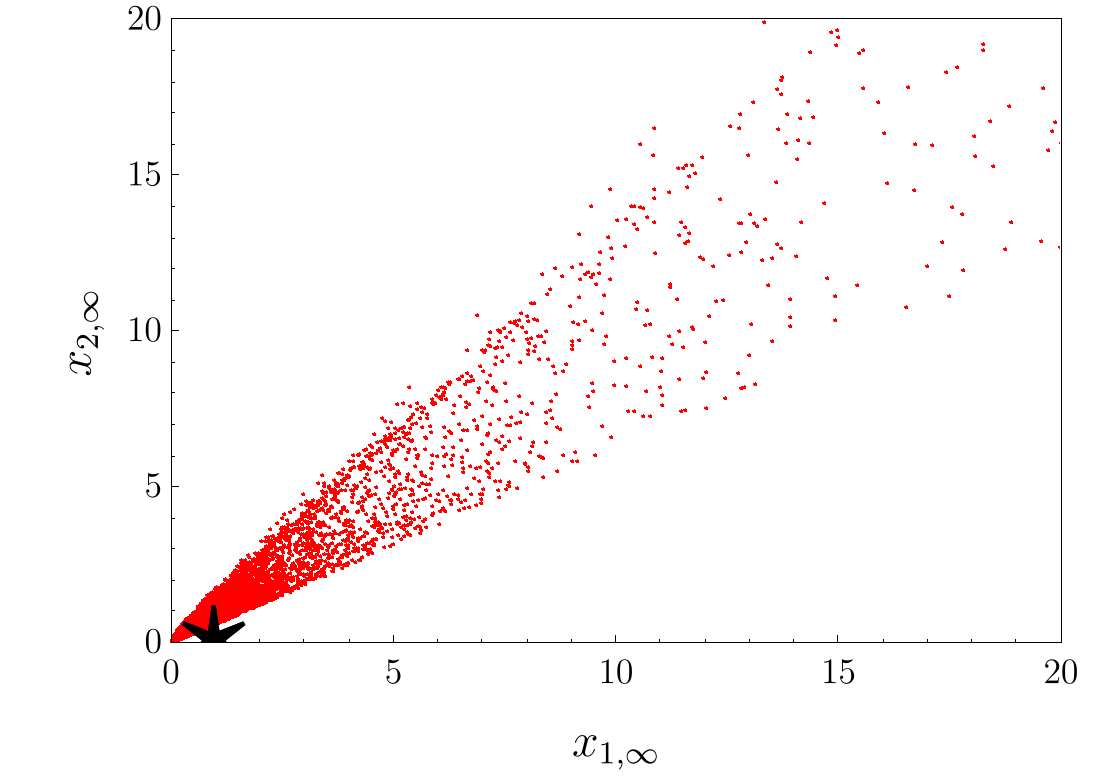}
  \caption{ \small Asymptotic values ${\bm x}_\infty\equiv{\bm x}|_{r=\infty}$ of solutions with boundary charge vector $\mathbf{q}=(1,1,0)\notin\calc_{\text{\tiny WH}}$.}
  \label{fig:model1inf110}
  \hfill
\end{subfigure}
\caption{\small Same scan as in Figure \ref{fig:model1} but with charges $\mathbf{q}=(1,0,1)$ and $\mathbf{q}=(1,1,0)$, as indicated in the captions of the plots. While the solution with vanishing $q_1$ charge is expected, the one with vanishing $q_2$ charge is apparently surprising as the points with $\ell_2=q_2 =0$ are at infinite distance. Indeed, a closer look at \ref{fig:model1domain110} reveals how getting closer to the point associated to the would-be homogeneous solution (indicated as star in the plot) the domain shrinks so that the the homogeneous solution is never actually reached, confirming that $\ell_{2,*} =0$ never delivers a valid configuration. This is a completely different behavior with respect to figure \ref{fig:model1domain101}, where the homogeneous solution has a ``dense'' neighborhood of valid solutions.
}
\label{fig:model1chargesOff}
\end{figure}

We can more explicitly test our expectation by considering the simple  models described in Section \ref{sec:Ftheory1} and in the Appendix \ref{app:Ftheory2}, in which ${\bm x}(\tau)$ should belong to $\mathbb{R}^2_{>0}$ and $\mathbb{R}^3_{>0}$ respectively. Consider first the  model of subsection  \ref{sec:Ftheory1}. Ignoring for the moment the sector associated with the dilaton \eqref{dualdilaton}, the cone of BPS instanton charges is generated by the effective divisors $E^1$ and $E^2$. Hence, in the basis of nef divisors $D^1,D^2$ the components of a BPS instanton charge  ${\bf q}=q_1 D^1+q_2D^2\in\calc_{\rm I}$ must satisfy $q_1+pq_2\geq 0$ and $q_2\geq 0$. On the other hand, ${\bf q}$ belongs to the subset $\calc^{\text{\tiny EFT}}_{\rm I}$ of  EFT instanton charges only if $q_1\geq 0$ and $q_2\geq 0$. For instance the charge vector ${\bf q}=E^2$, which corresponds to $(q_1,q_2)=(-p,1)$, is BPS but not EFT. Including back the dilaton, according to our general claim there should exist a large family of wormhole solutions for charges  in the set \eqref{calcWH}, in addition to the corresponding homogeneous wormholes which certainly exist. Recalling \eqref{model2EFTs}, we see that  $\calc_{\text{\tiny WH}}$ includes charge  vectors $(\hat q,q_1,q_2)$ with  $\hat q,q_2>0$ and $q_1\geq 0$. Let us first consider charge vectors which are in the interior of $\calp$, that is with $\hat q,q_1,q_2>0$.  
One can numerically integrate the equations \eqref{elleom} for different choices of the twisting constant $p>0$ (the case $p=0$ reducing to a trivial factorized model), charges $\hat q,q_1,q_2>0$ and initial values $(x_{*1},x_{*2})$. The results confirm our expectations, as  exemplified in Figs.\ \ref{fig:model1trajectories} and \ref{fig:model1}. We stress that the specific choice of the charges used in these plot has been made for purely visual reasons and we explicitly verified that our qualitative conclusions hold also for more general charges.
The plots show how, even if the allowed throat values $(x_{1,*},x_{2,*})$ are confined to a sharp specific sub-region, the corresponding asymptotic values $(x_{1,\infty },x_{2,\infty})$ spread all over $\mathbb{R}^2_{>0}$, as predicted by our general arguments.  The set $\calc_{\text{\tiny WH}}$ includes also  charge vectors with $q_1=0$ and $\hat q,q_2>0$, which are on the finite distance boundary $\ell_1=0$ of $\calp$, and we thus expect to find non-homogeneous wormhole solutions around the homogeneous one for these charges as well.  This is indeed what we find, as shown in the plots of Fig.\,\ref{fig:model1domain101} and \ref{fig:model1inf101}. Almost surprisingly, as shown in Fig.\,\ref{fig:model1domain110} and \ref{fig:model1inf110}, we also find solutions for charge vectors with  $\hat q,q_1>0$ and $q_2 =0$,  which are not included in $\calc_{\text{\tiny WH}}$ and for which no homogeneous solution is possible. A closer inspection of the plots reveals however how these solutions do not form a dense open set around the would-be homogeneous solution, with a spread distribution of end points $(x_{1,\infty },x_{2,\infty})$, but rather look like  the class of  ``accidental'' solutions of Fig.\  \ref{fig:model1}.

\begin{figure}[!htb]
\centering
\begin{subfigure}[t]{.49\textwidth}
  \centering
  \includegraphics[width=\textwidth]{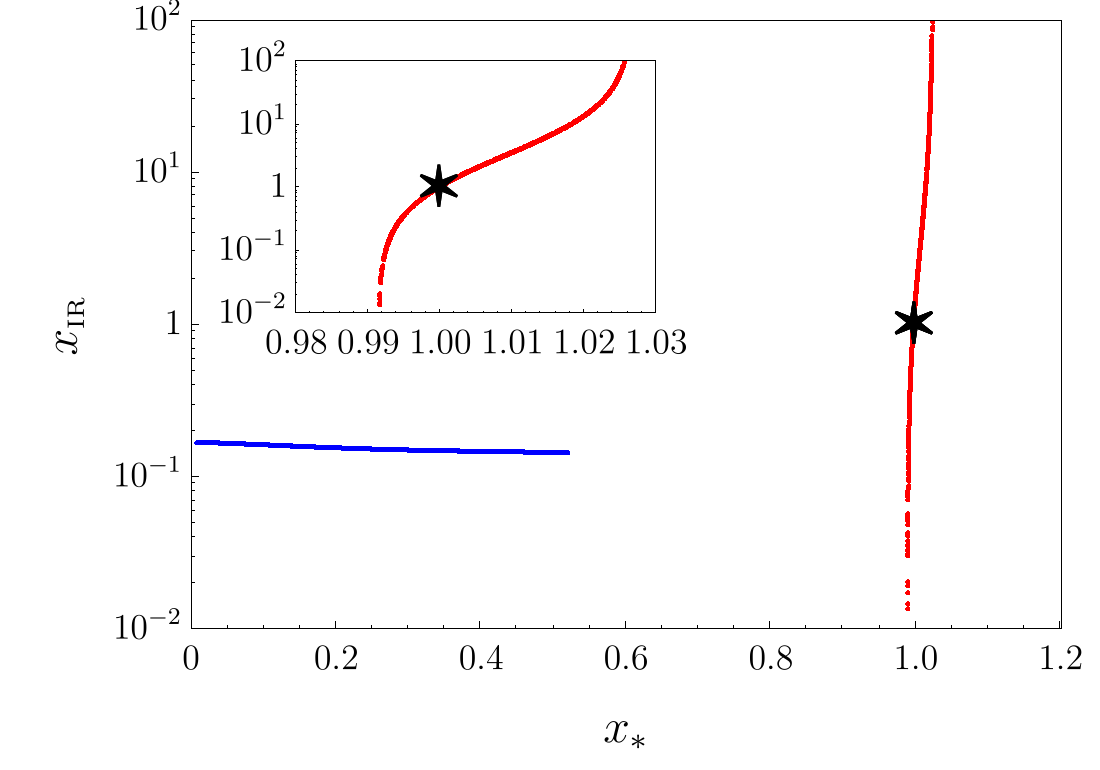}
  \caption{\small Scanned throat and asymptotic values of $x=\ell_1/\ell_2$.}
  \label{fig:model1degDomain}
\end{subfigure}%
\hfill
\begin{subfigure}[t]{.49\textwidth}
  \centering
  \includegraphics[width=\textwidth]{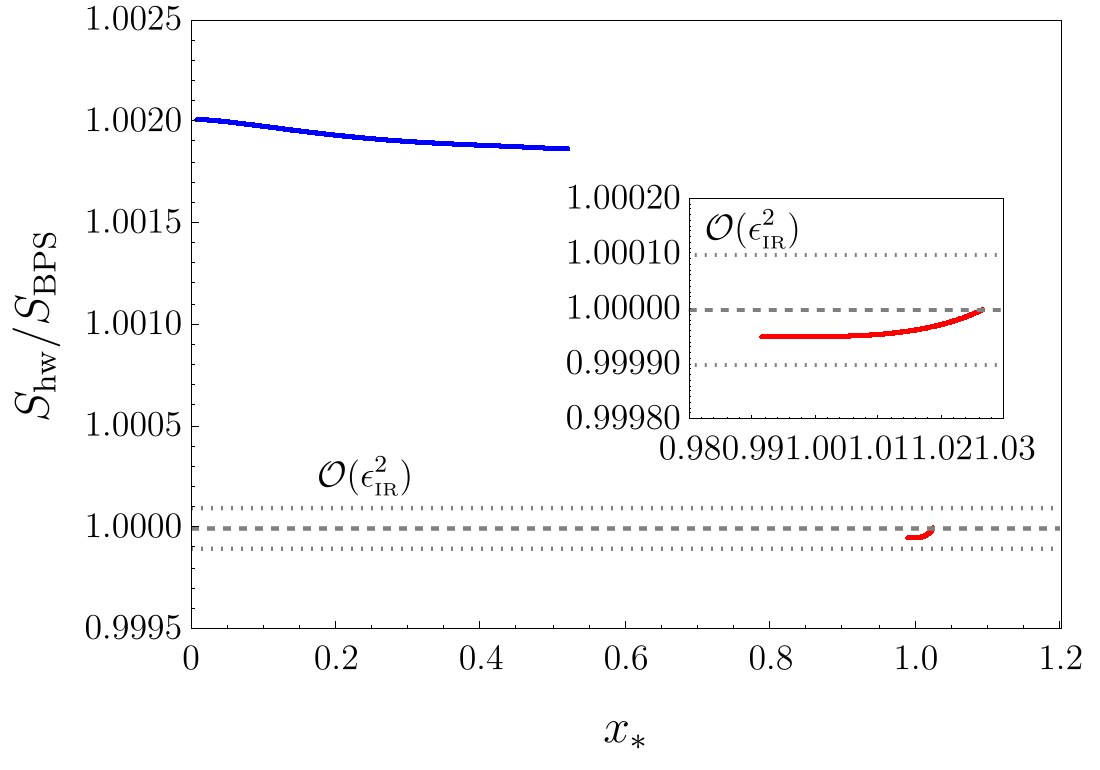}
  \caption{\small Ratio between the complete and the BPS action as a function of $x_*$.}
  \label{fig:model1degActions}
\end{subfigure}
\caption{\small 
Numerical scan obtained by integrating the equations of motion \eqref{elleom} with charge vector $\mathbf{q}=(1,1)$ in the F-theory model of Section \ref{sec:Ftheory1} without the dilaton and with random initial conditions for the dual saxions. The left panel compares the allowed throat values of $x_* =\ell_{1,*}/\ell_{2,*}$ to the asymptotic values $x_\text{\tiny IR}=\ell_{1,\text{\tiny IR}}/\ell_{2,\text{\tiny IR}}$ found by the scan, where acceptable solutions have been identified as the ones in which the dual saxions degenerate at a distance greater than $r_\text{\tiny IR}= 10L$ from the throat. The black star denotes the homogeneous  solution and the red dots in its neighborhood can be identified with the solutions predicted by the general perturbative argument, while the blue set on the left represents the usual accidental solutions. We note that, as expected, a tiny  $x_*$ interval of  $1$ is mapped to a much larger  $x_{\text{\tiny IR}}$ interval. The second plot shows the ratio $S_{\text{hw}}/S_{\text{BPS}}$ associated to the acceptable solutions, where $S_{\text{hw}}$ is obtained by cutting off the integration up to $r_\text{\tiny IR}$ as outlined in the main text and $S_{\text{BPS}}$ is given by \eqref{SlocBPS} . Around the homogeneous solution the two actions are compatible up to $\mathcal{O}(\epsilon_\text{\tiny IR}^2)$ terms, with $\epsilon_\text{\tiny IR}\equiv\Lambda_\text{\tiny IR}^2L^2$, as predicted by \eqref{apprSn=3}, while in the accidental region the deviation is slightly bigger (but still below $\mathcal{O}(\epsilon_\text{\tiny IR})$).} 
\label{fig:model1degenerate}
\end{figure}

It is also interesting to explicitly check the correspondence between fundamental EFT instantons and $n=3$ wormholes of section \ref{sec:n=3} beyond the homogeneous and perturbative regime. We can do this by keeping only the $\ell_a$ sector, which identifies an $n=3$ model, and performing the numerical scan. Applying the IR regularization outlined in Section \ref{sec:n=3}, which introduces an IR cutoff radius $r_\text{\tiny IR}=\Lambda^{-1}_\text{\tiny IR}$, we can test the correspondence  by comparing the complete numerical action of the IR regulated half-wormhole solutions to the associated BPS value \eqref{SlocBPS}. The result of the scan is reported in the plots of Fig.\   \ref{fig:model1degenerate}, where we  show both the usual spread in the asymptotic values associated to deformations around the homogeneous solutions and the comparison between the complete and the BPS action. If we introduce the expansion parameter $\epsilon_\text{\tiny IR}\equiv \Lambda^2_\text{\tiny IR}L^2\ll1$,  in the neighborhood of the homogeneous solutions the actions coincide up to $\mathcal{O}(\epsilon_\text{\tiny IR}^2)$, exactly as predicted by \eqref{apprSn=3}. Interestingly, a new region around  $\ell_1 =0$ also appears. This is not surprising as it closely resemble the situation of Figs.\ \ref{fig:model1} and \ref{fig:model1chargesOff}. This time the agreement between the complete and BPS action is roughly of $\mathcal{O}(\epsilon_\text{\tiny IR})$.

\begin{figure}[!htb]
\centering
\begin{subfigure}[t]{.33\textwidth}
  \centering
  \includegraphics[width=\textwidth]{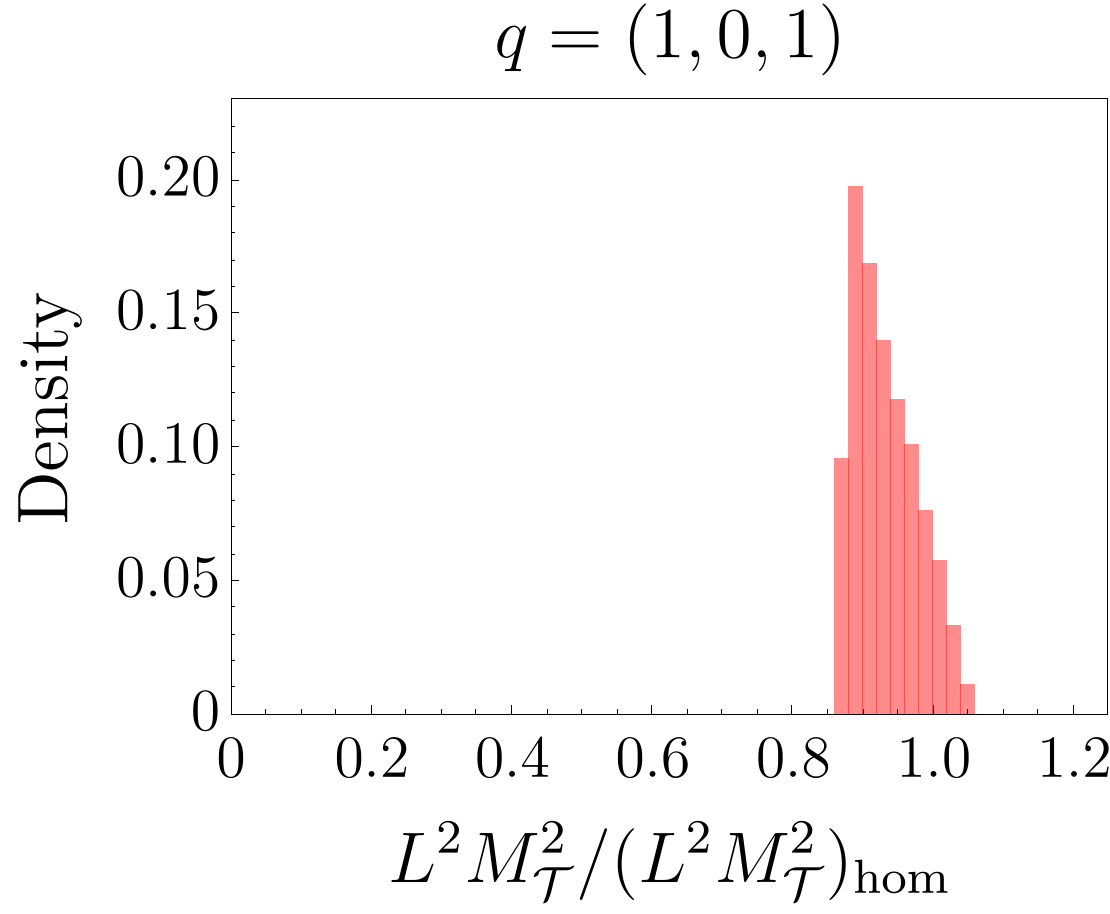}
  \caption{\small 
  $(L^2 M^2 _\mathcal{T} )_{\text{hom}} = 4/\sqrt{3}$.}
  \label{fig:hist101}
\end{subfigure}%
\hfill
\begin{subfigure}[t]{.33\textwidth}
  \centering
  \includegraphics[width=\textwidth]{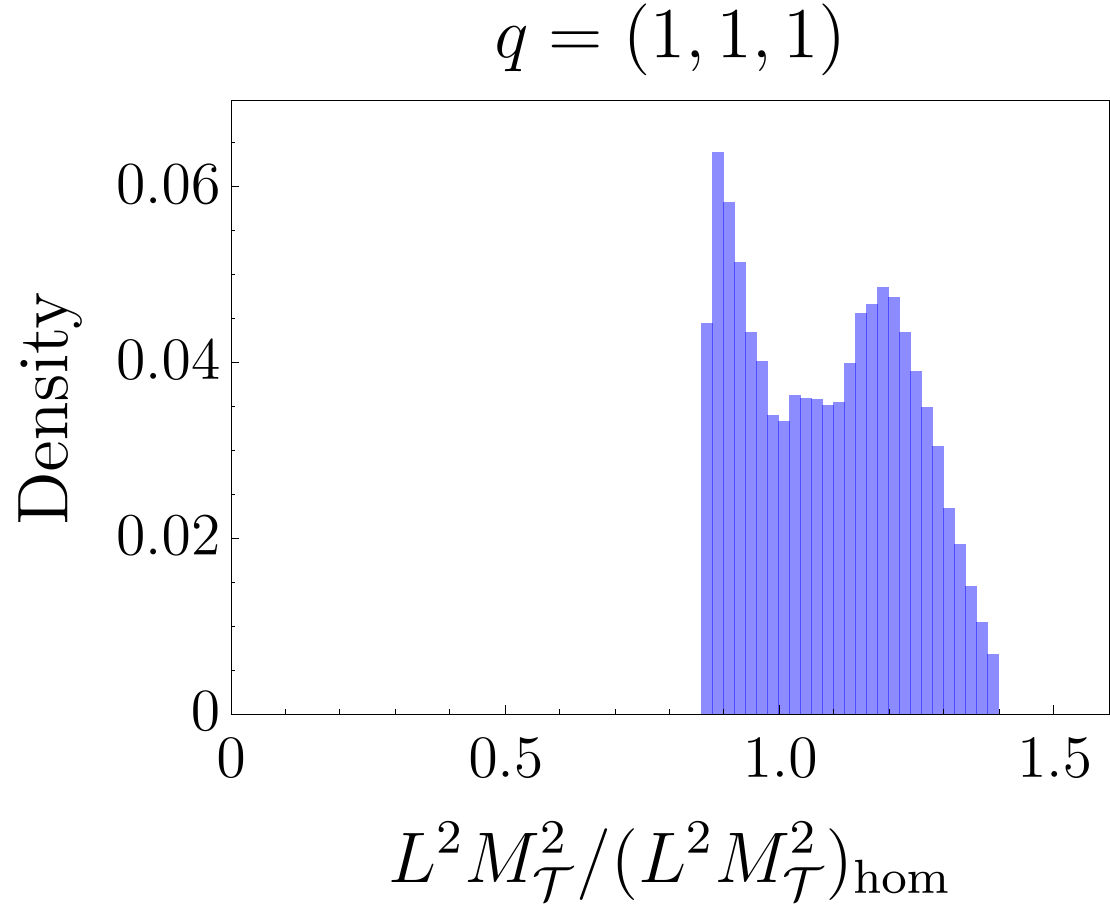}
  \caption{\small $(L^2 M^2 _{\calt})_{\text{hom}}= 4/\sqrt{3}$.}
  \label{fig:hist111}
\end{subfigure}
\begin{subfigure}[t]{.33\textwidth}
  \centering
  \includegraphics[width=\textwidth]{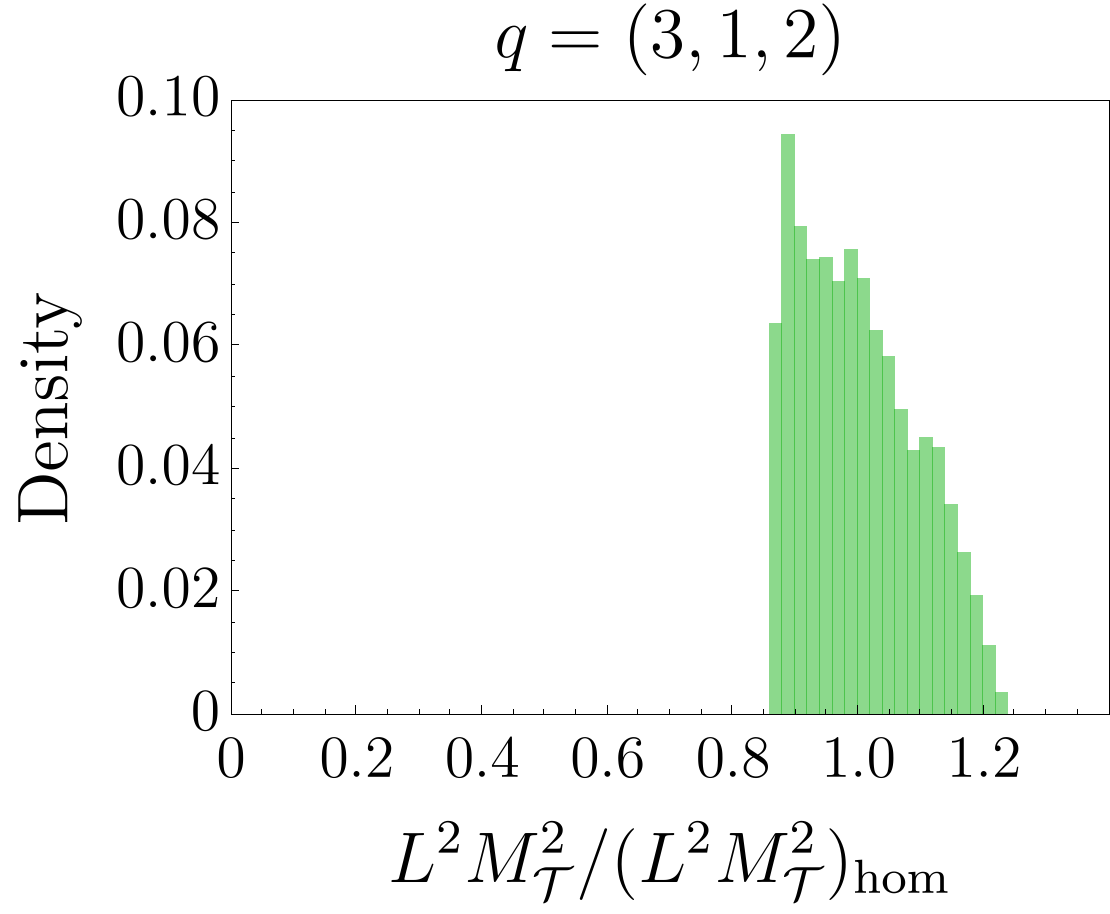}
  \caption{\small 
  $(L^2 M^2 _{\calt} )_{\text{hom}}$ $ = 8/\sqrt{3}$.}
  \label{fig:hist312}
\end{subfigure}
\caption{\small Density histograms of  $L^2M_{\calt}^2$ evaluated at the throat of  a set of non-homogeneous numerical wormhole solutions for the F-theory model of Section \ref{sec:Ftheory1}. The values of $L^2M_{\calt}^2$ have been normalized to the values of the associated homogeneous solution. The results clearly show how the values do not deviate substantially from the ones of the homogeneous solutions, guaranteeing $L^2 M_{\calt}^2\gtrsim 1$ for all these solutions, as required by  compatibility with the species scale.} 
\label{fig:histograms}
\end{figure}

A last crucial check regards the compatibility of our solutions with the energy bounds imposed by the dominant EFT string scale \eqref{SPbounddef}. In section \ref{sec:validityWH} we showed how on the homogeneous solution the condition $L^2 M_{\calt} ^2  \geq 1$ is always guaranteed if $\mathbf{q} \in \mathcal{C}_\text{WH}$, and argued that this should be the case also on more general solutions that deviate from the homogeneous one. We quantitatively verified this claim on a set of numerically obtained non-homogeneous solutions with charges $q=(1,0,1)$, $q=(1,1,1)$ and $q=(3,1,2)$. The results are reported in figure \ref{fig:histograms}, where we show the density histogram of $L^2 M^2 _{\calt}$ normalized to the one of the associated homogeneous solution. It is clearly visible how the ratio between the dominant EFT string mass squared ($M^2_{\calt}$) and the wormhole scale squared ($1/L^2$) does not deviate much from one, corresponding of the homogeneous solutions, guaranteeing the stability of these numerical solution and supporting our previous claim.

\begin{figure}[!htb]
\centering
\begin{subfigure}{.48\textwidth}
  \centering
  \includegraphics[width=\textwidth]{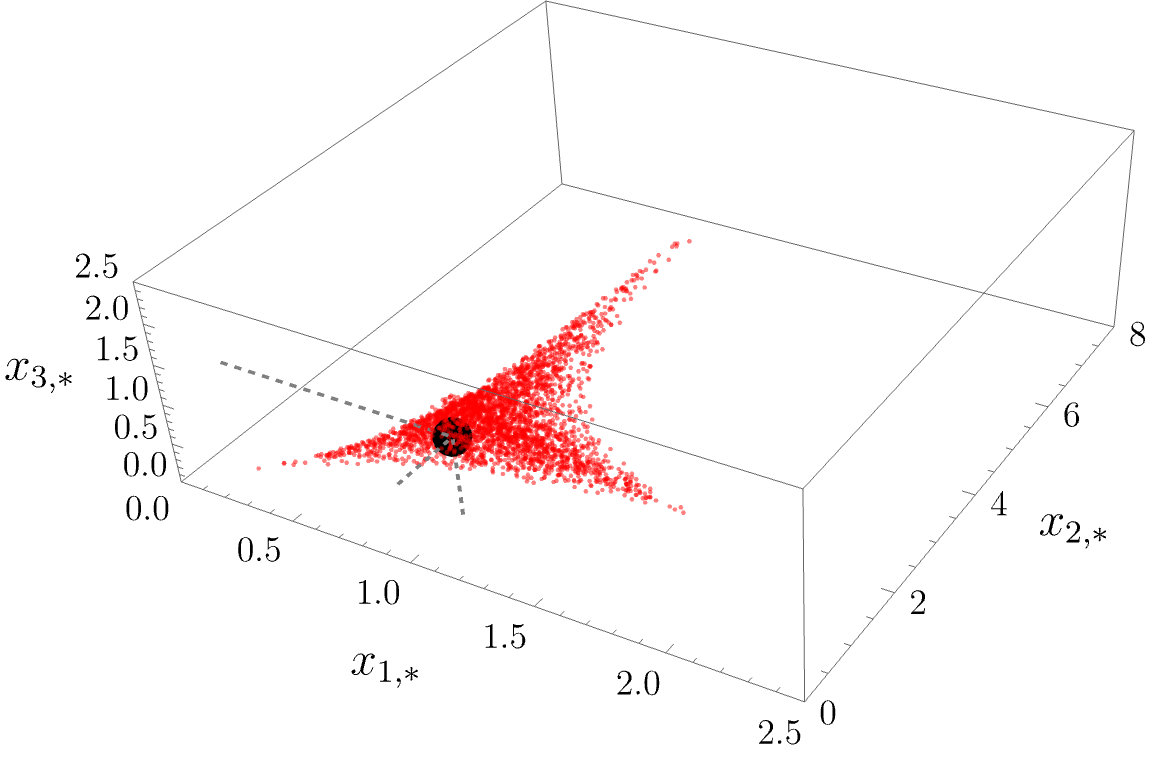}
  \caption{\small Scanned throat values ${\bm x}_*\equiv{\bm x}|_{r=L}$.}
  \label{fig:model2domain}
\end{subfigure}%
\hfill
\begin{subfigure}{.48\textwidth}
  \centering
  \includegraphics[width=\textwidth]{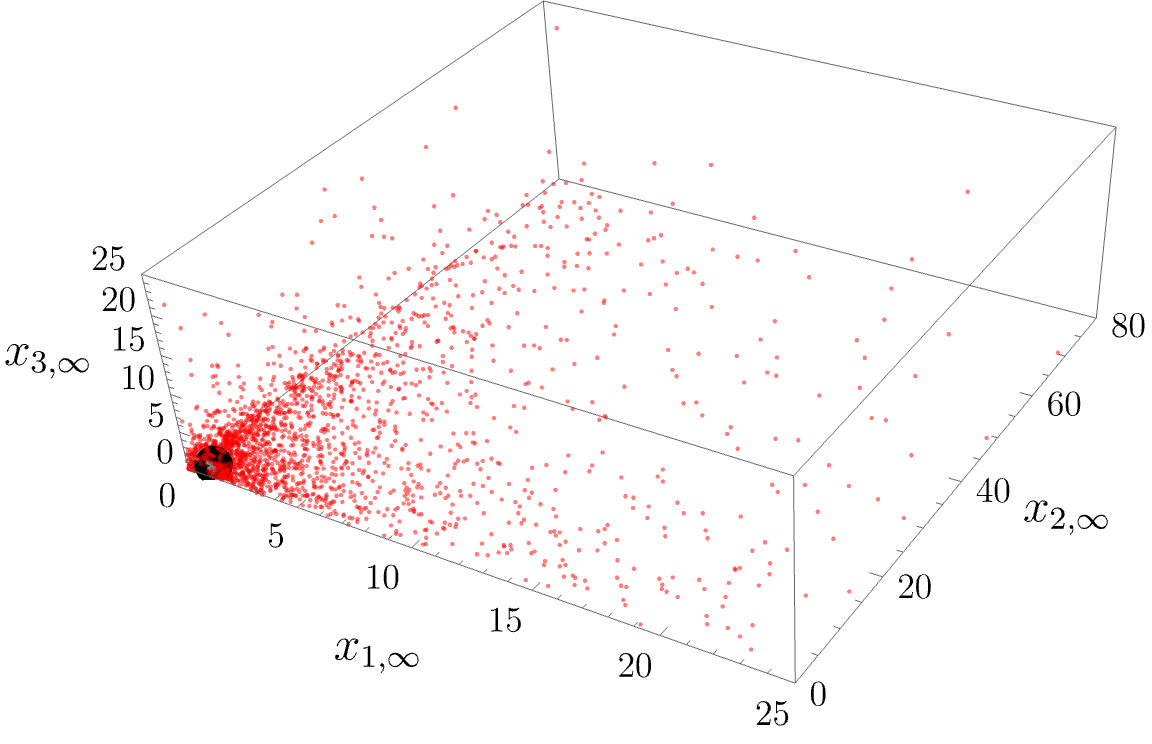}
  \caption{\small Corresponding asymptotic values ${\bm x}_\infty\equiv{\bm x}|_{r=\infty}$.}
  \label{fig:model2inf}
\end{subfigure}
\caption{\small Numerical scan obtained by integrating the equations of motion \eqref{elleom} with charge vector $\mathbf{q}=(1,1,1,1)$ in the $\mathbb{P}^1$ over $\mathbb{F}_p$ F-theory model of App.\ \ref{app:Ftheory2}, with $p=h=1$ and random initial conditions for the dual saxions. As in figure \ref{fig:model1}, the first plot contains the acceptable throat points ${\bm x}_*$ identified by the scan. The shape of the domain of acceptable solutions is clearly visible and generates a dense set around the homogeneous solution, indicated as a black ball connected by dashed gray lines for clarity. The second plot shows the asymptotic values ${\bm x}_\infty$ associated to the acceptable solutions, which spreads all over $\mathbb{R}^3_{>0}$, in agreement with our general argument.}
\label{fig:model2}
\end{figure}

The model described in the appendix \ref{app:Ftheory2} can be analyzed in the same way. In this case, in addition to the dilaton \eqref{dualdilaton}, we have a three-dimensional vector ${\bm\ell}=\ell_1 D^1+\ell_2D^2+\ell_3 D^3$ that takes values in the K\"ahler cone. So, with focus on this sector, any ${\bf q}\in\calc^{\text{\tiny EFT}}_{\rm I}$ can be identified with the generic nef divisor ${\bf q}=q_1D^1+q_2D^2+q_3D^3$, with $q_1,q_2,q_3\geq 0$. Again, the set $\calc_{\rm I}$ of BPS instanton charges is larger, and is generated by the effective divisors $E^{1,2,3}$ defined in \eqref{nefdivF1}. In particular, if $p,h>0$, the generators $E^1$ and $E^3$ identify BPS instanton charges $(q_1,q_2,q_3)=(1,-p,0)$ and $(q_1,q_2,q_3)=(-h,0,1)$, respectively, which do {\em not} belong to $\calc^{\text{\tiny EFT}}_{\rm I}$. Including the dilaton as in the previous model, the set \eqref{calcWH} includes charge vectors with $\hat q,q_3>0$ and $q_1,q_2\geq 0$. Again, we expect that for any choice of charges in $\calc_{\text{\tiny WH}}$, there should exist a large family of wormhole solutions reaching all possible asymptotic values $(x_{1,\infty},x_{2,\infty},x_{3,\infty})\in\mathbb{R}^3_{>0}$.  By numerically integrating \eqref{elleom} with initial conditions $(x_{1,* },x_{2,* },x_{3,*})\in\mathbb{R}^3_{>0}$ one gets another clear confirmation of our expectation, as evident from the plots in Fig.\ \ref{fig:model2}.

\subsubsection{Wormholes in heterotic  models}
\label{sec:WHst2}

EFT instantons and wormholes in the heterotic models of Section \ref{sec:het} can be discussed in a similar way. Hence we will be briefer, outlining only some relevant points. 

The set of BPS instanton charges $(q_0,{\bf q})\in \calc_{\rm I}$ is realized  by F1-strings wrapping effective curves ${\bf q}=q_a\Sigma^a$ and $q_0$ NS5-branes wrapping the internal space -- see \cite{Martucci:2022krl} for more details. The subset of EFT instanton charges $\calc_{\rm I}^{\text{\tiny EFT}}$ is obtained by imposing that the charges $(q_0,{\bf q})$ satisfy the same constraints \eqref{hetellcond} that must be obeyed by $(\ell_0,{\bm\ell})$.  
Let us first set $q_0=0$, so that \eqref{hetellcond} reduce to ${\bf q}=q_a\Sigma^a\in \calp^{\text{\tiny het}}_{\text{\tiny K}}$. The BPS instanton tower or sublattice WGCs of Section \ref{sec:EFTinst} translate into geometrical statements on the existence of irreducible effective curves that populate an infinite subset of $\calc_{\rm I}^{\text{\tiny EFT}}$ -- see \cite{Alim:2021vhs} for the corresponding statements in M-theory compactification to five dimensions.  Such EFT instantons correspond to F1-strings wrapping {\em movable} curves. As for the movable divisors encountered in Section \ref{sec:WHst1}, this implies that for some  $m\in\mathbb{Z}_{\geq 1}$ the charge $m{\bf q}$ can be represented by a curve that can  freely move along the  internal space. Hence, as already observed in F-theory models, EFT instantons correspond to microscopic configurations which can probe the entire compactification space. These EFT instanton charges do not belong to the set $\calc_{\text{\tiny WH}}$ associated with the complete perturbative regime described by $(\ell_0,{\bm\ell})$, since they have vanishing pairing with EFT string charges $(e^0,{\bf 0})$, $e^0>0$. Nevertheless, they characterize the restricted perturbative regime obtained by removing the EFT string charge $(1,{\bf 0})$ from the generators of $\Delta$,  realizing an example of the mechanism outlined in Footnote \ref{foot:CWH}.  In this regime, one can ``ignore'' the saxionic combination  $\hat s^0\equiv s^a-\frac12p_a s^s$ and the corresponding dual saxion $\ell_0$, and only the second term  of \eqref{hetK} remains relevant, whose homogeneity degree is $n=3$. Hence, the corresponding EFT instanton charges  can  be at best supported by marginally degenerate wormholes of the type discussed in Section \ref{sec:n=3}.  Since  EFT strings are realized by NS5-branes wrapping nef divisors \cite{Lanza:2021udy,Martucci:2022krl}, the set  \eqref{calcWH} corresponding to this restricted theory identifies 
movable curves ${\bf q}=q_a\Sigma^a$ which have non-vanishing (positive) intersection with any nef divisor.
Our general arguments imply that marginally degenerate wormholes  should exist for any such curve.  This certainly happens when ${\bf q}$ lies in the interior of $\calp^{\text{\tiny het}}_{\text{\tiny K}}$, and in these cases at least the homogeneous wormholes certainly exist. Note that these homogeneous  wormholes coincide with the string theory wormholes first constructed and studied  in \cite{Giddings:1989bq,Park:1990ep}.
We emphasize that this reasoning is based on an EFT which, as discussed in Section \ref{sec:het}, is fully reliable only if $s^0\gg p_as^a$. In this regime $\calt_*=M^2_{\text{\tiny P}}\ell_0$ determines a constant upper bound on the species scale, so that \eqref{LMTcond} does not hold for such wormholes. This is expected, since as observed above these charges do not belong to the set $\calc_{\text{\tiny WH}}$ associated with the full perturbative regime, which includes  $\ell_0$. In order to make these wormholes fully reliable, we should go beyond the $s^0\gg p_as^a$ regime, as in the toy model of Footnote \ref{foot:modelinfdist}, but unfortunately this is presently out of our reach. On the other hand, the quasi-BPSness  mentioned in Section \ref{sec:n=3}, and further discussed in \ref{sec:n=3physics} below, strongly suggests that these wormholes should survive away the $s^0\gg p_as^a$ regime.

We can then turn on a charge $q_0>0$, so that the EFT instanton uplifts to a configuration involving Euclidean NS5-branes wrapping the entire internal space. For generic choices of ${\bf q}=q_a\Sigma^a$ in the interior of $\calp^{\text{\tiny het}}_{\text{\tiny K}}$, the entire K\"ahler and kinetic potentials \eqref{hetK} and \eqref{hetKinpot} are relevant, which have homogeneity $n=4$. Hence there should exist corresponding homogeneous as well as non-homogeneous solutions, which are fully non-degenerate. Finally, one could also pick charges so that the relevant EFT has again $n=3$. Explicit examples could be obtained through heterotic/F-theory duality, which maps NS5 instantons to some  D3-brane instantons in F-theory -- see  for instance the detailed discussion of \cite{Palti:2020qlc} and Appendix \ref{app:hetFtheory}. As an example, a D3-brane wrapping the divisor $D^2$ in the F-theory model of Section \ref{sec:Ftheory1} corresponds to  an NS5-brane instanton in the dual heterotic description. By duality the relevant EFT instanton has homogeneity $n=3$ and admits marginally degenerate wormholes.  Other explicit examples could be obtained from  the heterotic/F-theory dictionary discussed in Appendix \ref{app:hetFtheory}. 

Finally we notice that we can adapt (part of) of the above discussion to wormholes associated with the K\"ahler moduli sector of IIA compactifications, either with orientifolds or not (and, in the latter case, with enhanced $\caln=2$ supersymmetry), and with the complex structure sector of IIB models by mirror symmetry. More general  wormholes of type IIA orientifold compactifications can be obtained by taking particular limits of the wormholes discussed in the next section.

\subsubsection{Wormholes in M-theory models}
\label{sec:WHst3}

Finally, we would like to briefly discuss how our general considerations apply to M-theory models compactified on $G_2$ manifolds -- see Section 6.5 of \cite{Lanza:2021udy} and Section 8 of \cite{Martucci:2022krl} for a summary of the relevant ingredients in the present context. The effective theory of these  models does not generically  admit an explicit description, basically because of the absence of a complex and K\"ahler structure on the internal space. One can nevertheless say something about the existence of corresponding wormholes. 

The information on the $G_2$-holonomy metric of the internal  seven-dimensional space $Y$ is completely encoded in the {\em associative} three-form $\Phi\in H^3(Y,\mathbb{R})$ \cite{Hitchin:2000jd}.   The saxions are then obtained by expanding $\Phi=s^i\omega_i$ in an appropriate cohomology basis $\omega_i\in H^3(Y,\mathbb{Z})$. Alternatively, the same information is encoded in the {\em coassociative} four-form $*_Y\Phi$, and its rescaled counterpart:
\be\label{hatPhi} 
\hat\Phi\equiv \frac{1}{2V_Y}*_Y\Phi\,,
\ee
where $V_Y$ is the volume of $Y$.
The dual saxions $\ell_i$ are then obtained from the expansion 
\be 
\hat\Phi=\ell_i\hat\omega^i\,,
\ee 
where $\hat\omega^i\in H^4(Y,\mathbb{Z})$ is the dual cohomology basis, such that $\int_Y\omega_i\wedge \hat\omega^j=\delta_i^j$. Both $\Phi$ and $\hat\Phi$ parametrize corresponding cones, which can be identified with the saxionic and the dual saxionic cone, respectively. BPS instantons uplift to M2-branes wrapping internal submanifolds $\Sigma_{\bf q}$ calibrated by $\Phi$. The corresponding charge vectors can be identified with the Poincar\'e dual four-forms ${\bf q}=[\Sigma]=q_i\hat\omega^i\in H^4(Y,\mathbb{Z})$. Hence, according to our general definition the cohomology class ${\bf q}$ identifies  an EFT  instanton if it admits, among its representative closed four-forms,  a coassociative four-form $\hat\Phi_{\bf q}$. Note that the existence of  calibrated submanifolds $\Sigma_{k{\bf q}}$ for any such ${\bf q}$ and some integer $k\geq 1$ is not obvious at all, and would be necessary in order to realize the BPS instanton tower or sublattice WGCs of Section \ref{sec:EFTinst}. 

The K\"ahler potential is given by the geometrical formula $K=-3\log\int_Y\Phi\wedge *_{\Phi}\Phi$ \cite{Beasley:2002db}, and then by \eqref{hatPhi}
the kinetic potential for the dual saxions takes the form \eqref{calfP} with  
\be\label{G2calf} 
\tilde P(\ell)=\left(\int_Y\hat\Phi\wedge *_{\hat\Phi}\hat\Phi\right)^3\,.
\ee
In these geometric formulas the Hodge-operators $*_{\Phi}$ and $*_{\hat\Phi}$ are implicitly defined by $\Phi$ and $\hat\Phi$, respectively. 
Even if it is generically hard to  make explicit the dependence of \eqref{G2calf} on the dual saxions, it is easy to see that it is homogeneous of degree $n=7$. Indeed, a rescaling $\ell_i\rightarrow \lambda\ell_i$ corresponds to $\hat\Phi\rightarrow \lambda\hat\Phi$. This  induces a rescaling $\d s^2_Y\rightarrow \lambda^{-\frac{2}{3}}\d s^2_Y$ of the corresponding metric, and hence $*_{\hat\Phi}\hat\Phi\rightarrow \lambda^{\frac{4}{3}} *_{\hat\Phi}\hat\Phi$. 

From our general discussion we conclude that whenever we pick ${\bf q}$ inside the dual saxionic cone  $\calp$, we certainly have ${\bf q}\in \calc_{\text{\tiny WH} }$ and there exists a corresponding homogeneous wormhole with $n=7$, as well as its non-homogeneous variations. We expect similar conclusions if ${\bf q}$ belongs to some finite field distance boundary of $\calp$. Charges ${\bf q}$ on some infinite distance boundary of $\calp$ can still correspond to sensible wormholes in restricted perturbative regimes, possibly with  $n<7$ -- see the discussion at the end of Section \ref{sec:univWH}. Furthermore, by considering $G_2$ compactifications admitting a weakly coupled IIA description \cite{Kachru:2001je}, one can immediately adapt these conclusions to type IIA compactifications on Calabi-Yau spaces with O6-planes.

\section{Low-energy implications}
\label{sec:phys}

In this section we would like to discuss some of the physical implications of wormholes. We mostly follow the standard literature \cite{Giddings:1987cg,Coleman:1988cy,Preskill:1988na}, see also \cite{Hebecker:2018ofv} for a review and a more complete list of references.

\begin{figure}[!t]
    \centering
     \begin{subfigure}{.48\textwidth}
    \centering
    \includegraphics[angle=0,width=230pt]{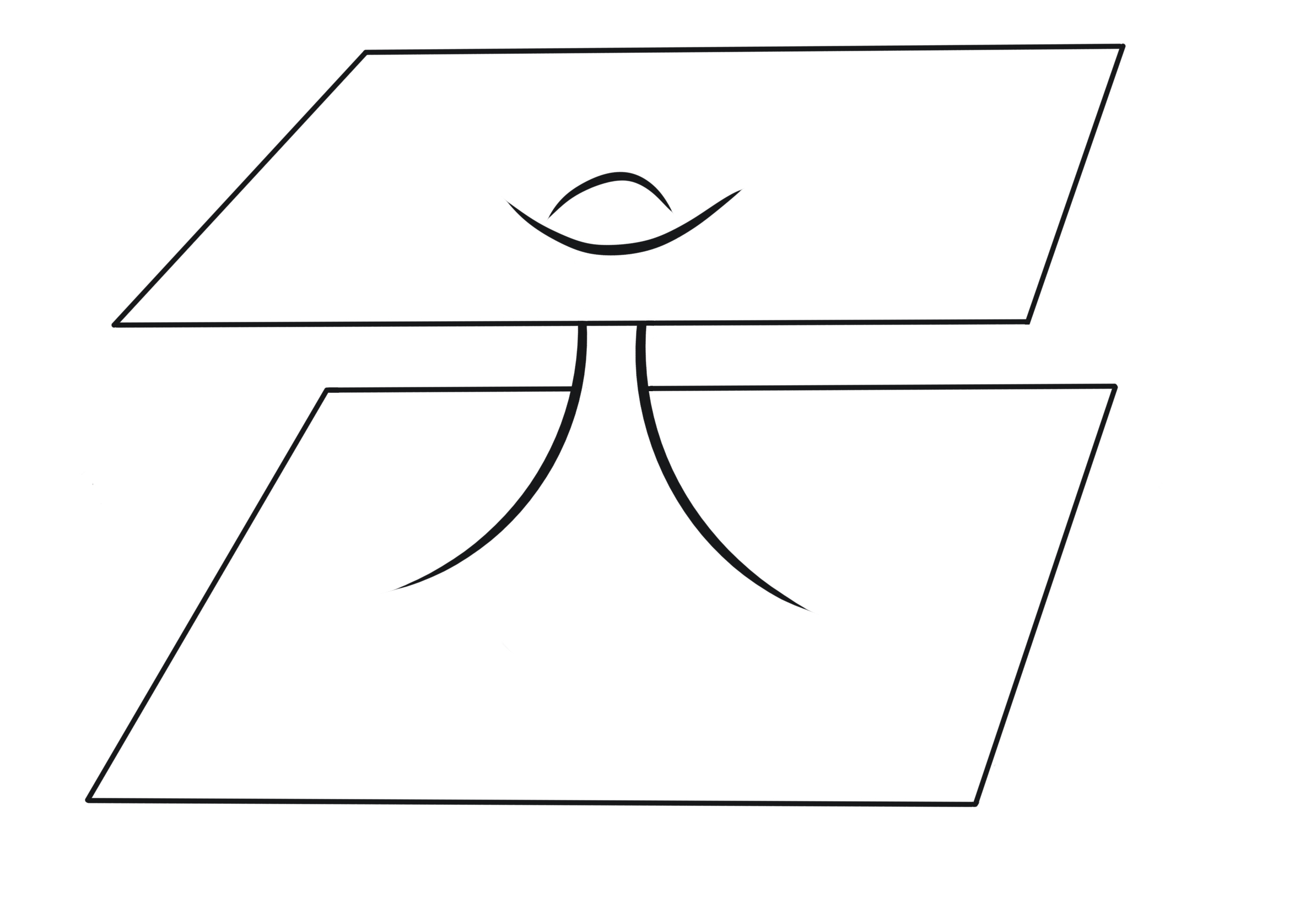}
    \caption{\small Wormhole connecting two different asymptotic spaces.}
    \label{fig:wormholePictureDifferentSpaces}
    \end{subfigure}
    \hfill
     \begin{subfigure}{.48\textwidth}
    \centering
    \begin{overpic}[angle=0,width=230pt]{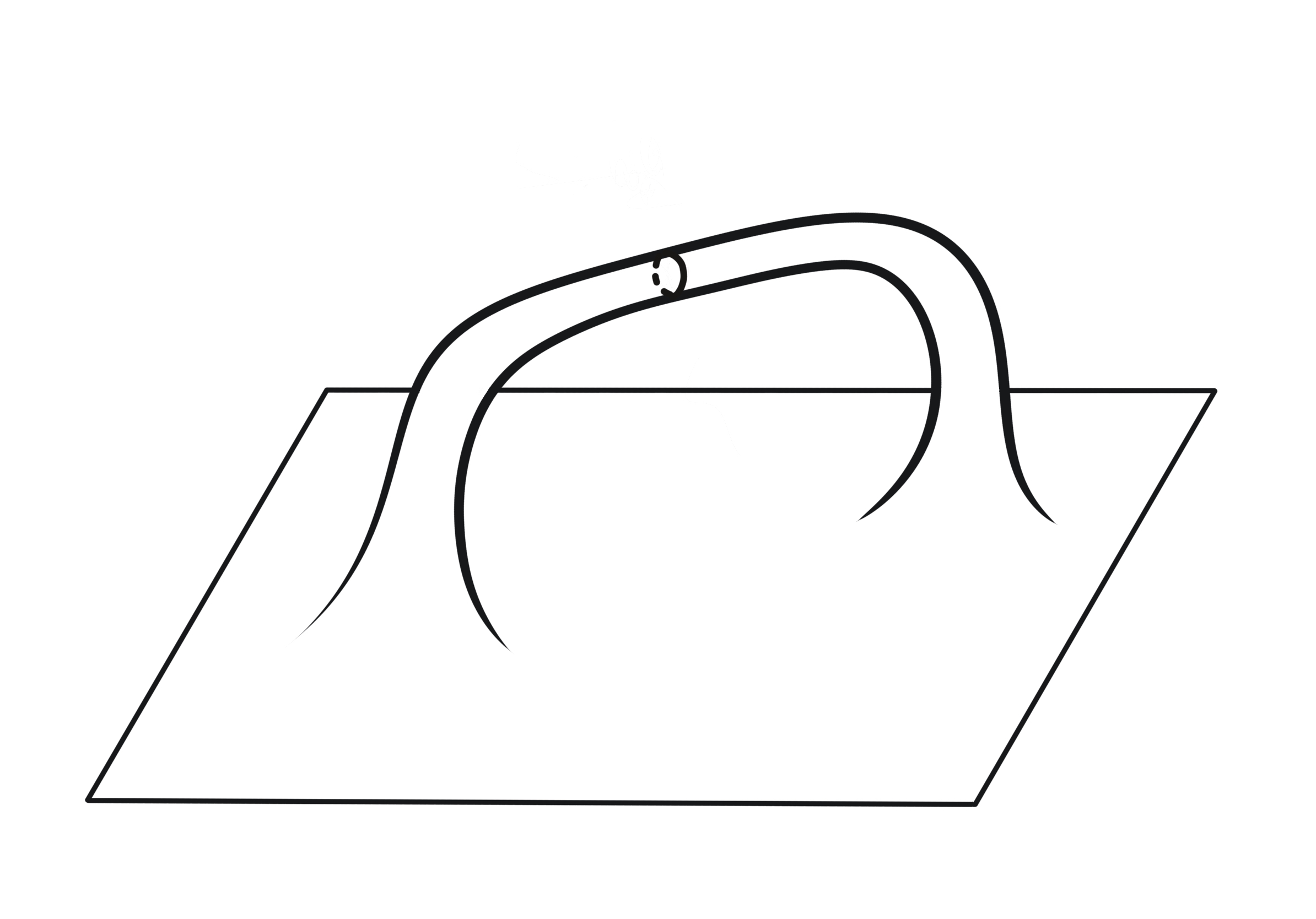}
    \put (49,53) {$L$}
    \end{overpic}
    \caption{\small Wormhole connecting two regions of the same asymptotic space.}
    \label{fig:wormholePictureSameSpace}
    \end{subfigure}
    \caption{\small Cartoons of a wormhole connecting  two different asymptotic spaces (left panel) and two regions of the same asymptotic space separated by a distance much greater than the wormhole neck radius $L$ (right panel).}
    \label{fig:wormholePicture}
\end{figure}

Consider a full (two-sided) non-extremal wormhole solution of the type discussed in the previous sections, illustrated in Fig.\ \ref{fig:wormholePictureDifferentSpaces}. One can view it as a conduit between two distinct asymptotically flat universes. Yet, the same solution also provides an accurate description of a short cut connecting two regions of the {\em same} background geometry separated by a distance much larger than the wormhole ``thickness'' $L$, as shown in figure \ref{fig:wormholePictureSameSpace}. In either case, at energy scales much smaller than $1/L$, wormholes can be ``integrated out'' and their effect codified in the appearance of appropriate interactions in an EFT defined at a new lower Wilsonian cutoff $\Lambda\ll 1/L$. These interactions are intrinsically non-local, and the leading order effects are captured by bi-local operators of the form \cite{Klebanov:1988eh,Preskill:1988na} 
\be\label{BKSlv}
\int \d^4x\sqrt{|g(x)|}\int \d^4y \sqrt{|g(y)|}~C^{IJ}\,\overline{\cal O}_I(x){\cal O}_J(y)\,,
\ee
with ${\cal O}_I$ local gauge-invariant operators and $C^{IJ}=\overline{C^{JI}}$ denoting a dimensionless 
non-degenerate constant matrix. In our setting, some properties of the operators $\calo_I(x)$ can be understood from general principles. Given a wormhole charge ${\bf q}\in \calc_{\text{\tiny WH}}$, one can pick a basis in which $\calo_I(y)$ represents the insertion of the corresponding half-wormhole, while $\overline\calo_I(x)$ represents the insertion of the anti-half-wormhole of charge $-{\bf q}$. Based on symmetry considerations alone, each $\calo_{I}(x)$ should be proportional to $e^{2\pi\ii q_i a^i(x)}$ and each $\overline\calo_{I}$ to $e^{-2\pi\ii q_i a^i}$. These phases precisely match the contribution of the boundary term \eqref{aHboundary} evaluated on the asymptotic boundary of each half-wormhole. Moreover, the entire contribution in \eqref{BKSlv} should be suppressed by $e^{-S_{\text{tot}}|_{\rm w}}$, where $S_{\text{tot}}|_{\rm w}$ denotes the \emph{real} part of the complete double-sided wormhole on-shell action (including possible higher-derivative terms). These two effects combine into the following universal dependence of the effective operators on the background fields 
\be\label{Ogen} 
{\cal O}_I\ \propto\  e^{-S_{\text{tot}}|_{\rm hw}}e^{2\pi\ii q_i a^i}\,,
\ee
where the action of half a wormhole $S_{\text{tot}}|_{\rm hw}\equiv \frac12 S_{\text{tot}}|_{\rm w}$ has made its appearance, as we anticipated above \eqref{Sw0}.

The bi-local contributions \eqref{BKSlv} can be rewritten in a manifestly local form by introducing the so-called Coleman's $\alpha$-parameters \cite{Coleman:1988cy}. Explicitly, \eqref{BKSlv} is equivalent to the insertion of the local terms 
\be\label{effcorrSlv} 
S_{\rm eff}=\int\d^4x\sqrt{|g|}\,\alpha^I\calo_I(x)+\text{c.c.}\,,
\ee
provided one includes the Gaussian integration 
\be\label{gaussianweightlv} 
 \int \d\alpha\,\d\bar\alpha\, e^{-\alpha^IC^{-1}_{IJ}\bar\alpha^J}
 \ee 
in the path integral. Corrections to \eqref{BKSlv}, as well as a non-trivial integration over otherwise disconnected geometries, affect the Gaussian weight but do not change the qualitative conclusion \cite{Klebanov:1988eh,Preskill:1988na}: wormholes can be viewed at low-energies as insertions of local operators provided an integration over seemingly innocuous ``parameters'' $\alpha^I$ is performed.

Wormholes are equivalently understood as non-perturbative tunneling processes in which ``baby universes'' are created or annihilated with amplitude $\sim e^{-S_{\text{tot}}|_{\rm hw}}$. The terms in \eqref{effcorrSlv} describe the creation and absorption of any number of baby universes by the perturbative vacuum state $|0\rangle$ we started from. The $\alpha^I$-parameters are expectation values of the creation-annihilation operators $\hat{\alpha}^I$ of such baby universes. One can then go to the so-called $\alpha$-vacua $|\alpha\rangle$, in which $\hat{\alpha}^I|\alpha\rangle=\alpha^I|\alpha\rangle$. The $\alpha$-vacua decohere and a given universe can be thought of as being in a superselection sector
labeled by a specific set of $\alpha^I$-parameters, which appear in the low-energy effective term \eqref{effcorrSlv}.

The presence of an Hilbert space of baby universes is one of the most striking, but also subtle, outcomes of Coleman's viewpoint on wormholes -- see also \cite{Marolf:2020xie} for an interesting recent revisitation of this interpretation. The associated integration over $\alpha^I$-parameters makes it impossible for the cluster decomposition principle to hold. This in general implies the non-factorization of correlation functions \cite{Arkani-Hamed:2007cpn}. In the context of the present paper, the fundamental quantum indeterminacy of the $\alpha$-parameters also appears in tension with experience from string theory, in which all couplings are dynamical and no free parameters exist, as also expected  for more general quantum gravity models \cite{McNamara:2020uza}.

There is yet another implication of the above logic that may appear puzzling -- see also the discussions in \cite{Heidenreich:2015nta,Hebecker:2016dsw}. It may seem to imply that no effective potential for our axions can be induced. In fact, according to \eqref{BKSlv}, complete wormholes subtract charge from $x$ and place it in another point $y$. An observer around $x$ would thus experience a {\emph{local}} violation of charge conservation, signaled by an anomaly in the Noether current:
\be\label{AnomalyWHlv}
\nabla_\mu J^\mu=\sum_{IJ}q_JC^{IJ}\left[\left(\int \overline{\cal O}_I\right){\cal O}_J-\overline{\cal O}_I\left(\int {\cal O}_J\right)\right].
\ee
Yet, {\emph{globally}} charge remains conserved.\footnote{This continues to hold when corrections to the dilute instanton gas approximation, in the form of wormhole and instanton interactions as described in \cite{Preskill:1988na}, are included.} In this situation, no effective potential for the axion can be generated. In order to induce such effect, the anomaly in \eqref{AnomalyWHlv} should have an overlap with the vacuum and a state of axions at rest, and this does not seem to happen here because the right-hand side of \eqref{AnomalyWHlv} is completely neutral. The only logical way to obtain an axion potential is if for some mysterious reason part of the wormholes did not bring back the charge to our Universe. In that case the symmetry between outgoing and incoming charge would be broken, and the vacuum would spontaneously break the shift symmetry. In the language of the $\alpha^I$-parameters, that peculiar condition is realized when some of the $\alpha^I$'s acquire a non-trivial vacuum expectation value, in conflict with what \eqref{gaussianweightlv} seems to indicate.

More generally, a natural way to avoid all the problems mentioned above is the existence of a huge gauge redundancy that identifies different $\alpha$ states,  implying that the Hilbert space of baby universes is in fact one-dimensional \cite{Marolf:2020xie,McNamara:2020uza}. In this case all $\alpha$ parameters would have unique values, controlled by some moduli fields. They would hence reduce to ordinary EFT couplings and no violation of factorization would take place. The Baby Universe Hypothesis \cite{McNamara:2020uza}  proposes that this is indeed what happens  in any consistent $d>3$-dimensional quantum gravity model. If correct, the Baby Universe Hypothesis would remove any arbitrariness in the $\alpha^I$-parameters appearing in \eqref{effcorrSlv},  allowing wormholes to break axionic symmetries, at least in principle.

In the following we assume that this ``natural selection'' of the $\alpha^I$-parameters actually takes place, but also provide arguments indicating that such selection must actually occur, at least in the case of $n=3$ non-extremal wormholes.

We begin with a discussion of the impact of the Gauss-Bonnet term on the low-energy theory (Section \ref{sec:phenoGB}). Subsequently we analyze more explicitly the structure of the low-energy EFT with $\Lambda\ll 1/L$. Specifically, the underlying supersymmetry imposes important restrictions on the wormhole-induced effective action: the local operators $\calo_{I}(x)$ should organize themselves in supersymmetric multiplets, and enter the low-energy effective action either through D-terms or F-terms. The a priori infinite series appearing in \eqref{effcorrSlv} is expected to be dominated by the  lowest derivative terms. We see in Section \ref{sec:extrFterms} that F-terms can be generated by EFT instantons (extremal wormholes), whereas regular wormholes can only generate D-term corrections to the K\"ahler potential at leading order (see Section \ref{sec:Dterms}). As we will see in Section \ref{sec:n=3physics}, the particular case of marginally degenerate wormholes ($n=3$) is associated with possible non-perturbatively generated superpotentials. Some phenomenological consequences for axion physics are discussed in Section \ref{sec:QFTpheno}.

\subsection{Large $N$ suppressions and Gauss-Bonnet}
\label{sec:phenoGB}

The two-derivative contribution to the half-wormhole action is bounded from below by \eqref{Swhsplit}, which codifies the combined gravitational-axionic contribution of the wormhole neck. Combining this observation with \eqref{boundonL}, which implies that any non-extremal wormhole solution can make sense only if $L^{-1}$ is smaller than the species scale evaluated at the wormhole's neck, and the consistency condition $M_{\text{sp}}\leq M_{\text{\tiny QG}}$ (see Section \ref{sec:perturbative} and \eqref{axiSC}), we obtain 
\be\label{minShw} 
S|_{\rm hw}\geq 
3\pi^3 \frac{M^2_{\text{\tiny P}}}{M_{\text{sp}}^2(s_*)}\geq \frac{3\pi}{16}N_{\text{sp}}\,
\ee
where $N_{\text{sp}}> N$ by construction. At the leading two-derivative level the low-energy operators \eqref{Ogen} are therefore suppressed by 
\be\label{supprOhw} 
e^{-S|_{\rm hw}}\leq e^{-\frac{3\pi}{16}N_{\text{sp}}}\,.
\ee
Even before asking whether F-terms or D-terms are actually induced or not (for that we will have to wait for Sections \ref{sec:extrFterms}, \ref{sec:Dterms}, \ref{sec:n=3physics}), we therefore see that wormhole effects must be very small. Clearly, the larger the number of degrees of freedom accessible to the QFT description, the smaller the UV cutoff, and the less relevant wormholes are.

Strictly speaking, \eqref{supprOhw} is an accurate estimate of the size of wormholes insertions only if a perturbative four-dimensional description can be extended all the way to the species scale. Presumably, this holds when the tower scale $M_{\text t}$ of Section \ref{sec:UVcutoff} corresponds to the string mass scale. In that case, generic higher dimensional operators are not expected to affect our estimate qualitatively because they are suppressed by powers of $1/(M_{\text{sp}} L)<1$. On the other hand, if $M_{\text t}$ denotes a KK mass, generically our four-dimensional description would be accurate only for $L>1/M_{\text{\tiny{KK}}}>1/M_{\text{sp}}$. Hence the purely four-dimensional contribution to the wormhole action would be $S|_{\rm hw}\geq 
3\pi^3 M^2_{\text{\tiny P}}/M_{\text{\tiny KK}}^2$ and the suppression of a wormhole insertion much more significant than shown in \eqref{supprOhw}. Of course, more interesting effects may potentially arise integrating in the KK resonances. For $L\ll 1/M_{\text{\tiny KK}}$ one may find higher-dimensional uplifts of our wormholes -- see \cite{Loges:2023ypl} for explicit examples of uplifted wormholes in string theory --  that potentially have smaller actions than in a strict four-dimensional regime with $L\gtrsim 1/M_{\text{\tiny KK}}$. A careful investigation of $d$-dimensional uplift of our wormholes would be necessary to confirm that, but this is beyond the scope of the present paper. In the following we will thus proceed using only results that are under our control. In that view \eqref{supprOhw} provides a conservative measure of the strength of non-perturbative gravitational effects in the axiverse.

While the previous discussion indicates that higher-derivative operators are generically not expected to change our conclusions qualitatively, 
as we emphasized in Section \ref{sec:semitop} this may not be the case for (semi-)topological terms. Let us thus discuss them. When considering wormhole physics, it turns out that the Pontryagin term is completely irrelevant. As discussed below \eqref{H3ansatz}, the Pontryagin operator  itself exactly vanishes by spherical symmetry on the wormhole solution \eqref{WHmetric}. More generally, the topological structure of a general non-spherically-symmetric wormhole configuration is such that the Pontryagin index (supplemented with the appropriate boundary term) vanishes. We can therefore safely ignore this term and focus on GB.

The potential relevance of the GB term for wormhole physics was emphasized in \cite{Giddings:1987cg,Kallosh:1995hi}. We find that indeed GB gives a {\emph{positive}} contribution to the on-shell action, at least if we are allowed to treat it as a perturbation. In such a case, at leading non-trivial order its effect on the equations of motion can be neglected when evaluating the on-shell action. Within this approximation $(S+S_{\text{\tiny GB}})|_{\text{hw}}$ is simply obtained by plugging in the leading order solution. Furthermore, in evaluating \eqref{GBterm2} on our solution we can practically ignore the asymptotic boundary term, since the wormhole metric \eqref{WHmetric} quickly becomes flat and so $Q\rightarrow Q_0$ as $r\rightarrow\infty$. Hence, the on-shell action \eqref{GBterm2} for the wormhole reduces to the purely bulk term 
\be\label{GBterm3}
\begin{aligned}
S_{\text{\tiny GB}}=-\int_\calm\sqrt{g}\, \gamma(\ell)\,E_{\text{\tiny GB}}\,.
\end{aligned}
\ee
Now, the GB density \eqref{GBdensity}  of a  wormhole metric \eqref{WHmetric} is negative definite since $E_{\text{\tiny GB}}|_{\text{wh}}= -{3 L^8}/({2\pi^2 r^{12}})$. So, denoting by $\gamma_{\rm min}$ the minimum value attained by $\gamma(\ell)$ along the dual saxion flow, we conclude that the on-shell Gauss-Bonnet term \eqref{GBterm3} of a semi-wormhole satisfies a lower bound
\be\label{GBlower0}
S_{\text{\tiny GB}}|_{\text{hw}}=-\int_{{\rm hw}}\sqrt{g}\,\gamma(\ell)\, E_{\text{\tiny GB}}\geq -\gamma_{\rm min}\int_{\rm hw}\sqrt{g}\, E_{\text{\tiny GB}}=\gamma_{\rm min}\,,
\ee
where in the last step we used the fact that  $\int_{\rm hw}\sqrt{g}\, E_{\text{\tiny GB}}=-1$, which we obtained by dividing by two the integral of $E_{\text{\tiny GB}}$ over the full wormholes of Fig.\,\ref{fig:wormholePicture}.\footnote{Recalling \eqref{Chern},  for asymptotically flat spaces with no finite distance boundaries we have $\int_{\calm}\sqrt{g}\, E_{\text{\tiny GB}}=\chi(\calm)-\chi(\calm_0)$, where $\calm$ denotes a general wormhole configuration with $S^3$ boundaries at asymptotically flat infinities and $\calm_0$ the corresponding (possibly reducible) flat space with the same asymptotic behavior, since the corresponding boundary terms in \eqref{Chern} cancel each other. If $\calm$ is as in Fig.\,\ref{fig:wormholePictureDifferentSpaces}, then $\chi(\calm)=0$ and $\chi(\calm_0)=\chi(\mathbb{E}_4\cup  \mathbb{E}_4)=2$, which gives $\int_{\calm}\sqrt{g}\, E_{\text{\tiny GB}}=-2$. Consistently, repeating the calculation for $\calm$  as in Fig.\,\ref{fig:wormholePictureSameSpace} one gets the same result.}

The r.h.s.\ of \eqref{GBlower0} reproduces the topological contribution considered in \cite{Giddings:1987cg}, where a constant $\gamma$ was assumed. Restricting our considerations to the controllable domain $\hat\Delta_\alpha$ and combining \eqref{GBlower0} and  \eqref{gammabound}, we get
\be\label{GBlower1} 
S_{\text{\tiny GB}}|_{\text{hw}}\geq\gamma_{\rm min}\geq \frac{N\pi}{6\alpha}\,,
\ee 
indicating that \eqref{Ogen} contains, on top of \eqref{supprOhw}, a further suppression of order
\be\label{OWHboundlv} 
e^{-S_{\text{\tiny GB}}|_{\text{hw}}}\leq e^{-\frac{N\pi}{6\alpha}}.
\ee
The condition \eqref{vafaSC} is equivalent to $S|_{\rm hw}\gtrsim S_{\text{\tiny GB}}|_{\rm hw}$, as required in a self-consistent derivative expansion. Hence, \eqref{OWHboundlv} only provides a correction to the dominant effect shown in \eqref{supprOhw}.\footnote{It is interesting to observe that the large suppression of  $e^{-\gamma}$ also guarantees  the weak-coupling regime of a possible third-quantization of the type proposed  in \cite{Giddings:1988wv}.}

In conclusion, it is not possible to precisely quantify the suppression of wormholes insertions, as the dominant contribution \eqref{supprOhw} depends on unknown details of the UV completion. Yet, the relation $S|_{\rm hw}\gtrsim S_{\text{\tiny GB}}|_{\rm hw}$ reminds us that a rough lower bound may be provided by the GB contribution alone. Interestingly, this itself is bounded from below by \eqref{GBlower1}, via a bound expressed solely in terms of quantities accessible to the low-energy observer. We can thus confidently claim that wormholes are substantially suppressed in the $N\gg 1$ limit. Even for moderately small couplings $\alpha=\pi/6\sim 0.5$ and just $N=100$ axions -- which is a quite natural number in stringy axiverse models -- the upper bound in \eqref{OWHboundlv} is of order $\sim 10^{-44}$. The actual suppression could be much stronger than \eqref{OWHboundlv}, though, since \eqref{gammabound} is generically quite conservative, as remarked below that equation. For instance, in the F-theory models of Section \ref{sec:FIIB} the argument in the exponential of  \eqref{OWHboundlv} must be corrected at least by a factor of six -- see \eqref{gammaboundF}. Furthermore, in the heterotic models of Section \eqref{sec:het}, the lower bound on the GB coefficient is even stronger. The numerical investigation summarized in Fig.\ \ref{fig:GBvsN} confirms our analytic expectations and also suggests that the GB coefficient $\gamma(s)$ can in fact scale with higher powers of $N$ as soon as $N>100$. Wormhole effects are evidently very suppressed in the axiverse.

\subsection{Extremal wormholes and F-terms}
\label{sec:extrFterms}

We now turn to a more explicit discussion of the leading low-energy terms that can be induced by wormholes. We begin with the BPS extremal wormholes introduced in Section \ref{sec:extremalBPS}. Conceptually, those objects appear to be qualitatively different from non-extremal wormholes, since they are more naturally interpreted as configurations sourced by localized EFT instantons and do not entail the introduction of $\alpha^I$-parameters nor baby universes. Nevertheless, the present discussion will serve as a reference for our subsequent investigation of non-extremal wormholes.

Our main concern here is to understand whether extremal wormholes are described at low energies via superpotential or K\"ahler terms. A necessary condition for a superpotential term to be generated is that the extremal wormhole configuration admits exactly two fermionic zero-modes. These would correspond to  't Hooft vertices with two fermion insertions, which is something that only a superpotential term can provide. As shown in Appendix \ref{app:WHsusy1}, extremal wormholes break two of the four supercharges preserved by the underlying flat spacetime. The corresponding two Goldstino-like zero-modes \eqref{extrzeromodes} are contained in the chiralinos $\bar\chi^{i\,\dot\alpha}$, the partners of $\bar t^i$. One would therefore expect extremal wormholes to be able to induce an effective superpotential at low energies. Let us elaborate on this possibility.

Because of the two zero-modes, a single extremal wormhole can only generate non-vanishing contributions to correlation functions involving at most two $\bar\chi^{i\,\dot\alpha}$ in excess over $\chi^i_\alpha$. We thus focus on the two-point function $\langle\bar\chi^{i\dot\alpha}(x)\bar\chi^{j\dot\beta}(y)\rangle$. We estimate it in a UV theory sourced by \eqref{SlocBPS} ignoring for the moment possible subtleties related to the insertion of the fundamental instanton (we will come back to this shortly). Explicitating only the integration over the zero-modes -- the massive modes can only affect the overall factor --  we schematically get 
\be 
\begin{aligned}\label{chi^2}
\langle\bar\chi^{i\,\dot\alpha}(x)\bar\chi^{j\,\dot\beta}(y)\rangle
\propto
&\ e^{-S_{\rm BPS}+2\pi\ii\langle{\bf q},{\bm a}_\infty \rangle }\int \d^4x_0\int \d^2 \theta\,[\bar\chi_{(0)}^{i\,\alpha}(x-x_0)]^{\dot\alpha}\theta_\alpha\,[\bar\chi_{(0)}^{j\,\beta}(y-x_0)]^{\dot\beta}\theta_\beta\\
\sim &\ \,e^{\frac{1}{2}K_\infty}e^{2\pi\ii\langle{\bf q},{\bm t}_\infty\rangle}{\calg}^{im}_\infty q_m{\calg}^{jn}_\infty q_n\int \d^4x_0\,[S_{\rm F}(x-x_0)]^{\dot\alpha}_\alpha[S_{\rm F}(y-x_0)]^{\dot\beta}_\beta\varepsilon^{\alpha\beta}
\end{aligned}
\ee
with the complete complex BPS on-shell action \eqref{complBPSlocaction} appearing as an overall factor. In deriving \eqref{chi^2} we replaced $\bar\chi^{i\,\dot\alpha}(x)=[\bar\chi_{(0)}^{i\,\beta}(x-x_0)]^{\dot\alpha}\theta_\beta+\cdots$ (the dots refer to non-zero-modes), where the zero-mode content of the field is parametrized by the wavefunction defined in \eqref{extrzeromodes}, with $x_0$ and $\theta_\alpha$ denoting the associated bosonic and Grassmanian parameters. In the final step we used the fact that in the limit $|x-x_0|,|y-x_0|\to\infty$ the asymptotic dependence of the Goldstino wavefunction is precisely the same as that of the Feynman's propagator $S_{\rm F}$, see \eqref{asympzerom}, whereas the overall factor of $e^{K_\infty/2}$ is due to its normalization. 

It is now easy to show that the above correlator can be reproduced, at distances sufficiently far from the singularity, by a supergravity F-term (we adopt the conventions of \cite{Wess:1992cp})
\be\label{extrFterms} 
\int\d^2\theta\, 2\cale\, W_{\bf q}\,
\ee 
where
\be\label{BPSsuperpotential}
 W_{\bf q}=\cala_{\bf q}\, M^3_{\text{\tiny  P}}\,e^{2\pi\ii\langle{\bf q},{\bm t}\rangle} \,,
\ee
and $\cala_{\bf q}$ includes the contribution of the non-zero-modes.  The superpotential \eqref{BPSsuperpotential} induces an effective fermionic vertex $e^{K/2}(\del_i\del_j W_{\bf q})\chi^{i}_\alpha\chi^j_\beta\varepsilon^{\alpha\beta}$, and observing that $\langle\chi^i_\alpha(x_0)\bar\chi^{j\dot\alpha}(x)\rangle\propto \calg^{ij} [S_{\rm F}(x-x_0)]^{\dot\alpha}_\alpha$, we see that a single insertion of \eqref{BPSsuperpotential} would precisely reproduce the structure in \eqref{chi^2}, provided ${\bm t}_\infty$ is identified with the classical background in the effective field theory. In analogy to \cite{Affleck:1983mk} we can thus conclude that at low energies the BPS extremal wormhole of charge ${\bf q}\in\calc^\text{\tiny EFT}_{\rm I}$ generates the superpotential \eqref{BPSsuperpotential}. BPS extremal anti-wormholes of charge $-{\bf q}\in\calc^\text{\tiny EFT}_{\rm I}$ would instead generate the complex conjugate of \eqref{extrFterms}. See \cite{Rey:1989xj} for a discussion along similar lines.

Before jumping to the conclusion that extremal wormholes can generate superpotentials at low energies, we should however address the subtlety we alluded to earlier. Recall that extremal wormholes are singular and that a localized UV-sensitive contribution \eqref{SlocBPS} had to be included in our argument. Additional contributions may be present, though. Indeed, fundamental instantons generically carry additional internal degrees of freedom, which can in principle contribute both zero and non-zero fermionic modes. The presence of additional fermionic zero-modes may invalidate our calculation and for example reveal that fundamental instantons actually induce multi-fermion and higher-derivative F-terms as in \cite{Beasley:2004ys,Beasley:2005iu}, as opposed to superpotential terms like those in \eqref{extrFterms}. Whether or not this does indeed occur can only be determined with some knowledge of the microscopic UV completion of the four-dimensional model. Let us therefore consider some known examples in string theory.

In string theory models the fundamental instantons sourcing  extremal wormholes uplift to various types of branes wrapping internal cycles, which support various types of localized world-volume  degrees of freedom -- see for instance \cite{Dine:1986zy,Dine:1987bq,Becker:1995kb,Witten:1996bn,Harvey:1999as,Witten:1999eg} and \cite{Blumenhagen:2009qh} for a review. Extra fermionic zero-modes are known to appear for example as Goldstinos of an ``accidental'' enhanced supersymmetry felt by  microscopic brane configurations that are ``strongly'' isolated, in the sense that these brane configurations as well as any multiple thereof  are isolated and can probe only some local internal geometry. However, the string theory realizations  described in Section \ref{sec:WHst} clearly indicate that such strongly isolated branes correspond to {\em non}-EFT instanton charges, whereas  an EFT instanton charge (or an appropriate multiple thereof) corresponds to a non-isolated brane that can explore the entire compactification space and in principle ``detect'' its global minimally supersymmetric structure. Thus, our EFT considerations should not be affected by  additional Goldstino-like fermionic zero-modes associated with a local enhanced supersymmetry. 

More concretely, in the F-theory models described in Sections \ref{sec:FtheoryP3}, \ref{sec:Ftheory1} and \ref{sec:WHst1}, two simple upliftings of EFT instantons are provided by Euclidean D3-branes wrapping the hyperplane divisor in $\mathbb{P}^3$ and the vertical divisor in a $\mathbb{P}^1$ fibration over $\mathbb{P}^2$. Even if at first sight these branes contain extra fermionic (and bosonic) zero-modes, those are actually lifted by the inclusion of world-volume fluxes \cite{Bianchi:2011qh} and/or the interaction with background D3-branes \cite{Palti:2020qlc}. As a result, in these explicit examples the localized contribution associated to the EFT instanton is known not to bring extra fermionic zero-modes. By F-theory/heterotic duality, the F-theory mechanism must have heterotic counterparts. As discussed in Section \ref{sec:WHst2}, in the large volume perturbative regime part of the EFT instantons correspond to F1-strings wrapping movable curves, which are either non-isolated or admit a non-isolated multiple. Our remarks then fit well with the early observation \cite{Dine:1987bq} that non-isolated genus zero world-sheet instantons satisfy all necessary conditions to contribute to the superpotential. 

Note that non-EFT instantons may also  generate superpotentials of the form \eqref{BPSsuperpotential}. In fact, many string theory examples of brane instanton  superpotentials are generated by strongly isolated branes -- see e.g.\ the review \cite{Blumenhagen:2009qh}. However, according to the above observations and to the discussion of Section \ref{sec:EFTinst}, these non-perturbative effects may be interpreted as being non-gravitational in nature.

These considerations suggest that extremal wormholes generated by EFT instantons should dominantly match onto effective superpotentials like \eqref{BPSsuperpotential} at low energies. More precisely, because the cleanest examples of \cite{Palti:2020qlc} correspond to extremal wormholes in perturbative sectors of homogeneity degree $n=3$, this conclusion should at least apply to that subclass of extremal wormholes. This expectation fits well with the Supersymmetric Genericity Conjecture of \cite{Palti:2020qlc}, which implies that in an $\caln=1$ four-dimensional theory of quantum gravity a superpotential term should vanish only if the theory is related to a higher supersymmetric one. In our context,  the $n=3$ EFT instantons appear as the natural responsible for the realization of this conjecture, at least for $t^i$ chiral sectors that cannot get superpotential terms at tree level, as for instance induced by the three-form potentials mentioned in Footnote \ref{foot:fluxsup}. Note that the Supersymmetric Genericity Conjecture is quantum gravitational in nature, which resonates well with the already emphasized gravitational nature of the EFT instantons, and the fact that their existence can be guaranteed by invoking the BPS instanton tower or sublattice WGCs of Section \ref{sec:EFTinst}.  These speculations would also imply that the low energy theory contains a sum $\sum_{\bf q}W_{\bf q}$ of contributions like \eqref{extrFterms}, for all  charges populated by EFT instantons with $n=3$, and that according to BPS instanton tower/sublattice WGCs this sum is infinite.

In Subsection \ref{sec:n=3physics}, we will see how these considerations can be generalized to, and further supported by, the marginally degenerate wormholes discussed in Section \ref{sec:n=3}.

\subsection{Wormhole D-terms}
\label{sec:Dterms}

Let us now consider the everywhere non-degenerate wormholes (i.e.\ those with $n>3$). They do not preserve any of the four supersymmetries of the asymptotic Minkowski vacuum, and as such they are expected to have four corresponding Goldstino zero-modes. In the case of homogeneous wormholes of Section \ref{sec:univWH}, such zero-modes are explicitly constructed in Appendix \ref{app:WHsusy2}. A computation like \eqref{chi^2} now vanishes identically: no F-term can be generated. Instead, one can check that it is correlators of the type $\langle\bar\chi\bar\chi\chi\chi\rangle$ that can be reproduced at large distances by a local four-fermion operator, since all zero-modes have wavefunctions that asymptotically behave as the free Feynman propagator $S_{\rm F}$ for any $n>3$, as one can check by expanding  \eqref{eq:appSUSY:subextrHomogeneous} for $r\gg L$. The case $n=3$, as usual, should be treated separately, and will be considered  in Section \ref{sec:n=3physics}.

The leading terms in the low-energy  contributions \eqref{effcorrSlv} should then organize themselves into a supergravity D-term providing a correction $\Delta K_{\bf q}$ of the K\"ahler potential.  Following the same procedure used to derive \eqref{chi^2}, or equivalently adopting the arguments  of  \cite{Anglin:1992ym}, one finds that $\Delta K_{\bf q}$ should be given by the superfield extension of the operator \eqref{Ogen}. Specifying to the homogeneous solution \eqref{ellsol} we have   
 \be\label{nonpertK} 
\Delta K_{\bf q}= \cala'_{\bf q}\, \left(e^{2\pi\ii \langle {\bf q},{\bm t}\rangle}+e^{-2\pi\ii \langle {\bf q},\overline{\bm t}\rangle}\right)e^{-2\pi f(\Im t)}\,,
 \ee
where $\cala'_{\bf q}$ is another unknown normalization constant containing also loop effects, $t^i=a^i+is^i$ should of course be considered as chiral superfields, and from the on-shell  action \eqref{Sw01} we read
\be\label{fimt} 
2\pi f(\Im t)\equiv S|_{\text{hw}}-2\pi \langle {\bf q},{\bm s}\rangle=\left[\sin\left(\frac{\pi}{2}\sqrt{\frac3n}\right)-1\right]2\pi\langle {\bf q},\Im{\bm t}\rangle\,.
\ee
This expression is reliable as long as ${\bf q}$ is large, $n>3$, and $s^i$ is aligned with $\calf^{i}({\bf q})$. For smaller charges and/or non-homogeneous saxionic wormholes $f(\Im t)$ may of course be different. Still, according to the general estimates given at the beginning of this section, any non-perturbative correction like \eqref{nonpertK} should be extremely tiny within the regime of validity of the EFT, and certainly much smaller than the expected {\em perturbative} corrections to the leading K\"ahler potential \eqref{KP(s)}. However, even if tiny, (7.16) represents a breaking of the axionic shift symmetries, which cannot  be broken by perturbative corrections.

As already emphasized, the two derivative action \eqref{Sw01}, which is always accurate in the large ${\bf q}$ limit, is {\em lower} than the BPS action of an EFT instanton, or of a multi BPS instanton configuration carrying the same charge. We have argued in \ref{sec:BPSbound} that it is precisely because of a finite size throat that this happens. We are thus tempted to interpret non-extremal half-wormholes as non-BPS bound states of fundamental instantons describing throats that open up around the EFT instantons themselves. In this view, the throat-opening would effectively integrate out whatever localized internal degrees of freedom the EFT instantons may have. Furthermore, being non-BPS bound states, non-extremal wormholes would induce D-terms like \eqref{nonpertK}, rather than F-terms as in Section \ref{sec:extrFterms}. Non-extremal wormholes are thus qualitatively different than EFT instantons, but may simultaneously be intimately connected. This view is consistent with the ten-dimensional uplifts of similar wormholes recently constructed in \cite{Loges:2023ypl}, that were interpreted in terms of intersecting Euclidean branes. In the following subsection we will further elaborate on the connection between wormholes and instantons.

\subsection{Marginally degenerate wormholes and superpotentials}
\label{sec:n=3physics}

It is now time to discuss the physics associated with the marginally degenerate wormholes ($n=3$). In Section \ref{sec:n=3} we have seen how, slightly away from the wormhole's neck, such configurations have approximately the same form as extremal BPS wormholes associated with EFT instantons of the same charge. Moreover, after an appropriate IR regularization their action \eqref{apprSn=3} looks approximately like the instanton local term \eqref{SlocBPS} that complements extremal wormholes. These observations strongly suggest a deep connection between wormholes with $n=3$ and fundamental EFT instantons. In this section we would like to elaborate upon this connection. We first motivate the apparently ad hoc IR regularization introduced in Section \ref{sec:n=3} and subsequently discuss how, extending to our general framework a mechanism proposed in \cite{Giddings:1989bq,Park:1990ep}, marginally degenerate wormholes may be able to generate superpotential terms as those discussed in Section \ref{sec:extrFterms}.

In order to unclutter the notation, here  the IR-regularizing cutoff will be  simply denoted as $\Lambda$, rather than $\Lambda_{\text{\tiny IR}}$ as in Section  \ref{sec:n=3}.  This choice is also consistent with the fact that in this section $\Lambda$ plays the role of the Wilsonian cutoff of the IR effective theory that one obtains by ``integrating out'' the wormhole contributions, as discussed at the beginning of this section. One should then keep in mind that in this section $\Lambda$ is much lower than, and  should not be confused with, the UV cutoff of the initial EFT in which the wormhole solutions are derived.  

\begin{figure}[!t]
    \centering
    \begin{subfigure}{.48\textwidth}
    \begin{overpic}[angle=0,width=230pt]{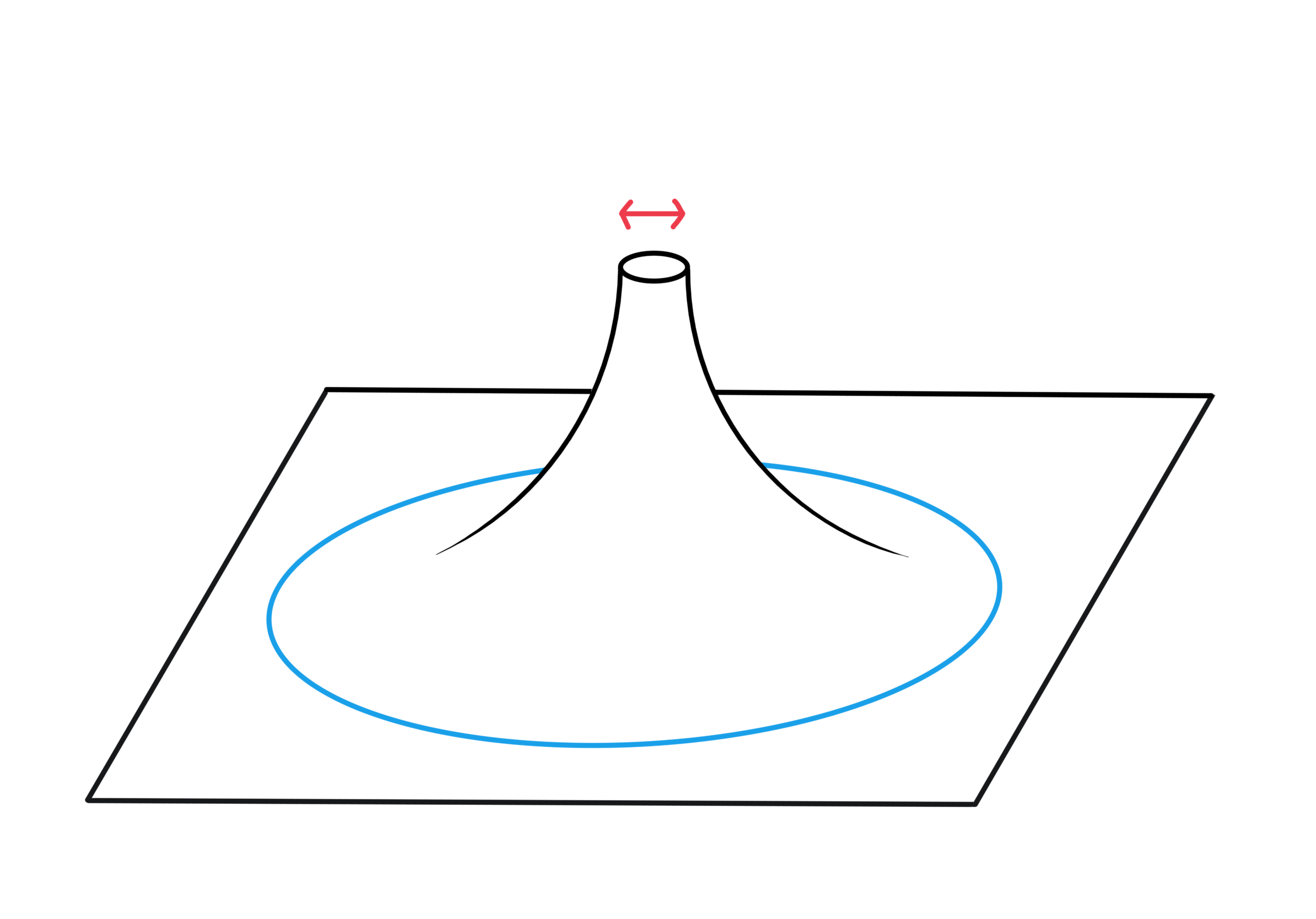}
    \put (48.1,55.9) {${\color{red}L}$}
    \end{overpic}
    \end{subfigure}
    \hfill
    \begin{subfigure}{.48\textwidth}
    \centering
    \includegraphics[angle=0,width=230pt]{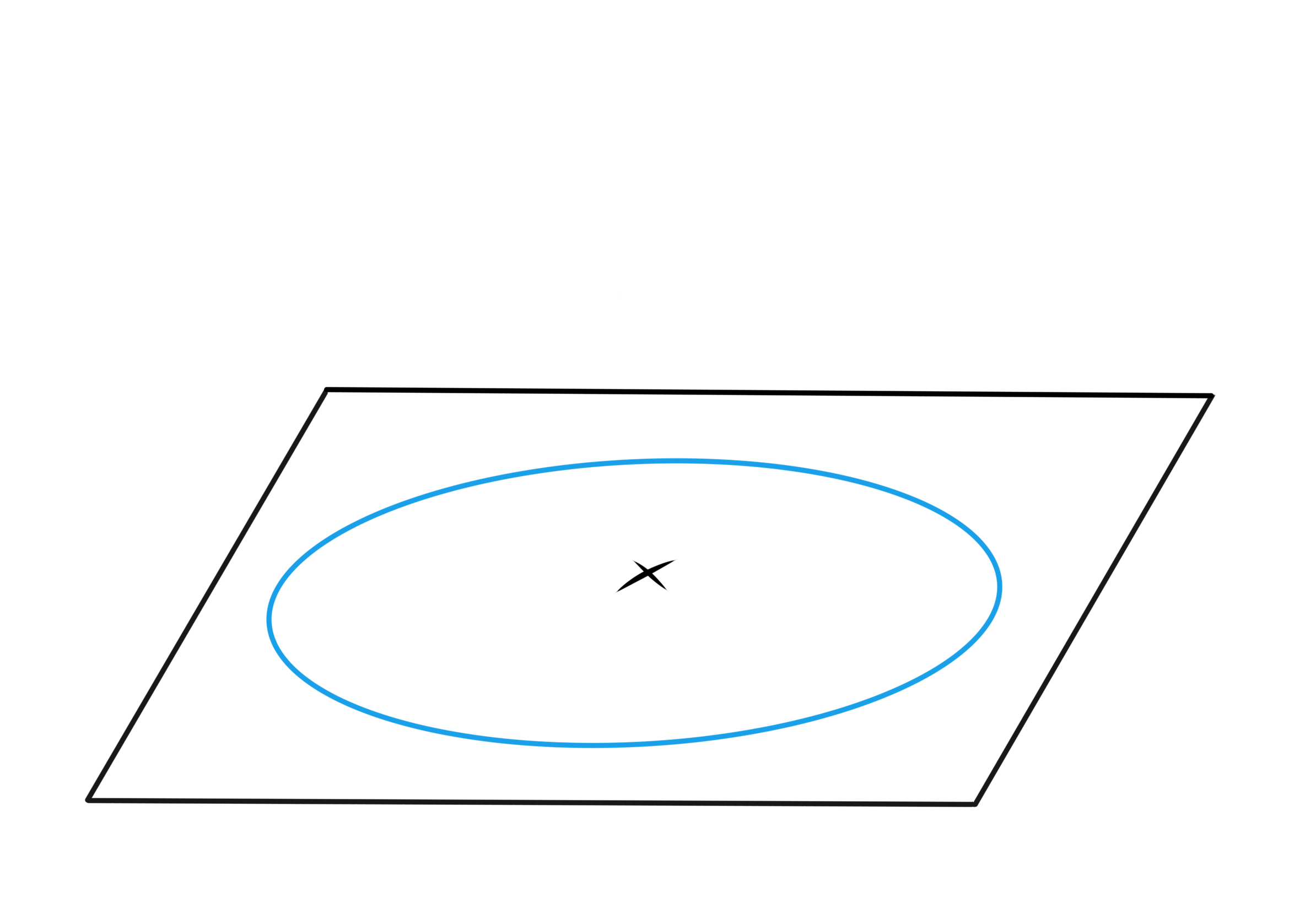}
    \end{subfigure}
    \caption{Pictorial representation of a marginally degenerate half-wormhole surrounded by a radius at the IR cutoff $1/\Lambda\gg L$ (left panel). Sending $L\to0$ one formally obtains a fundamental BPS instanton insertion (right panel).}
    \label{fig:prova}
\end{figure}

The regularization prescription adopted in Section \ref{sec:n=3} may be interpreted as identifying approximate wormhole solutions of the equations of motion, following \cite{Coleman:1989zu}. One cuts a ball of radius $r_\Lambda\equiv 1/\Lambda$ out of a flat Euclidean background and dual saxions with constant background, and replaces it with the IR-regularized half-wormhole, smoothly gluing the fields at the common boundary three-sphere (see the left of Fig.\ \ref{fig:prova}). In our case, the value of the background dual saxions on the  boundary three-sphere is given by ${\bm\ell}_{\text{\tiny IR}}={\bm\ell}(r_\Lambda)$. This cut-and-paste procedure gives a configuration that is an approximate saddle-point whose action receives non-negligible contributions only from the region $r\leq r_\Lambda$. The result is essentially the regularized action \eqref{apprSn=3} with $\Lambda_{\text{\tiny IR}}\rightarrow\Lambda$, and the dependence of the associated effective operators \eqref{Ogen} on the background fields is thus also expected to be controlled by the same ``on-shell" action \cite{Abbott:1989jw,Coleman:1989zu}.\footnote{In principle this regularization procedure may be applied to the everywhere non-degenerate  $n\geq 4$ wormholes as well. However, in those cases the background fields quickly reach their asymptotic value, cf.\ \eqref{tildellr}, and one can hence identify ${\bm\ell}_\Lambda$ with ${\bm\ell}_\infty$, as we implicitly did in our treatment of Section \ref{sec:Dterms} (see also the comments at the end of Section \ref{sec:effd=1}).}

A natural interpretation of the IR regularization of Section \ref{sec:n=3} is to formally view it as a prescription to ``integrate out" the wormhole down to energy scales $\Lambda\ll1/L$ \cite{Abbott:1989jw,Coleman:1989zu}. Suppose we are interested in computing some correlation function at distances $\gg L$. We can take into account the effect of the wormhole in two equivalent ways. Either we compute the full path integral for our initial EFT, including the wormhole and its fluctuations around it, or we use a low-energy effective description defined below $\Lambda$ where the geometry is essentially flat and the wormhole is replaced by a set of effective local operators ${\cal O}_{\text{\tiny IR}}$ like those in \eqref{Ogen}. We may call these the ``UV" and ``IR" descriptions of our theory. In a semi-classical expansion, the first, ``UV" theory gives a factor $e^{-S|_{\text{hw}}}$, where $S|_{\text{hw}}$ is precisely the wormhole action we would compute in our initial EFT \eqref{kinetic}. The second gives ${\cal O}_{\text{\tiny IR}} e^{-S|_{\text{hw}}^{\rm IR}}$, where $S|_{\text{hw}}^{\rm IR}$ is the on-shell action for the ``IR" theory in the presence of the operator insertion. Because by definition the low-energy degrees of freedom are precisely the same, the long-distance $r>1/\Lambda$ contributions to the actions exactly match; they only differ at distances smaller and of order the matching scale $1/\Lambda$. Therefore, equating the results of the two descriptions, we learn the effective operator must depend on the background fields as follows \cite{Abbott:1989jw} 
\be
{\cal O}_{\text{\tiny IR}}\propto e^{-(S|_{\text{hw}}-S|_{\text{hw}}^{\rm IR})}=e^{-S|_{\text{hw}}^\Lambda}.
\ee
The final exponential contains the expression \eqref{apprSn=3}. The effect of the marginally degenerate wormhole, once renormalization-group evolved down to $\Lambda\ll1/L$, is thus captured by effective local operators suppressed by the exponential of our IR-regularized action \eqref{apprSn=3}. The agreement between \eqref{apprSn=3} and \eqref{SlocBPS} is not a mere coincidence, as one may have thought by our analysis in Section \ref{sec:n=3}: up to corrections of order $\Lambda^4L^4$ the effect of a marginally degenerate wormhole is {\emph{exactly equivalent}} to that of a fundamental BPS instanton of the same charge.

If what we are arguing is correct, then the results of Section \ref{sec:extrFterms} would tell us that marginally degenerate wormholes can induce superpotentials at low energies, as opposed to their non-degenerate ($n>3$) siblings. At the very least this conclusion should hold when $\Lambda^4L^4\to0$, where the solution formally coincides with that of a fundamental instanton.

The claim that wormholes can induce F-terms was first made in \cite{Giddings:1989bq,Park:1990ep}. At a first sight, there is an apparent obstruction: for any $L\neq0$ the $n=3$ wormholes break all supersymmetries and therefore have additional Goldstino-like zero-modes in the chiral partners $\chi^i$ of $t^i$, see Appendix \ref{app:n=3}. This would naively suggest that such configurations should generate D-terms like those of Section \ref{sec:Dterms}, rather than superpotential terms. Interestingly, also this conclusion is too rushed. Let us actually ask if our $n=3$ wormhole describes four-fermion interactions at low energies $\Lambda\ll1/L$. As mentioned in Section \ref{sec:Dterms}, a procedure analogous to the one carried out in \eqref{chi^2} can be applied to evaluate four-point functions of the form $\langle\bar\chi\bar\chi\chi\chi\rangle$. Yet, here that result cannot be interpreted as the insertion of a local operator involving four free fermions, since the contribution of the wavefunctions of the extra zero-modes vanishes away from the wormhole faster than a free Feynman propagator. The reason is that the additional Goldstino-like zero-modes present in the marginally degenerate wormhole are localized around the throat, where supersymmetry is completely broken. Indeed, we have seen in \eqref{moregenn=3} that the throat is the only region where the $n=3$ configuration departs from the extremal solution by corrections of order $\calo(L^4/r^4)$, and as a consequence that must also be the only region where the profiles of the zero-modes of the two backgrounds can differ. This is indeed the case. In conclusion, the fact that for $n=3$ the two extra zero-modes are localized at the throat prevents us from obtaining a four-fermion operator at low energies. Still, their presence seems to obstruct the generation of a superpotential as well. How can we reconcile this with our earlier observation that marginally degenerate wormholes are BPS instantons in disguise?

Well, as emphasized around \eqref{n=3Larb}, $n=3$ wormholes appear in an entire family of solutions with arbitrary $L$. The extra zero-modes of such solutions would certainly be relevant for any $L$ within the calculable regime. But the profiles of the special wormholes obtained in the formal limit $L\to0$  disappear behind the singularity. If we could ignore curvature singularities, the extra zero-modes of such wormholes would in some sense become invisible to our EFT path integral. The subclass of $n=3$ wormholes with $L\to0$ would then have effectively two zero-modes like the BPS solution, and hence the ability to generate a superpotential as in Section \ref{sec:extrFterms}, as proposed in \cite{Giddings:1989bq,Park:1990ep}. The limit $L\to0$ is visually seen passing from the left to the right of Fig.\ \ref{fig:prova}.

The arguments we just presented are essentially a rephrasing, in our more general setting, of those of \cite{Giddings:1989bq,Park:1990ep}. To see this, take our $n=3$ solution and, instead of regularizing it with an IR cutoff at $r_\Lambda$,  adopt a different regularization (see Appendix \ref{app:n=3} for more details). Namely, multiply the kinetic potential $\calf(\ell)$ by $(1+\varepsilon)$, where $\varepsilon>0$ is a very small constant, effectively replacing $n=3$ with $n=3(1+\varepsilon)>3$. This produces an everywhere non-degenerate wormhole profile, with a precise relation between its IR value $\tilde{\ell}_\infty$ and its UV value $\tilde{\ell}_*$ (see \eqref{tildellr}). Using \eqref{Ln=3}, and at leading non-trivial order in $\varepsilon$, this reads
\be\label{lIRlUV}
\tilde\ell_\infty=\frac\pi4 \varepsilon \tilde\ell_*=\frac{\varepsilon}{8  M^2_{\text{\tiny P}}L^2}\,.
\ee
One finally picks $\varepsilon$ so that ${\bm\ell}_\infty$ in the $\varepsilon$-regularized theory coincides with the ${\bm\ell}_\Lambda$ of our IR-regularized theory. This way we have found a different regularization of our profile where, if needed, the limit $r\to\infty$ may be taken. The papers \cite{Giddings:1989bq,Park:1990ep} suggest then that the contribution of the $n=3$ wormholes to the IR effective theory can be computed by taking the limit $\varepsilon\rightarrow 0$ of the $\varepsilon$-regularized wormholes while keeping ${\bm\ell}_\infty$ fixed. From \eqref{lIRlUV} we see that this prescription is actually equivalent to the limit $L\to0$ we alluded to earlier. As in the case discussed above, also the $\varepsilon$-regularized wormholes have extra Goldstino-like zero-modes $\delta\chi^i_\alpha$ localized around their throat. Taking $\varepsilon\rightarrow 0$, or equivalently $L\to0$, the extra Goldstino-like zero-modes are effectively pushed beyond the EFT and, as such, they are no more integration variables in the EFT path integral.

In conclusion, the reasoning of \cite{Park:1990ep} applied to our context implies that marginally degenerate wormholes generate a superpotential of the form \eqref{BPSsuperpotential}. Assuming that the $L\rightarrow 0$ limit sensibly connects these wormholes to the corresponding fundamental instantons, from this we can also infer that the $n=3$ EFT instantons do not contain any extra fermionic zero-mode and hence do generate \eqref{BPSsuperpotential}, as speculated in Section \ref{sec:extrFterms}. The correspondence between $n=3$ wormholes and EFT instantons has additional important implications. It says that the wormhole $\alpha$-parameters, whose presence  has been for simplicity suppressed, must be fully determined by an EFT instanton calculation in the UV complete theory, which has no free parameters at the onset. This would represent a concrete manifestation of the Baby Universe Hypothesis \cite{McNamara:2020uza}. Moreover, a non-trivial wormhole-induced superpotential tells us that the associated $\alpha$-parameters are generically non-vanishing, realizing the Supersymmetric Genericity Conjecture of \cite{Palti:2020qlc}. (Note that, in some non-generic cases, the UV-complete computation may still give vanishing $\alpha$-parameters.)

Another interesting consequence of  the wormhole argument is that a superpotential of the form \eqref{BPSsuperpotential} should be generated  for {\em any} charge ${\bf q}\in\calc_{\text{\tiny WH}}$ corresponding to an $n=3$ sector. This conclusion may be extended to ${\bf q}\in\calc_{\rm S}^{\text{\tiny EFT}}-\calc_{\text{\tiny WH}}$ by considering restricted perturbative regimes, as described in Footnote \ref{foot:CWH} and for instance exemplified in Section \ref{sec:WHst2}. The picture that emerges is that fundamental  EFT instantons carrying the same charges ${\bf q}$, or at least a multiple thereof (possibly selected by some stricter $\calh_{3,i}$ quantization rule), should exist as well, consistently with the BPS instanton tower/sublattice WGCs of Section \ref{sec:EFTinst}.

Finally, as shown in \eqref{apprSn=3} in the above $L\rightarrow 0$ limit the on-shell action reduces to the BPS one, $S|_{\rm hw}\rightarrow S_{\text{BPS}}=2\pi\langle {\bf q},{\bm s}_{\text{\tiny IR}}\rangle$. Since  ${\bf q}\in\calc_{\text{\tiny WH}}$, by definition it satisfies $\langle {\bf q}, {\bf e}\rangle \geq 1$ for any ${\bf e}\in \calc_{\rm S}^{\text{\tiny EFT}}$, see  \eqref{calcWH}. We can now run the same argument used to go from \eqref{gammabound1} to \eqref{tildeCs} and conclude that    $\langle {\bf q},{\bm s}_{\text{\tiny IR}}\rangle\geq N/\alpha$ for ant ${\bm s}_{\text{\tiny IR}}\in\hat\Delta_\alpha$, and therefore: 
\be 
e^{-S_{\text{BPS}}}\leq e^{-\frac{2\pi}{\alpha}N},~~~~~~~~~~({\bm s}_{\text{\tiny IR}}\in\hat\Delta_\alpha~~~\&~~~{\bf q}\in\calc_{\text{\tiny WH}})\,.
\ee
We see that also the contribution of the marginally degenerate wormholes is strongly suppressed in perturbative regimes associated  with $N\gg 1$ (s)axions.

\subsection{Implications for purely axionic models}
\label{sec:QFTpheno}

The symmetry-breaking mediated by non-perturbative effects can potentially induce effective axion potentials $V_{\text{eff}}(a)$ at low energies. Quantifying the impact of these effects is crucial in models for inflation and for the QCD axion approach to the strong CP problem, where even tiny corrections can have dramatic phenomenological consequences.

The physics of axionic wormholes  has been largely studied in the literature from an EFT perspective -- see for instance the  review papers  \cite{Hebecker:2018ofv,Alvey:2020nyh} and references therein. From an agnostic low-energy point of view, the axions appearing in such bottom-up constructions can either be a subgroup of our $a^i$'s, whose saxionic partners have been stabilized and acquired a mass, or can be completely independent degrees of freedom, say arising from the breaking of accidental compact symmetries linearly realized in the EFT. We will refer to the former as scenarios of ``fundamental axions" and the latter as scenarios of ``QFT axions". In the following we ask what lessons can be drawn about symmetry-breaking in such scenarios from our results. Our key assumption is that any such low-energy axion model admits, at sufficiently high energies, an intermediate  ${\caln=1}$ axiverse description as in Sections  \ref{sec:N=1models} and \ref{sec:regimesEFT} (plus the necessary additional degrees of freedom).

As a preliminary observation we recall that, in a supersymmetric setup like the one considered in this paper, an $\calo(e^{-S|_{\text{hw}}})$ violation of axionic symmetries is encoded in the sum, over all wormhole charges, of K\"ahler or superpotential corrections of the form \eqref{nonpertK} and \eqref{BPSsuperpotential}. In the supersymmetric regime, then, the effective axion potential at momenta $\ll1/L$ is controlled at least by a factor of order $\calo(e^{-2S|_{\text{hw}}})$. If, on the other hand, supersymmetry is softly broken at a scale $M_{\text{\tiny{SSB}}}\ll M_{\text{sp}}$, then the effective axion potential may also receive corrections of order $(M_{\text{\tiny{SSB}}}/ M_{\text{sp}})^p e^{-S|_{\text{hw}}}$ for some $p>0$. Hence, the effective axion potential is always more suppressed than just $\calo(e^{-S|_{\text{hw}}})$. Let us next estimate what $S|_{\text{hw}}$ can be in scenarios with fundamental or QFT axions.

Consider first the case of fundamental axions. That is, assume the EFT contains a number $\leq N$ of approximately massless fundamental axions, precisely like the ones discussed in our paper, while the saxions are instead stabilized by a SUSY-breaking potential $V_{\rm s}$. The presence of $V_{\rm s}$ introduces a correction to \eqref{1dS} in the form of a new effective potential for the (dual) saxions 
\be\label{effSAX}
V_{\bf q}=-\frac12\|{\bf q}\|^2-\pi^2r^6M^2_{\text{\tiny P}}V_{\rm s}.
\ee
The saxion potential $V_{\rm s}$ generates an IR threshold mass $M_{\text{\tiny IR}}$, as anticipated around \eqref{IRcutoff}. Approximate, purely axionic (non-extremal) wormholes  may be constructed at distances $r\gg1/M_{\text{\tiny IR}}$, where the saxions are fixed at their asymptotic values and the only relevant dynamical degrees of freedom are the axions --  see \cite{Montero:2015ofa,Bachlechner:2015qja} for discussions on multi-axion generalizations of the axionic wormhole of \cite{Giddings:1987cg}. Their action $S|_{\text{hw}}=3\pi^3 M_{\text{\tiny P}}^2L^2$ is characterized by curvature lengths $L\geq1/M_{\text{\tiny IR}}$, and according to \eqref{supprOhw} the associated wormholes would be very inefficient sources of symmetry breaking in the controllable regime $M_{\text{\tiny IR}}\ll M_{\text{sp}}$ because $S|_{\text{hw}}\gg N_{\text{sp}}$. In the more interesting regime $L\leq1/M_{\text{\tiny IR}}$, wormholes inevitably excite the saxionic degrees of freedom via \eqref{effSAX} -- see \cite{Andriolo:2022rxc,Jonas:2023ipa} for  explicit examples in simple dilatonic models -- and the results of our paper directly apply. Hence, in the more interesting cases the effect is estimated as discussed in Section \ref{sec:phenoGB}.

Axions may also originate as approximate Nambu-Goldstone bosons of some linearly realized accidental compact symmetry. In that case one can build wormholes with those ``QFT axions",  the associated saxionic radial modes, and possibly other degrees of freedom as well, whereas the fundamental saxions and axions we have been considering here remain mere silent spectators. One can now envision QFT models in which the radial modes shrink as one approaches the wormhole throat, resulting in small on-shell actions (see e.g.\ \cite{Kallosh:1995hi}). But small actions are a clear indication of sizable quantum effects, and that may signal we are outside the perturbative domain. Fortunately, irrespective of that we can confidently claim that wormhole effects cannot get larger than indicated in \eqref{OWHboundlv}. Our bound on the GB coefficient is indeed largely insensitive to the details of the wormhole configuration and impervious to those violations of perturbativity. Barring unnatural cancellations due to other higher derivative terms, then, the suppression \eqref{OWHboundlv} should apply to QFT wormholes as well, of course provided the UV completion is ultimately well described by an ${\cal N}=1$ axiverse.

\section{Conclusions}
\label{sec:conclusions}

 In this paper we explored various aspects of $\caln=1$ axiverse models containing an arbitrary number $N$ of fundamental axions, adopting the general framework developed in \cite{Lanza:2020qmt,Lanza:2021udy,Lanza:2022zyg} and taking into account various quantum gravity constraints.

We proposed the upper bound \eqref{SPbound} on the ultimate (moduli-dependent) UV cutoff of any EFT description of gravity, the so-called species scale. Our upper limit is set by the dominant EFT string scale \eqref{SPbounddef}, which is a perturbatively exact quantity depending on four-dimensional data already available at the two-derivative level. Our proposal \eqref{SPbound} has been checked in a number of string theory models and compared to another upper bound on the species scale suggested in \cite{vandeHeisteeg:2022btw,vandeHeisteeg:2023dlw}, finding very good agreement.

The approximate axionic shift symmetries that characterize our EFTs are explicitly broken by two classes of non-perturbative effects. The first class consists in short distance effects due to fundamental instantons. The second class is associated to non-perturbative effects within the EFT, part of which are of gravitational nature and potentially due to axionic wormholes. A significant portion of our paper was dedicated to the study of axionic wormholes in $\caln=1$ axiverse models,  and of their relation with fundamental instantons.

Our axiverse models support a large class of extremal and non-extremal wormhole solutions, which carry specific sets of axionic charges. The extremal ones are BPS and singular, but their singularity has a clear interpretation as due to the insertion of fundamental BPS  instantons. Among the fundamental BPS  instantons, a special role is played by the {\em EFT instantons}  \cite{Lanza:2021udy}. We argued, both within the macroscopic EFT description and by discussing microscopic string theory realizations,   that EFT instantons are intrinsically gravitational in nature. The analogy with five-dimensional back holes  suggests that EFT instantons satisfy an axion version of the BPS tower or sublattice  WGC proposed in \cite{Alim:2021vhs}. 
 
Interestingly, a slightly restricted subset of EFT instanton charges always support {\em homogeneous} non-extremal wormhole solutions as well, which describe $N$ saxions moving along a specific direction in the saxionic space. For the same set of charges, more general solutions  can also be constructed, both perturbatively, as small deformations of the homogeneous configurations, and numerically. Crucial to the existence of our homogeneous wormhole solutions is the homogeneity degree $n$ of the function $P(s)$ characterizing the saxion K\"ahler potential \eqref{KP(s)}. With $n>3$ the solutions are regular everywhere and in principle capable of inducing at low energy exponentially suppressed symmetry-breaking corrections to the K\"ahler potential. These wormholes may be interpreted as non-BPS bound states of EFT instantons and may capture, in the low-energy theory, some of their  physical effects. Wormholes with $n<3$, instead, degenerate at a small distance from their throat  and no relevant semi-classical effects can be associated to them. 

The non-extremal configurations with $n=3$ are degenerate only asymptotically and deserved a separate discussion. We revisited and extended a proposal of \cite{Park:1990ep}, which implies that these marginally degenerate wormholes should generate effective symmetry-breaking superpotentials. Moreover, we argued that these configurations can be more directly interpreted as a low-energy manifestation of EFT instantons, thus identifying a clearer link between two a priori independent classes of non-perturbative effects. The combination of these considerations has far-reaching consequences. On the one hand, it implies that EFT instantons corresponding to $n=3$ sectors are generically expected to generate superpotential terms, compatibly with the Supersymmetric Genericity Conjecture \cite{Palti:2020qlc}. On the other hand, it indicates that Coleman's $\alpha$-parameters of the $n=3$ wormholes  should be somehow fixed by the UV complete description of the corresponding EFT instantons, providing a concrete realization of the Baby Universe Hypothesis  \cite{McNamara:2020uza}. Actually, according to the Baby Universe Hypothesis the same ``natural selection" of the $\alpha$-parameters should also take place for the $n>3$ regular wormholes and the corresponding D-terms. We leave a more in depth exploration of these interesting speculations for the future.

We showed that the effect of non-extremal wormholes associated to  $\caln=1$ axiverses is always very suppressed. For non-degenerate wormholes, the basic reason is that the species scale is expected to decrease parametrically with increasing $N$, hence there is a maximal curvature mass scale below which the EFT predictions are reliable, and this suggests that the wormhole on-shell action should be at least of order $N$. One way to quantify this suppression is by taking advantage of a lower bound we established on the coefficient of the Gauss-Bonnet operator: in the EFT domain of validity, its value is always positive and growing with at least a power of $N$. For $n=3$ marginally degenerate wormholes one can equally show that the on-shell action is enhanced by $N$. As a result, we can robustly conclude that  in the axiverse the effect of  non-extremal wormholes is suppressed by powers of $e^{-N}$. 
Despite their tiny impact on the axiverse dynamics, non-extremal wormholes remain very interesting and concrete laboratories for the study of quantum gravity and of its low-energy consequences.


\vspace{1cm}

\centerline{\large\em Acknowledgments}

\vspace{0.5cm}

\noindent We thank M.~Bianchi, V.~Cagioni, D.~Cassani, L.~McAllister, F.~Marchesano, J.~Moritz, G.~Shiu, J.~Stout, A.~Strominger, C.~Vafa, I.~Valenzuela, D.~van de Heisteeg, T.~Van Riet, T.~Weigand, and M.~Wiesner for useful discussions.  LM thanks the Jefferson Physical Laboratory of Harvard University, and the members of the Harvard Swampland Initiative for hospitality during the final stage of this work. This work was supported in part by the Italian MUR Departments of Excellence grant 2023-2027 ``Quantum Frontiers”, by the MUR-PRIN contract 2022YZ5BA2  -  Effective Quantum Gravity,  and by the MIUR-PRIN contract 2017CC72MK003. AV acknowledges the Swiss National
Science Foundation (SNF) for funding through the Eccellenza Professorial Fellowship “Flavor Physics at the High Energy Frontier” project number 186866. The work of LV was partly supported by the Italian MIUR under contract 202289JEW4 (Flavors: dark and intense), the Iniziativa Specifica “Physics at the Energy, Intensity, and Astroparticle Frontiers” (APINE) of Istituto Nazionale di Fisica Nucleare (INFN), and the European Union’s Horizon 2020 research and innovation programme under the Marie Sklodowska-Curie grant agreement No 860881-HIDDeN.
\vspace{2cm}

\newpage

\centerline{\LARGE \bf Appendix}
\vspace{0.5cm}


\appendix

\section{Naive Dimensional Analysis}
\label{app:2pis}

In this appendix we review the concept of Naive Dimensional Analysis (NDA) to explain the structure of the derivative expansion in our EFT, justify the definitions \eqref{strPS1011} of the strong coupling scales in string and M-theory, and derive \eqref{axiSC} and \eqref{GBNDA}. In the process we will clarify the origin of the ``geometric" $2\pi$-factors  that appear throughout the paper; all other factors of order unity cannot be detected by NDA arguments and will be ignored. NDA was first used in the context of QCD by S.~Weinberg in \cite{Weinberg:1978kz} and later formalized by \cite{Manohar:1983md} (see also \cite{Cohen:1997rt} and references therein and, e.g.\ \cite{Chacko:1999hg} for a $d$-dimensional analysis). Its application to string theory models is not as popular.

For clarity we anticipate here the proxy for the strong coupling scale of a $d$-dimensional gravitational theory that we found to apply to all scenarios of interest to us:
\be\label{strongCoupld}
\widehat M_{(d)}^{d-2}\equiv(2\pi)^{\lfloor\frac{d}2\rfloor}{M}_{(d)}^{d-2}.
\ee
Here $M_{(d)}$ denotes the $d$-dimensional Planck scale which, to avoid unnecessary $1/2$ factors, in this appendix is defined via~\footnote{$M_{(4)}$ is thus related to the $M_{\text{\tiny P}}$ used in the rest of the paper via $M^2_{\text{\tiny P}}=2 M^2_{(4)}$.}
\be\label{PlanckdAPPA}
{\cal L}_{(d)}\supset \sqrt{-g_{(d)}}M_{(d)}^{d-2}R_{(d)}. 
\ee
The quantity $\widehat M_{(d)}$ identifies the highest possible scale at which a $d$-dimensional gravitational EFT can be extrapolated. If a parametrically large number $N_{(d)}$ of $d$-dimensional degrees of freedom is present, \eqref{strongCoupld} must be further reduced by a factor of $1/N_{(d)}$, so \eqref{strongCoupld} is to be interpreted as an upper bound. Eq.\ \eqref{strPS1011} shows the ten- and eleven-dimensional incarnations of \eqref{strongCoupld}, respectively.

\subsection*{The strong coupling scale}

Imagine we match a $d$-dimensional EFT to its UV completion at the lowest energy threshold ${\mathtt M}_{(d)}$ of the UV theory. Using the terminology of Section \eqref{sec:UVcutoff}, ${\mathtt M}_{(d)}$ denotes the $d$-dimensional version of the species scale. A simple $\hbar$ power counting reveals that the matching procedure must result in the following effective Lagrangian (we focus on gravity, but our arguments are completely general)
\be\label{NDAd}
{\cal L}_{(d)}=
\frac{{\mathtt M}_{(d)}^{d}}{{\mathtt g}_{(d)}^2}\sqrt{-g_{(d)}}\left\{
\frac{R_{(d)}}{{\mathtt M}_{(d)}^2}+c_2\frac{R_{(d)}^2}{{\mathtt M}_{(d)}^4}+c_3\frac{R_{(d)}^3}{{\mathtt M}_{(d)}^6}+\cdots
\right\}
\ee
where $c_{2,3,\cdots}$ are pure numbers at most of order unity that depend on ratios of mass scales and/or ratios of couplings. The strength of the Einstein-Hilbert term at momenta $p\sim {\mathtt M}_{(d)}$ is measured by the EFT coupling ${\mathtt g}_{(d)}$, which has units of $[{\mathtt g}_{(d)}]=[1/\sqrt\hbar]$. Denoting by $N_{(d)}$ the number of $d$-dimensional degrees of freedom below ${\mathtt M}_{(d)}$, a typical loop integral with momenta $p\sim {\mathtt M}_{(d)}$ is estimated to be of order 
\be\label{strongCoupling}
{\mathtt g}_{(d)}^2N_{(d)}\int \frac{\d\Omega_d}{(2\pi)^d}=\frac{2\, {\mathtt g}_{(d)}^2N_{(d)}}{(4\pi)^{\frac{d}{2}}\Gamma\left(d/2\right)}\sim\frac{{\mathtt g}_{(d)}^2}{(2\pi)^{\lfloor\frac{d+1}2\rfloor}}
\ee
with $\lfloor\cdots\rfloor$ the floor function. In the last step of \eqref{strongCoupling} we kept track of the powers of $2\pi$ but neglected a factor $N_{(d)}/(d-2)!!$ for $d={\text{even}}$ or $2N_{(d)}/(d-2)!!$ for $d={\text{odd}}$. It turns out that in the higher-dimensional scenarios we consider such factors are of order one. They will hence be neglected along with all other factors of order unity because NDA cannot keep track of them. The symbol $\sim$ will be used when our NDA prescription of ignoring numbers of order unity is employed.

Conventionally, a coupling is called maximally strong when loop effects are of order unity \cite{Cohen:1997rt,Chacko:1999hg}. According to \eqref{strongCoupling}, in our scenarios this occurs when ${\mathtt g}_{(d)}^2\sim{\mathtt g}_{(d)}^2|_{\text{max}}\equiv(2\pi)^{\lfloor\frac{d+1}2\rfloor}$.  
Now, from \eqref{NDAd} it follows that the $d$-dimensional Planck scale can be expressed in terms of the matching scale and the EFT coupling. Recalling our convention \eqref{PlanckdAPPA} we have $M_{(d)}^{d-2}={{\mathtt M}_{(d)}^{d-2}}/{{\mathtt g}_{(d)}^2}$. This says that, for fixed $M_{(d)}$, the higher the matching scale the stronger the coupling. The maximally allowed value ${\mathtt M}_{(d)}^2|_{\text{max}}$ is obtained with the maximally strong coupling ${\mathtt g}_{(d)}^2|_{\text{max}}$ identified above:
\be\label{MQGMd}
\left.{\mathtt M}_{(d)}^{d-2}\right|_{\text{max}}\equiv(2\pi)^{\lfloor\frac{d+1}2\rfloor}M_{(d)}^{d-2}.
\ee
This expression coincides with \eqref{strongCoupld} for any even dimensionality, but is slightly larger when $d$ is odd. We will see that \eqref{MQGMd} does indeed represent a good proxy for strong coupling in ten-dimensional string theories, but can be slightly refined for eleven-dimensional M-theory. A perturbative $d$-dimensional theory must always satisfy ${\mathtt g}_{(d)}^2\ll{\mathtt g}_{(d)}^2|_{\text{max}}$ so as to ensure that the description remains perturbative all the way up to the matching scale. For the particular case $d=4$ this logic explains the $2\pi$'s in Section \ref{sec:regimesEFT}. In general the ratio ${\mathtt M}_{(d)}^{d-2}/{\mathtt M}_{(d)}^{d-2}|_{\text{max}}$ parametrizes the expansion parameter on which the coefficients $c_{2,4,\cdots}$ depend.

To check the consistency of these considerations let us first consider type IIA string theory at not-too-large dilaton. 
At weak string coupling the matching scale ${\mathtt M}_{(10)}$ in \eqref{NDAd} corresponds to the string mass scale in the Einstein frame:
\begin{align}
    {\mathtt M}_{(10)}^2=2\pi T_{\rm F1}=\frac{(2\pi)^2}{l_{(10)}^2}e^{\phi/2}.
\end{align}
As the string coupling $g_{\rm s}=e^\phi$ increases, ${\mathtt M}_{(10)}$ grows and eventually is expected to reach the value ${\mathtt M}_{(10)}|_{\text{max}}$ at which both the UV theory and the EFT are strongly coupled. Recalling \eqref{S-10-theory} we have $M_{(10)}^8=2\pi/l_{(10)}^8$, which combined with \eqref{MQGMd} gives
\begin{align}
{\mathtt M}_{(10)}^2|_{\text{max}}=\,&(2\pi)^{\frac54}M_{(10)}^2=\frac{(2\pi)^{\frac32}}{l_{(10)}^2}.
\end{align}
To verify that ${\mathtt M}_{(10)}|_{\text{max}}$ gives also a good measure of the scale at which the string coupling gets strong we inspect the coefficients of the $R_{(10)}^4$ operators in the EFT. The latter can be arranged as shown in \eqref{NDAd} with coefficients (see for instance \cite{Grimm:2017okk}, and references therein)
\begin{align}
    c_4|_{\text{Type IIA}}=\,&\zeta(3)\,T_1+\frac{\pi^2e^{2\phi}}{3}T_2=\zeta(3)T_1+\frac{1}{12}\frac{{\mathtt M}_{(10)}^8}{{\mathtt M}_{(10)}^8|_{\text{max}}}T_2,
\end{align}
where $T_1$ and $T_2$ are appropriate tensor structures contracting the indices of the Riemann tensors. Consistently with our expectation, the loop corrections become of the same order as the tree-level terms when $(2\pi e^\phi)^2={\mathtt M}_{(10)}^8/{\mathtt M}_{(10)}^8|_{\text{max}}$ gets of order one. We can therefore conclude that ${\mathtt M}_{(10)}|_{\text{max}}$ is a physically reasonable estimate of the strong coupling scale in Type IIA string theory, at least for $2\pi e^\phi={\mathtt M}_{(10)}^4/{\mathtt M}_{(10)}^4|_{\text{max}}<1$, and that $2\pi e^\phi$ is an appropriate expansion parameter. For ten-dimensional string theory the strong coupling scale introduced in Section \ref{sec:UVcutoff} can thus be taken to be $\widehat M_{(10)}={\mathtt M}_{(10)}|_{\text{max}}$, as anticipated in \eqref{strPS1011}.

In the eleven-dimensional description of M-theory  
the approximation in \eqref{strongCoupling} is again justified. The matching scale (or, equivalently, spieces scale)  ${\mathtt M}_{(11)}$ should correspond to a strong coupling scale, but the story here is a little more subtle than before because of the presence of branes. We will argue that it is more appropriate to identify the strong coupling scale of M-theory with \eqref{strongCoupld} rather than \eqref{MQGMd}. While in Type IIA the strongest coupling at the matching scale is set by gravity, from which \eqref{MQGMd} follows, in M-theory it is not a priori evident which coupling is the strongest. Its action contains the standard Einstein-Hilbert term as well as M2-branes and  M5-branes. The coefficients of these terms are completely fixed by a number of independent arguments. The purely gravitational contributions are given by 
\be\label{MtheoryAct}
\frac{2\pi}{l_{(11)}^9}\int R_{(11)}+\frac{2\pi}{l_{(11)}^3}\int_{\text{M2}}{\text{vol}}_{\text{M2}}+\frac{2\pi}{l_{(11)}^6}\int_{\text{M5}}{\text{vol}}_{\text{M5}}\,.
\ee
Let us estimate the maximal strong coupling scale for each term in turn. The one for the Einstein-Hilbert interaction can be obtained via the logic leading to \eqref{MQGMd}. Combining with our definition \eqref{S-M-theory} one gets ${\mathtt M}_{(11)}^9|_{\text{max,EH}}=(2\pi)^7/l_{(11)}^9$. The world-volume theory supported by a $p$-brane of tension $T_p$  is  effectively described by a $p+1$-dimensional action with coupling ${\mathtt g}_{(p+1)}^2= {\mathtt M}_{(11)}^{p+1}|_{\text{Mp}}/T_p$. The corresponding maximal coupling scale can be determined in a way completely analogous to what done for the bulk gravitational theory in \eqref{strongCoupling}, though now the dimensionality is not that of spacetime. Ignoring again factors of order one coming from the number of degrees of freedom living on the branes, for the M5 brane that logic gives ${\mathtt M}_{(11)}^6|_{\text{max,M5}}=(2\pi)^4/l_{(11)}^6$ and for the M2 brane we have ${\mathtt M}_{(11)}^3|_{\text{max,M2}}=(2\pi)^3/l_{(11)}^3$. The actual regime of strong coupling of M-theory is presumably controlled by the smallest of these scales. According to our NDA estimate, that is the one associated to the M5 brane. It is hence natural to take
\be\label{MQG-Mtheory}
{\mathtt M}_{(11)}=\widehat M_{(11)}=\frac{(2\pi)^{2/3}}{l_{(11)}}.
\ee
This is smaller than ${\mathtt M}_{(11)}|_{\text{max}}$ given in \eqref{MQGMd} by just a factor $(2\pi)^{1/9}$. While such difference is very small, we are tempted to take our estimate seriously because NDA can only identify the geometric powers of $2\pi$, and according to a $2\pi$ counting \eqref{MQG-Mtheory} should be our candidate scale. 

Our choice \eqref{MQG-Mtheory} can also be justified by an alternative criterion, which is essentially semiclassical and may be formulated as follows: a  $d$-dimensional theory is said to be maximally strong at $p\sim {\mathtt M}_{(d)}$ if its action  is one (in $\hbar=1$ units) when evaluated on spherically symmetric configurations controlled by distance and curvature scales $\sim1/{\mathtt M}_{(d)}$. With this new prescription the matching scale of a strongly-coupled theory would always be set by \eqref{strongCoupld} rather than \eqref{MQGMd}, so the new criterion would agree with the one discussed earlier only for even dimensionalities. Interestingly, when we apply the new criterion to M-theory one finds that 
all three terms in \eqref{MtheoryAct} become strong at the very same scale \eqref{MQG-Mtheory}. This coincidence seems to support our identification.

An empirical non-trivial check that the powers of $2\pi$ suggested by \eqref{MQG-Mtheory} have some truth in them is obtained noting that according to \eqref{NDAd} we should expect the coefficients of the $R_{(11)}^4$ terms in M-theory to be proportional to 
\be
\frac{M_{(11)}^{9}}{{\widehat M}_{(11)}^6}=\frac{M_{(11)}^{3}}{(2\pi)^{10/3}}
\ee
up to a pure number. Amusingly, the powers of $2\pi$ obtained from that relation match exactly those obtained in explicit calculations, as for instance collected in \cite{Grimm:2017okk} and \cite{vandeHeisteeg:2023dlw}. 

\subsection*{The four-dimensional theory}

Finally, we would like to see how the Lagrangian \eqref{NDAd} appears to a four-dimensional observer at energy scales $<M_{\text{\tiny KK}}$ below the Kaluza-Klein mass scale. We look for gravitational solutions with line element as in \eqref{lineelement} and proceed as follows. We first derive the dimensionally reduced four-dimensional version of \eqref{NDAd} with all the KK resonances integrated in. Subsequently we integrate out the KK resonances to obtain an EFT at the scale $M_{\text{\tiny KK}}$. 

The dimensionally-reduced four-dimensional EFT for the graviton zero-mode and KK resonances is, taking care of the appropriate factors of $e^{2A}$, a Lagrangian completely analogous to \eqref{NDAd}: 
\be\label{NDA4}
\left.{\cal L}\right|_{\Lambda<M_{\text{sp}}}=
\frac{M_{\text{sp}}^4}{{\mathtt g}^2}\sqrt{-g}\left\{\frac{R}{M_{\text{sp}}^2}+c_2\frac{R^2}{M_{\text{sp}}^4}+c_3\frac{R^3}{M_{\text{sp}}^6}+\cdots+{\text{KK-resonances}}\right\}
\ee
where 
\be\label{fourdimMC}
M_{\text{sp}}^2= e^{2A}{\mathtt M}_{(d)}^2,~~~~~~~~~~~~~~~~~~{\mathtt g}^2=\frac{{\mathtt g}_{(d)}^2}{{\mathtt V}_{(d-4)}},
\ee
and ${\mathtt V}_{(d-4)}={\mathtt M}_{(d)}^{d-4}\int \d^{d-4}y~\sqrt{g_{(d-4)}}$. The matching scale, converted to the four-dimensional Einstein frame, is what we called species scale in Section \ref{sec:UVcutoff}. The quantity ${\mathtt V}_{(d-4)}$ is the volume normalized in units of the fundamental scale ${\mathtt M}^{-1}_{(d)}$ in $d$-dimensions. It is related to the number of four-dimensional KK degrees of freedom below ${\mathtt M}_{(d)}$ by Weyl's asymptotic formula (see for instance \cite{Ivrii}) 
\be\label{NKK}
N_{\text{\tiny KK}}\simeq\frac{N_{(d)}\Omega_{(d-4)}{\mathtt V}_{d-4}}{(d-4)(2\pi)^{d-4}}=\frac{(d-2)\, N_{(d)}{\mathtt V}_{d-4}}{2(4\pi)^{\frac{d}{2}-2}\Gamma\left(d/2\right)}
\,.
\ee
If we identify $N_{\text{\tiny KK}}$ with the number of four-dimensional degrees of freedom, then the 't Hooft expansion parameter in the reduced four-dimensional theory is  
\be\label{coupliKK}
\frac{{\mathtt g}^2N_{\text{\tiny KK}}}{(2\pi)^2}\simeq (d-2)\,\frac{2 \,{\mathtt g}_{(d)}^2\, N_{(d)}}{(4\pi)^{\frac{d}{2}}\Gamma\left(d/2\right)}
\,.
\ee
Up to to the  $(d-2)$ factor, this is precisely the $d$-dimensional loop estimate appearing in  \eqref{strongCoupling}. This simply reminds us that a weakly-coupled $d$-dimensional description necessarily leads to a weakly-coupled four-dimensional reduction. Now, following the convention in \eqref{PlanckdAPPA}, \eqref{NDA4} leads to the identification  $M_{\text{sp}}={\mathtt g}M_{(4)}$ which, combined with \eqref{coupliKK} and adopting the same approximation as in \eqref{strongCoupling}, reads 
\be\label{MGQapp}
M_{\text{sp}}^2\sim\frac{{\mathtt g}_{(d)}^{2}}{(2\pi)^{\lfloor\frac{d+1}2\rfloor}}\frac{(2\pi M_{(4)})^2}{N_{\text{\tiny KK}}}\,.
\ee
This relation says that the species scale seen by the four-dimensional observer satisfies an upper bound $M_{\text{sp}}\lesssim 2\pi M_{(4)}/\sqrt{N_{\text{\tiny KK}}}\sim 2\pi M_{\text{\tiny P}}/\sqrt{N_{\text{\tiny KK}}}$, which we recognize to be \eqref{axiSC} whenever the large number of species can be identified with the number of KK modes. This is just the statement that the species scale and the maximally strong coupling scale \eqref{MQGMd} coincide only when the expansion parameter appearing on the r.h.s.\ of \eqref{strongCoupling} is one.

The next step consists in integrating out the KK modes to obtain a four-dimensional EFT at scales $\lesssim M_{\text{\tiny KK}}$. In carrying out this final step we note that the tree-level exchange of a KK resonance $\Phi$ can only be relevant when linear couplings of the form $\Phi J$ are present in \eqref{NDA4}. According to the discussion in Section \ref{sec:UVcutoff}, neither Einstein-Hilbert nor Gauss-Bonnet are corrected by tree-level KK exchange in string theory compactifications. Moreover, loops cannot introduce additional powers of $M_{\text{sp}}^2/M_{\text{\tiny KK}}^2$ because controlled by irrelevant couplings. Hence we deduce that neither the Einstein-Hilbert term nor the Gauss-Bonnet term are qualitatively affected by KK exchange.~\footnote{On the other hand, Ricci squared interactions (as well as higher powers of the curvature) may receive corrections already at tree-level.} Ignoring numbers of order unity we thus conclude that below the KK scale our four-dimensional EFT formally reads
\be\label{NDA4EFT}
\left.{\cal L}\right|_{\Lambda<M_{\text{\tiny KK}}}=
M_{(4)}^2M_{\text{sp}}^2\sqrt{-g}\left\{\frac{R}{M_{\text{sp}}^2}+c'_2\frac{R^2}{M_{\text{sp}}^4}+c'_3\frac{R^3}{M_{\text{sp}}^6}+\cdots\right\}.
\ee
Let us consider the Riemann squared operator, which is particularly relevant to us because related to the GB term. Exploiting its symmetry properties we observe that $R_{abcd}R^{abcd}=4\sum_{a<b, c<d}R_{abcd}R^{abcd}$. Therefore it is reasonable to expect a combinatorial factor of $1/4$ in front of the Wilson coefficient. We thus take the $c_2$ associated to the Riemann squared operator in \eqref{NDAd} to be $c_2=c_{\text{GB}}/4$ with $c_{\text{GB}}$ of order one. In addition, we just argued that $c'_{\text{GB}}\sim c_{\text{GB}}$ (whereas, typically, $c'_{R^2}\sim  c_{R^2}M_{\text{sp}}^2/M_{\text{\tiny KK}}^2$). Using now our conventions in \eqref{GBdensity} and \eqref{GRtheta}, the coefficient of the four-dimensional Gauss-Bonnet term in \eqref{NDA4EFT} reads as shown in \eqref{GBNDA}.

\section{Matching in heterotic/F-theory dual models}
\label{app:hetFtheory}

In this section we  more explicitly show how the K\"ahler/kinetic potentials \eqref{hetK} and \eqref{FFK} of heterotic/F-theory dual models match in an appropriate limit. F-theory compactifications on a $\mathbb{P}^1$ fibered space $X$ is dual to a heterotic compactification on an elliptically fibered Calabi-Yau three-fold $\hat X$ over the same  base two-fold $B$. We assume $B$ to be weak-Fano, and then to support smooth elliptic fibrations \cite{Halverson:2015jua}. By definition, such a base has a nef and big  anti-canonical divisor  $\overline K_B$. Let us denote it as
\be 
c_B\equiv \overline K_B\in {\rm Nef}^1(B)\,.
\ee
The bigness condition is equivalent to the self-intersection positivity  $c_B\cdot c_B>0$. Let us also introduce two dual bases of 2-cycles $c_\alpha,\tilde c^\alpha\in H_2(B,\mathbb{Z})$, such that $c_a\cdot\tilde c^\beta=\delta_\alpha^\beta$.

\subsection{F-theory side}

 The space $X$ can be described as a projectivized bundle \cite{Friedman:1997yq}
\be 
X_{\text{F}}=\mathbb{P}(\calo_B\oplus \call)\,,
\ee
where the line  bundle $\call$ on the base $B$  determines the twist of the $\mathbb{P}^1$-fibration. We denote by $\pi:X\rightarrow P$ the corresponding projection map.  Introducing fibral projective coordinate $[x:y]$, we can identify the global section $\sigma: B\rightarrow X$, defined by $x=0$, and a corresponding effective divisor $S=\sigma_{*}B\subset X$. The cone of effective divisors ${\rm Eff}^1(X)\simeq \calc_{\rm I}$ is generated by $S$ and the divisors of the form $E=\pi^*(e)$, where $e\in {\rm Eff}_1(B)$, i.e.\ $e$ is an effective curve in $B$. We will also assume that the line bundle $\call$ is positive,  that is 
\be 
c_\call\equiv c_1(\call) \equiv p_\alpha\tilde c^\alpha
\in {\rm Eff}_1(B)\,,
\ee 
where again Poincar\'e duality is implicit. 
The cone ${\rm Nef}^1(X)$ of nef divisors is generated by the divisor
 \be
 H=S+\pi^*c_\call
 \ee
and the divisors of the form $\pi^*c$, where $c\in {\rm Nef}^1(B)\simeq {\rm Mov}_1(B)$.  

 We can expand the dual saxion vector ${\bm\ell}$ in the basis of four-cycles provided by $H$ and the four-cycles $\pi^*\tilde c^\alpha$:
\be 
{\bm\ell}=\ell_0 H+\ell_\alpha \pi^*\tilde c^\alpha\,.
\ee
Then, according to our general discussion of Section \ref{sec:FIIB}, the dual saxionic domain is defined by $\ell_0\geq 0$ and $\ell_\alpha \tilde c^\alpha\in {\rm Nef}^1(B)_{\mathbb{R}}$. A simple example is provided  by the model of Section \ref{sec:Ftheory1}, which corresponds to $B=\mathbb{P}^2$, $(D^1,H)=(D^1,D^2)$ and the redefinition $(\ell_1,\ell_0)\rightarrow (\ell_1,\ell_2)$. 

 The basis  $(H,\pi^*\tilde c^\alpha)\in H_4(X,\mathbb{Z})$ is dual to the basis $(C_{\rm F},C_\alpha)$, where  $C_{\rm F}$  is the $\mathbb{P}^1$-fiber $C_{\rm F}$ and $C_\alpha\equiv \sigma_*c_\alpha$. The saxions $s^i=(s^0,s^\alpha)$  can be obtained by expanding ${\bm s}\in {\rm Mov}_1(X)$ in this basis:
\be\label{P1fibs} 
{\bm s}=s^0C_{\rm F}+s^\alpha C_\alpha=(s^0-p_\alpha s^\alpha) C_{\rm F}+s^\alpha H\cdot\pi^*c_\alpha\,,
\ee
where we have used $C_\alpha=S\cdot \pi^* c_\alpha=H\cdot\pi^*c_\alpha-p_\alpha C_{\rm F}$. 
The cone of movable curves is generated by $C_{\rm F}$ and curves of the form $H\cdot\pi^*(c)$, with $c\in {\rm Nef}^1(B)$.  Hence, the saxionic cone is defined by the conditions 
\be\label{saxFH} 
s^0\geq p_\alpha s^\alpha\quad,\quad s^\alpha c_\alpha\in {\rm Nef}^1(B)\,. 
\ee

The kinetic potential \eqref{FFK}  takes the form
\be 
\calf=\log\kappa({\bm\ell},{\bm\ell},{\bm\ell})=\log\left[\cali^{\alpha\beta}p_\alpha p_\beta\ell_0^3+3\cali^{\alpha\beta}p_\alpha\ell_\beta\ell_0^2+3\cali^{\alpha\beta}\ell_\alpha\ell_\beta\ell_0\right]\,,
\ee
where $\cali^{\alpha\beta}\equiv \tilde c^\alpha\cdot\tilde c^\beta$.
Applying \eqref{elldef2}, we get 
\be 
s^0=\frac{3(\cali^{\alpha\beta}p_\alpha p_\beta\,\ell_0^2+2\cali^{\alpha\beta}p_\alpha\ell_\beta\ell_0+\cali^{\alpha\beta}\ell_\alpha\ell_\beta)}{2\kappa({\bm\ell},{\bm\ell},{\bm\ell})}\quad,\quad s^\alpha=\frac{3(\cali^{\alpha\beta} p_\beta\ell_0^2+2\cali^{\alpha\beta}\ell_\beta\ell_0)}{2\kappa({\bm\ell},{\bm\ell},{\bm\ell})}\,.
\ee
 We now take the limit $\ell_0\ll |\ell_\alpha|$ limit, in which 
\be 
\calf\simeq \log\ell_0+\log\left[ \cali^{\alpha\beta}(\ell_\alpha+\frac12p_\alpha\ell_0) (\ell_\beta+\frac12p_\beta\ell_0)\right]\,,
\ee
up to a term of order $\calo(\ell^2_0/|\ell_\alpha|^2)$, and an irrelevant additive constant. In this limit we have 
\be 
s^0-\frac12 p_\alpha s^\alpha\simeq \frac{1}{2\ell_0}\gg 1\quad,\quad s^\alpha\simeq \frac{\cali^{\alpha\beta}(\ell_\beta+\frac12 p_\beta\ell_0)}{\cali^{\gamma\delta}(\ell_\gamma+\frac12p_\gamma\ell_0) (\ell_\delta+\frac12p_\delta\ell_0)}\,.
\ee
and the K\"ahler potential is then given by
\be\label{redKpot} 
K\simeq -\log\left(s^0-\frac12 p_\alpha s^\alpha\right)-\log\left(\cali_{\alpha\beta}s^\alpha s^\beta\right)\,,
\ee
where $\cali_{\alpha\beta}=c_\alpha\cdot c_\beta$ is the inverse of $\cali^{\alpha\beta}$.

\subsection{Heterotic side}

The elliptically fibered Calabi-Yau $\hat\pi:\hat X\rightarrow B$ admits the global section $\hat\sigma:B\rightarrow \hat X$. One can then identify the nef divisor $\hat H=\sigma_*B+\pi^*c_B$. 
The K\"ahler saxionic vector ${\bm s}$ appearing in \eqref{npDelta} can then be expanded as follows
\be 
{\bm s}=\hat s \hat H+s^\alpha \pi^*c_\alpha\,,
\ee
so that the saxionic cone condition ${\bm s}\in\overline{\calk(\hat X)}$ reads  $\hat s\geq 0$ and $s^\alpha c_\alpha\in {\rm Nef}^1(B)$. The corresponding contribution to the K\"ahler potential \eqref{hetK} is 
\be 
-\log\kappa({\bm s},{\bm s},{\bm s})=-\log\left[(c_B\cdot c_B)\hat s^3+3(c_B\cdot c_\alpha)s^\alpha\hat s^2+3(c_\alpha\cdot c_\beta)s^\alpha s^\beta\hat s\right]
\ee
The saxion $\hat s$ measures the volume of the elliptic fiber. Let us apply these results to the models of Section \ref{sec:het},  taking  the limit in which the fiber is of stringy size, and hence much smaller than the base. This describes a restricted perturbative regime parametrized only by the saxion $s^0$ and the base K\"ahler  saxions $s^\alpha$. In this limit  $p_as^a=\hat p\hat s+p_\alpha s^\alpha\simeq p_\alpha s^\alpha$, and the K\"ahler potential \eqref{hetK} becomes
\be 
-\log(s^0-\frac12p_as^a)-\log\kappa({\bm s},{\bm s},{\bm s})\simeq -\log(s^0-\frac12p_\alpha s^\alpha)-\log\left(\cali_{\alpha\beta}s^\alpha s^\beta\right)+\ldots \,,
\ee
which indeed matches \eqref{redKpot}, provided we identify the constants $p_\alpha$ appearing in these two different settings. This identification was already proposed in \cite{Martucci:2022krl}, and is also consistent with the identification of the saxionic cone conditions \eqref{saxFH} and \eqref{npDelta}.

\section{Other tests of the species scale bound}
\label{app:SCtests}

In this appendix we discuss other non-trivial examples in string theory compactifications where we can test explicitly our proposal on the relevant energy scales formulated in Section \ref{sec:UVcutoff}. The reader can see this as a natural addition to Section \ref{sec:stringtheorymodels}. In particular, we are going to explore the case of a $\mathbb{P}^1$ fibration over a Hirzebruch surface $\mathbb{F}_p$ in F-theory, and delve a bit deeper in the various K\"ahler moduli limits of heterotic/IIA models. 

\subsection{Another F-theory model: $\mathbb{P}^1$ fibration over $\mathbb{F}_p$} 
\label{app:Ftheory2}

Another class of simple F-theory models where we can check our proposal is obtained by choosing $X$ to be a $\mathbb{P}^1$ fibration over the Hirzebruch surface $\mathbb{F}_p$,  which in turn can be described as a  $\mathbb{P}^1$ fibration over $\mathbb{P}^1$, specified by the integer $p\geq 0$. This model has been recently discussed  in a closely related framework by \cite{Cota:2022yjw}, to which we refer for more details. For our purposes it is sufficient to restrict to a $\mathbb{P}^1$ fibration over $\mathbb{F}_p$ specified a single non-negative integers $h\in\mathbb{Z}_{\geq 0}$,\footnote{\label{foot:fibr} Following \cite{Friedman:1997yq}, the $\mathbb{P}^1$ fibration could be defined in terms of a line bundle $\call=h d_1+t d_2$, where $d_1,d_2$ are elementary nef divisors over $\mathbb{F}_p$. In particular, their  intersection numbers are given by the coefficients of $\cali(\mathbb{F}_p)=pd_1^2+d_1d_2$. In the notation of \cite{Cota:2022yjw}, our integers $(h,t)$  correspond to $(s,t)$. Here we are restricting to fibrations with $t=0$.} and to identify the relevant cones of divisors and curves and their intersection numbers. 

The cone of effective divisors is simplicial and is generated by three effective divisors $E^a$, $a=1,2,3$.
These three effective divisors can be roughly regarded as twisted products of the  possible pairs of the three $\mathbb{P}^1$'s involved in the geometry. We can also introduce  a basis of nef divisors $D^a$, with respect to which
\be\label{nefdivF1} 
E^1=D^1-pD^2\,,\quad E^2=D^2\,,\quad E^3=D^3-hD^1\,.
\ee 
The divisors $D^a$  generate all the other nef divisors, as well as the K\"ahler cone. Hence, by using these  divisors in the expansion \eqref{JexpF} the K\"ahler cone corresponds to $v_1,v_2,v_3>0$. The triple intersections are given by the coefficients of the formal object 
\be 
\begin{aligned}
\cali(X)&=D^1D^2D^3+p(D^1)^2D^3+hp D^1(D^3)^2+hD^2(D^3)^2+h^2p(D^3)^3\,.
\end{aligned}
\ee
By using the expansion ${\bm \ell}=\ell_aD^a$, the kinetic potential  \eqref{FFK} becomes
\be\label{ex1calf}
\calf_{\text{\tiny K}}=\log\kappa({\bm \ell},{\bm \ell},{\bm \ell})=\log\left(6\ell_1\ell_2\ell_3+3p \ell_1^2\ell_3+3hp\ell_1\ell_3^2+3h\ell_2\ell_3^2+h^2p\ell_3^3\right)\,.
\ee

The condition that $J$ belongs to the K\"ahler cone is equivalent to $\ell_a>0$. In order to understand the complete (dual) saxionic domain, we have to consider the saxions $s^a=\frac{3\kappa^{abc}\ell_b\ell_c}{2\kappa({\bm \ell},{\bm \ell},{\bm \ell})}$, which identify the $\mathbb{R}$-effective curves:
\be\label{sellF} 
{\bm s}=s^a\Sigma_a=\frac{3\,{\bm\ell}\cdot{\bm\ell}}{2\kappa({\bm \ell},{\bm \ell},{\bm \ell})}\,,
\ee
where  $\Sigma_a$ are effective curves 
\be 
\Sigma_1=E^2\cdot E^3\quad,\quad \Sigma_2=E^1\cdot E^3\quad,\quad \Sigma_3=E^1\cdot E^2\,,
\ee
which generate the whole cone of effective curves ${\rm Eff}_1(X)$ and are dual to the nef divisors $D^1$: $D^a\cdot\Sigma_b=\delta^a_b$. On the other hand,  the saxionic cone can be identified with the cone of movable curves -- see \eqref{sF} -- which is generated by the (effective) movable curves 
\be\label{movcurve2} 
\hat\Sigma_1=D^2\cdot D^3=\Sigma_1+h\Sigma_3\,,\quad \hat\Sigma_2=D^1\cdot D^3=p\Sigma_1+\Sigma_2+hp\Sigma_3\,,\quad \hat\Sigma_3=D^1\cdot D^2=\Sigma_3\,,
\ee
which are dual to the effective divisors $E^a$: $\hat\Sigma_a\cdot E^b=\delta_a^b$. Hence, if we use the expansion ${\bm s}=\hat s^a\hat\Sigma_a$, the saxionic cone is defined by the positivity conditions $\hat s^a>0$. By
using \eqref{nefdivF1} 
and  \eqref{sellF} we can compute the components $\hat s^a=E^a\cdot{\bm s}=\frac{3E^a \cdot{\bm\ell}\cdot{\bm\ell}}{2\kappa({\bm\ell},{\bm\ell},{\bm\ell})}$, getting
\be\label{hatsell} 
\hat s^1=\frac{6\ell_2\ell_3}{2\kappa({\bm\ell},{\bm\ell},{\bm\ell})}\quad,\quad \hat s^2=\frac{3(2\ell_1\ell_3+h\,\ell_3^2)}{2\kappa({\bm\ell},{\bm\ell},{\bm\ell})}\quad,\quad \hat s^3=\frac{3(2\ell_1\ell_2+p\,\ell_1^2)}{2\kappa({\bm\ell},{\bm\ell},{\bm\ell})}\,.
\ee
The saxionic cone condition $\hat s^a> 0$ is clearly satisfied if $\ell_a>0$.\footnote{Note that, if we consider fibrations of the kind described in footnote \ref{foot:fibr} with $t>0$, in general the image of the cone $\{\ell_a>0\}$ under the map \eqref{hatsell}  does not cover the entire saxionic cone $\Delta$  -- see   \cite{Cota:2022yjw} -- as discussed in general in Section \ref{sec:FIIB}. This issue is absent if we set $t=0$.} 

\begin{figure}[!htb]
    \centering
    \includegraphics[width=0.5\textwidth]{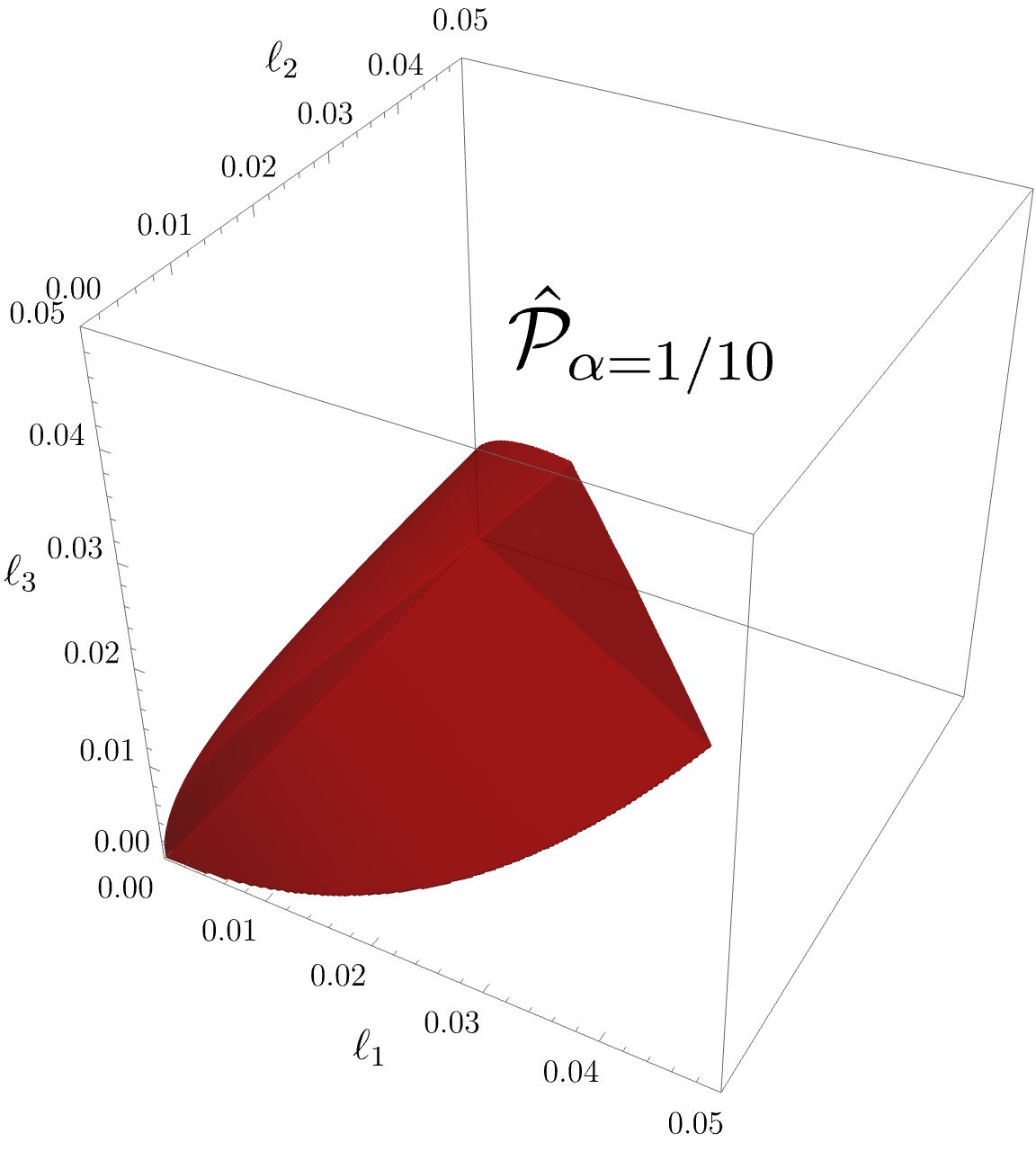}
    \caption{\small Dual saxionic convex hull $\hat{\mathcal{P}}_\alpha$ for the F-theory model $\mathbb{P}^1$ over $\mathbb{F}_p$. The plot has been drawn with the reference value $\alpha =1/10$. $\hat{\mathcal{P}}_\alpha$ corresponds to the red region.}
    \label{fig:model2domainP}
\end{figure}

The $\calp_{\text{\tiny K}}$ boundaries  can again be characterized in terms of tensionless string limits. 
The set $\calc^{\text{\tiny EFT}}_{\rm S}$ of EFT string charges is generated by the movable curves \eqref{movcurve2} and hence, in the basis $\Sigma_a$, we identify the following corresponding tensions:
\be\label{model3EFTtens} \calt_{\hat\Sigma_1}=M^2_{\text{\tiny P}}(\ell_1+h\ell_3)\,,\quad \calt_{\hat\Sigma_2}=M^2_{\text{\tiny P}}(\ell_2+p\ell_1+ hp\ell_3)\,,\quad \calt_{\hat\Sigma_3}=M^2_{\text{\tiny P}}\ell_3\,.
\ee
We then see that, assuming $p,h>0$, $\calt_{\hat\Sigma_3}=0$ on the two-dimensional boundary component $\{\ell_3=0\}$, $\calt_{\hat\Sigma_1}=0$ on the one dimensional boundary component $\{\ell_1=\ell_3=0\}$, while  $\calt_{\hat\Sigma_2}=0$ at the tip $\{\ell_1=\ell_2=\ell_3=0\}$. These boundary components are at infinite field distance. On the other hand, the BPS but non-EFT strings of charges $\Sigma_1$ and $\Sigma_2$ have tensions  $\calt_{\Sigma_1}=M^2_{\text{\tiny P}}\ell_1$ and $\calt_{\Sigma_2}=M^2_{\text{\tiny P}}\ell_2$, which vanish on the  boundaries $\ell_1=0$ and $\ell_2=0$, respectively, which are at finite field distance (if $\ell_3>0$). More precisely, they correspond to the finite distance $\Delta$ boundaries $\hat s^3=0$ and $\hat s^1=0$, respectively, while $\ell_3\rightarrow 0$ corresponds  to the infinite distance limit $\hat s^3\rightarrow \infty$. Viceversa, a limit $\hat s^2\rightarrow 0$ (with fixed $\hat s^1,\hat s^3$)  corresponds to a limit $\ell_2\rightarrow \infty$.  So, as in subsection \ref{sec:Ftheory1}, while the saxionic convex hull is simply given by $\hat\Delta_\alpha=\{\hat s^a\geq \frac1\alpha\}$, its dual saxionic counterpart $\hat\calp_\alpha$ is more complicated
-- see figure \ref{fig:model2domainP}.

The anti-canonical divisor is 
\be
\begin{aligned}
\overline K_X&=2E^3+(2+h)E^1+(2+p+ph)E^2\\
&=2D^3+(2-h)D^1+(2-p)D^2\,,
\end{aligned}
\ee
which, combined with \eqref{gammas} and \eqref{FtildeC}, gives 
\be\label{model3gamma} 
\gamma(s)=\pi\left[2\hat s^3+(2+h)\hat s^1+(2+p+ph)\hat s^2\right]\,.
\ee
 Hence 
\be 
\gamma(s)|_{\hat\Delta_\alpha}\geq \frac{\pi(6+p+h+ph)}\alpha\,,
\ee 
which is again  stronger than \eqref{gammabound} with $N=3$.
If for instance $p,h\geq 1$ and $\alpha\leq \frac1{10}$, we get the lower bound  $\gamma(s)|_{\hat\Delta_\alpha}> 282$.

\subsection{Energy scales in the $\mathbb{P}^1$ over $\mathbb{F}_p$ model}

The tensions \eqref{model3EFTtens} are associated to the generators of the cone of EFT string charges.
From \eqref{model3EFTtens}, and recalling that $\ell_a>0$, is clear that $\calt_{\hat\Sigma_3}$ is always the lowest of the three when $p,h>0$, and can thus be identified with the dominant EFT string scale \eqref{SPbounddef} 
\be\label{Fboundmax1} 
M^2_{\calt}=2\pi \calt_{\hat\Sigma_3}=2\pi M^2_{\text{\tiny P}}\ell_3\,.
\ee
Noticing that
\be 
\calt_{\hat\Sigma_3}\leq  \calt_{\hat\Sigma_1}\leq \calt_{\hat\Sigma_2}\,,
\ee
the limit  $\calt_{\hat\Sigma_3}\ll \calt_{\hat\Sigma_1}$, or equivalently $\ell_3\ll\ell_1$, corresponds to the weak string  coupling limit of a dual heterotic model, and $\calt_{\hat\Sigma_3}$ represents the corresponding critical string tension. In this regime $(2\pi \calt_{\hat\Sigma_3})^{1/2}$ can be identified with the species scale, and \eqref{SPbound} is saturated. 

If instead $\calt_{\hat\Sigma_3}\simeq \calt_{\hat\Sigma_1}$ the species scale may be given by the quantum gravity scale \eqref{FtheorySC}, which for the present model reads
\be\label{app:10dscale}
\begin{aligned}
M^2_{\text{\tiny QG}} &=  \sqrt{2\pi\left(2\ell_1 \ell_2 \ell_3 + p\,\ell_1^2 \ell_3 + hp \ell_1\ell_3^2 + h\ell_2\ell_3^2 + \frac{1}{3} h^2 p \ell_3^3\right)}\,M^2_{\text{\tiny P}}\\
&=\sqrt{\frac1{2\pi}\left[\frac{\ell_1}{\ell_3}\left( p\,\ell_1 +2\ell_2\right) + h\left(p \ell_1 + \ell_2 + \frac{1}{3} h p \ell_3\right)\right]}\,M^2_{\calt}\, .
\end{aligned}
\ee 
Imposing the constraint \eqref{TMP} on $\calt_{\hat\Sigma_2}$ gives  $p\ell_1 + \ell_2 + hp\ell_3 <2\pi$, which  in turn implies $p\ell_1 + \ell_2 <2\pi$. 
Applying these two inequalities we find
\be\label{model2Min} 
M^2_{\text{\tiny QG}}<\sqrt{\frac{\ell_1}{\ell_3}\left(2-p\frac{\ell_1}{2\pi}\right)+h\left(1-hp\frac{\ell_3}{3\pi}\right)}\,M^2_{\calt}\,.
\ee
 Since 
\be 
\frac{\calt_{\hat\Sigma_1}}{\calt_{\hat\Sigma_3}} = \frac{\ell_1}{\ell_3} + h > h \, ,
\ee
the assumption $\calt_{\hat\Sigma_3}\simeq \calt_{\hat\Sigma_1}$ requires that $h \sim \calo(1)$ and $\ell_1/\ell_3 \lesssim 1 $. Hence \eqref{model2Min} implies that $M^2_{\text{\tiny QG}}\lesssim M^2_{\calt}$, up to a factor of order one, compatibly with \eqref{SPbound}. 
The precise order-one factor depends on the dual saxions, and may be both smaller as well as (slightly) greater than one. One may then reverse the logic, and use \eqref{SPbound} as concrete criterion to more precisely identify  the species scale, as the smallest one between  \eqref{Fboundmax1} and \eqref{app:10dscale}.

We can also compare \eqref{Fboundmax1} to \eqref{vafaSC}. By \eqref{model3gamma}, in the present model we have
\be 
 M^2_{\gamma} = \frac{4\pi M^2_{\text{\tiny P}}}{2\hat s^3+(2+h)\hat s^1+(2+p+ph)\hat s^2}\, .
\ee
Recalling \eqref{hatsell} we then  obtain
\be
\frac{M^2_{\calt}}{M^2_{\gamma}}=\frac14\left[2+\frac{12\ell_2\ell_3+6(2+p)\ell_1\ell_3 + (6h+3ph+ ph^2 )\ell_3^2}{6\ell_1\ell_2 + 3p\ell_1^2 + 3hp\ell_1 \ell_3 + 3h\ell_2\ell_3 + ph^2 \ell_3^2}\right]\,.
\ee
We can see this ratio is bounded from below and from above such that it is always an $\calo(1)$ quantity
\be 
\frac{1}{2} \leq \frac{M^2_{\calt}}{M^2_{\gamma}} \leq \frac{1}{2} + \operatorname{max}\left\{ \frac{1}{h}, \frac{1}{4}+ \frac{3(2+p)}{4ph} \right\} \leq \frac{3}{4} + \frac{3(2+p)}{4ph} \leq 3 \,.
\ee 
This confirms that these two upper bounds on the species scale are of the same order.

\subsection{Asymptotic tests of our species scale bound}
\label{app:N=2tensions}

In this appendix we would like to provide further evidence for the validity of the bound \eqref{SPbound}. We will study the heterotic models discussed in Section \ref{sec:het}   in various EFT string limits, assuming that all EFT string charges satisfy $e^ap_a\geq 0$ as in Section \ref{sec:het}. Notice that the results in the case $p_a=0$ apply also to  $\caln=2$ models obtained from type IIA Calabi-Yau compactifications, or from the mirror type IIB models.  

The infinite distance limit associated with an EFT string  charge vector $(e^0,{\bf e})$ is given by the $\sigma\rightarrow \infty$ limit of  the saxionic flow
\be\label{EFTlimit} 
s^0=s^0_0+e^0\,\sigma\quad,\quad {\bm s}(\sigma)={\bm s}_0+{\bf e}\,\sigma\,.
\ee
  Let us assume that ${\bf e}$ is an elementary charge, and that $e^0$ is the minimal one compatible with the saxionic cone \eqref{npDelta}: $e^0=p_ae^a$. According to the discussion of  Section \ref{sec:UVcutoff}, along the limit \eqref{EFTlimit} the species scale is either given by $\sqrt{2\pi\calt_{\bf e}}$, if  $w_{\bf e}=1$, or by $M_{\text{\tiny QG}}$, if $w_{\bf e}\geq 2$.   Some of these limits will exit the M-theory/heterotic (or IIA) geometric regime and would require a dualization to an alternative description. While \eqref{SPbound} is automatically satisfied and saturated if $w_{\bf e}=1$, in the $w_{\bf e}\geq 2$ case the identification of the appropriate decompactification frame, and of the corresponding $M_{\text{\tiny QG}}$, may  not be obvious. In such cases, we will rather apply \eqref{axiSC} as a short-cut to estimate  $M_{\text{\tiny QG}}$, and verify \eqref{SPbound}. We will hence need to identify the relevant light masses that can appear in these limits.  

Consider first  the KK scale along the M-theory interval/circle. Taking into account the ansatz \eqref{11to10} and Einstein frame rescaling \eqref{scalingMth}, the corresponding KK mass is given by
\be\label{D0mass}
M_{\text{\tiny KK}}^2 = \frac{(2\pi)^2 e^{2A}}{l_{(11)}^2e^{\frac43\phi}} = \frac{6\pi M^2_{\text{\tiny P}}}{\kappa({\bm s},{\bm s},{\bm s})}\, .
\ee
We will also encounter towers of light states corresponding to M2-branes multiply wrapped around curves $\calc$. Using again  \eqref{scalingMth}, and the relation $\d s^2_{\text{\tiny M}}(X)=e^{-\frac23\phi}\d s^2_{\text{st}}(X)$ between the M-theory and the string frame Calabi-Yau metric, their minimal mass is given by: 
\be\label{D2mass}
M_{\text{\tiny M2}}^2 = \frac{(2\pi)^2 e^{2A}  V({\calc})^2}{l_{(11)}^2} = \frac{6\pi M^2_{\text{\tiny P}} V_{\rm st}(\calc)^2}{\kappa({\bm s},{\bm s},{\bm s})}\, ,
\ee 
where $V_{\rm st}(\calc)={\bm s}\cdot \calc=s^a D_a\cdot\calc$ is the string frame volume of $\calc$. As we will see, there will also appear  additional light towers of KK-states associated to the internal Calabi-Yau.  

Which towers of states should be considered in computing the quantum gravity scale along these EFT string limits depends on whether $p_a e^a=0$ or $p_a e^a>0$, and on the intersection properties of the nef divisor  ${\bf e}=e^aD_a$. As remarked in \cite{Lanza:2021udy} and \cite{Martucci:2022krl}, following  \cite{Lee:2019wij}, the latter can grouped into three cases. We will then separately discuss these three possibilities, for both $p_a e^a=0$ and $p_a e^a>0$, in turn.

\subsubsection{EFT string limits with $p_ae^a=0$}
\label{app:casespe=0}

Let us first assume that $e^0=p_ae^a=0$. The saxion $s^0$,  the average  M-theory  Calabi-Yau volume $ V(X)=s^0-\frac12p_as^a$, and the tension $\calt_*$ defined in \eqref{eq:HetTensionsa} remain  constant along the flow \eqref{EFTlimit}. On the other hand, \eqref{hets0} implies that
\be\label{hetdilaton} 
e^{2\phi}=\frac{\kappa({\bm s},{\bm s},{\bm s})}{6(s^0-\frac12p_as^a)}\,,
\ee
which, as we will see, diverges as some positive  power of $\sigma$. 
Furthermore \eqref{eq:HetTensionsb} reduces to the first contribution appearing on its r.h.s.\ and  determines, in the limit $\sigma\rightarrow \infty$, the dominant EFT string scale $M_\calt$ defined in \eqref{SPbounddef}:
\be\label{MTpe0} 
M^2_\calt=\frac{3\pi\kappa({\bf e},{\bm s},{\bm s})}{\kappa({\bm s},{\bm s},{\bm s})}\, M^2_{\text{\tiny P}}\,.
\ee
Following \cite{Lanza:2021udy}, we can then distinguish three cases. Note that, while it is not a priori obvious  how to concretely realize the condition $p_ae^a=0$ in heterotic compactifications, if we set $p_a\equiv 0$ the following results apply anyway to $\caln=2$ type IIA compactifications over $X$. 

\bigskip

{\em{\large Case 1: $\kappa({\bf e},{\bf e},{\bf e})>0$}}

\smallskip

\noindent In this case, 
 evaluating \eqref{D0mass} and \eqref{MTpe0} along \eqref{EFTlimit} and taking the limit $\sigma\rightarrow\infty$, one gets $M^2_\calt \simeq \frac{3\pi M^2_{\text{\tiny P}}}{\sigma}$ and  $M_{\text{\tiny KK}}^2 \simeq \frac{6\pi M^2_{\text{\tiny P}}}{\kappa({\bf e},{\bf e},{\bf e}) \sigma^3}$. Moreover, from \eqref{hetdilaton} we see that   $e^{2\phi}$ diverges as $\sigma^3$, and then the M-theory description is more appropriate. This  is a $w=3$ EFT string limit \cite{Lanza:2021udy}. The Calabi-Yau KK mass-squared scales as $\sigma^{-1}$ and  is parametrically higher than $M_{\text{\tiny KK}}$. This limit corresponds to a partial M-theory decompactification, and the species scale should then be given by the quantum  gravity scale \eqref{hetMQG}, which asymptotically behaves as 
\be\label{case1MQG} 
M^2_{\text{\tiny QG}}\simeq \left[\frac{12\pi}{(s^0_0-\frac12p_as^a_0)^2\kappa({\bf e},{\bf e},{\bf e})}\right]^\frac13\frac{M^2_{\text{\tiny P}}}{2\sigma}\simeq  \frac{M^2_\calt}{\left[18\pi^2(s^0_0-\frac12p_as^a_0)^2\kappa({\bf e},{\bf e},{\bf e})\right]^\frac13}\,.
\ee
Our saxionic convex hull conditions \eqref{hethatdelta} requires $s^0_0\geq \frac1{\alpha}+p_as_0^a$, with  $\alpha\leq 2\pi$. Hence \eqref{case1MQG} implies that $M_{\text{\tiny QG}}$ and $M_\calt$ scale asymptotically in the same way, but with $M_{\text{\tiny QG}}<M_\calt$, hence realizing \eqref{SPbound} with $M_{\rm sp}=M_{\text{\tiny QG}}$.  

\bigskip

{\em\large Case 2: $\kappa({\bf e},{\bf e},{\bf e})=0$ but $\kappa({\bf e},{\bf e},{\bf e}')> 0$ for some ${\bf e}'\in{\rm Nef}_{\mathbb{Z}}(X)$}\label{app:CASE2}

\smallskip

\noindent In this limit, the Calabi-Yau base $X$ can be seen as a $T^2$ fibration over a base two-fold $B$, with the non-trivial curve $\calc = {\bf e} \cdot {\bf e}$ being a multiple of the $T^2$ fiber. From \eqref{MTpe0} and \eqref{D0mass} get  $M^2_\calt \simeq 2\pi M^2_{\text{\tiny P}}/ \sigma$ and $M_{\text{\tiny KK}}^2 \simeq 2\pi M^2_{\text{\tiny P}}/[\kappa({\bf e},{\bf e},{\bm s}_0) \sigma^2]$, realizing a $w=2$ EFT string limit. The M-theory volume   $ V(\calc)=e^{-2\phi/3}V_{\rm st}(\calc)=e^{-2\phi/3}\kappa({\bf e},{\bf e},{\bm s})$ of $\calc$ vanishes as $\sigma^{-2/3}$, while the M-theory volume of base two-fold $B$ grows as $\sigma^{2/3}$. Hence  
the base KK mass-squared scales as $1/\sigma$, and is again much heavier than $M_{\text{\tiny KK}}$. On the other hand, since we can identify $\calc$ with an integral  $k$-multiple of the $T^2$ fiber, the mass  \eqref{D2mass} of an M2-brane on the $T^2$ fiber is asymptotically given by $M_{\text{\tiny M2}}^2 \simeq 2\pi M^2_{\text{\tiny P}}\kappa({\bf e},{\bf e},{\bm s}_0)/(k^2\sigma^2)$, and hence scales with the same rate of $M^2_{\text{\tiny KK}}$. According to the ESC \cite{Lee:2019wij}, there should exist some dual description in which this limit corresponds to a decompactification limit, which then realizes the species scale as a corresponding quantum gravity scale. Indeed, as in the F-theory limit of elliptically fibered M-theory compactifications, while one looses the two directions of the shrinking  $T^2$ fiber, the above light M2 states should correspond to KK modes of a new emergent compact direction. Therefore, the dual compactification space should roughly be  given by a (possibly twisted) product of the M-theory interval/circle,  the new emergent circle, and the base $B$. The corresponding KK species can then be approximately considered as multiplicative --  
see for instance  \cite{Castellano:2021mmx} for a discussion on 
multiplicative species.
More precisely, we  estimate the species scale by applying \eqref{axiSC} with~\footnote{According to Weyl's rule, which can be extracted from Eq.\ \eqref{NKK}, the number of KK resonances associated with an $n$-dimensional space with mass below the species scale  is well approximated by the formula $N_{(n)}\sim {\mathtt V}_{n}/(2\pi)^{\lfloor(n+1)/2 \rfloor}$, up to  numerical factors which will not alter our main conclusion. On the other hand, in the present setting, we have ${\mathtt V}_6\simeq {\mathtt V}_{1,{\text{\tiny KK}}}{\mathtt V}_{1,{\text{\tiny M2}}}{\mathtt V}_{4,B}$ and $N_{\rm sp}\simeq N_{6}N_{(10)}$. Combining these estimates we obtain \eqref{limitsNsp}.
}
\be\label{limitsNsp} 
N_{\rm sp}\simeq  2\pi N_{\text{\tiny KK}}N_{\text{\tiny M2}} N_BN_{(10)}=\frac{2\pi M^2_{\text{sp}}}{M_{\text{\tiny KK}}M_{\text{\tiny M2}}}\,  N_B N_{(10)}
\ee
where $N_{\text{\tiny KK}}$, $N_{\text{\tiny M2}}$,  $N_B$ denote the numbers of  KK modes with mass below the species scale, corresponding to each geometric (local) factor, and $N_{(10)}$ denotes the number of light degrees of freedom in the dual ten-dimensional theory. 
Using \eqref{limitsNsp}
in \eqref{axiSC}, with the above asymptotic values of $M_{\text{\tiny KK}}$ and $M_{\text{\tiny M2}}$, we get 
\be\label{case211}
M^2_{\rm sp}\simeq \frac{1}{k\sqrt{N_BN_{(10)}}}\,\frac{2\pi M^2_{\text{\tiny P}}}{\sigma}\simeq \frac{\,M^2_\calt}{k\sqrt{N_BN_{(10)}}}\,.
\ee
Since in the limit the base volume grows, we expect $N_B\gg 1$. Moreover, in string theory  $N_{(10)}$ is  significantly larger than one too. Hence $M^2_{\rm sp}< M^2_\calt$, in agreement with \eqref{SPbound}.

\bigskip

{\em\large Case 3: $\kappa({\bf e},{\bf e},{\bf e}')=0$ for any ${\bf e}'\in{\rm Nef}_{\mathbb{Z}}(X)$}

\smallskip

\noindent In this case we have ${\bf e} \cdot {\bf e} = 0$ and in this limit the Calabi-Yau $X$ can be seen as a K3 or $T^4$ fibration over a $\mathbb{P}^1$, where the fiber can be identified with the divisor ${\bf e}=e^aD_a$. Along this EFT string limit, which has scaling weight $w=1$, the M-theory volume of the base $\mathbb{P}^1$ grows as $\sigma^\frac23$, while the M-theory volume of K3 or $T^4$ decreases as $\sigma^{-\frac23}$. 
By the ESC  \cite{Lee:2019wij} there should exist a dual description in which this EFT string limit is a tensionless critical string limit. Hence in this limit  $M_{\rm sp}=M_\calt$ and \eqref{SPbound} is satisfied and saturated.

\subsubsection{EFT string limits with $p_ae^a\geq 1$}

Looking at the asymptotic behavior of \eqref{eq:HetTensions} along \eqref{EFTlimit}, it is clear that if $p_ae^a\geq 1$ the dominant EFT string scale is always given by the EFT string tension \eqref{eq:HetTensionsa}:
\be\label{MTpe>1} 
M_\calt^2=2\pi \calt_*=\frac{\pi M^2_{\text{\tiny P}}}{(s^0-\frac12 p_as^a)}\simeq \frac{2\pi M^2_{\text{\tiny P}}}{p_ae^a}\,\frac1\sigma\,.
\ee
Differently from what happened in Section \ref{app:casespe=0}, now the average M-theory Calabi-Yau volume  diverges, ${V}(X) = s^0 - \frac{1}{2} p_a s^a\simeq \frac{1}{2} p_a e^a \sigma$,  and this also affects the asymptotic behavior of the dilaton \eqref{hetdilaton}.

We can then proceed   discussing in turn the three cases already considered as in in Section \ref{app:casespe=0}, 
following \cite{Martucci:2022krl}.

\bigskip

{\em{\large Case 1: $\kappa({\bf e},{\bf e},{\bf e})>0$}}

\smallskip

\noindent In this case both the M-theory interval and the Calabi-Yau volume diverge. Hence the species scale $M_{\rm sp}$ is given by \eqref{hetMQG}. Evaluating \eqref{MhetTQG} along the flow \eqref{EFTlimit}, in the limit $\sigma\rightarrow\infty$ we get $M^2_\calt/M^2_{\text{\tiny QG}}\propto \sigma^{2/3}$, and then \eqref{SPbound} is certainly satisfied.

\bigskip

{\em\large Case 2: $\kappa({\bf e},{\bf e},{\bf e})=0$ but $\kappa({\bf e},{\bf e},{\bf e}')> 0$ for some ${\bf e}'\in{\rm Nef}_{\mathbb{Z}}(X)$}

\smallskip

\noindent 
This limit has scaling weight $w=2$ \cite{Martucci:2022krl}. As already discussed in Section \ref{app:casespe=0}, the Calabi-Yau $X$ can be regarded as a $T^2$ fibration over a base two-fold $B$. In M-theory, the $T^2$ fiber has volume $ V(T^2)=V(\calc)/k=\frac1ke^{-2\phi/3}\kappa({\bf e},{\bf e},{\bm s})\propto \sigma^{-1/3}$, where $k\geq 1$ is the number of $T^2$ fibers contained in $\calc$, and hence the base $B$ volume grows as $\sigma^{4/3}$. From \eqref{D0mass} and \eqref{D2mass} we get the same asymptotic behaviors found in the $p_ae^a=0$ case:   $M_{\text{\tiny KK}}^2 \simeq 2\pi M^2_{\text{\tiny P}}/[\kappa({\bf e},{\bf e},{\bf s}_0)\sigma^2]$ and  $M_{\text{\tiny M2}}^2 \simeq 2\pi M^2_{\text{\tiny P}}\kappa({\bf e},{\bf e},{\bm s}_0)/(k^2\sigma^2)$. Now we cannot use \eqref{hetMQG} and \eqref{MhetTQG}, but we can proceed as in the discussion of Case 2 in Section \ref{app:casespe=0}, and use \eqref{axiSC} as estimate of the species scale, with $N_{\rm sp}$ as in \eqref{limitsNsp}. The first estimate in \eqref{case211} still holds.  Recalling \eqref{MTpe>1}, it  can be rewritten as 
\be 
M^2_{\rm sp}= \frac{p_a e^a}{k\sqrt{N_BN_{(10)}}}\, M^2_\calt\,.
\ee
Recalling that the base $B$ decompactifies  faster than in the $p_ae^a=0$ case, we expect an even larger $N_B\gg 1$, implying that \eqref{SPbound} is again satisfied. 

\bigskip

{\em\large Case 3: $\kappa({\bf e},{\bf e},{\bf e}')=0$ for any ${\bf e}'\in{\rm Nef}_{\mathbb{Z}}(X)$}

\smallskip

This is a $w=2$ limit \cite{Martucci:2022krl}. As discussed in Section \ref{app:casespe=0}, $X$ is a  K3/$T^4$ fibration over $\mathbb{P}^1$, and we can identify the fiber with the nef divisor ${\bf e}=e^aD_a$. Eq.~\eqref{hetdilaton} implies that the dilaton tends to the constant value $e^{2\phi}\simeq  \kappa({\bf e},{\bm s}_0,{\bm s}_0)/(p_a e^a)$. Similarly, the M-theory volume  of the K3/$T^4$ fiber is asymptotically constant too: $V_{\bf e}=\frac12e^{-\frac{4}{3}\phi}\kappa({\bf e},{\bm s},{\bm s})\simeq  \frac12 (p_ae^a)^{\frac23}\kappa({\bf e},{\bm s}_0,{\bm s}_0)^{\frac13}$. 
 Hence the $\mathbb{P}^1$ base diverges as $\sigma$, the heterotic/M-theory EFT description remains valid, and the discussion of Section \ref{sec:hetscales} applies. In particular, if the asymptotic value of  $2\pi e^\phi$ is smaller than one, then $M_{\rm sp}$ asymptotically coincides with \eqref{MTpe>1}, and saturates the bound \eqref{SPbound}. If instead it is larger than one, then  $M_{\rm sp}$ is given by \eqref{hetMQG}
 and \eqref{MhetTQG} implies that the bound  \eqref{SPbound} is satisfied.

\section{Wormholes and supersymmetry}
\label{app:WHsusy}

In this paper we have focused on asymptotically flat wormholes, regarded as quantum excitations of a supersymmetric Minkowski vacuum preserving $\caln=1$ supersymmetry.  In general a wormhole  breaks, at least partially,  the corresponding four supercharges. In this appendix we discuss at a more quantitative level this breaking and the  presence of corresponding Goldstino-like fermionic zero-modes.

We will use the four- and two-component spinor notations. In four-component notation the Majorana supersymmetry generators split into  $\epsilon=\epsilon_{\text{\tiny L}}+\epsilon_{\text{\tiny R}}$, with $\gamma_5\epsilon_{\text{\tiny L}}=\epsilon_{\text{\tiny L}}$ and $\gamma_5\epsilon_{\text{\tiny R}}=-\epsilon_{\text{\tiny R}}$. The relation with the two-component notation \cite{Wess:1992cp} is given by
\be\label{fourtworepr} 
\gamma^{a}=\ii\left(\begin{array}{cc} 0 & \sigma^a \\
\bar\sigma^a & 0
\end{array}\right)\,,\quad \gamma_5=-\ii \gamma^{\underline{0123}}=\left(\begin{array}{cc} -\mathds{1} & 0 \\
0 & \mathds{1}
\end{array}\right)\,,\quad \epsilon=\left(\begin{array}{c} \epsilon_{\text{\tiny R}}  \\
\epsilon_{\text{\tiny L}}
\end{array}\right)=\left(\begin{array}{c} 
\epsilon_{\alpha}\\ \bar\epsilon^{\dot\alpha}  
\end{array}\right)\,.
\ee
where $(\sigma^a)_{\alpha\dot\beta}=(-\mathds{1},\vec\sigma)$ and $(\bar\sigma^a)^{\dot\alpha\beta}=(-\mathds{1},-\vec\sigma)$. We use Latin letters from the middle of the alphabet ($m,n,\ldots$)  to denote curved indices, and from the beginning of the alphabet ($a,b,\ldots$) to denote flat indices. If necessary, to avoid further ambiguities, we underline flat indices. In two-component notation the Majorana condition relates left- and right-moving components: 
\be\label{Majcond}
\bar\epsilon_{\dot\alpha}\equiv \varepsilon_{\dot\alpha\dot\beta}\bar\epsilon^{\dot\beta}=(\epsilon_\alpha)^*\,.
\ee

In four-component notation the supersymmetry transformations of gravitino and chiralinos (the supersymmetric partners of $t^i$) are \cite{Wess:1992cp} 
\begin{subequations}\label{fermitr} 
\begin{align}
\delta\psi_m&=\Big[\nabla_m-\frac\ii2\Im\left(\del_mt^i\del_i K\right) \gamma_5\Big]\epsilon\,,\label{fermitra}\\ 
\delta\chi^i_{\text{\tiny R}}&=\sqrt{2}\,\del_m t^i \gamma^m \epsilon_{\text{\tiny L}}\,,\quad \delta\chi^{i}_{\text{\tiny L}}=\sqrt{2}\,\del_m \bar t^i \gamma^m \epsilon_{\text{\tiny R}}\,.\label{fermitrb}
\end{align}
\end{subequations}
In this paper we restrict to K\"ahler potential that depend on $t^i=a^i+\ii s^i$ only through the saxions $s^i=\Im t^i$. Recalling \eqref{elldef}, \eqref{fermitra} can be rewritten as 
\be\label{gravtr} 
\delta\psi_m=\Big(\nabla_m-\frac\ii2\,\ell_i\del_m a^i\, \gamma_5\Big)\epsilon\,.
\ee

In the Wick-rotated Euclidean formulation $\gamma^{\underline{4}}=\gamma_{\underline{4}}=\ii\gamma^{\underline{0}}$ and, correspondingly, $\sigma^{\underline{4}}=\sigma_{\underline{4}}=\ii\sigma^{\underline{0}}=-\ii\mathds{1}$
and $\bar\sigma^{\underline{4}}=\bar\sigma_{\underline{4}}=\ii\bar\sigma^{\underline{0}}=-\ii\mathds{1}$ whereas $\vec\sigma$ remain unchanged, so that the representation \eqref{fourtworepr} formally still holds. Moreover, the left- and right-moving components of the Majorana spinors become independent fields and are not related by complex conjugation anymore, but must be regarded as analytic continuation of the Lorentzian ones. In particular, the  Majorana condition  \eqref{Majcond} must be relaxed.

\subsection{Extremal BPS wormholes}
\label{app:WHsusy1}

We first focus on the extremal wormholes of Section \ref{sec:extremalBPS}. These are characterized by the following BPS-like equations \cite{Lanza:2021udy}:
\be\label{BPSH3ell} 
\calh_{3,i}=-M_{\text{\tiny P}}^2*\d\ell_i\,.
\ee
The metric is flat and, in the case of the extremal wormhole sourced by a fundamental instanton centered at $x_0$, the Bianchi identity \eqref{H3BI} must be corrected by a localized term: $\d\calh_{3,i}=2\pi\delta_{4}(x_0)$. Hence, the dual saxions profiles must satisfy $\d*\d\ell_i=-2\pi M_{\text{\tiny P}}^{-2}\delta_{4}(x_0)$, which  is  solved  by
\be\label{multicenter} 
\ell_i=\ell_i^\infty+\frac{q_{i}}{2\pi M^2_{\text{\tiny P}}\,|x-x_0|^2}\,,
\ee
where $|x|^2\equiv \delta_{mn}x^mx^n$. The solution \eqref{multicenter} can be immediately generalized to multi-centered ones, sourced by multiple fundamental instantons.   Given  the Euclidean axion/two-form duality relation 
\be\label{Euclduality} 
\d a^i=\frac{\ii}{M_{\text{\tiny P}}^2}\calg^{ij}*\calh_{3,j}\,,
\ee
in terms of the axions $a^i$ the BPS conditions \eqref{BPSH3ell} reads\footnote{Recall that in Euclidean space the axions dual to a 2-form become purely imaginary fields.}
\be\label{axionBPS} 
\d a^i=\ii\calg^{ij}\d\ell_j
=-\ii\d s^i\,.
\ee
By using this and  \eqref{elldef}, and the fact that the metric is flat, the gravitino transformation \eqref{gravtr} on the multicenter extremal wormhole solution becomes 
\be\label{Killing0} 
\delta\psi_m=(\del_m+\frac14 \del_mK\gamma_5)\epsilon=e^{-\frac14 K\gamma_5}\del_m\left(e^{\frac14 K\gamma_5}\epsilon\right)\,.
\ee
We then see that the gravitino remains invariant, $\delta\psi_m=0$, under the supersymmetry transformations
\be\label{Killing1} 
\epsilon=e^{-\frac14(K-K_\infty)\gamma_5}\eta\quad\Leftrightarrow\quad \epsilon_\alpha=e^{\frac14(K-K_\infty)}\eta_\alpha\quad,\quad \bar\epsilon^{\dot\alpha}=e^{-\frac14(K-K_\infty)}\bar\eta^{\dot\alpha}\,,
\ee
with $\eta_\alpha$ and $\bar\eta^{\dot\alpha}$ independent  constant anticommuting spinors. We have chosen the integration constant $e^{\frac14 K_\infty\gamma_5}$, where $K_\infty\equiv K(s_\infty)$,  so that $\epsilon\rightarrow \eta$ at infinite spatial distance. The four constant components $(\eta_\alpha,\bar\eta^{\dot\alpha})$  parametrize four independent global supersymmetry transformations which leave the gravitino invariant -- see also \cite{Huebscher:2009bp}.  

Let us now turn  to the chiralino transformations \eqref{fermitrb}. We first notice that the BPS condition \eqref{axionBPS} is equivalent to $\d t^i=0$, while $\d\bar t^i=-2\ii\d s^i$. This implies that
\be\label{deltachiBPS} 
\delta\chi^i_\alpha=0\,,\quad \delta\bar\chi^{\dot\alpha \,i}=2\sqrt{2}e^{\frac14(K-K_\infty)}\del_m s^i(\bar\sigma^m)^{\dot\alpha\beta}\eta_\beta\,,
\ee
under the transformations generated by \eqref{Killing1}. We see that the extremal wormhole solutions preserve the two supersymmetries generated by $\bar\eta^{\dot\alpha}$, while break the ones generated by $\eta_{\alpha}$. In particular,
because $\eta_{\alpha}$ acts non-trivially on the background, the configuration 
\be\label{zero-modesBPS}
\bar\chi^{\dot\alpha \,i}(x)\equiv [\bar\chi^{i\,\beta}_{(0)}(x-x_0)]^{\dot\alpha}\eta_\beta\,,
\ee
with ($\beta=1,2$)
\begin{align}\label{extrzeromodes} 
[\bar\chi^{i\,\beta}_{(0)}(x-x_0)]^{\dot\alpha}
={\cal N}\,e^{\frac14(K-K_\infty)}(\bar\sigma^m)^{\dot\alpha\beta}\del_m s^i\,,
\end{align}
corresponds to zero-modes of the linearized equations of motion around the background. Note that the location of the fundamental instanton $x_0$ is a free parameter. The normalization constant $\mathcal{N}$ is fixed by
\be\label{normcondLV} 
\delta^{\beta\beta'}=\int\d^4 x\, \calg_{ij}\left\{[\bar\chi^{i\,\beta}_{(0)}]^{\dot\alpha}\right\}^\dagger[\bar\chi^{j\,\beta'}_{(0)}]^{\dot\alpha}.
\ee
Since the integral should be restricted to the controllable regime $|x-x_0|^2\geq \Lambda^{-2}$, the normalization constant $\caln$ in general satisfies ${\cal N}\propto \Lambda\,e^{K_\infty/4}$ up to a dimensionless function of $M_{\text{\tiny P}}^2/{\Lambda^2}$.
Furthermore, as $\partial_ms^i\sim M_{\text{\tiny P}}^{-2}\calg^{ij}_\infty q_j\,\del_m |x-x_{0}|^{-2}$ for large  $|x-x_0|^2$,
the zero-modes \eqref{zero-modesBPS} behave as the components of Feynman's fermionic propagator in the asymptotic vacuum:
\begin{align}\label{asympzerom} 
[\bar\chi^{i\,\beta}_{(0)}]^{\dot\alpha}
\propto {\cal N}\,e^{\frac14(K-K_\infty)}\frac{\calg^{ij} q_j}{2\pi M_{\text{\tiny P}}^2}\,(\bar\sigma^m)^{\dot\alpha\beta}\del_m \frac{1}{|x-x_{0}|^2} \propto \varepsilon^{\beta\alpha}q_j\calg^{ij}[S_{\rm F} (x-x_{0})]_{\alpha}^{\dot\alpha} \, .
\end{align}

\subsection{Non-extremal wormholes}
\label{app:WHsusy2}

 In order to study the fermionic zero-modes of regular non-extremal wormholes it is convenient, following \cite{Anglin:1992ym}, to introduce a new dimensionless radial coordinate 
\be\label{ydef} 
y\equiv 2\pi M^2_{\text{\tiny P}}L^2 \tau\quad~~~\Leftrightarrow\quad~~~ \cos y=\frac{L^2}{r^2}  \,.
\ee
Note that $y\in(-\frac\pi2,\frac\pi2)$. 
We recall that $\tau$ is related to the three-sphere radius $r$ by \eqref{rtaurel}. Then $e^{-2A}=\sin^2y$ and
the metric \eqref{WHmetric} becomes
\be\label{ymetric} 
\d s^2=L^2\left(\frac{\d y^2}{4\cos^3 y}+\frac{\d\Omega^2}{\cos y}\right)\,.
\ee
Let us pick the vielbein  $e^a=(e^{\underline{4}},e^{\underline{i}})$, with
\be 
e^{\underline{4}}=\frac{L}{2(\cos y)^{\frac32}}\d y\quad,\quad e^{\underline{i}}=\frac{L}{\sqrt{\cos y}}\,\hat e^{\underline{i}}\,,
\ee
where $e^{\underline{i}}$ is a vielbein for a three-sphere of radius $r=1$. The components of the spin connection $\omega^a{}_b=\omega^a{}_{bm}\d y^m$ are
\be\label{spincon} 
\omega^{\underline{i}}{}_{\underline{j}}=\hat\omega^{\underline{i}}{}_{\underline{j}}\quad,\quad \omega^{\underline{i}}{}_{\underline{4}}=\sin y\,e^{\underline{i}}\,.
\ee

By combing  \eqref{Euclduality} and \eqref{H3ansatz} we have
\be\label{daEucl}
\d a^i=-\frac{\ii \calg^{ij}q_j}{2\pi M^2_{\text{\tiny P}}L^2}\d y\,.
\ee
Plugging \eqref{daEucl} inside \eqref{gravtr} we get 
\be\label{gravtr1}
\delta\psi_m=\left(\nabla_m-\frac{\langle {\bf q},{\bm s}\rangle }{4\pi M^2_{\text{\tiny P}}L^2}\delta^y_m\gamma_5\right)\epsilon\,.
\ee
Let us now gauge-fix the local supersymmetry by imposing the transverse gravitino gauge
\be\label{gravitinofix} 
\gamma^m\psi_m=0\,,
\ee 
where $\gamma_m=e_m^a\gamma_a$. Imposing that \eqref{gravitinofix} is preserved under \eqref{gravtr1} and using \eqref{spincon}, we get the following condition on $\epsilon$:
\be\label{epsilonconst} 
2\cos y\,\del_y\epsilon +\gamma^{\underline{4}}\hat\gamma^i\hat\nabla_i\epsilon+\frac32\sin y\,\epsilon=\frac{\langle {\bf q},{\bm s}\rangle }{2\pi M^2_{\text{\tiny P}}L^2}\cos y\,\gamma_5\epsilon\,.
\ee
We would like to determine the form of $\epsilon$ imposing that it becomes asymptotically covariantly constant  on one of the two sides of the wormhole: $\nabla_m\epsilon|_{y=\pm\frac\pi2}=0$. Since 
\be 
\nabla_i\epsilon=\hat\nabla_i\epsilon+\frac12\sin y\,\hat\gamma_i\gamma_{\underline{4}}\epsilon\,,
\ee
we get the asymptotic conditions 
\be 
\hat\nabla_i\epsilon|_{\pm\frac\pi2}=\mp\frac12\hat\gamma_i\gamma_{\underline{4}}\epsilon|_{y=\pm\frac\pi2}\,.
\ee
Given the $O(4)$ symmetry, these conditions should in fact be satisfied by $\epsilon$ at any $y\in(-\frac\pi2, \frac\pi2)$. We must then consider two possibilities 
\be\label{asymepsilon} 
\hat\nabla_i\epsilon_\pm=\mp\frac12\hat\gamma_i\gamma_{\underline{4}}\epsilon_\pm\,.
\ee
So $\epsilon_+$ can asymptotically  tend   to a non-vanishing covariantly constant spinor on the first half-wormhole, $y\in [0,\frac\pi2)$, while it should quickly vanish in the second half, for $y\rightarrow -\frac\pi2$. Viceversa, $\epsilon_-$ tends  to a covariantly constant spinor on the second half-wormhole, $y\in (-\frac\pi2,0]$ and must be quickly vanishing for $y\rightarrow \frac\pi2$. In the following we will focus on $\epsilon_+$, omitting the subscript for simplicity. (The corresponding results for $\epsilon_-$ can be immediately obtained by inverting the role of the two half-wormholes.) 

So, taking into account \eqref{asymepsilon}, \eqref{epsilonconst} becomes 
\be\label{epsilonconst1} 
2\cos y\,\del_y\epsilon +\frac32(\sin y-1)\,\epsilon=\frac{\langle {\bf q},{\bm s}\rangle }{2\pi M^2_{\text{\tiny P}}L^2}\cos y\,\gamma_5\epsilon\,.
\ee
The solution to this equation can be written in the form
\be
\epsilon=\mathbf{M}(y)\,\eta\quad~~~~\text{with}\quad~~~~~ \mathbf{M}(y)=M_1(y)+ M_2(y)\gamma_5\,,
\ee
where $\eta$ is an $y$-independent spinor satisfying the same equation \eqref{asymepsilon} of $\epsilon_+$: $\hat\nabla_i\eta=-\frac12\hat\gamma_i\gamma_{\underline{4}}\eta$. We will impose that ${\bf M}(y=\frac\pi2)=\mathds{1}$, so that $\eta$ represents the asymptotic covariantly constant spinor.  The condition \eqref{epsilonconst1} translates into the following equation for the matrix $\mathbf{M}(y)$:
\be 
\del_y\log\mathbf{M}=-\frac34\tan \left(\frac{y}2-\frac\pi4\right)+\frac{\langle {\bf q},{\bm s}\rangle }{4\pi M^2_{\text{\tiny P}}L^2}\,\gamma_5\,.
\ee
This can be integrated into 
\be\label{genMform} 
\mathbf{M}(y)=\left[\cos \left(\frac{y}2-\frac\pi4\right)\right]^{\frac32}\exp\left[-\frac{\gamma_5 }{4\pi M^2_{\text{\tiny P}}L^2}\int^{\frac\pi2}_y \d\tilde y\,\langle {\bf q},{\bm s}(\tilde y)\rangle\right]\,,
\ee
where we have fixed the integration constants by imposing ${\bf M}(y)|_{y=\frac\pi2}=\mathds{1}$.
The right- and left-handed components $(\epsilon_\alpha,\bar\epsilon^{\dot\alpha})$ are then given by 
\be\label{epsilongens}
\epsilon_\alpha=\eta_\alpha\left[\cos \left(\frac{y}2-\frac\pi4\right)\right]^{\frac32}e^{f(y)}\,,\quad\bar\epsilon^{\dot\alpha}=\bar\eta^{\dot\alpha}\left[\cos \left(\frac{y}2-\frac\pi4\right)\right]^{\frac32}e^{-f(y)}\,,
\ee
with 
\be 
f(y)\equiv \frac{1}{4\pi M^2_{\text{\tiny P}}L^2}\int^{\frac\pi2}_y \d\tilde y\,\langle {\bf q},{\bm s}(\tilde y)\rangle\,.
\ee
The common cosine  factor implies that all components of $\epsilon$ quickly vanish as one approaches $y=-\frac\pi2$, as expected. The exponential factors instead give an enhancement for $\epsilon_\alpha$ and a further suppression for $\bar\epsilon^{\dot\alpha}$.  
One can get an idea of the possible form of $e^{f(y)}$ by restricting to the homogeneous wormholes of  Section \ref{sec:univWH}. In terms of the $y$ radial coordinate, \eqref{ellsol} becomes 
\be\label{tildelly} 
\tilde\ell(y)=\frac{1}{2\pi M^2_\text{\tiny P} L^2}\sqrt{\frac{n}{3}}\cos\left(\sqrt{\frac{3}{n}}\,y\right)\,.
\ee
From \eqref{tildelbound}, \eqref{linfstar} and \eqref{Ltau}, we get
\be\label{efhom} 
e^{f(y)}=\left[\frac{\tan\left(\frac\pi4-\frac12\sqrt{\frac{3}n}\,y\right)}{\tan\left(\frac\pi4-\frac\pi4\sqrt{\frac{3}n}\right)}\right]^{\frac{n}4}\,.
\ee

Using \eqref{epsilongens} inside \eqref{fermitr} one gets the corresponding (unnormalized) Goldstino zero-modes,
\begin{align}
\begin{aligned}
\delta\chi^i_{\text{\tiny R}}&= \frac{ 2\sqrt{2}}{L} \,\del_y t^i \left(\cos y \right) ^{3/2} \sigma^4 \epsilon_{\text{\tiny L}}\,,\\
\delta\chi^{i}_{\text{\tiny L}}&=\frac{ 2\sqrt{2}}{L} \,\del_y \bar t^i \left(\cos y \right)^{3/2} \bar\sigma^4 \epsilon_{\text{\tiny R}}\,.
\end{aligned}
\label{eq:appSUSY:subextrChiralinos}
\end{align}
These explicitly show that, in general, no supersymmetry is preserved by non-extremal wormhole configurations. We can further verify this by inserting the profile of $t^i$ on the homogeneous solution into \eqref{eq:appSUSY:subextrChiralinos}, obtaining
\begin{align}
\begin{aligned}
\delta\chi^i_{\text{\tiny R}}&=-\text{i} \frac{6\sqrt{2}\pi M^2_{\text{\tiny P}} L}{n}  \left[1-\sin \left( \sqrt{ \frac{3}{n}} y \right) \right]  \frac{ \left[ \cos y \cos \left( \frac{y}{2}-\frac{\pi}{4} \right) \right]^{3/2} }{ \left[ \cos \left( \sqrt{\frac{3}{n}} y\right) \right]^2} e^{-f(y)} \, \sigma^4 \eta_{\text{\tiny L}}\,, \\
\delta\chi^{i}_{\text{\tiny L}}&=-\text{i}\frac{6\sqrt{2}\pi M^2_{\text{\tiny P}} L}{n} \left[1+\sin \left( \sqrt{ \frac{3}{n}} y \right) \right]  \frac{ \left[ \cos y \cos \left( \frac{y}{2}-\frac{\pi}{4} \right) \right]^{3/2} }{ \left[ \cos \left( \sqrt{\frac{3}{n}} y\right) \right]^2}  e^{+f(y)} \, \bar{\sigma}^{ 4} \eta_{\text{\tiny R}}\,, \\
\end{aligned}
\label{eq:appSUSY:subextrHomogeneous}
\end{align}
with $e^{f(y)}$ as in \eqref{efhom}. For $n>3$ these expressions are regular and behave as $\sim r^{-3}$ as we approach the asymptotic flat space, like a free Feynman propagator analogously to \eqref{asympzerom}. The case $n=3$ corresponds to the marginally degenerated case and will be discussed below.

\subsection{Marginally degenerate wormholes}
\label{app:n=3}

In Section \eqref{sec:n=3} we have regularized the $n=3$ marginally degenerate 
 wormholes by introducing  an IR cutoff $r_\text{\tiny IR}$.  As discussed in Section \ref{sec:n=3physics}, this prescription receives a natural justification if one aims at finding the effective operator encoding the wormhole effects at distances larger than $\Lambda^{-1}_\text{\tiny IR}\gg L$. In 
Section \ref{sec:n=3}
we have also observed how the marginally degenerate wormholes look very similar to extremal BPS wormholes carrying the same charge. The deviation from the extremal case is concentrated around the wormhole's neck. It is then clear that, at radii slightly larger than $L$,  the  marginally degenerate wormhole should admit, to very good approximation, two Killing spinors $\bar\epsilon^{\dot\alpha}$ of the form  \eqref{Killing1}. This approximation of course breaks down as we approach the throat. However, the corresponding zero-modes $\delta\chi^i_{\text{\tiny R}}$ are necessarily localized around the wormhole throat and quickly vanish as we approach $r_{\text{\tiny IR}}$.  

In order to more explicitly  check this qualitative expectation, it is convenient to use the alternative (smooth) regularization $\calf=\log\tilde P\rightarrow (1+\varepsilon)\log\tilde P$, with $\varepsilon\ll 1$, which amounts to setting
\be 
n=3\left(1+\varepsilon\right)\,.
\ee
In order to illustrate the form of the zero-modes, let us focus for simplicity on the homogeneous case. The regularization implies a finite
\be 
\tilde\ell_\infty=\frac\pi4 \varepsilon \tilde\ell_*\,.
\ee
Recalling the discussion in section \ref{sec:n=3} we may pick  $\varepsilon=\frac{4}{\pi}\epsilon_\Lambda =\frac{4}{\pi} \Lambda_\text{\tiny IR} ^2 L^2$ so that we can identify  $\tilde\ell_\infty$ with $\tilde\ell_\Lambda$, formally identifying a correspondence between the $\Lambda_\text{\tiny IR}$ and $\varepsilon$ regularization procedures. In this way, Eq.\ \eqref{efhom} becomes
\be 
e^{-f(y,\varepsilon)}\simeq \left[ \frac{\pi\varepsilon}{8}\tan\left(\frac\pi4+\frac{y}{2\sqrt{1+\varepsilon}}\right) +\calo(\varepsilon^2) \right]^{\frac34}\,.
\ee
so that $\delta \chi_{\text{\tiny R}}, \delta \chi_{\text{\tiny L}} $ of \eqref{eq:appSUSY:subextrHomogeneous} are now well defined in the whole domain of $y$. The quantitative expectation that the modes associated to $\delta \chi_{\text{\tiny R}}$ become unobservable far from the wormhole throat can then be verified by inspecting the functional dependence on $y$ of the regularized wavefunctions associated to the zero-modes. In particular, their ratio is given by
\begin{align}
\left[ \frac{1-\sin\left(\frac{y}{\sqrt{1+\varepsilon}} \right) }{1+\sin\left(\frac{y}{\sqrt{1+\varepsilon}} \right) } \right]  e^{-2f(y,\varepsilon)}.
\end{align}
Taking $y=\pi/2$, the above expression becomes
\begin{align}
    \left[ \frac{1-\sin\left(\frac{\pi}{2\sqrt{1+\varepsilon}} \right) }{1+\sin\left(\frac{\pi}{2\sqrt{1+\varepsilon}} \right) } \right]e^{-2f(\pi/2,\varepsilon)}  \approx \frac{\pi ^2 \varepsilon^2}{64} = \frac{\Lambda_\text{\tiny IR} ^4 L^4}{4}\,,
\end{align}
where in the last step we exploited the correspondence between the two regularization procedures. Already at distances $1/\Lambda_\text{\tiny IR}= 5L$ this gives a relative $\calo( 10^{-4})$ relative suppression. Hence from a IR viewpoint only the zero-mode $\delta \chi_{\text{\tiny L}}$ is observable. At $y=-\pi/2$ the situation is reversed, with the zero-mode $\delta \chi_{\text{\tiny L}}$ being suppressed in the IR. This is consistent with the identification of the $y=-\pi/2$ side as an anti-extremal BPS wormhole conserving the opposite supercharges.

\bibliographystyle{jhep}
\bibliography{references}

\providecommand{\href}[2]{#2}\begingroup\raggedright\begin{thebibliography}{100}

\bibitem{Peccei:1977hh}
R.~D. Peccei and H.~R. Quinn, \emph{{CP Conservation in the Presence of
  Instantons}}, \href{https://doi.org/10.1103/PhysRevLett.38.1440}{\emph{Phys.
  Rev. Lett.} {\bfseries 38} (1977) 1440}.

\bibitem{Weinberg:1977ma}
S.~Weinberg, \emph{{A New Light Boson?}},
  \href{https://doi.org/10.1103/PhysRevLett.40.223}{\emph{Phys. Rev. Lett.}
  {\bfseries 40} (1978) 223}.

\bibitem{Wilczek:1977pj}
F.~Wilczek, \emph{{Problem of Strong $P$ and $T$ Invariance in the Presence of
  Instantons}}, \href{https://doi.org/10.1103/PhysRevLett.40.279}{\emph{Phys.
  Rev. Lett.} {\bfseries 40} (1978) 279}.

\bibitem{Preskill:1982cy}
J.~Preskill, M.~B. Wise and F.~Wilczek, \emph{{Cosmology of the Invisible
  Axion}}, \href{https://doi.org/10.1016/0370-2693(83)90637-8}{\emph{Phys.
  Lett. B} {\bfseries 120} (1983) 127}.

\bibitem{Abbott:1982af}
L.~F. Abbott and P.~Sikivie, \emph{{A Cosmological Bound on the Invisible
  Axion}}, \href{https://doi.org/10.1016/0370-2693(83)90638-X}{\emph{Phys.
  Lett. B} {\bfseries 120} (1983) 133}.

\bibitem{Dine:1982ah}
M.~Dine and W.~Fischler, \emph{{The Not So Harmless Axion}},
  \href{https://doi.org/10.1016/0370-2693(83)90639-1}{\emph{Phys. Lett. B}
  {\bfseries 120} (1983) 137}.

\bibitem{Hu:2000ke}
W.~Hu, R.~Barkana and A.~Gruzinov, \emph{{Cold and fuzzy dark matter}},
  \href{https://doi.org/10.1103/PhysRevLett.85.1158}{\emph{Phys. Rev. Lett.}
  {\bfseries 85} (2000) 1158}
  [\href{https://arxiv.org/abs/astro-ph/0003365}{{\ttfamily
  astro-ph/0003365}}].

\bibitem{Frieman:1995pm}
J.~A. Frieman, C.~T. Hill, A.~Stebbins and I.~Waga, \emph{{Cosmology with
  ultralight pseudo Nambu-Goldstone bosons}},
  \href{https://doi.org/10.1103/PhysRevLett.75.2077}{\emph{Phys. Rev. Lett.}
  {\bfseries 75} (1995) 2077}
  [\href{https://arxiv.org/abs/astro-ph/9505060}{{\ttfamily
  astro-ph/9505060}}].

\bibitem{Svrcek:2006yi}
P.~Svrcek and E.~Witten, \emph{{Axions In String Theory}},
  \href{https://doi.org/10.1088/1126-6708/2006/06/051}{\emph{JHEP} {\bfseries
  06} (2006) 051} [\href{https://arxiv.org/abs/hep-th/0605206}{{\ttfamily
  hep-th/0605206}}].

\bibitem{Arvanitaki:2009fg}
A.~Arvanitaki, S.~Dimopoulos, S.~Dubovsky, N.~Kaloper and J.~March-Russell,
  \emph{{String Axiverse}},
  \href{https://doi.org/10.1103/PhysRevD.81.123530}{\emph{Phys. Rev. D}
  {\bfseries 81} (2010) 123530}
  [\href{https://arxiv.org/abs/0905.4720}{{\ttfamily 0905.4720}}].

\bibitem{McAllister:2023vgy}
L.~McAllister and F.~Quevedo, \emph{{Moduli Stabilization in String Theory}},
  \href{https://arxiv.org/abs/2310.20559}{{\ttfamily 2310.20559}}.

\bibitem{Giddings:1987cg}
S.~B. Giddings and A.~Strominger, \emph{{Axion Induced Topology Change in
  Quantum Gravity and String Theory}},
  \href{https://doi.org/10.1016/0550-3213(88)90446-4}{\emph{Nucl. Phys. B}
  {\bfseries 306} (1988) 890}.

\bibitem{Hebecker:2018ofv}
A.~Hebecker, T.~Mikhail and P.~Soler, \emph{{Euclidean wormholes, baby
  universes, and their impact on particle physics and cosmology}},
  \href{https://doi.org/10.3389/fspas.2018.00035}{\emph{Front. Astron. Space
  Sci.} {\bfseries 5} (2018) 35}
  [\href{https://arxiv.org/abs/1807.00824}{{\ttfamily 1807.00824}}].

\bibitem{Demirtas:2018akl}
M.~Demirtas, C.~Long, L.~McAllister and M.~Stillman, \emph{{The Kreuzer-Skarke
  Axiverse}}, \href{https://doi.org/10.1007/JHEP04(2020)138}{\emph{JHEP}
  {\bfseries 04} (2020) 138}
  [\href{https://arxiv.org/abs/1808.01282}{{\ttfamily 1808.01282}}].

\bibitem{Demirtas:2021gsq}
M.~Demirtas, N.~Gendler, C.~Long, L.~McAllister and J.~Moritz, \emph{{PQ
  Axiverse}},  \href{https://arxiv.org/abs/2112.04503}{{\ttfamily 2112.04503}}.

\bibitem{Lanza:2020qmt}
S.~Lanza, F.~Marchesano, L.~Martucci and I.~Valenzuela, \emph{{Swampland
  Conjectures for Strings and Membranes}},
  \href{https://arxiv.org/abs/2006.15154}{{\ttfamily 2006.15154}}.

\bibitem{Lanza:2021udy}
S.~Lanza, F.~Marchesano, L.~Martucci and I.~Valenzuela, \emph{{The EFT stringy
  viewpoint on large distances}},
  \href{https://doi.org/10.1007/JHEP09(2021)197}{\emph{JHEP} {\bfseries 09}
  (2021) 197} [\href{https://arxiv.org/abs/2104.05726}{{\ttfamily
  2104.05726}}].

\bibitem{Lanza:2022zyg}
S.~Lanza, F.~Marchesano, L.~Martucci and I.~Valenzuela, \emph{{Large Field
  Distances from EFT strings}},  in \emph{{21st Hellenic School and Workshops
  on Elementary Particle Physics and Gravity}}, 5, 2022,
  \href{https://arxiv.org/abs/2205.04532}{{\ttfamily 2205.04532}}.

\bibitem{Dvali:2007hz}
G.~Dvali, \emph{{Black Holes and Large N Species Solution to the Hierarchy
  Problem}}, \href{https://doi.org/10.1002/prop.201000009}{\emph{Fortsch.
  Phys.} {\bfseries 58} (2010) 528}
  [\href{https://arxiv.org/abs/0706.2050}{{\ttfamily 0706.2050}}].

\bibitem{Dvali:2007wp}
G.~Dvali and M.~Redi, \emph{{Black Hole Bound on the Number of Species and
  Quantum Gravity at LHC}},
  \href{https://doi.org/10.1103/PhysRevD.77.045027}{\emph{Phys. Rev. D}
  {\bfseries 77} (2008) 045027}
  [\href{https://arxiv.org/abs/0710.4344}{{\ttfamily 0710.4344}}].

\bibitem{Dvali:2009ks}
G.~Dvali and D.~Lust, \emph{{Evaporation of Microscopic Black Holes in String
  Theory and the Bound on Species}},
  \href{https://doi.org/10.1002/prop.201000008}{\emph{Fortsch. Phys.}
  {\bfseries 58} (2010) 505} [\href{https://arxiv.org/abs/0912.3167}{{\ttfamily
  0912.3167}}].

\bibitem{Dvali:2010vm}
G.~Dvali and C.~Gomez, \emph{{Species and Strings}},
  \href{https://arxiv.org/abs/1004.3744}{{\ttfamily 1004.3744}}.

\bibitem{Palti:2019pca}
E.~Palti, \emph{{The Swampland: Introduction and Review}},
  \href{https://doi.org/10.1002/prop.201900037}{\emph{Fortsch. Phys.}
  {\bfseries 67} (2019) 1900037}
  [\href{https://arxiv.org/abs/1903.06239}{{\ttfamily 1903.06239}}].

\bibitem{vanBeest:2021lhn}
M.~van Beest, J.~Calder\'on-Infante, D.~Mirfendereski and I.~Valenzuela,
  \emph{{Lectures on the Swampland Program in String Compactifications}},
  \href{https://arxiv.org/abs/2102.01111}{{\ttfamily 2102.01111}}.

\bibitem{Grana:2021zvf}
M.~Gra\~na and A.~Herr\'aez, \emph{{The Swampland Conjectures: A Bridge from
  Quantum Gravity to Particle Physics}},
  \href{https://doi.org/10.3390/universe7080273}{\emph{Universe} {\bfseries 7}
  (2021) 273} [\href{https://arxiv.org/abs/2107.00087}{{\ttfamily
  2107.00087}}].

\bibitem{Agmon:2022thq}
N.~B. Agmon, A.~Bedroya, M.~J. Kang and C.~Vafa, \emph{{Lectures on the string
  landscape and the Swampland}},
  \href{https://arxiv.org/abs/2212.06187}{{\ttfamily 2212.06187}}.

\bibitem{Beasley:2004ys}
C.~Beasley and E.~Witten, \emph{{New instanton effects in supersymmetric QCD}},
  \href{https://doi.org/10.1088/1126-6708/2005/01/056}{\emph{JHEP} {\bfseries
  01} (2005) 056} [\href{https://arxiv.org/abs/hep-th/0409149}{{\ttfamily
  hep-th/0409149}}].

\bibitem{Beasley:2005iu}
C.~Beasley and E.~Witten, \emph{{New instanton effects in string theory}},
  \href{https://doi.org/10.1088/1126-6708/2006/02/060}{\emph{JHEP} {\bfseries
  0602} (2006) 060} [\href{https://arxiv.org/abs/hep-th/0512039}{{\ttfamily
  hep-th/0512039}}].

\bibitem{Coleman:1988cy}
S.~R. Coleman, \emph{{Black Holes as Red Herrings: Topological Fluctuations and
  the Loss of Quantum Coherence}},
  \href{https://doi.org/10.1016/0550-3213(88)90110-1}{\emph{Nucl. Phys. B}
  {\bfseries 307} (1988) 867}.

\bibitem{Marolf:2020xie}
D.~Marolf and H.~Maxfield, \emph{{Transcending the ensemble: baby universes,
  spacetime wormholes, and the order and disorder of black hole information}},
  \href{https://doi.org/10.1007/JHEP08(2020)044}{\emph{JHEP} {\bfseries 08}
  (2020) 044} [\href{https://arxiv.org/abs/2002.08950}{{\ttfamily
  2002.08950}}].

\bibitem{McNamara:2020uza}
J.~McNamara and C.~Vafa, \emph{{Baby Universes, Holography, and the
  Swampland}},  \href{https://arxiv.org/abs/2004.06738}{{\ttfamily
  2004.06738}}.

\bibitem{Giddings:1989bq}
S.~B. Giddings and A.~Strominger, \emph{{String Wormholes}},
  \href{https://doi.org/10.1016/0370-2693(89)91651-1}{\emph{Phys. Lett. B}
  {\bfseries 230} (1989) 46}.

\bibitem{Park:1990ep}
Y.~Park, M.~Srednicki and A.~Strominger, \emph{{Wormhole induced supersymmetry
  breaking in string theory}},
  \href{https://doi.org/10.1016/0370-2693(90)90335-4}{\emph{Phys. Lett. B}
  {\bfseries 244} (1990) 393}.

\bibitem{ArkaniHamed:2006dz}
N.~Arkani-Hamed, L.~Motl, A.~Nicolis and C.~Vafa, \emph{{The String landscape,
  black holes and gravity as the weakest force}},
  \href{https://doi.org/10.1088/1126-6708/2007/06/060}{\emph{JHEP} {\bfseries
  06} (2007) 060} [\href{https://arxiv.org/abs/hep-th/0601001}{{\ttfamily
  hep-th/0601001}}].

\bibitem{Ooguri:2006in}
H.~Ooguri and C.~Vafa, \emph{On the geometry of the string landscape and the
  swampland},
  \href{https://doi.org/10.1016/j.nuclphysb.2006.10.033}{\emph{Nucl.Phys.B}
  {\bfseries 766} (2007) 21}
  [\href{https://arxiv.org/abs/hep-th/0605264}{{\ttfamily hep-th/0605264}}].

\bibitem{Townsend:1979js}
P.~K. Townsend and P.~van Nieuwenhuizen, \emph{{Anomalies, Topological
  Invariants and the {Gauss-Bonnet} Theorem in Supergravity}},
  \href{https://doi.org/10.1103/PhysRevD.19.3592}{\emph{Phys. Rev. D}
  {\bfseries 19} (1979) 3592}.

\bibitem{Cecotti:1985mf}
S.~Cecotti, S.~Ferrara, L.~Girardello, M.~Porrati and A.~Pasquinucci,
  \emph{{Matter Coupling in Higher Derivative Supergravity}},
  \href{https://doi.org/10.1103/PhysRevD.33.2504}{\emph{Phys. Rev. D}
  {\bfseries 33} (1986) 2504}.

\bibitem{Cecotti:1987mr}
S.~Cecotti, S.~Ferrara, L.~Girardello, A.~Pasquinucci and M.~Porrati,
  \emph{{Matter Coupled Supergravity With {Gauss-Bonnet} Invariants: Component
  Lagrangian and Supersymmetry Breaking}},
  \href{https://doi.org/10.1142/S0217751X88000734}{\emph{Int. J. Mod. Phys. A}
  {\bfseries 3} (1988) 1675}.

\bibitem{Bonora:2013rta}
L.~Bonora and S.~Giaccari, \emph{{Weyl transformations and trace anomalies in
  N=1, D=4 supergravities}},
  \href{https://doi.org/10.1007/JHEP08(2013)116}{\emph{JHEP} {\bfseries 08}
  (2013) 116} [\href{https://arxiv.org/abs/1305.7116}{{\ttfamily 1305.7116}}].

\bibitem{Martucci:2022krl}
L.~Martucci, N.~Risso and T.~Weigand, \emph{{Quantum Gravity Bounds on N=1
  Effective Theories in Four Dimensions}},
  \href{https://arxiv.org/abs/2210.10797}{{\ttfamily 2210.10797}}.

\bibitem{Wess:1992cp}
J.~Wess and J.~Bagger, \emph{{Supersymmetry and supergravity}}. Princeton
  University Press, Princeton, NJ, USA, 1992.

\bibitem{Zwiebach:1985uq}
B.~Zwiebach, \emph{{Curvature Squared Terms and String Theories}},
  \href{https://doi.org/10.1016/0370-2693(85)91616-8}{\emph{Phys. Lett. B}
  {\bfseries 156} (1985) 315}.

\bibitem{Lindstrom:1983rt}
U.~Lindstrom and M.~Rocek, \emph{{Scalar Tensor Duality and N=1, N=2 Nonlinear
  Sigma Models}},
  \href{https://doi.org/10.1016/0550-3213(83)90638-7}{\emph{Nucl. Phys. B}
  {\bfseries 222} (1983) 285}.

\bibitem{Lanza:2019xxg}
S.~Lanza, F.~Marchesano, L.~Martucci and D.~Sorokin, \emph{{How many fluxes fit
  in an EFT?}}, \href{https://doi.org/10.1007/JHEP10(2019)110}{\emph{JHEP}
  {\bfseries 10} (2019) 110}
  [\href{https://arxiv.org/abs/1907.11256}{{\ttfamily 1907.11256}}].

\bibitem{Blumenhagen:2009qh}
R.~Blumenhagen, M.~Cvetic, S.~Kachru and T.~Weigand, \emph{{D-Brane Instantons
  in Type II Orientifolds}},
  \href{https://doi.org/10.1146/annurev.nucl.010909.083113}{\emph{Ann.Rev.Nucl.Part.Sci.}
  {\bfseries 59} (2009) 269} [\href{https://arxiv.org/abs/0902.3251}{{\ttfamily
  0902.3251}}].

\bibitem{Demirtas:2019lfi}
M.~Demirtas, C.~Long, L.~McAllister and M.~Stillman, \emph{{Minimal Surfaces
  and Weak Gravity}},
  \href{https://doi.org/10.1007/JHEP03(2020)021}{\emph{JHEP} {\bfseries 03}
  (2020) 021} [\href{https://arxiv.org/abs/1906.08262}{{\ttfamily
  1906.08262}}].

\bibitem{Long:2021lon}
C.~Long, A.~Sheshmani, C.~Vafa and S.-T. Yau, \emph{{Non-Holomorphic Cycles and
  Non-BPS Black Branes}},
  \href{https://doi.org/10.1007/s00220-022-04587-4}{\emph{Commun. Math. Phys.}
  {\bfseries 399} (2023) 1991}
  [\href{https://arxiv.org/abs/2104.06420}{{\ttfamily 2104.06420}}].

\bibitem{Vilenkin:2000jqa}
A.~Vilenkin and E.~S. Shellard, \emph{{Cosmic Strings and Other Topological
  Defects}}. Cambridge University Press, 7, 2000.

\bibitem{Katz:2020ewz}
S.~Katz, H.-C. Kim, H.-C. Tarazi and C.~Vafa, \emph{{Swampland Constraints on
  5d $\mathcal{N}=1$ Supergravity}},
  \href{https://doi.org/10.1007/JHEP07(2020)080}{\emph{JHEP} {\bfseries 07}
  (2020) 080} [\href{https://arxiv.org/abs/2004.14401}{{\ttfamily
  2004.14401}}].

\bibitem{Etheredge:2022opl}
M.~Etheredge, B.~Heidenreich, S.~Kaya, Y.~Qiu and T.~Rudelius,
  \emph{{Sharpening the Distance Conjecture in diverse dimensions}},
  \href{https://doi.org/10.1007/JHEP12(2022)114}{\emph{JHEP} {\bfseries 12}
  (2022) 114} [\href{https://arxiv.org/abs/2206.04063}{{\ttfamily
  2206.04063}}].

\bibitem{Lee:2019wij}
S.-J. Lee, W.~Lerche and T.~Weigand, \emph{{Emergent strings from infinite
  distance limits}}, \href{https://doi.org/10.1007/JHEP02(2022)190}{\emph{JHEP}
  {\bfseries 02} (2022) 190}
  [\href{https://arxiv.org/abs/1910.01135}{{\ttfamily 1910.01135}}].

\bibitem{Castellano:2022bvr}
A.~Castellano, A.~Herr\'aez and L.~E. Ib\'a\~nez, \emph{{The emergence proposal
  in quantum gravity and the species scale}},
  \href{https://doi.org/10.1007/JHEP06(2023)047}{\emph{JHEP} {\bfseries 06}
  (2023) 047} [\href{https://arxiv.org/abs/2212.03908}{{\ttfamily
  2212.03908}}].

\bibitem{Marchesano:2022axe}
F.~Marchesano and L.~Melotti, \emph{{EFT strings and emergence}},
  \href{https://doi.org/10.1007/JHEP02(2023)112}{\emph{JHEP} {\bfseries 02}
  (2023) 112} [\href{https://arxiv.org/abs/2211.01409}{{\ttfamily
  2211.01409}}].

\bibitem{Blumenhagen:2023yws}
R.~Blumenhagen, A.~Gligovic and A.~Paraskevopoulou, \emph{{The emergence
  proposal and the emergent string}},
  \href{https://doi.org/10.1007/JHEP10(2023)145}{\emph{JHEP} {\bfseries 10}
  (2023) 145} [\href{https://arxiv.org/abs/2305.10490}{{\ttfamily
  2305.10490}}].

\bibitem{Basile:2023blg}
I.~Basile, D.~Lust and C.~Montella, \emph{{Shedding black hole light on the
  emergent string conjecture}},
  \href{https://arxiv.org/abs/2311.12113}{{\ttfamily 2311.12113}}.

\bibitem{vandeHeisteeg:2022btw}
D.~van~de Heisteeg, C.~Vafa, M.~Wiesner and D.~H. Wu, \emph{{Moduli-dependent
  Species Scale}},  \href{https://arxiv.org/abs/2212.06841}{{\ttfamily
  2212.06841}}.

\bibitem{vandeHeisteeg:2023ubh}
D.~van~de Heisteeg, C.~Vafa and M.~Wiesner, \emph{{Bounds on Species Scale and
  the Distance Conjecture}},
  \href{https://arxiv.org/abs/2303.13580}{{\ttfamily 2303.13580}}.

\bibitem{Cribiori:2023ffn}
N.~Cribiori, D.~Lust and C.~Montella, \emph{{Species entropy and
  thermodynamics}}, \href{https://doi.org/10.1007/JHEP10(2023)059}{\emph{JHEP}
  {\bfseries 10} (2023) 059}
  [\href{https://arxiv.org/abs/2305.10489}{{\ttfamily 2305.10489}}].

\bibitem{Calderon-Infante:2023uhz}
J.~Calder\'on-Infante, M.~Delgado and A.~M. Uranga, \emph{{Emergence of species
  scale black hole horizons}},
  \href{https://doi.org/10.1007/JHEP01(2024)003}{\emph{JHEP} {\bfseries 01}
  (2024) 003} [\href{https://arxiv.org/abs/2310.04488}{{\ttfamily
  2310.04488}}].

\bibitem{vandeHeisteeg:2023dlw}
D.~van~de Heisteeg, C.~Vafa, M.~Wiesner and D.~H. Wu, \emph{{Species Scale in
  Diverse Dimensions}},  \href{https://arxiv.org/abs/2310.07213}{{\ttfamily
  2310.07213}}.

\bibitem{Castellano:2023aum}
A.~Castellano, A.~Herr\'aez and L.~E. Ib\'a\~nez, \emph{{On the Species Scale,
  Modular Invariance and the Gravitational EFT expansion}},
  \href{https://arxiv.org/abs/2310.07708}{{\ttfamily 2310.07708}}.

\bibitem{Castellano:2023stg}
A.~Castellano, I.~Ruiz and I.~Valenzuela, \emph{{A Universal Pattern in Quantum
  Gravity at Infinite Distance}},
  \href{https://arxiv.org/abs/2311.01501}{{\ttfamily 2311.01501}}.

\bibitem{Castellano:2023jjt}
A.~Castellano, I.~Ruiz and I.~Valenzuela, \emph{{Stringy Evidence for a
  Universal Pattern at Infinite Distance}},
  \href{https://arxiv.org/abs/2311.01536}{{\ttfamily 2311.01536}}.

\bibitem{Cota:2022yjw}
C.~F. Cota, A.~Mininno, T.~Weigand and M.~Wiesner, \emph{{The asymptotic Weak
  Gravity Conjecture for open strings}},
  \href{https://doi.org/10.1007/JHEP11(2022)058}{\emph{JHEP} {\bfseries 11}
  (2022) 058} [\href{https://arxiv.org/abs/2208.00009}{{\ttfamily
  2208.00009}}].

\bibitem{Denef:2008wq}
F.~Denef, \emph{{Les Houches Lectures on Constructing String Vacua}},
  {\emph{Les Houches} {\bfseries 87} (2008) 483}
  [\href{https://arxiv.org/abs/0803.1194}{{\ttfamily 0803.1194}}].

\bibitem{Weigand:2018rez}
T.~Weigand, \emph{{F-theory}}, {\emph{PoS} {\bfseries TASI2017} (2018) 016}
  [\href{https://arxiv.org/abs/1806.01854}{{\ttfamily 1806.01854}}].

\bibitem{fulger2016zariski}
M.~Fulger and B.~Lehmann, \emph{Zariski decompositions of numerical cycle
  classes},  2016.

\bibitem{boucksom2013pseudo}
S.~Boucksom, J.-P. Demailly, M.~Paun and T.~Peternell, \emph{The
  pseudo-effective cone of a compact k{\"a}hler manifold and varieties of
  negative kodaira dimension}, {\emph{Journal of Algebraic Geometry} {\bfseries
  22} (2013) 201}.

\bibitem{cox2011toric}
D.~Cox, J.~Little and H.~Schenck, \emph{Toric Varieties}, Graduate studies in
  mathematics. American Mathematical Soc., 2011.

\bibitem{miyaoka1985chern}
Y.~Miyaoka, \emph{The chern classes and kodaira dimension of a minimal
  variety}, {\emph{Algebraic geometry, Sendai} {\bfseries 10} (1985) 449}.

\bibitem{Horava:1995qa}
P.~Horava and E.~Witten, \emph{{Heterotic and type I string dynamics from
  eleven-dimensions}},
  \href{https://doi.org/10.1016/0550-3213(95)00621-4}{\emph{Nucl. Phys. B}
  {\bfseries 460} (1996) 506}
  [\href{https://arxiv.org/abs/hep-th/9510209}{{\ttfamily hep-th/9510209}}].

\bibitem{Horava:1996ma}
P.~Horava and E.~Witten, \emph{{Eleven-dimensional supergravity on a manifold
  with boundary}},
  \href{https://doi.org/10.1016/0550-3213(96)00308-2}{\emph{Nucl. Phys. B}
  {\bfseries 475} (1996) 94}
  [\href{https://arxiv.org/abs/hep-th/9603142}{{\ttfamily hep-th/9603142}}].

\bibitem{Witten:1996mz}
E.~Witten, \emph{{Strong coupling expansion of Calabi-Yau compactification}},
  \href{https://doi.org/10.1016/0550-3213(96)00190-3}{\emph{Nucl. Phys. B}
  {\bfseries 471} (1996) 135}
  [\href{https://arxiv.org/abs/hep-th/9602070}{{\ttfamily hep-th/9602070}}].

\bibitem{Blumenhagen:2006ux}
R.~Blumenhagen, S.~Moster and T.~Weigand, \emph{{Heterotic GUT and standard
  model vacua from simply connected Calabi-Yau manifolds}},
  \href{https://doi.org/10.1016/j.nuclphysb.2006.06.005}{\emph{Nucl. Phys. B}
  {\bfseries 751} (2006) 186}
  [\href{https://arxiv.org/abs/hep-th/0603015}{{\ttfamily hep-th/0603015}}].

\bibitem{Demirtas:2022hqf}
M.~Demirtas, A.~Rios-Tascon and L.~McAllister, \emph{{CYTools: A Software
  Package for Analyzing Calabi-Yau Manifolds}},
  \href{https://arxiv.org/abs/2211.03823}{{\ttfamily 2211.03823}}.

\bibitem{Loges:2022nuw}
G.~J. Loges, G.~Shiu and N.~Sudhir, \emph{{Complex saddles and Euclidean
  wormholes in the Lorentzian path integral}},
  \href{https://doi.org/10.1007/JHEP08(2022)064}{\emph{JHEP} {\bfseries 08}
  (2022) 064} [\href{https://arxiv.org/abs/2203.01956}{{\ttfamily
  2203.01956}}].

\bibitem{Jonas:2023qle}
C.~Jonas, G.~Lavrelashvili and J.-L. Lehners, \emph{{Stability of axion-dilaton
  wormholes}}, \href{https://doi.org/10.1103/PhysRevD.109.086022}{\emph{Phys.
  Rev. D} {\bfseries 109} (2024) 086022}
  [\href{https://arxiv.org/abs/2312.08971}{{\ttfamily 2312.08971}}].

\bibitem{Hertog:2024nys}
T.~Hertog, S.~Maenaut, B.~Missoni, R.~Tielemans and T.~Van~Riet,
  \emph{{Stability of Axion-Saxion wormholes}},
  \href{https://arxiv.org/abs/2405.02072}{{\ttfamily 2405.02072}}.

\bibitem{Coleman:1989zu}
S.~R. Coleman and K.-M. Lee, \emph{{Wormholes made without massless matter
  fields}}, \href{https://doi.org/10.1016/0550-3213(90)90149-8}{\emph{Nucl.
  Phys. B} {\bfseries 329} (1990) 387}.

\bibitem{Arkani-Hamed:2007cpn}
N.~Arkani-Hamed, J.~Orgera and J.~Polchinski, \emph{{Euclidean wormholes in
  string theory}},
  \href{https://doi.org/10.1088/1126-6708/2007/12/018}{\emph{JHEP} {\bfseries
  12} (2007) 018} [\href{https://arxiv.org/abs/0705.2768}{{\ttfamily
  0705.2768}}].

\bibitem{Gibbons:1976ue}
G.~W. Gibbons and S.~W. Hawking, \emph{{Action Integrals and Partition
  Functions in Quantum Gravity}},
  \href{https://doi.org/10.1103/PhysRevD.15.2752}{\emph{Phys. Rev. D}
  {\bfseries 15} (1977) 2752}.

\bibitem{Farakos:2017jme}
F.~Farakos, S.~Lanza, L.~Martucci and D.~Sorokin, \emph{{Three-forms in
  Supergravity and Flux Compactifications}},
  \href{https://doi.org/10.1140/epjc/s10052-017-5185-y}{\emph{Eur. Phys. J.}
  {\bfseries C77} (2017) 602}
  [\href{https://arxiv.org/abs/1706.09422}{{\ttfamily 1706.09422}}].

\bibitem{Bandos:2018gjp}
I.~Bandos, F.~Farakos, S.~Lanza, L.~Martucci and D.~Sorokin,
  \emph{{Three-forms, dualities and membranes in four-dimensional
  supergravity}}, \href{https://doi.org/10.1007/JHEP07(2018)028}{\emph{JHEP}
  {\bfseries 07} (2018) 028}
  [\href{https://arxiv.org/abs/1803.01405}{{\ttfamily 1803.01405}}].

\bibitem{Bielleman:2015ina}
S.~Bielleman, L.~E. Ibanez and I.~Valenzuela, \emph{{Minkowski 3-forms, Flux
  String Vacua, Axion Stability and Naturalness}},
  \href{https://doi.org/10.1007/JHEP12(2015)119}{\emph{JHEP} {\bfseries 12}
  (2015) 119} [\href{https://arxiv.org/abs/1507.06793}{{\ttfamily
  1507.06793}}].

\bibitem{Myers:1987yn}
R.~C. Myers, \emph{{Higher Derivative Gravity, Surface Terms and String
  Theory}}, \href{https://doi.org/10.1103/PhysRevD.36.392}{\emph{Phys. Rev. D}
  {\bfseries 36} (1987) 392}.

\bibitem{Chern45}
S.~shen Chern, \emph{On the curvatura integra in a riemannian manifold},
  {\emph{Annals of Mathematics} {\bfseries 46} (1945) 674}.

\bibitem{Eguchi:1980jx}
T.~Eguchi, P.~B. Gilkey and A.~J. Hanson, \emph{{Gravitation, Gauge Theories
  and Differential Geometry}},
  \href{https://doi.org/10.1016/0370-1573(80)90130-1}{\emph{Phys. Rept.}
  {\bfseries 66} (1980) 213}.

\bibitem{Gibbons:1979xm}
G.~W. Gibbons and S.~W. Hawking, \emph{{Classification of Gravitational
  Instanton Symmetries}},
  \href{https://doi.org/10.1007/BF01197189}{\emph{Commun. Math. Phys.}
  {\bfseries 66} (1979) 291}.

\bibitem{VanRiet:2020pcn}
T.~Van~Riet, \emph{{Instantons, Euclidean wormholes and AdS/CFT}},
  \href{https://doi.org/10.22323/1.376.0121}{\emph{PoS} {\bfseries CORFU2019}
  (2020) 121} [\href{https://arxiv.org/abs/2004.08956}{{\ttfamily
  2004.08956}}].

\bibitem{Harlow:2022ich}
D.~Harlow, B.~Heidenreich, M.~Reece and T.~Rudelius, \emph{{Weak gravity
  conjecture}}, \href{https://doi.org/10.1103/RevModPhys.95.035003}{\emph{Rev.
  Mod. Phys.} {\bfseries 95} (2023) 035003}
  [\href{https://arxiv.org/abs/2201.08380}{{\ttfamily 2201.08380}}].

\bibitem{Kallosh:1995hi}
R.~Kallosh, A.~D. Linde, D.~A. Linde and L.~Susskind, \emph{{Gravity and global
  symmetries}}, \href{https://doi.org/10.1103/PhysRevD.52.912}{\emph{Phys. Rev.
  D} {\bfseries 52} (1995) 912}
  [\href{https://arxiv.org/abs/hep-th/9502069}{{\ttfamily hep-th/9502069}}].

\bibitem{Andriolo:2022rxc}
S.~Andriolo, G.~Shiu, P.~Soler and T.~Van~Riet, \emph{{Axion wormholes with
  massive dilaton}},
  \href{https://doi.org/10.1088/1361-6382/ac8fdc}{\emph{Class. Quant. Grav.}
  {\bfseries 39} (2022) 215014}
  [\href{https://arxiv.org/abs/2205.01119}{{\ttfamily 2205.01119}}].

\bibitem{Jonas:2023ipa}
C.~Jonas, G.~Lavrelashvili and J.-L. Lehners, \emph{{Zoo of axionic
  wormholes}}, \href{https://doi.org/10.1103/PhysRevD.108.066012}{\emph{Phys.
  Rev. D} {\bfseries 108} (2023) 066012}
  [\href{https://arxiv.org/abs/2306.11129}{{\ttfamily 2306.11129}}].

\bibitem{Cheong:2023hrj}
D.~Y. Cheong, S.~C. Park and C.~S. Shin, \emph{{Effective Theory Approach for
  Axion Wormholes}},  \href{https://arxiv.org/abs/2310.11260}{{\ttfamily
  2310.11260}}.

\bibitem{Rey:1989xj}
S.-J. Rey, \emph{{The Confining Phase of Superstrings and Axionic Strings}},
  \href{https://doi.org/10.1103/PhysRevD.43.526}{\emph{Phys. Rev. D} {\bfseries
  43} (1991) 526}.

\bibitem{Gibbons:1995vg}
G.~W. Gibbons, M.~B. Green and M.~J. Perry, \emph{{Instantons and seven-branes
  in type IIB superstring theory}},
  \href{https://doi.org/10.1016/0370-2693(95)01565-5}{\emph{Phys. Lett. B}
  {\bfseries 370} (1996) 37}
  [\href{https://arxiv.org/abs/hep-th/9511080}{{\ttfamily hep-th/9511080}}].

\bibitem{Mayr:1996sh}
P.~Mayr, \emph{{Mirror symmetry, N=1 superpotentials and tensionless strings on
  Calabi-Yau four folds}},
  \href{https://doi.org/10.1016/S0550-3213(97)00196-X}{\emph{Nucl. Phys. B}
  {\bfseries 494} (1997) 489}
  [\href{https://arxiv.org/abs/hep-th/9610162}{{\ttfamily hep-th/9610162}}].

\bibitem{Alim:2021vhs}
M.~Alim, B.~Heidenreich and T.~Rudelius, \emph{{The Weak Gravity Conjecture and
  BPS Particles}}, \href{https://doi.org/10.1002/prop.202100125}{\emph{Fortsch.
  Phys.} {\bfseries 69} (2021) 2100125}
  [\href{https://arxiv.org/abs/2108.08309}{{\ttfamily 2108.08309}}].

\bibitem{Arkani-Hamed:2006emk}
N.~Arkani-Hamed, L.~Motl, A.~Nicolis and C.~Vafa, \emph{{The String landscape,
  black holes and gravity as the weakest force}},
  \href{https://doi.org/10.1088/1126-6708/2007/06/060}{\emph{JHEP} {\bfseries
  06} (2007) 060} [\href{https://arxiv.org/abs/hep-th/0601001}{{\ttfamily
  hep-th/0601001}}].

\bibitem{Coleman:1989ky}
S.~R. Coleman and K.-M. Lee, \emph{{Escape From the Menace of the Giant
  Wormholes}}, \href{https://doi.org/10.1016/0370-2693(89)91705-X}{\emph{Phys.
  Lett. B} {\bfseries 221} (1989) 242}.

\bibitem{Coleman:1990tz}
S.~R. Coleman and K.-M. Lee, \emph{{Big Wormholes and Little Interactions}},
  \href{https://doi.org/10.1016/0550-3213(90)90263-D}{\emph{Nucl. Phys. B}
  {\bfseries 341} (1990) 101}.

\bibitem{kawamata1988crepant}
Y.~Kawamata, \emph{Crepant blowing-up of 3-dimensional canonical singularities
  and its application to degenerations of surfaces}, {\emph{Annals of
  Mathematics} {\bfseries 127} (1988) 93}.

\bibitem{Palti:2020qlc}
E.~Palti, C.~Vafa and T.~Weigand, \emph{{Supersymmetric Protection and the
  Swampland}}, \href{https://doi.org/10.1007/JHEP06(2020)168}{\emph{JHEP}
  {\bfseries 06} (2020) 168}
  [\href{https://arxiv.org/abs/2003.10452}{{\ttfamily 2003.10452}}].

\bibitem{Hitchin:2000jd}
N.~J. Hitchin, \emph{{The geometry of three-forms in six and seven
  dimensions}}, {\emph{J. Diff. Geom.} {\bfseries 55} (2000) 547}
  [\href{https://arxiv.org/abs/math/0010054}{{\ttfamily math/0010054}}].

\bibitem{Beasley:2002db}
C.~Beasley and E.~Witten, \emph{A note on fluxes and superpotentials in m
  theory compactifications on manifolds of g(2) holonomy},
  \href{https://doi.org/10.1088/1126-6708/2002/07/046}{\emph{JHEP} {\bfseries
  07} (2002) 046} [\href{https://arxiv.org/abs/hep-th/0203061}{{\ttfamily
  hep-th/0203061}}].

\bibitem{Kachru:2001je}
S.~Kachru and J.~McGreevy, \emph{{M theory on manifolds of G(2) holonomy and
  type IIA orientifolds}},
  \href{https://doi.org/10.1088/1126-6708/2001/06/027}{\emph{JHEP} {\bfseries
  06} (2001) 027} [\href{https://arxiv.org/abs/hep-th/0103223}{{\ttfamily
  hep-th/0103223}}].

\bibitem{Preskill:1988na}
J.~Preskill, \emph{{Wormholes in Space-time and the Constants of Nature}},
  \href{https://doi.org/10.1016/0550-3213(89)90592-0}{\emph{Nucl. Phys. B}
  {\bfseries 323} (1989) 141}.

\bibitem{Klebanov:1988eh}
I.~R. Klebanov, L.~Susskind and T.~Banks, \emph{{Wormholes and the Cosmological
  Constant}}, \href{https://doi.org/10.1016/0550-3213(89)90538-5}{\emph{Nucl.
  Phys. B} {\bfseries 317} (1989) 665}.

\bibitem{Heidenreich:2015nta}
B.~Heidenreich, M.~Reece and T.~Rudelius, \emph{{Sharpening the Weak Gravity
  Conjecture with Dimensional Reduction}},
  \href{https://doi.org/10.1007/JHEP02(2016)140}{\emph{JHEP} {\bfseries 02}
  (2016) 140} [\href{https://arxiv.org/abs/1509.06374}{{\ttfamily
  1509.06374}}].

\bibitem{Hebecker:2016dsw}
A.~Hebecker, P.~Mangat, S.~Theisen and L.~T. Witkowski, \emph{{Can
  Gravitational Instantons Really Constrain Axion Inflation?}},
  \href{https://doi.org/10.1007/JHEP02(2017)097}{\emph{JHEP} {\bfseries 02}
  (2017) 097} [\href{https://arxiv.org/abs/1607.06814}{{\ttfamily
  1607.06814}}].

\bibitem{Loges:2023ypl}
G.~J. Loges, G.~Shiu and T.~Van~Riet, \emph{{A 10d construction of Euclidean
  axion wormholes in flat and AdS space}},
  \href{https://doi.org/10.1007/JHEP06(2023)079}{\emph{JHEP} {\bfseries 06}
  (2023) 079} [\href{https://arxiv.org/abs/2302.03688}{{\ttfamily
  2302.03688}}].

\bibitem{Giddings:1988wv}
S.~B. Giddings and A.~Strominger, \emph{{Baby Universes, Third Quantization and
  the Cosmological Constant}},
  \href{https://doi.org/10.1016/0550-3213(89)90353-2}{\emph{Nucl. Phys. B}
  {\bfseries 321} (1989) 481}.

\bibitem{Affleck:1983mk}
I.~Affleck, M.~Dine and N.~Seiberg, \emph{{Dynamical Supersymmetry Breaking in
  Supersymmetric QCD}},
  \href{https://doi.org/10.1016/0550-3213(84)90058-0}{\emph{Nucl. Phys. B}
  {\bfseries 241} (1984) 493}.

\bibitem{Dine:1986zy}
M.~Dine, N.~Seiberg, X.~G. Wen and E.~Witten, \emph{{Nonperturbative Effects on
  the String World Sheet}},
  \href{https://doi.org/10.1016/0550-3213(86)90418-9}{\emph{Nucl. Phys. B}
  {\bfseries 278} (1986) 769}.

\bibitem{Dine:1987bq}
M.~Dine, N.~Seiberg, X.~G. Wen and E.~Witten, \emph{{Nonperturbative Effects on
  the String World Sheet. 2.}},
  \href{https://doi.org/10.1016/0550-3213(87)90383-X}{\emph{Nucl. Phys. B}
  {\bfseries 289} (1987) 319}.

\bibitem{Becker:1995kb}
K.~Becker, M.~Becker and A.~Strominger, \emph{{Five-branes, membranes and
  nonperturbative string theory}},
  \href{https://doi.org/10.1016/0550-3213(95)00487-1}{\emph{Nucl.Phys.}
  {\bfseries B456} (1995) 130}
  [\href{https://arxiv.org/abs/hep-th/9507158}{{\ttfamily hep-th/9507158}}].

\bibitem{Witten:1996bn}
E.~Witten, \emph{{Nonperturbative superpotentials in string theory}},
  \href{https://doi.org/10.1016/0550-3213(96)00283-0}{\emph{Nucl.Phys.}
  {\bfseries B474} (1996) 343}
  [\href{https://arxiv.org/abs/hep-th/9604030}{{\ttfamily hep-th/9604030}}].

\bibitem{Harvey:1999as}
J.~A. Harvey and G.~W. Moore, \emph{{Superpotentials and membrane instantons}},
   \href{https://arxiv.org/abs/hep-th/9907026}{{\ttfamily hep-th/9907026}}.

\bibitem{Witten:1999eg}
E.~Witten, \emph{{World sheet corrections via D instantons}},
  \href{https://doi.org/10.1088/1126-6708/2000/02/030}{\emph{JHEP} {\bfseries
  02} (2000) 030} [\href{https://arxiv.org/abs/hep-th/9907041}{{\ttfamily
  hep-th/9907041}}].

\bibitem{Bianchi:2011qh}
M.~Bianchi, A.~Collinucci and L.~Martucci, \emph{{Magnetized E3-brane
  instantons in F-theory}},
  \href{https://doi.org/10.1007/JHEP12(2011)045}{\emph{JHEP} {\bfseries 1112}
  (2011) 045} [\href{https://arxiv.org/abs/1107.3732}{{\ttfamily 1107.3732}}].

\bibitem{Anglin:1992ym}
J.~R. Anglin and R.~C. Myers, \emph{{Wormholes and supersymmetry}},
  \href{https://doi.org/10.1103/PhysRevD.46.2469}{\emph{Phys. Rev. D}
  {\bfseries 46} (1992) 2469}
  [\href{https://arxiv.org/abs/hep-th/9206072}{{\ttfamily hep-th/9206072}}].

\bibitem{Abbott:1989jw}
L.~F. Abbott and M.~B. Wise, \emph{{Wormholes and Global Symmetries}},
  \href{https://doi.org/10.1016/0550-3213(89)90503-8}{\emph{Nucl. Phys. B}
  {\bfseries 325} (1989) 687}.

\bibitem{Alvey:2020nyh}
J.~Alvey and M.~Escudero, \emph{{The axion quality problem: global symmetry
  breaking and wormholes}},
  \href{https://doi.org/10.1007/JHEP01(2021)032}{\emph{JHEP} {\bfseries 01}
  (2021) 032} [\href{https://arxiv.org/abs/2009.03917}{{\ttfamily
  2009.03917}}].

\bibitem{Montero:2015ofa}
M.~Montero, A.~M. Uranga and I.~Valenzuela, \emph{{Transplanckian axions!?}},
  \href{https://doi.org/10.1007/JHEP08(2015)032}{\emph{JHEP} {\bfseries 08}
  (2015) 032} [\href{https://arxiv.org/abs/1503.03886}{{\ttfamily
  1503.03886}}].

\bibitem{Bachlechner:2015qja}
T.~C. Bachlechner, C.~Long and L.~McAllister, \emph{{Planckian Axions and the
  Weak Gravity Conjecture}},
  \href{https://doi.org/10.1007/JHEP01(2016)091}{\emph{JHEP} {\bfseries 01}
  (2016) 091} [\href{https://arxiv.org/abs/1503.07853}{{\ttfamily
  1503.07853}}].

\bibitem{Weinberg:1978kz}
S.~Weinberg, \emph{{Phenomenological Lagrangians}},
  \href{https://doi.org/10.1016/0378-4371(79)90223-1}{\emph{Physica A}
  {\bfseries 96} (1979) 327}.

\bibitem{Manohar:1983md}
A.~Manohar and H.~Georgi, \emph{{Chiral Quarks and the Nonrelativistic Quark
  Model}}, \href{https://doi.org/10.1016/0550-3213(84)90231-1}{\emph{Nucl.
  Phys. B} {\bfseries 234} (1984) 189}.

\bibitem{Cohen:1997rt}
A.~G. Cohen, D.~B. Kaplan and A.~E. Nelson, \emph{{Counting 4 pis in strongly
  coupled supersymmetry}},
  \href{https://doi.org/10.1016/S0370-2693(97)00995-7}{\emph{Phys. Lett. B}
  {\bfseries 412} (1997) 301}
  [\href{https://arxiv.org/abs/hep-ph/9706275}{{\ttfamily hep-ph/9706275}}].

\bibitem{Chacko:1999hg}
Z.~Chacko, M.~A. Luty and E.~Ponton, \emph{{Massive higher dimensional gauge
  fields as messengers of supersymmetry breaking}},
  \href{https://doi.org/10.1088/1126-6708/2000/07/036}{\emph{JHEP} {\bfseries
  07} (2000) 036} [\href{https://arxiv.org/abs/hep-ph/9909248}{{\ttfamily
  hep-ph/9909248}}].

\bibitem{Grimm:2017okk}
T.~W. Grimm, K.~Mayer and M.~Weissenbacher, \emph{{Higher derivatives in Type
  II and M-theory on Calabi-Yau threefolds}},
  \href{https://doi.org/10.1007/JHEP02(2018)127}{\emph{JHEP} {\bfseries 02}
  (2018) 127} [\href{https://arxiv.org/abs/1702.08404}{{\ttfamily
  1702.08404}}].

\bibitem{Ivrii}
V.~Ivrii, \emph{{100 years of Weyl's law}},
  \href{https://doi.org/10.48550/arXiv.1608.03963}{\emph{Bulletin of
  Mathematical Sciences (Springer)} {\bfseries Volume 6, Issue 3} (October
  2016) 379}.

\bibitem{Halverson:2015jua}
J.~Halverson and W.~Taylor, \emph{{$ {\mathrm{\mathbb{P}}}^1 $-bundle bases and
  the prevalence of non-Higgsable structure in 4D F-theory models}},
  \href{https://doi.org/10.1007/JHEP09(2015)086}{\emph{JHEP} {\bfseries 09}
  (2015) 086} [\href{https://arxiv.org/abs/1506.03204}{{\ttfamily
  1506.03204}}].

\bibitem{Friedman:1997yq}
R.~Friedman, J.~Morgan and E.~Witten, \emph{{Vector bundles and F theory}},
  \href{https://doi.org/10.1007/s002200050154}{\emph{Commun. Math. Phys.}
  {\bfseries 187} (1997) 679}
  [\href{https://arxiv.org/abs/hep-th/9701162}{{\ttfamily hep-th/9701162}}].

\bibitem{Castellano:2021mmx}
A.~Castellano, A.~Herr\'aez and L.~E. Ib\'a\~nez, \emph{{IR/UV mixing, towers
  of species and swampland conjectures}},
  \href{https://doi.org/10.1007/JHEP08(2022)217}{\emph{JHEP} {\bfseries 08}
  (2022) 217} [\href{https://arxiv.org/abs/2112.10796}{{\ttfamily
  2112.10796}}].

\bibitem{Huebscher:2009bp}
M.~Huebscher, P.~Meessen and T.~Ortin, \emph{{Domain walls and instantons in
  N=1, d=4 supergravity}},
  \href{https://doi.org/10.1007/JHEP06(2010)001}{\emph{JHEP} {\bfseries 06}
  (2010) 001} [\href{https://arxiv.org/abs/0912.3672}{{\ttfamily 0912.3672}}].

\end{thebibliography}\endgroup

\end{document}